\documentclass{aastex63}

\usepackage{amsmath}
\usepackage[title]{appendix}

\usepackage{xcolor, microtype}
\definecolor{twitterblue}{RGB}{64,153,255}
\definecolor{linkcolor}{rgb}{0.1216,0.4667,0.7059}

\usepackage{ mathrsfs }
\usepackage[T1]{fontenc}
\usepackage{hyperref}
\usepackage{comment}

\newcommand{\rearth}{$R_{\oplus}$}
\newcommand{\mearth}{$M_{\oplus}$}
\newcommand{\rsun}{$R_{\odot}$}
\newcommand{\msun}{$M_{\odot}$}

\newcommand{\epicthreefiveninebalias}{K2-371~b}

\newcommand{\epicfourfivefivebalias}{K2-388~b}

\newcommand{\epicthreenineeightbalias}{K2-392~b}

\newcommand{\epicfiveeightzerobalias}{K2-370~b}

\newcommand{\epicfiveninesevenbalias}{K2-402~b}

\newcommand{\epicfourninezerobalias}{K2-408~b}

\newcommand{\epicthreeninefivealias}{K2-409}
\newcommand{\epicthreeninefivebalias}{K2-409~b}

\newcommand{\epicthreeonefourbalias}{K2-391~b}

\newcommand{\epiceightninefivebalias}{K2-382~b}

\newcommand{\epiczerooneonesixbalias}{K2-365~b}

\newcommand{\epiceightthreeninealias}{K2-381}
\newcommand{\epiceightthreeninebalias}{K2-381~b}
\newcommand{\epiceightthreeninecalias}{K2-381~c}
\newcommand{\epiceightthreeninedalias}{K2-381~d}

\newcommand{\epictwosevenninebalias}{K2-404~b}

\newcommand{\epicsixfiveonealias}{K2-407}
\newcommand{\epicsixfiveonebalias}{K2-407~b}
\newcommand{\epicsixfiveonecalias}{K2-407~c}

\newcommand{\epiconefourzeroalias}{K2-399}
\newcommand{\epiconefourzerobalias}{K2-399~b}

\newcommand{\epicninetwoeightbalias}{K2-405~b}

\newcommand{\epicthreeninetwoalias}{K2-396}
\newcommand{\epicthreeninetwobalias}{K2-396~b}
\newcommand{\epicthreeninetwocalias}{K2-396~c} 

\newcommand{\epiczerofoureightalias}{K2-389}
\newcommand{\epiczerofoureightbalias}{K2-389~b}
\newcommand{\epiczerofoureightcalias}{K2-389~c}

\newcommand{\epicsixeighttwoalias}{K2-368}
\newcommand{\epicsixeighttwobalias}{K2-368~b}
\newcommand{\epicsixeighttwocalias}{K2-368~c}
\newcommand{\epicsixeighttwodalias}{K2-368~d}

\newcommand{\epicfoursevensevenbalias}{K2-385~b}

\newcommand{\epictwothreethreebalias}{K2-387~b}

\newcommand{\epictwoseventwoalias}{K2-384}
\newcommand{\epictwoseventwobalias}{K2-384~b} 
\newcommand{\epictwoseventwocalias}{K2-384~c} 
\newcommand{\epictwoseventwodalias}{K2-384~d} 
\newcommand{\epictwoseventwoealias}{K2-384~e} 
\newcommand{\epictwoseventwofalias}{K2-384~f} 

\newcommand{\epiceightoneninealias}{K2-395}
\newcommand{\epiceightoneninebalias}{K2-395~b}
\newcommand{\epiceightoneninecalias}{K2-395~c}

\newcommand{\epictwooneonesixalias}{K2-374}
\newcommand{\epictwooneonesixbalias}{K2-374~b}
\newcommand{\epictwooneonesixcalias}{K2-374~c}

\newcommand{\epicfiveonefouralias}{K2-401}
\newcommand{\epicfiveonefourbalias}{K2-401~b}

\newcommand{\epiceighttwoeightalias}{K2-378}
\newcommand{\epiceighttwoeightbalias}{K2-378~b}

\newcommand{\epicsixsixfouralias}{K2-366}
\newcommand{\epicsixsixfourbalias}{K2-366~b}

\newcommand{\epictwoeightsixalias}{EPIC 206317286}
\newcommand{\epictwoeightsixbalias}{EPIC 206317286~b} 
\newcommand{\epictwoeightsixcalias}{EPIC 206317286~c}

\newcommand{\epicninenineninealias}{K2-372}
\newcommand{\epicninenineninebalias}{K2-372~b}

\newcommand{\epicthreeoneeightalias}{K2-375}
\newcommand{\epicthreeoneeightbalias}{K2-375~b}

\newcommand{\epiconeoneeightalias}{K2-377}
\newcommand{\epiconeoneeightbalias}{K2-377~b}

\newcommand{\epiczerosixninealias}{K2-379}
\newcommand{\epiczerosixninebalias}{K2-379~b}

\newcommand{\epicfivefourfivealias}{K2-380}
\newcommand{\epicfivefourfivebalias}{K2-380~b}

\newcommand{\epicsixzerotwoalias}{K2-383}
\newcommand{\epicsixzerotwobalias}{K2-383~b}

\newcommand{\epiceightsevenfouralias}{K2-386}
\newcommand{\epiceightsevenfourbalias}{K2-386~b}

\newcommand{\epiczerofourninealias}{K2-393}
\newcommand{\epiczerofourninebalias}{K2-393~b}

\newcommand{\epiczerofourzeroalias}{K2-394}
\newcommand{\epiczerofourzerobalias}{K2-394~b}

\newcommand{\epicninesevensevenalias}{K2-390}
\newcommand{\epicninesevensevenbalias}{K2-390~b}

\newcommand{\epiczerozerothreealias}{K2-397}
\newcommand{\epiczerozerothreebalias}{K2-397~b}

\newcommand{\epicthreefivezeroalias}{K2-398}
\newcommand{\epicthreefivezerobalias}{K2-398~b}
\newcommand{\epicthreefivezerocalias}{K2-398~c}

\newcommand{\epicthreefivethreealias}{K2-403}
\newcommand{\epicthreefivethreebalias}{K2-403~b}

\newcommand{\epicfoursevenonealias}{K2-406}
\newcommand{\epicfoursevenonebalias}{K2-406~b}

\newcommand{\epicnineeightonealias}{K2-367}
\newcommand{\epicnineeightonebalias}{K2-367~b}

\newcommand{\epicninefivesevenalias}{K2-369}
\newcommand{\epicninefivesevenbalias}{K2-369~b}

\newcommand{\epiczerofivefouralias}{K2-373}
\newcommand{\epiczerofivefourbalias}{K2-373~b}

\newcommand{\epicthreeeightthreealias}{K2-376}
\newcommand{\epicthreeeightthreebalias}{K2-376~b}

\newcommand{\epicthreezerosevenalias}{K2-400}
\newcommand{\epicthreezerosevenbalias}{K2-400~b}

\newcommand{\totalval}{60} 
\newcommand{\totalsys}{46}
\newcommand{\totalchecked}{91} 
\newcommand{\totalcheckedsystems}{78} 

\graphicspath{{./figures/}{}}

\received{\today}
\revised{\today}
\accepted{\today}
\submitjournal{AJ}

\shorttitle{Scaling K2: \totalval\ New Exoplanets}

\shortauthors{Christiansen et al. (2021)}

\begin{document}

\title{Scaling \emph{K2}. V. Statistical Validation of \totalval\ New Exoplanets From K2 Campaigns 2--18}

\correspondingauthor{Jessie L. Christiansen}
\email{christia@ipac.caltech.edu}

\author[0000-0002-8035-4778]{Jessie L. Christiansen}
\affiliation{Caltech/IPAC-NASA Exoplanet Science Institute, Pasadena, CA 91125, USA}

\author[0000-0002-6673-8206]{Sakhee Bhure}
\affiliation{Caltech/IPAC-NASA Exoplanet Science Institute, Pasadena, CA 91125, USA}

\author[0000-0003-1848-2063]{Jon K. Zink}\thanks{NASA Hubble Fellow}
\affiliation{Department of Astronomy, California Institute of Technology, Pasadena, CA 91125, USA}

\author[0000-0003-3702-0382]{Kevin K. Hardegree-Ullman}
\affiliation{Department of Astronomy, The University of Arizona, Tucson, AZ 85721, USA}

\author{Britt Duffy Adkins}
\affiliation{Sol Price School of Public Policy, University of Southern California, Los Angeles, CA 90089, USA}


\author{Christina Hedges}
\affiliation{Bay Area Environmental Research Institute, P.O. Box 25, Moffett Field, CA 94035, USA}
\affiliation{NASA Ames Research Center, Moffett Field, CA, USA}

\author{Timothy D. Morton}
\affiliation{Department of Physics and Astronomy, University of Southern California, Los Angeles, CA 90089, USA}


\author[0000-0001-6637-5401]{Allyson Bieryla}
\affiliation{Center for Astrophysics \textbar \ Harvard \& Smithsonian, 60 Garden Street, Cambridge, MA 02138, USA}

\author[0000-0002-5741-3047]{David R. Ciardi}
\affiliation{Caltech/IPAC-NASA Exoplanet Science Institute, Pasadena, CA 91125, USA}

\author{William D. Cochran}
\affiliation{McDonald Observatory, The University of Texas, Austin, TX 78712, USA}

\author[0000-0001-8189-0233]{Courtney D. Dressing}
\affiliation{Department of Astronomy, University of California, Berkeley, CA 94720, USA}

\author{Mark E. Everett} %
\affiliation{NSF’s Optical Infrared Astronomy Research Laboratory, 950 North Cherry Avenue Tucson, AZ 85719, USA}

\author{Howard Isaacson}
\affiliation{Department of Astronomy, University of California Berkeley, Berkeley CA 94720, USA}
\affiliation{Centre for Astrophysics, University of Southern Queensland, Toowoomba, QLD, Australia}

\author{John H. Livingston} 
\affiliation{Department of Astronomy, University of Tokyo, 7-3-1 Hongo, Bunkyo-ku, Tokyo 113-0033, Japan}

\author[0000-0002-0619-7639]{Carl Ziegler} 
\affiliation{Department of Physics, Engineering and Astronomy, Stephen F. Austin State University, 1936 North St, Nacogdoches, TX 75962, USA}


\author{Perry Berlind}
\affiliation{Center for Astrophysics | Harvard \& Smithsonian, 60 Garden St, Cambridge, MA, 02138, USA}

\author{Michael L. Calkins}
\affiliation{Center for Astrophysics | Harvard \& Smithsonian, 60 Garden St, Cambridge, MA, 02138, USA}

\author{Gilbert A. Esquerdo}
\affiliation{Center for Astrophysics | Harvard \& Smithsonian, 60 Garden St, Cambridge, MA, 02138, USA}

\author{David W. Latham}
\affiliation{Center for Astrophysics | Harvard \& Smithsonian, 60 Garden St, Cambridge, MA, 02138, USA}


\author{Michael Endl}
\affiliation{McDonald Observatory, The University of Texas, Austin, TX 78712}

\author{Phillip J. MacQueen}
\affiliation{McDonald Observatory, The University of Texas, Austin, TX 78712}


\author{Benjamin J. Fulton}
\affiliation{Caltech/IPAC-NASA Exoplanet Science Institute, Pasadena, CA 91125}

\author{Lea A. Hirsch}
\affiliation{Kavli Center for Particle Astrophysics and Cosmology, Stanford University, Stanford, CA 94305, USA}

\author{Andrew W. Howard}
\affiliation{Department of Astronomy, California Institute of Technology, Pasadena, CA 91125, USA}

\author[0000-0002-3725-3058]{Lauren M. Weiss}
\affiliation{Department of Physics, University of Notre Dame, Notre Dame, IN 46556, USA}


\author{Bridgette E. Allen}
\affiliation{University of Wisconsin-Stout, Menomonie, WI 54751, USA}
\affiliation{Caltech/IPAC-NASA Exoplanet Science Institute, Pasadena, CA 91125, USA}

\author{Arthur Berberyann}
\affiliation{College of the Canyons, Santa Clarita, CA 91355, USA}
\affiliation{Caltech/IPAC-NASA Exoplanet Science Institute, Pasadena, CA 91125, USA}

\author{Krys N. Ciardi}
\affiliation{Rhode Island College, Providence, RI 02908, USA}
\affiliation{Caltech/IPAC-NASA Exoplanet Science Institute, Pasadena, CA 91125, USA}

\author{Ava Dunlavy}
\affiliation{University of California, Santa Cruz, Santa
Cruz CA 95065, USA}
\affiliation{Caltech/IPAC-NASA Exoplanet Science Institute, Pasadena, CA 91125, USA}

\author{Sofia H. Glassford}
\affiliation{University of California, Davis, Davis, CA 95616, USA}
\affiliation{Caltech/IPAC-NASA Exoplanet Science Institute, Pasadena, CA 91125, USA}


\author{Fei Dai} 
\affiliation{Division of Geological and Planetary Sciences, 1200 E. California Boulevard, Pasadena, CA, 91125, USA}

\author{Teruyuki Hirano}
\affiliation{Astrobiology Center, 2-21-1 Osawa, Mitaka, Tokyo 181-8588, Japan}
\affiliation{National Astronomical Observatory of Japan, NINS, 2-21-1 Osawa, Mitaka, Tokyo 181-8588, Japan}
\affiliation{Department of Astronomical Science, School of Physical Sciences, The Graduate University for Advanced Studies (SOKENDAI), 2-21-1, Osawa, Mitaka, Tokyo, 181-8588, Japan}

\author[0000-0002-6510-0681]{Motohide Tamura} 
\affiliation{Department of Astronomy, University of Tokyo, 7-3-1 Hongo, Bunkyo-ku, Tokyo 113-0033, Japan}
\affiliation{Astrobiology Center, 2-21-1 Osawa, Mitaka-shi, Tokyo 181-8588, Japan}
\affiliation{National Astronomical Observatory, 2-21-1 Osawa, Mitaka-shi, Tokyo 181-8588, Japan}




\author{Charles Beichman}
\affiliation{Caltech/IPAC-NASA Exoplanet Science Institute, Pasadena, CA 91125}

\author{Erica J. Gonzales}
\affiliation{Department of Physics, University of Notre Dame, 225 Nieuwland Science Hall, Notre Dame IN 46556, USA}
\affiliation{University of California, Santa Cruz, Santa
Cruz CA 95065, USA}

\author[0000-0001-5347-7062]{Joshua E. Schlieder}
\affiliation{Exoplanets and Stellar Astrophysics Laboratory, Code 667, NASA Goddard Space Flight Center, Greenbelt, MD 20771}


\author[0000-0001-7139-2724]{Thomas Barclay}
\affiliation{University of Maryland, Baltimore County, 1000 Hilltop Cir, Baltimore, MD 21250, USA}
\affiliation{NASA Goddard Space Flight Center, 8800 Greenbelt Rd, Greenbelt, MD 20771, USA}

\author[0000-0002-1835-1891]{Ian J. M. Crossfield}
\affiliation{Department of Physics and Astronomy, University of Kansas, Lawrence, KS 66045}

\author[0000-0002-0388-8004]{Emily A. Gilbert}
\affiliation{Department of Astronomy and Astrophysics, University of
Chicago, 5640 S. Ellis Ave, Chicago, IL 60637, USA}
\affiliation{University of Maryland, Baltimore County, 1000 Hilltop Circle, Baltimore, MD 21250, USA}
\affiliation{The Adler Planetarium, 1300 South Lakeshore Drive, Chicago, IL 60605, USA}
\affiliation{NASA Goddard Space Flight Center, 8800 Greenbelt Road, Greenbelt, MD 20771, USA}

\author{Elisabeth C. Matthews}
\affiliation{Observatoire de l’Université de Genève, Chemin des Maillettes 51, 1290 Versoix, Switzerland}


\author[0000-0002-8965-3969]{Steven Giacalone}
\affiliation{Department of Astronomy, University of California, Berkeley, CA 94720}


\author[0000-0003-0967-2893]{Erik A. Petigura}
\affiliation{Department of Physics and Astronomy, University of California, Los Angeles, CA 90095}



\begin{abstract}

The NASA \emph{K2} mission, salvaged from the hardware failures of the \emph{Kepler} telescope, has continued \emph{Kepler's} planet-hunting success. It has revealed nearly 500 transiting planets around the ecliptic plane, many of which are the subject of further study, and over 1000 additional candidates. Here we present the results of an ongoing project to follow-up and statistically validate new \emph{K2} planets, in particular to identify promising new targets for further characterization. By analyzing the reconnaissance spectra, high-resolution imaging, centroid variations, and statistical likelihood of the signals of \totalchecked\ candidates, we validate \totalval\ new planets in \totalsys\ systems. These include: a number of planets amenable to transmission spectroscopy  (\epictwoseventwofalias, \epictwothreethreebalias, \epicninesevensevenbalias, \epicthreefivethreebalias, and \epicthreefivezerocalias), emission spectroscopy (\epicthreefiveninebalias, \epicfiveeightzerobalias, and \epiconefourzerobalias), and both (\epicninetwoeightbalias\ and \epicfoursevenonebalias); several systems with planets in or close to mean motion resonances (\epiceightthreeninealias, \epicthreefivezeroalias) including a compact, TRAPPIST-1-like system of five small planets orbiting a mid-M dwarf (\epictwoseventwoalias); an ultra-short period sub-Saturn in the hot Saturn desert (\epiconefourzerobalias); and a super-Earth orbiting a moderately bright ($V=11.93$), metal-poor ([Fe/H]=$-0.579\pm0.080$) host star (\epicfourninezerobalias). In total we validate planets around 4 F stars, 26 G stars, 13 K stars, and 3 M dwarfs. In addition, we provide a list of 37 vetted planet candidates that should be prioritized for future follow-up observation in order to be confirmed or validated.

\end{abstract}



\section{Introduction}

The NASA \emph{K2} mission operated from 2014--2018, observing 18 full Campaigns along the ecliptic plane \citep{Howell14}. During each Campaign of 51--89 days, the 1-m telescope acquired images every 30 minutes on between 10,000 and 30,000 stars brighter than $V\sim16$. From these observations, nearly 500 transiting planets have been either confirmed or statistically validated, and an additional $\sim$1,000 candidates await confirmation\footnote{From the NASA Exoplanet Archive as of December 10, 2021. \url{https://exoplanetarchive.ipac.caltech.edu}}. While the NASA \emph{Kepler} mission, which stared at one carefully chosen field of view for four years, produced a largely homogeneous data set designed for occurrence rate studies, the \emph{K2} fields and targets were much more heterogeneous. The \emph{K2} Campaigns span a range of Galactic latitudes ($-65^{\circ}$ to $+65^{\circ}$) and longitudes (0$^{\circ}$ to 360$^{\circ}$), wildly varying the likelihood for contamination of the target star flux by background stars in a given pixel. The properties of the target stars themselves were not well constrained, being based in large part on photometric colors \citep{hub16}; some populations, such as cool stars, were subsequently characterized at higher fidelity \citep{Dressing2017,Dressing2019}. In addition, the final target lists for each Campaign were compiled from lists provided by guest observers, with stellar populations that varied systematically from Campaign to Campaign as projects with different scientific aims were developed or completed. Finally, unlike \textit{Kepler}, there was no single, standard data processing pipeline that was producing time series and signal detections that could be characterized for statistical studies. 

The Scaling \emph{K2} project has had as its primary goal the production of a catalog of \emph{K2} planet candidates that were uniformly detected and characterized, with stellar and planet parameters that were calculated to reduce the impacts of systematic biases in the data, and using a pipeline for which we could measure the detection efficiency (completeness) and reliability. In Paper I \citep{har20}, we published our uniformly derived catalog of stellar parameters for the \emph{K2} targets, using the subset of targets with both photometric colors and spectroscopic parameters as a training set for a machine learning algorithm to classify the much larger number of targets with only photometric colors. In Paper II \nocite{zin20a} (Zink et al. 2020a), we presented a pilot study of \emph{K2} Campaign 5, where we described our fully automated pipeline and released the first publicly available \emph{K2} transit candidate vetting code, EDI-Vetter. In Paper III \nocite{zin20b} (Zink et al. 2020b), we found comparable occurrence rates from \emph{K2} Campaign 5 and \emph{Kepler} in the same parameter space, validating the pipeline and vetting algorithms. Finally, in Paper IV \citep{Zink2021}, we presented our full, uniform \emph{K2} planet candidate catalog from Campaigns 1--8 and 10--18 (Campaign 9 being a specialized microlensing campaign that was unsuitable for our transit search). Here, we validate a significant number of planet candidates from those works and other previously published \emph{K2} planet candidate catalogs. In Section \ref{sec:validation} we describe the validation process, including the follow-up observations that were acquired, and the \texttt{vespa} and \texttt{centroid} analyses. In Section \ref{sec:planets} we discuss the newly validated planets, and highlight some of the more interesting new planetary systems. Finally, in Section \ref{sec:candidates} we briefly discuss the candidates that did not pass our validation thresholds, and highlight additional candidates that passed visual inspection but require additional follow-up observations to validate.

\section{Candidate Validation}
\label{sec:validation}

The planet candidate catalog presented in Paper IV contained 747 transit signals that were classified by the EDI-Vetter vetting algorithm as planet candidates, 366 of which were newly reported. In this work, we acquire and analyze additional observations and attempt to statistically validate as bona fide planets those planet candidates that had passed our EDI-Vetter vetting thresholds. Removing the 273 known planets, we visually inspected the phase-folded, detrended light curves produced in Paper IV for each of the 474 candidates to identify the most promising candidates for additional follow-up. Light curves that showed signals which appeared by eye to be caused by stellar variability, eclipsing binaries, or instrumental systematics, as compared to planet candidates, were discarded (and are noted as `FP' in that catalog). For each of the 329 remaining candidates, we retrieved the following information: (i) the Renormalized Unit Weight Error (RUWE) score from \emph{Gaia} EDR3, to flag potential binary stars \citep{GaiaDR3}; (ii) the status on the NASA Exoplanet Archive, to determine whether it was a new candidate, whether it had been previously deemed a false positive by another publication, or whether there were additional published candidates in the same system that our pipeline had missed; and (iii) the follow-up observations and observing notes that had been uploaded to ExoFOP\footnote{\url{https://exofop.ipac.caltech.edu/}}, if any, to determine whether it had already been deemed a false positive, and if not, to assess whether any additional follow-up data were needed. In a number of cases we obtained additional follow-up observations to complement the data that had already been archived. 

In order to consider whether a given candidate was ready to be validated, we required a single-lined reconnaissance spectrum, to mitigate the presence of spectroscopic binaries, and no evidence of additional stars in the high-resolution image, to constrain the presence of bound companions or chance alignments of the target star with more distant background stars. We note that while in many cases, the indication of additional stars in the spectroscopy, high-resolution imaging, or RUWE value may not actually disprove the existence of a planet, disentangling the true source of the transiting signal in these cases and deriving accurate stellar and planet parameters is beyond the scope of this work. We leave these as planet candidates for now, noting the presence of companions where observed, and highlighting them as interesting future cases to solve. We gathered the required observations to test validation of \totalchecked\ planet candidates around \totalcheckedsystems\ stars; we list these stars and their properties in Table \ref{tab:stellarpars}. The stellar parameters are drawn from Paper I, to provide a uniform source of stellar (and planetary) parameters in this work, however we note that the planet candidate catalog presented in Paper IV is designed to be agnostic to the stellar parameters, and to allow users to update the parameters to their desired values at will. The planet parameters are fit as in Paper IV, using the \texttt{batman} and \texttt{emcee} Python packages \citep{kre15,goo10,for13} and assuming circular orbits to measure the posterior distributions of the orbital period, transit midpoint, the planet-to-star radius ratio, the transit impact parameter, and the semi-major axis to stellar radius ratio. As described in Paper IV, quadratic limb-darkening parameters are fixed to values derived using the ATLAS model coefficients for the Kepler bandpasses \citep{cla12}, in concert with the Paper I stellar parameters.

\subsection{Follow-up Observations}

\subsubsection{Hale/PHARO and Keck/NIRC2}

High-resolution adaptive optics (AO) imaging were obtained with the PHARO instrument \citep{Hayward2001} on the Palomar 200-inch Hale telescope, and the NIRC2 instrument on the Keck II telescope. The PHARO observations were obtained using the P3K natural guide star AO system \citep{Dekany2013} in standard 5-point quincunx dither patterns. The dither positions were each observed three times, offset in position from each other by 0.500 arcsec for a total of 15 frames. The NIRC2 observations were obtained using the natural guide star AO system (Wizinowich et al. 2000) in the standard 3-point dither pattern that is used to avoid the noisier left lower quadrant of the detector. The dither pattern was repeated twice, with each dither offset from the previous dither by 0.500 arcsec. \citet{Schlieder2021} presents additional details on the NIRC2 follow-up observing program. Both the PHARO and NIRC2 observations were typically obtained in the narrow-band Br$-\gamma$ filter ($\lambda_o$ = 2.1686; $\Delta\lambda$ = 0.0326~$\mu$m).

The PHARO and NIRC2 data were processed and analyzed with a custom set of IDL tools. Flat fields were produced using a median average of dark subtracted sky frames. Sky frames were generated from the median average of the dithered science frames; each science image was then sky-subtracted and flat-fielded. The reduced science frames were combined into a single combined image using a intra-pixel interpolation that conserves flux, shifts the individual dithered frames by the appropriate fractional pixels, and median-coadds the frames.

\subsubsection{Gemini/DSSI and WIYN/NESSI}


Speckle imaging observations were obtained using the instruments DSSI (Horch et al. 2012) at the Gemini South 8-m telescope on Cerro Pachon and NESSI \citep{Scott2018} at the WIYN 3.5-m telescope on Kitt Peak. Both instruments collimate the incoming beam from the telescope, split it into red and blue channels with a dichroic beamsplitter and pass the beams through relatively narrow-band filter before focusing each on a separate Andor Ixon EMCCD camera. Data were taken in time series composed of 1000 frames apiece with individual exposure times of 40~ms for WIYN/NESSI and 60~ms for Gemini/DSSI. The images themselves were sub-array readouts of 256$\times$256 pixels, corresponding to a 4.6$^{\prime\prime}\times$4.6$^{\prime\prime}$ field-of-view (0.018$^{\prime\prime}$/pixel) for NESSI and a 2.8$^{\prime\prime}\times$2.8$^{\prime\prime}$ field-of-view (0.011$^{\prime\prime}$/pixel) for DSSI. The number of 1000-frame image sets taken per star depended on target brightness and observing conditions, but 3--10 image sets were typically obtained for our science targets. The observing strategy was very similar with both instruments. Science targets were observed close in time with nearby bright stars selected to represent point sources. The data were reduced following the procedures described in \citet{Howell2011}.

\subsubsection{SOAR}

We searched for close stellar companions to EPIC~211830293 with speckle imaging on the 4.1-m Southern Astrophysical Research (SOAR) telescope \citep{Tokovinin2018} on 18 March 2019 UT, observing in the visible Cousins I-band. The observations were each sensitive to an approximately 5-magnitude fainter star at an angular distance of $1^{\prime\prime}$ from the targets, and the target was cleared of bright companions at angular separations within $\sim3^{\prime\prime}$. 

\subsubsection{Keck/HIRES}

We collected high-resolution spectra with Keck/HIRES on Maunakea from 2014 Dec to 2017 Sep. Those spectra collected for reconnaissance purposes have a resolution of 60,000, a typical signal-to-noise ratio of 40/pixel at 550~nm, and a median seeing on Keck/HIRES of 1.0$^{\prime\prime}$. Those stars fainter than $V\sim10$ were observed with the C2 decker ($0.87^{\prime\prime}\times14.0^{\prime\prime}$), allowing for efficient removal of night sky emission lines. The spectra were reduced with the standard pipeline of the California Planet Search \citep{Howard2010}.

\subsubsection{FLWO/TRES}

The Tillinghast Reflector Echelle Spectrograph (TRES) is mounted on the 1.5m telescope at the Fred Lawrence Whipple Observatory (FLWO) in Arizona, USA. TRES is a fiber-fed, optical spectrograph with a R$\sim$44,000. The TRES spectra were extracted as described in \citet{Buchhave2010} and were visually inspected to determine that the spectra are single-lined. 

\subsubsection{McDonald/TS23}

The McDonald Observatory reconnaissance spectra were obtained with the Tull Coude Spectrometer \citep{Tull1995} of the McDonald Observatory 2.7m Harlan J. Smith Telescope. This is a cross-dispersed echelle white-pupil spectrograph, which gives spectral resolving power of 60,000 in its ``TS23'' setup, which was used for these observations. We used an exposure meter to achieve our target signal-to-noise ratio of $\sim$25/pixel. Data were reduced and the spectra were extracted using standard CCD techniques. 

\subsubsection{IRTF/SpeX and Hale/TripleSpec}

Medium resolution reconnaissance spectra were obtained using SpeX on the NASA Infrared Telescope Facility (IRTF), and TripleSpec on the Palomar 200-inch Hale telescope. The SpeX observations were conducted in SXD mode, using the 0.3$^{\prime\prime}\times$15$^{\prime\prime}$ slit to obtain $R\approx2000$ spectra. The observations were taken after the SpeX upgrade in 2014 and cover 0.7--2.55~$\mu$m. The TripleSpec observations were obtained using the fixed 1$^{\prime\prime}\times$30$^{\prime\prime}$ slit, with simultaneous coverage from 1.0--2.4~$\mu$m at a resolution $R=2500-2700$. The SpeX spectra were reduced using the publicly available \texttt{Spextool} pipeline \citep{Cushing2004}, and the TripleSpec spectra using a specialized version of \texttt{Spextool} adapted for using with TripleSpec data (available upon request from M. Cushing). The spectra were corrected for telluric contamination using the \texttt{xtellcor} package (Vacca et al. 2003), which is included in both versions of the Spextool pipeline. Additional details on the observing mode and data reduction can be found in \citet{Dressing2017, Dressing2019}.

\subsection{Validation with \texttt{vespa}}

The many thousands of \emph{Kepler} candidates detected early in the mission quickly out-paced the capacity of the dedicated network of follow-up telescopes to confirm them. This bottleneck ushered in the era of statistical validation, where the likelihood that a given transit signal around a given target star is explained by a planet is compared to the likelihood that it is explained by a suite of astrophysical false positives. These could include, for instance, eclipsing binaries, background/blended eclipsing binaries, and hierarchical eclipsing stellar systems. A number of codes were developed to analyze \emph{Kepler} candidates, including \texttt{BLENDER} \citep{Torres2011}, \texttt{PASTIS} \citep{Diaz2014}, and \texttt{vespa} \citep{Morton2016}. The latter of these is a publicly available Python package\footnote{\url{https://github.com/timothydmorton/VESPA}} that has been used to validate over a thousand \emph{Kepler} planets \citep{Morton2016}, and has been used in the validation of the vast majority of \emph{K2} planets to date \citep{Montet2015,Crossfield2016,Sinukoff2016,Dressing2017,Livingston2018,Mayo2018,Heller2019,Castro2020}.\\

In this analysis we use \texttt{vespa} version 0.6, and for each candidate, calculate the false positive probability (FPP) that the candidate transit signal is due to a bona fide planet. These values are given in Table \ref{tab:validation}. We set a threshold of $<1$\% to consider a planet validated. As inputs, \texttt{vespa} uses stellar parameters (position, parallax, photometry, $T_{\rm eff}$, log$g$, and [Fe/H]), transit signal parameters (period, $Rp/Rs$), the detrended light curve, and any available contrast curves from the high-resolution follow-up imaging. The latter are useful for constraining the potential parameter space within which nearby or blended eclipsing binaries could have remained undetected. In each case, we calculate the maximum radius for contaminating background sources from the size of the aperture used to generate the photometry, and the maximum allowed secondary eclipse depth using $0.1\times(Rp/Rs)^2$. Since \texttt{vespa} treats each candidate in multi-planet systems independently, we separately calculate and apply a multiplicity boost for these candidates. This boost, first introduced in \citet{Lissauer2012} for \emph{Kepler} and first adapted for \emph{K2} by \citet{Sinukoff2016}, relies on the fact that the conceivable false positive scenarios become significantly less probable when there are multiple transit signals in the same light curve; the likelihood, for instance, that two eclipsing binary systems are coincident along the line of sight is much lower than the likelihood of a star hosting two transiting planets, given what we know about the occurrence rate of multi-planet systems. Following the method of \citet{Sinukoff2016}, we calculated a multiplicity boost for each \emph{K2} campaign. The boost for Campaign $c$ is given by $X_{c}=F_{\rm{multi},c}/F_{\rm{cand},c}$, where $F_{\rm{cand},c}$ is the fraction of targets in the campaign with planet candidates, and $F_{\rm{multi},c}$ is the fraction of planet candidate hosts in the Campaign with more than one planet candidate. The calculated $X_{c}$ values are given in Table \ref{tab:multiplicity}; we did not calculate values for Campaigns 11 or 17 as there are no confirmed or candidate multi-planet systems in those campaigns. We calculated a different value for each campaign due to the varying Galactic latitudes and longitudes, which dramatically alter the stellar density in the field of view, and the likelihood of a given pixel containing a potentially contaminating eclipsing binary.

\subsection{Centroid Testing}
\label{sec:centroidingtest}

One area in which \emph{K2} observations were significantly degraded compared to \emph{Kepler} observations was pointing stability. In order for \emph{K2} to point accurately, it relied on the two remaining functional reaction wheels balancing against the radiation pressure from the Sun. This created an unstable equilibrium, where the spacecraft was continuously rolling slightly out of position and then being adjusted back to its original pointing. Unlike \emph{Kepler}, which achieved sub-pixel pointing precision \citep{Haas2010}, \emph{K2} was hampered by drift on the order of $\sim$1 pixel over several hours \citep[see][]{Saunders2019}. The subsequent increased sampling of the intrapixel and interpixel sensitivity variations, and time-dependent flux loss in the optimal aperture, significantly increased the correlated noise in the \emph{K2} light curves, on a timescale unfortunately comparable to typical transit durations.
The motion in the light curves was substantially corrected by several teams \citep[e.g.][]{van14, lug16, Aigran2016}.

For the exquisitely precise \emph{Kepler} pointing, measuring the change in position of the center of light from a target star (the centroid) during transit was a powerful tool for discriminating between true planet candidates and background or nearby eclipsing binaries \citep{Bryson2013}. For \emph{K2} however, the significantly worse pointing stability has thus far largely precluded the centroid test from being implemented (see \citet{kos19} for some success). The lack of a centroid test has led to the retraction of a number of previously validated \emph{K2} planets which were found to be eclipsing binaries within the same pixel \citep{Cabrera2017}. Here we use the new implementation of a centroid test for unstable/moving data from the \texttt{vetting}\footnote{https://github.com/ssdatalab/vetting} Python package \citep{vetting}, which accounts for the motion of the spacecraft.

In brief, the approach of this centroid test is to: (i) find the weighted average position of the source inside the pipeline aperture in the $x$ and $y$ dimension; (ii) correct these centroids for motion by detrending against the \texttt{POS\_CORR} arguments from the pipeline processed TPF (which estimate the position of the source on the detector) using the same Self Flat Fielding approach used by \cite{van14} for \emph{K2} data; (iii) remove a Gaussian-smoothed, long-term trend with a width of 21 cadences to remove long term drifts due to velocity aberration; and (iv) employ a student-t test to assess the likelihood of the distribution of centroids during transit being drawn from the distribution of centroids out of transit. This likelihood can then be used to discount false positives. Here, we used a threshold of $p=0.05$ to define a significant difference in the two distributions; candidates with $p<0.05$ were deemed to have a significant offset between their in- and out-of-transit centroids and were not considered validated. The threshold was chosen after performing the centroid test on known confirmed planets and false positives. 

\begin{longrotatetable}
\begin{deluxetable*}{p{0.6in}p{0.6in}p{0.7in}p{0.7in}p{0.7in}p{0.75in}p{0.34in}p{0.34in}p{0.34in}p{0.8in}p{1in}}
\tablewidth{700pt}
\tablecaption{Stellar parameters of the candidate systems as derived in Paper I; the uncertainties on $T_{\rm eff}$, log$g$, and [Fe/H] are root-mean-square uncertainties from the machine learning process. The instrument key is as follows: HIRES = Keck/HIRES; TRES = FLWO/TRES; TS23 = McDonald/TS23; TSpec = Hale/TripleSpec; SpeX = IRTF/Spex; PHARO = Hale/PHARO; NIRC2 = Keck/NIRC2, NIRI = Gemini/NIRI; DSSI = Gemini/DSSI; NESSI = WIYN/NESSI; HRCam = SOAR/HRCam. \label{tab:stellarpars}}
\tabletypesize{\scriptsize}
\tablehead{
\colhead{EPIC ID} & 
\colhead{$T_{\rm eff}$} & 
\colhead{$R_s$} &
\colhead{$M_s$} &
\colhead{log$g$} & 
\colhead{[Fe/H]} &
\colhead{$V$} & 
\colhead{$K$} &
\colhead{RUWE} &
\colhead{Spectrum} &
\colhead{Imaging} \\
\colhead{} & 
\colhead{[K]} & 
\colhead{[$R_{\odot}$]} &
\colhead{[$M_{\odot}$]} & 
\colhead{[dex]} &
\colhead{[dex]} &
\colhead{[mag]} &
\colhead{[mag]} &
\colhead{} & 
\colhead{} & 
\colhead{} 
}
\startdata
204750116 & $5561\pm138$ & $1.051^{+0.065}_{-0.061}$ & $0.823^{+0.369}_{-0.25}$ & $4.313\pm0.15$ & $-0.021\pm0.235$ & 11.37 & 9.62 & 1.10 & TRES & NIRC2\\
205029914 & $5755\pm138$ & $1.126^{+0.071}_{-0.065}$ & $0.835^{+0.376}_{-0.253}$ & $4.257\pm0.151$ & $-0.312\pm0.198$ & 12.29 & 9.71 & 1.07 & TRES$\bullet$ & NIRC2$\dagger$ \\
205111664 & $5601\pm138$ & $0.94^{+0.065}_{-0.058}$ & $0.825^{+0.366}_{-0.255}$ & $4.406\pm0.15$ & $-0.216\pm0.235$ & 12.49 & 9.88 & 1.10 & TRES & NIRC2 \\
205944181 & $5266\pm138$ & $0.852^{+0.057}_{-0.052}$ & $0.985^{+0.436}_{-0.306}$ & $4.568\pm0.15$ & $0.119\pm0.235$ & 12.60 & 10.65 & 0.94 & TRES, HIRES & NIRC2, NESSI\\
205957328 & $5300\pm138$ & $0.797^{+0.051}_{-0.05}$ & $0.981^{+0.438}_{-0.299}$ & $4.63\pm0.15$ & $0.053\pm0.235$ & 12.65 & 10.64 & 0.92 & TRES & NIRC2\\
205999468 & $5016\pm138$ & $0.754^{+0.052}_{-0.047}$ & $0.93^{+0.415}_{-0.29}$ & $4.65\pm0.15$ & $-0.046\pm0.235$ & 12.93 & 11.01 & 1.05 & TRES, HIRES & PHARO \\
206055981 & $4348\pm128$ & $0.625^{+0.041}_{-0.046}$ & $0.756^{+0.231}_{-0.347}$ & $4.726\pm0.150$ & $-0.351\pm0.235$ & 13.80 & 10.96 & 0.97 & HIRES, SpeX & NESSI \\
206135682 & $4663\pm138$ & $0.663^{+0.047}_{-0.045}$ & $0.746^{+0.334}_{-0.232}$ & $4.667\pm0.15$ & $-0.25\pm0.235$ & 13.54 & 11.04 & 1.01 & TSpec & PHARO \\
206146957 & $5682\pm138$ & $0.881^{+0.054}_{-0.049}$ & $0.735^{+0.334}_{-0.224}$ & $4.418\pm0.15$ & $-0.234\pm0.235$ & 11.79 & 9.93 & 0.90 & TRES & PHARO \\
206192335 & $5433\pm138$ & $0.787^{+0.051}_{-0.046}$ & $0.794^{+0.346}_{-0.248}$ & $4.542\pm0.15$ & $-0.201\pm0.235$ & 12.31 & 	10.25 & 0.80 & TRES, HIRES & NIRC2$\dagger$, NIRI$\dagger$, PHARO$\dagger$\\
206317286 & $4752\pm138$ & $0.758^{+0.052}_{-0.057}$ & $0.832^{+0.253}_{-0.367}$ & $4.596\pm0.15$ & $0.019\pm0.235$ & 14.05 & 11.63 & 0.97 & SpeX & NIRC2 \\
211399359 & $4925\pm138$ & $0.736^{+0.052}_{-0.049}$ & $0.958^{+0.44}_{-0.298}$ & $4.683\pm0.15$ & $-0.107\pm0.235$ & 14.64 & 12.39 & 1.03 & HIRES & NIRI, NIRC2\\
211428897 & $3780\pm93$ & $0.408^{+0.012}_{-0.012}$ & $0.405^{+0.009}_{-0.009}$ & $4.824\pm0.027$ & $-0.088\pm0.170$ & 14.09 & 9.62 & 1.07 & SpeX, TSpec, TS23 & NESSI, NIRC2, DSSI \\
211490999 & $5479\pm27$ & $0.922^{+0.018}_{-0.018}$ & $0.847^{+0.101}_{-0.088}$ & $4.437\pm0.045$ & $-0.064\pm0.025$ & 13.60 & 11.87 & 1.06 & HIRES, TS23 & NESSI, NIRI \\
211539054 & $6138\pm138$ & $1.486^{+0.083}_{-0.079}$ & $1.158^{+0.499}_{-0.349}$ & $4.156\pm0.15$ & $0.049\pm0.235$ & 10.47 & 9.16 & 0.90 & TRES & PHARO \\
211711685 & $5509\pm21$ & $0.919^{+0.016}_{-0.015}$ & $0.802^{+0.073}_{-0.068}$ & $4.414\pm0.035$ & $0.044\pm0.02$ & 12.55 & 10.84 & 0.94 & TRES & NIRC2 \\
211965883 & $4128\pm138$ & $0.646^{+0.05}_{-0.043}$ & $0.725^{+0.326}_{-0.224}$ & $4.678\pm0.15$ & $-0.229\pm0.235$ & 14.63 & 11.36 & 1.08 & TSpec & NESSI \\
212006318 & $5822\pm138$ & $1.543^{+0.104}_{-0.095}$ & $1.119^{+0.492}_{-0.344}$ & $4.107\pm0.15$ & $-0.091\pm0.235$ & 13.04 & 11.56 & 1.04 & TRES & PHARO, NESSI\\
212351868 & $6051\pm138$ & $2.481^{+0.145}_{-0.135}$ & $1.936^{+0.856}_{-0.596}$ & $3.934\pm0.15$ & $-0.038\pm0.235$ & 9.99 & 8.72 & 0.96 & HIRES$\ast$ & NIRC2, NESSI\\
212530118 & $4200\pm138$ & $0.691^{+0.053}_{-0.049}$ & $0.821^{+0.371}_{-0.258}$ & $4.673\pm0.15$ & $-0.008\pm0.235$ & 14.05 & 10.99 & 0.98 & SpeX & NESSI \\
212575828 & $4823\pm138$ & $0.732^{+0.058}_{-0.052}$ & $0.923^{+0.425}_{-0.291}$ & $4.674\pm0.15$ & $-0.206\pm0.235$ & 15.79 & 13.39 & 1.05 & TSpec & DSSI \\
212585579 & $5896\pm138$ & $1.086^{+0.065}_{-0.063}$ & $0.811^{+0.356}_{-0.252}$ & $4.276\pm0.15$ & $-0.163\pm0.235$ & 12.81 & 11.29 & 1.15 & TRES & NESSI, DSSI \\
212730483 & $4409\pm138$ & $0.71^{+0.053}_{-0.048}$ & $0.807^{+0.366}_{-0.256}$ & $4.642\pm0.15$ & $-0.088\pm0.235$ & 13.45 & 10.71 & 0.92 & SpeX & NESSI \\
212797028 & $5980\pm138$ & $1.869^{+0.115}_{-0.107}$ & $1.209^{+0.537}_{-0.373}$ & $3.978\pm0.15$ & $0.009\pm0.235$ & 13.28 & 11.54 & 0.87 & HIRES & NIRC2, DSSI, NESSI\\
213817056 & $5051\pm138$ & $0.831^{+0.058}_{-0.054}$ & $0.991^{+0.457}_{-0.309}$ & $4.595\pm0.151$ & $-0.324\pm0.198$ & 13.17 & 10.46 & 1.20 & TRES & NIRC2 \\
214173069 & $4603\pm138$ & $0.751^{+0.056}_{-0.052}$ & $0.856^{+0.375}_{-0.265}$ & $4.615\pm0.15$ & $-0.045\pm0.235$ & 13.21 & 10.51 & 0.81 & TRES & DSSI \\
214419545 & $5911\pm138$ & $1.32^{+0.078}_{-0.072}$ & $0.993^{+0.416}_{-0.302}$ & $4.194\pm0.15$ & $-0.049\pm0.235$ & 11.65 & 9.68 & 0.72 & TRES & NIRC2\\
216892056 & $3705\pm138$ & $0.471^{+0.014}_{-0.014}$ & $0.472^{+0.011}_{-0.011}$ & $4.767\pm0.027$ & $0.016\pm0.235$ & 13.50 & 9.13 & 1.20 & TRES & NIRI, DSSI \\
217192839 & $4473\pm138$ & $0.675^{+0.052}_{-0.047}$ & $0.754^{+0.337}_{-0.239}$ & $4.653\pm0.15$ & $-0.244\pm0.235$ & 13.01 & 10.30 & 1.02 & TRES, HIRES & NIRI, DSSI \\
217977895 & $5452\pm138$ & $0.811^{+0.053}_{-0.047}$ & $0.922^{+0.401}_{-0.284}$ & $4.585\pm0.15$ & $-0.027\pm0.235$ & 12.98 & 11.05 & 0.95 & TRES & DSSI \\
218668602 & $5129\pm138$ & $0.804^{+0.055}_{-0.051}$ & $1.061^{+0.459}_{-0.327}$ & $4.655\pm0.15$ & $0.187\pm0.235$ & 12.68 & 10.46 & 0.93 & TRES & NIRC2 \\
220221272 & $3623\pm138$ & $0.348^{+0.011}_{-0.01}$ & $0.33^{+0.009}_{-0.009}$ & $4.874\pm0.028$ & $0.068\pm0.198$ & 14.26$^{\dagger}$ & 11.29 & 1.31 & SpeX & NESSI\\
220294712 & $6145\pm43$ & $1.206^{+0.034}_{-0.033}$ & $1.024^{+0.194}_{-0.163}$ & $4.286\pm0.07$ & $-0.116\pm0.04$ & 12.34 & 10.97 & 1.00 & HIRES, TS23, TRES & NIRC2, PHARO$\dagger$, DSSI, NESSI\\
220400100 & $4648\pm138$ & $0.728^{+0.055}_{-0.048}$ & $0.832^{+0.374}_{-0.258}$ & $4.632\pm0.15$ & $-0.014\pm0.235$ & 13.60 & 11.25 & 1.00 & TS23 & NESSI \\
220459477 & $4844\pm138$ & $0.761^{+0.057}_{-0.052}$ & $0.927^{+0.412}_{-0.286}$ & $4.64\pm0.15$ & $-0.224\pm0.235$ & 14.65 & 12.24 & 0.96 & TSpec & NESSI \\
220510874 & $5749\pm22$ & $0.97^{+0.023}_{-0.022}$ & $0.911^{+0.088}_{-0.083}$ & $4.424\pm0.036$ & $0.074\pm0.02$ & 13.17 & 11.53 & 0.90 & TS23 & NESSI \\
220571481 & $6017\pm138$ & $1.139^{+0.068}_{-0.063}$ & $0.888^{+0.382}_{-0.271}$ & $4.272\pm0.15$ & $-0.077\pm0.235$ & 13.24 & 11.66 & 0.95 & TS23 & NESSI \\
220696233 & $3572\pm138$ & $0.585^{+0.018}_{-0.018}$ & $0.57^{+0.051}_{-0.048}$ & $4.66\pm0.046$ & $0.126\pm0.235$ & 16.22 & 12.29 & 1.08 & SpeX & PHARO, NESSI\\
226042826 & $5140\pm138$ & $0.976^{+0.067}_{-0.062}$ & $1.478^{+0.655}_{-0.462}$ & $4.625\pm0.151$ & $0.035\pm0.198$ & 11.79 & 9.75 & 0.98 & TRES & DSSI \\
245943455 & $5367\pm138$ & $0.897^{+0.058}_{-0.053}$ & $0.842^{+0.376}_{-0.259}$ & $4.456\pm0.15$ & $0.124\pm0.235$ & 12.82 & 10.93 & 1.09 & HIRES, TRES & NIRI, PHARO, NESSI\\
245955351 & $5436\pm138$ & $0.812^{+0.051}_{-0.049}$ & $0.918^{+0.406}_{-0.275}$ & $4.585\pm0.15$ & $-0.061\pm0.235$ & 13.19 & 11.34 & 0.97 & HIRES, TRES & NIRI$\dagger$, PHARO$\dagger$, NESSI\\
245991048 & $5782\pm138$ & $1.084^{+0.067}_{-0.061}$ & $0.802^{+0.357}_{-0.246}$ & $4.273\pm0.15$ & $-0.034\pm0.235$ & 12.30 & 10.18 & 1.06 & HIRES, TRES & NESSI, PHARO, NIRI\\
245995977 & $5558\pm138$ & $0.966^{+0.059}_{-0.056}$ & $0.789^{+0.351}_{-0.239}$ & $4.368\pm0.15$ & $-0.116\pm0.235$ & 13.57 & 11.72 & 0.95 & HIRES & NIRI, PHARO\\
246074314 & $5543\pm138$ & $0.566^{+0.112}_{-0.076}$ & $0.757^{+0.341}_{-0.24}$ & $4.973\pm0.15$ & $-0.543\pm0.278$ & 12.36$^{\dagger}$ & 10.71 & 0.97 & TRES & NESSI \\
246084398 & $5726\pm138$ & $1.058^{+0.07}_{-0.065}$ & $0.841^{+0.37}_{-0.258}$ & $4.314\pm0.15$ & $-0.238\pm0.235$ & 12.82 & 11.24 & 1.19 & TRES & NESSI\\
246429049 & $5678\pm138$ & $1.071^{+0.065}_{-0.06}$ & $0.881^{+0.381}_{-0.269}$ & $4.321\pm0.15$ & $0.076\pm0.235$ & 11.84 & 10.28 & 1.04 & TRES & NESSI \\
210797580 & $5377\pm138$ & $0.971^{+0.065}_{-0.057}$ & $0.984^{+0.428}_{-0.303}$ & $4.454\pm0.15$ & $0.018\pm0.235$ & 11.11 & 9.24 & 0.99 & HIRES, TRES & NESSI, NIRC2\\
246876040 & $4938\pm138$ & $0.631^{+0.044}_{-0.04}$ & $0.478^{+0.214}_{-0.147}$ & $4.515\pm0.15$ & $-0.388\pm0.235$ & 12.72 & 9.57 & 1.11 & TRES & NESSI \\
246891819 & $5007\pm138$ & $0.783^{+0.054}_{-0.051}$ & $0.952^{+0.421}_{-0.293}$ & $4.628\pm0.15$ & $-0.033\pm0.235$ & 14.66 & 11.37 & 1.07 & TSpec, HIRES & NIRC2 \\
246947582 & $4692\pm138$ & $1.14^{+0.086}_{-0.079}$ & $1.232^{+0.548}_{-0.386}$ & $4.413\pm0.15$ & $0.037\pm0.235$ & 15.79 & 11.22 & 1.08 & TSpec & NIRC2$\dagger$ \\
246953392 & $5236\pm138$ & $0.92^{+0.062}_{-0.057}$ & $0.954^{+0.429}_{-0.293}$ & $4.488\pm0.15$ & $0.237\pm0.235$ & 13.24 & 10.81 & 0.91 & TRES & NESSI \\
247164043 & $6265\pm138$ & $1.272^{+0.067}_{-0.066}$ & $0.988^{+0.419}_{-0.293}$ & $4.223\pm0.15$ & $-0.088\pm0.235$ & 9.48 & 8.23 & 1.06 & TRES & NIRC2 \\
247383003 & $5387\pm138$ & $0.966^{+0.06}_{-0.057}$ & $0.937^{+0.406}_{-0.285}$ & $4.435\pm0.15$ & $0.11\pm0.235$ & 11.83 & 9.95 & 1.29 & TRES & NIRC2, NESSI \\
247698108 & $4993\pm138$ & $4.259^{+0.338}_{-0.314}$ & $1.249^{+0.585}_{-0.385}$ & $3.279\pm0.15$ & $-0.325\pm0.235$ & 14.68 & 10.78 & 0.97 & HIRES & NIRC2 \\
247724061 & $5496\pm154$ & $1.289^{+0.093}_{-0.085}$ & $1.041^{+0.468}_{-0.322}$ & $4.235\pm0.151$ & $-0.1\pm0.275$ & 12.03 & 10.45 & 1.21 & TRES & NESSI \\
248222323 & $5895\pm138$ & $1.427^{+0.085}_{-0.079}$ & $1.128^{+0.5}_{-0.347}$ & $4.184\pm0.15$ & $0.085\pm0.235$ & 12.41 & 9.47 & 1.19 & TRES & NIRC2 \\
248463350 & $5946\pm20$ & $1.239^{+0.032}_{-0.03}$ & $0.945^{+0.09}_{-0.083}$ & $4.227\pm0.034$ & $-0.257\pm0.019$ & 13.15 & 11.72 & 1.19 & HIRES, TRES & NIRC2, NESSI\\
248472140 & $5690\pm38$ & $1.541^{+0.048}_{-0.047}$ & $0.783^{+0.131}_{-0.114}$ & $3.956\pm0.063$ & $0.252\pm0.036$ & 13.03 & 11.38 & 5.89 & TRES, HIRES & NIRC2, NESSI \\
248518307 & $3578\pm138$ & $0.382^{+0.012}_{-0.012}$ & $0.374^{+0.009}_{-0.009}$ & $4.847\pm0.028$ & $-0.051\pm0.235$ & 15.12 & 10.41 & 1.05 & TSpec & NESSI \\
248527514 & $4213\pm138$ & $0.675^{+0.051}_{-0.046}$ & $0.87^{+0.395}_{-0.269}$ & $4.719\pm0.15$ & $-0.098\pm0.235$ & 14.20 & 11.11 & 1.09 & TSpec & NESSI \\
248621597 & $5716\pm138$ & $1.28^{+0.089}_{-0.082}$ & $0.944^{+0.419}_{-0.29}$ & $4.197\pm0.15$ & $-0.378\pm0.235$ & 13.03 & 11.38 & 0.96 & TRES & NESSI \\
248639411 & $4626\pm49$ & $0.799^{+0.024}_{-0.023}$ & $0.87^{+0.186}_{-0.15}$ & $4.573\pm0.08$ & $0.072\pm0.047$ & 13.51 & 	11.01 & 1.05 & HIRES, TSpec & NIRC2, NESSI\\
248740016 & $5872\pm138$ & $1.025^{+0.059}_{-0.058}$ & $0.863^{+0.381}_{-0.258}$ & $4.352\pm0.15$ & $-0.074\pm0.235$ & 11.15 & 9.64 & 0.89 & TRES & NESSI\\
248758353 & $5511\pm138$ & $0.886^{+0.056}_{-0.052}$ & $0.784^{+0.34}_{-0.241}$ & $4.439\pm0.15$ & $-0.162\pm0.235$ & 12.36 & 10.58 & 0.90 & TRES & NESSI \\
248827616 & $6216\pm138$ & $1.75^{+0.106}_{-0.1}$ & $1.566^{+0.695}_{-0.474}$ & $4.147\pm0.15$ & $-0.184\pm0.235$ & 11.92 & 10.36 & 0.98 & HIRES, TRES & NIRC2 \\
248861279 & $3748\pm138$ & $0.552^{+0.016}_{-0.016}$ & $0.552^{+0.015}_{-0.015}$ & $4.695\pm0.027$ & $0.005\pm0.235$ & 14.56 & 10.75 & 1.05 & TSpec & NIRC2, NESSI \\
248874928 & $4927\pm138$ & $0.765^{+0.054}_{-0.05}$ & $1.034^{+0.459}_{-0.314}$ & $4.685\pm0.15$ & $0.132\pm0.235$ & 12.86 & 10.55 & 1.05 & TRES, HIRES & NIRC2, NESSI \\
249223471 & $5784\pm138$ & $0.961^{+0.058}_{-0.054}$ & $0.798^{+0.339}_{-0.245}$ & $4.374\pm0.15$ & $-0.038\pm0.235$ & 9.47 & 8.03 & 1.05 & TRES, HIRES & DSSI\\
249403651 & $5563\pm138$ & $0.927^{+0.058}_{-0.052}$ & $0.881^{+0.39}_{-0.267}$ & $4.45\pm0.15$ & $0.014\pm0.235$ & 11.97 & 10.21 & 1.08 & TRES & DSSI \\
249816490 & $5335\pm138$ & $0.784^{+0.051}_{-0.047}$ & $0.813^{+0.368}_{-0.249}$ & $4.559\pm0.15$ & $-0.171\pm0.235$ & 11.93 & 9.98 & 1.10 & TRES & DSSI \\
249827330 & $5631\pm138$ & $1.783^{+0.115}_{-0.109}$ & $0.949^{+0.423}_{-0.289}$ & $3.916\pm0.15$ & $-0.201\pm0.235$ & 12.94 & 11.13 & 1.04 & TRES & DSSI \\
249865296 & $6117\pm138$ & $1.319^{+0.088}_{-0.08}$ & $1.03^{+0.469}_{-0.314}$ & $4.212\pm0.15$ & $-0.197\pm0.235$ & 12.96 & 11.43 & 1.14 & TRES & DSSI \\
249924395 & $5742\pm138$ & $1.228^{+0.077}_{-0.072}$ & $0.886^{+0.374}_{-0.271}$ & $4.204\pm0.15$ & $-0.036\pm0.235$ & 12.76 & 11.04 & 1.05 & TRES & DSSI \\
211732116 & $5789\pm21$ & $0.907^{+0.018}_{-0.017}$ & $0.731^{+0.067}_{-0.061}$ & $4.386\pm0.035$ & $-0.292\pm0.02$ & 12.80 & 11.29 & 0.95 & TRES & NIRC2 \\
211830293 & $5278\pm20$ & $1.851^{+0.045}_{-0.044}$ & $0.811^{+0.077}_{-0.07}$ & $3.812\pm0.033$ & $0.168\pm0.018$ & 12.13 & 10.21 & 0.96 & TRES$\bullet$ & HRCam \\
212222383 & $5937\pm138$ & $1.149^{+0.067}_{-0.062}$ & $0.908^{+0.394}_{-0.281}$ & $4.275\pm0.15$ & $-0.176\pm0.235$ & 10.42 & 9.01 & 0.86 & TRES & PHARO \\
212705192 & $5778\pm138$ & $2.155^{+0.135}_{-0.128}$ & $1.54^{+0.679}_{-0.477}$ & $3.958\pm0.15$ & $-0.265\pm0.235$ & 12.20 & 10.28 & 1.11 & TRES$\diamond$ & NESSI \\
211784767 & $6044\pm138$ & $1.58^{+0.096}_{-0.088}$ & $1.288^{+0.554}_{-0.39}$ & $4.149\pm0.15$ & $-0.075\pm0.235$ & 11.97 & 10.56 & 1.29 & TRES & PHARO \\
\enddata
\tablenotetext{\dagger}{Contaminating star/s detected in high-resolution imaging or spectroscopy.}
\tablenotetext{\bullet}{Radial velocities indicate stellar mass companion.}
\tablenotetext{\ast}{Too high $v$sin$i$ to detect possible stellar companions in spectrum.}
\tablenotetext{\diamond}{Double-lined spectroscopic binary.}
\end{deluxetable*}
\end{longrotatetable}


\begin{longrotatetable}
\begin{deluxetable*}{lllllllllll}
\tablewidth{700pt}
\tablecaption{Validation parameters of the candidate systems. HEBs: the calculated probability that the signal is caused by a hierarchical eclipsing binary; HEBs ($2\times$P): the same for a binary with twice the measured orbital period. EBs: the calculated probability that the signal is caused by an eclipsing binary; EBs ($2\times$P): the same for an eclipsing binary with twice the period. BEBs: the calculated probability that the signal is caused by a background eclipsing binary; BEBs: ($2\times$P) the same for a background eclipsing binary with twice the period. Planets: the calculated probability that the signal is caused by a bona fide planet. FPP: the final \texttt{vespa} false positive probability. \label{tab:validation}}
\tabletypesize{\scriptsize}
\tablehead{
\colhead{Candidate ID} & 
\colhead{Period} & 
\colhead{HEBs} & 
\colhead{HEBs ($2\times$P)} & 
\colhead{EBs} & 
\colhead{EBs ($2\times$P)} & 
\colhead{BEBs} & 
\colhead{BEBs ($2\times$P)} & 
\colhead{Planets} & 
\colhead{FPP} & 
\colhead{Centroid $p$-value}  \\ 
\colhead{} & 
\colhead{[d]} &
\colhead{} &
\colhead{} & 
\colhead{} & 
\colhead{} & 
\colhead{} & 
\colhead{} & 
\colhead{} &
\colhead{} &
\colhead{} 
}
\startdata
\emph{Validated} & & & & & & & & & & \\
204750116.01 & 23.4481724 & $2.54 \times 10^{-12}$ & $1.31 \times 10^{-08}$ & $3.02 \times 10^{-04}$ & $7.05 \times 10^{-04}$ & $5.73 \times 10^{-04}$ & $2.48 \times 10^{-05}$ & 0.998 & $1.60 \times 10^{-03}$ & 0.5219 \\
205111664.01 & 15.937039 & $3.72 \times 10^{-06}$ & $3.10 \times 10^{-06}$ & $4.23 \times 10^{-05}$ & $8.59 \times 10^{-05}$ & $1.64 \times 10^{-06}$ & $3.81 \times 10^{-06}$ & 1.000 & $1.40 \times 10^{-04}$ $^{\dagger}$ & 0.4962 \\
206055981.01 & 20.6450985 & $1.13 \times 10^{-05}$ & $2.00 \times 10^{-07}$ & $6.47 \times 10^{-04}$ & $1.87 \times 10^{-05}$ & $4.34 \times 10^{-05}$ & $7.21 \times 10^{-38}$ & 0.999 & $7.21 \times 10^{-04}$ $^{\dagger}$ & 0.5842 \\
206135682.01 & 5.02559754 & $1.77 \times 10^{-05}$ & $1.85 \times 10^{-04}$ & $3.69 \times 10^{-05}$ & $5.19 \times 10^{-05}$ & $7.43 \times 10^{-05}$ & $1.83 \times 10^{-04}$ & 0.999 & $1.90 \times 10^{-05}$ $^{\ast}$ & 0.2070 \\
206135682.02 & 20.2010847 & $1.26 \times 10^{-08}$ & $3.36 \times 10^{-08}$ & $7.63 \times 10^{-06}$ & $1.23 \times 10^{-05}$ & $1.73 \times 10^{-04}$ & $1.10 \times 10^{-04}$ & 1.000 & $3.44 \times 10^{-05}$ $^{\ast}$ & 0.3671 \\
206135682.03 & 9.66018618 & $1.42 \times 10^{-04}$ & $3.94 \times 10^{-07}$ & $9.19 \times 10^{-06}$ & $2.28 \times 10^{-07}$ & $2.38 \times 10^{-04}$ & $2.06 \times 10^{-06}$ & 1.000 & $2.46 \times 10^{-05}$ $^{\ast}$ & 0.1385 \\
206146957.01 & 5.76161216 & $2.5 \times 10^{-06}$ & $1.44 \times 10^{-05}$ & $9.31 \times 10^{-07}$ & $1.24 \times 10^{-05}$ & $0.0 \times 10^{+00}$ & $0.0 \times 10^{+00}$ & 1.000 & $3.02 \times 10^{-05}$ $^{\dagger}$ & 0.4584 \\
206317286.01 & 17.5154723 & $7.57 \times 10^{-66}$ & $1.44 \times 10^{-51}$ & $3.25 \times 10^{-05}$ & $1.11 \times 10^{-03}$ & $7.81 \times 10^{-05}$ & $8.07 \times 10^{-05}$ & 0.999 & $8.16 \times 10^{-05}$ $^{\ast}$ & 0.1811 \\
211399359.01 & 3.11490475 & $8.14 \times 10^{-65}$ & $9.25 \times 10^{-30}$ & $8.13 \times 10^{-173}$ & $1.58 \times 10^{-104}$ & $1.89 \times 10^{-13}$ & $2.09 \times 10^{-82}$ & 1.000 & $1.89 \times 10^{-13}$ & 0.1241 \\
211490999.01 & 9.84407732 & $5.48 \times 10^{-14}$ & $5.9 \times 10^{-18}$ & $3.23 \times 10^{-03}$ & $1.57 \times 10^{-04}$ & $0.00 \times 10^{+00}$ & $0.00 \times 10^{+00}$ & 0.997 & $3.39 \times 10^{-03}$ $^{\dagger}$ & 0.4752 \\ 
211539054.01 & 11.0206598 & $3.81 \times 10^{-07}$ & $3.16 \times 10^{-08}$ & $1.76 \times 10^{-03}$ & $3.73 \times 10^{-04}$ & $0.00 \times 10^{+00}$ & $0.00 \times 10^{+00}$ & 0.998 & $2.13 \times 10^{-03}$ $^{\dagger}$ & 0.3779 \\
212006318.01 & 14.453895 & $2.43 \times 10^{-05}$ & $1.39 \times 10^{-05}$ & $9.96 \times 10^{-05}$ & $3.97 \times 10^{-03}$ & $0.00 \times 10^{+00}$ & $1.55 \times 10^{-07}$ & 0.996 & $4.11 \times 10^{-03}$ $^{\dagger}$ & 0.1458 \\
212530118.01 & 12.8322355 & $6.79 \times 10^{-05}$ & $2.48 \times 10^{-14}$ & $9.12 \times 10^{-04}$ & $2.01 \times 10^{-05}$ & $1.37 \times 10^{-04}$ & $2.16 \times 10^{-07}$ & 0.999 & $1.14 \times 10^{-03}$ $^{\dagger}$ & 0.5774 \\
212575828.01 & 2.06043843 & $5.11 \times 10^{-05}$ & $2.31 \times 10^{-05}$ & $6.49 \times 10^{-04}$ & $2.23 \times 10^{-05}$ & $1.67 \times 10^{-04}$ & $0.00 \times 10^{+00}$ & 0.999 & $9.12 \times 10^{-04}$ $^{\dagger}$ & 0.4968 \\
214173069.01 & 8.7779358 & $1.54 \times 10^{-05}$ & $1.19 \times 10^{-06}$ & $1.20 \times 10^{-03}$ & $1.12 \times 10^{-03}$ & $1.41 \times 10^{-03}$ & $1.63 \times 10^{-03}$ & 0.995 & $5.38 \times 10^{-03}$ $^{\dagger}$ & 0.8487 \\
214419545.01 & 9.4013125 & $2.21 \times 10^{-05}$ & $2.24 \times 10^{-07}$ & $5.19 \times 10^{-05}$ & $1.94 \times 10^{-06}$ & $8.61 \times 10^{-03}$ & $1.42 \times 10^{-04}$ & 0.991 & $8.82 \times 10^{-03}$ & 0.6540 \\
217192839.01 & 16.0346559 & $7.26 \times 10^{-06}$ & $7.51 \times 10^{-08}$ & $6.96 \times 10^{-05}$ & $4.15 \times 10^{-06}$ & $2.41 \times 10^{-45}$ & $3.69 \times 10^{-62}$ & 1.000 & $1.46 \times 10^{-05}$ $^{\dagger\ast}$ & 0.3334 \\
217192839.02 & 7.938933355 & $3.01 \times 10^{-03}$ & $5.22 \times 10^{-05}$ & $3.93 \times 10^{-02}$ & $1.55 \times 10^{-03}$ & $2.08 \times 10^{-19}$ & $3.32 \times 10^{-42}$ & 0.956 & $8.22 \times 10^{-03}$ $^{\dagger\ast}$ & 0.2215 \\
217192839.03 & 26.803023 & $3.52 \times 10^{-04}$ & $1.43 \times 10^{-05}$ & $5.92 \times 10^{-03}$ & $1.16 \times 10^{-04}$ & $0.00 \times 10^{+00}$ & $0.00 \times 10^{+00}$ & 0.994 & $1.16 \times 10^{-03}$ $^{\dagger\ast}$ & 0.2891 \\
217977895.01 & 21.7001564 & $4.33 \times 10^{-10}$ & $4.05 \times 10^{-20}$ & $4.6 \times 10^{-04}$ & $3.18 \times 10^{-07}$ & $2.16 \times 10^{-03}$ & $9.68 \times 10^{-42}$ & 0.997 & $2.62 \times 10^{-03}$ $^{\dagger}$ & 0.3721 \\
218668602.01 & 1.86596186 & $6.93 \times 10^{-05}$ & $2.16 \times 10^{-06}$ & $4.41 \times 10^{-04}$ & $1.53 \times 10^{-06}$ & $1.53 \times 10^{-03}$ & $7.84 \times 10^{-05}$ & 0.998 & $2.12 \times 10^{-03}$ & 0.1391 \\
220221272.01 & 13.6274904 & $2.03 \times 10^{-04}$ & $6.59 \times 10^{-05}$ & $4.25 \times 10^{-04}$ & $8.15 \times 10^{-04}$ & $1.37 \times 10^{-05}$ & $7.62 \times 10^{-07}$ & 0.998 & $1.12 \times 10^{-04}$ $^{\dagger\ast}$ & 0.5460 \\%
220221272.02 & 6.679581962 & $3.59 \times 10^{-04}$ & $3.79 \times 10^{-04}$ & $6.58 \times 10^{-04}$ & $2.84 \times 10^{-03}$ & $7.76 \times 10^{-07}$ & $1.76 \times 10^{-05}$ & 0.996 & $3.14 \times 10^{-04}$ $^{\dagger\ast}$ & 0.5203 \\
220221272.03 & 4.194766393 & $1.6 \times 10^{-03}$ & $4.52 \times 10^{-04}$ & $4.10 \times 10^{-03}$ & $2.69 \times 10^{-03}$ & $8.53 \times 10^{-06}$ & $8.68 \times 10^{-07}$ & 0.991 & $6.57 \times 10^{-04}$ $^{\dagger\ast}$ & 0.1135 \\
220221272.04 & 9.715042543 & $2.66 \times 10^{-03}$ & $5.22 \times 10^{-03}$ & $1.14 \times 10^{-02}$ & $3.06 \times 10^{-02}$ & $2.25 \times 10^{-05}$ & $1.16 \times 10^{-05}$ & 0.950 & $3.85 \times 10^{-03}$ $^{\dagger\ast}$ & 0.4207 \\
220221272.05 & 2.231527238 & $8.26 \times 10^{-04}$ & $8.01 \times 10^{-04}$ & $9.14 \times 10^{-03}$ & $1.13 \times 10^{-02}$ & $1.25 \times 10^{-07}$ & $5.02 \times 10^{-06}$ & 0.978 & $7.20 \times 10^{-03}$ $^{\dagger\ast}$ & 0.6315 \\
220459477.01 & 2.3808672 & $6.28 \times 10^{-08}$ & $1.26 \times 10^{-19}$ & $2.83 \times 10^{-13}$ & $5.47 \times 10^{-20}$ & $1.08 \times 10^{-05}$ & $4.86 \times 10^{-12}$ & 1.000 & $1.09 \times 10^{-05}$ $^{\dagger}$ & 0.1107 \\
220510874.01 & 7.4732234 & $3.07 \times 10^{-205}$ & $1.99 \times 10^{-98}$ & $1.09 \times 10^{-13}$ & $3.34 \times 10^{-05}$ & $1.01 \times 10^{-23}$ & $2.61 \times 10^{-14}$ & 1.000 & $3.34 \times 10^{-05}$ $^{\dagger}$ & 0.6971 \\
220696233.01 & 28.7353275 & $1.50 \times 10^{-257}$ & $2.87 \times 10^{-122}$ & $3.25 \times 10^{-10}$ & $7.11 \times 10^{-15}$ & $8.07 \times 10^{-12}$ & $2.98 \times 10^{-62}$ & 1.000 & $3.33 \times 10^{-10}$ $^{\dagger}$ & 0.3918 \\
245943455.01 & 6.33906865 & $8.97 \times 10^{-51}$ & $3.45 \times 10^{-65}$ & $8.02 \times 10^{-08}$ & $1.07 \times 10^{-12}$ & $0.00 \times 10^{+00}$ & $0.00 \times 10^{+00}$ & 1.000 & $8.02 \times 10^{-08}$ $^{\dagger}$ & 0.0938 \\
245991048.01 & 8.5835664 & $1.14 \times 10^{-04}$ & $3.95 \times 10^{-05}$ & $2.03 \times 10^{-03}$ & $2.31 \times 10^{-03}$ & $0.00 \times 10^{+00}$ & $0.00 \times 10^{+00}$ & 0.996 & $6.41 \times 10^{-05}$ $^{\dagger\ast}$ & 0.2457 \\
245991048.02 & 20.8513367 & $9.04 \times 10^{-05}$ & $5.44 \times 10^{-07}$ & $3.86 \times 10^{-03}$ & $8.12 \times 10^{-05}$ & $2.02 \times 10^{-225}$ & $0.00 \times 10^{+00}$ & 0.996 & $5.75 \times 10^{-05}$ $^{\dagger\ast}$ & 0.2053 \\
245995977.01 & 3.31279234 & $5.36 \times 10^{-08}$ & $3.58 \times 10^{-12}$ & $2.22 \times 10^{-08}$ & $1.10 \times 10^{-08}$ & $0.00 \times 10^{+00}$ & $0.00 \times 10^{+00}$& 1.000 & $8.68 \times 10^{-08}$ $^{\dagger}$ & 0.6208 \\
246074314.01 & 4.62265408 & $5.42 \times 10^{-05}$ & $1.08 \times 10^{-07}$ & $2.12 \times 10^{-05}$ & $3.21 \times 10^{-07}$ & $2.28 \times 10^{-05}$ & $1.78 \times 10^{-245}$ & 1.000 & $9.87 \times 10^{-05}$ $^{\dagger}$ & 0.3581 \\
246084398.01 & 15.3987228 & $8.04 \times 10^{-06}$ & $3.99 \times 10^{-08}$ & $2.07 \times 10^{-03}$ & $3.04 \times 10^{-06}$ & $3.38 \times 10^{-05}$ & $6.88 \times 10^{-09}$ & 0.998 & $2.12 \times 10^{-03}$ $^{\dagger}$ & 0.5614 \\
246429049.01 & 10.4131807 & $1.30 \times 10^{-04}$ & $5.69 \times 10^{-05}$ & $6.61 \times 10^{-03}$ & $2.03 \times 10^{-03}$ & $8.16 \times 10^{-05}$ & $9.30 \times 10^{-05}$ & 0.991 & $9.00 \times 10^{-03}$ $^{\dagger}$ & 0.3512 \\
210797580.01 & 2.1408398 & $2.10 \times 10^{-04}$ & $4.68 \times 10^{-07}$ & $1.90 \times 10^{-03}$ & $8.92 \times 10^{-08}$ & $1.24 \times 10^{-03}$ & $1.99 \times 10^{-24}$ & $9.97 \times 10^{-01}$ & $1.77 \times 10^{-03}$ $^{\dagger}$ & 0.4204 \\
246876040.01 & 5.09571665 & $2.12 \times 10^{-08}$ & $1.82 \times 10^{-08}$ & $8.35 \times 10^{-22}$ & $9.61 \times 10^{-14}$ & $4.38 \times 10^{-04}$ & $2.50 \times 10^{-203}$ & 1.000 & $4.38 \times 10^{-04}$ $^{\dagger}$ & 0.2143 \\
246891819.01 & 4.80335174 & $1.21 \times 10^{-14}$ & $7.23 \times 10^{-12}$ & $2.15 \times 10^{-13}$ & $1.34 \times 10^{-08}$ & $0.00 \times 10^{+00}$ & $0.00 \times 10^{+00}$ & 1.000 & $6.35 \times 10^{-10}$ $^{\dagger\ast}$ & 0.4413 \\
246891819.02 & 8.49128189 & $1.42 \times 10^{-05}$ & $6.90 \times 10^{-06}$ & $1.12 \times 10^{-04}$ & $2.34 \times 10^{-04}$ & $0.00 \times 10^{+00}$ & $0.00 \times 10^{+00}$ & 1.000 & $1.74 \times 10^{-05}$ $^{\dagger\ast}$ & 0.4640 \\
246953392.01 & 0.67386241 & $5.32 \times 10^{-04}$ & $4.92 \times 10^{-04}$ & $3.04 \times 10^{-03}$ & $4.24 \times 10^{-03}$ & $2.34 \times 10^{-04}$ & $1.83 \times 10^{-04}$ & 0.991 & $4.16 \times 10^{-04}$ $^{\dagger\ast}$ & 0.4124 \\
246953392.02 & 25.7605085 & $2.83 \times 10^{-10}$ & $1.48 \times 10^{-11}$ & $1.40 \times 10^{-04}$ & $4.29 \times 10^{-15}$ & $8.03 \times 10^{-05}$ & $2.00 \times 10^{-76}$ & 1.000 & $1.07 \times 10^{-05}$ $^{\dagger\ast}$ & 0.3421 \\
247383003.01 & 3.57232573 & $1.03 \times 10^{-04}$ & $1.07 \times 10^{-05}$ & $6.81 \times 10^{-03}$ & $1.06 \times 10^{-03}$ & $0.00 \times 10^{+00}$ & $0.00 \times 10^{+00}$ & 0.992 & $7.99 \times 10^{-03}$ $^{\dagger}$ & 0.5226 \\
248463350.01 & 6.39302548 & $1.08 \times 10^{-03}$ & $9.01 \times 10^{-04}$ & $6.17 \times 10^{-03}$ & $2.33 \times 10^{-02}$ & $0.00 \times 10^{+00}$ & $0.00 \times 10^{+00}$ & 0.969 & $4.80 \times 10^{-04}$ $^{\dagger\ast}$ & 0.3000 \\
248463350.02 & 18.787839 & $5.67 \times 10^{-05}$ & $4.65 \times 10^{-04}$ & $3.32 \times 10^{-05}$ & $1.56 \times 10^{-02}$ & $0.00 \times 10^{+00}$ & $0.00 \times 10^{+00}$ & 0.984 & $2.43 \times 10^{-04}$ $^{\dagger\ast}$ & 0.3974 \\
248472140.01 & 0.75997781 & $1.54 \times 10^{-04}$ & $1.24 \times 10^{-07}$ & $2.66 \times 10^{-04}$ & $3.6 \times 10^{-04}$ & $0.00 \times 10^{+00}$ & $0.00 \times 10^{+00}$ & 0.999 & $7.80 \times 10^{-04}$ $^{\dagger}$ & 0.1608 \\
248518307.01 & 3.86505292 & $8.36 \times 10^{-05}$ & $1.22 \times 10^{-04}$ & $5.2 \times 10^{-04}$ & $4.29 \times 10^{-04}$ & $6.66 \times 10^{-05}$ & $1.28 \times 10^{-04}$ & 0.999 & $1.35 \times 10^{-03}$ $^{\dagger}$ & 0.5332 \\
248527514.01 & 6.29309867 & $9.99 \times 10^{-04}$ & $3.01 \times 10^{-05}$ & $2.45 \times 10^{-03}$ & $2.17 \times 10^{-04}$ & $7.93 \times 10^{-05}$ & $9.35 \times 10^{-07}$ & 0.996 & $3.78 \times 10^{-03}$ $^{\dagger}$ & 0.3051 \\
248621597.01 & 17.2747396 & $5.62 \times 10^{-06}$ & $1.77 \times 10^{-06}$ & $4.11 \times 10^{-09}$ & $1.94 \times 10^{-04}$ & $2.09 \times 10^{-06}$ & $3.45 \times 10^{-06}$ & 1.000 & $2.07 \times 10^{-04}$ $^{\dagger}$ & 0.4934 \\
248758353.01 & 33.5899786 & $4.19 \times 10^{-11}$ & $2.49 \times 10^{-09}$ & $1.03 \times 10^{-14}$ & $1.05 \times 10^{-05}$ & $2.13 \times 10^{-11}$ & $0.00 \times 10^{+00}$ & 1.000 & $1.05 \times 10^{-05}$ $^{\dagger}$ & 0.6589 \\
248861279.01 & 13.1153647 & $3.91 \times 10^{-05}$ & $2.88 \times 10^{-63}$ & $5.71 \times 10^{-04}$ & $2.52 \times 10^{-04}$ & $0.00 \times 10^{+00}$ & $0.00 \times 10^{+00}$ & 0.999 & $8.62 \times 10^{-04}$ $^{\dagger}$ & 0.4292 \\
248874928.01 & 3.43545277 & $1.06 \times 10^{-07}$ & $4.42 \times 10^{-06}$ & $1.47 \times 10^{-14}$ & $1.24 \times 10^{-10}$ & $0.00 \times 10^{+00}$ & $0.00 \times 10^{+00}$ & 1.000 & $4.53 \times 10^{-06}$ $^{\dagger}$ & 0.5057 \\
249223471.01 & 22.5494062 & $1.01 \times 10^{-06}$ & $8.33 \times 10^{-07}$ & $1.53 \times 10^{-04}$ & $1.14 \times 10^{-03}$ & $6.18 \times 10^{-08}$ & $3.69 \times 10^{-06}$ & 0.999 & $1.30 \times 10^{-03}$ $^{\dagger}$ & 0.6741 \\
249403651.01 & 4.94190653 & $2.91 \times 10^{-04}$ & $6.29 \times 10^{-06}$ & $6.75 \times 10^{-04}$ & $3.79 \times 10^{-05}$ & $4.11 \times 10^{-04}$ & $2.08 \times 10^{-08}$ & 0.999 & $3.23 \times 10^{-05}$ $^{\dagger\ast}$ & 0.0826 \\
249403651.02 & 9.22452977 & $1.52 \times 10^{-04}$ & $1.01 \times 10^{-05}$ & $2.35 \times 10^{-03}$ & $1.20 \times 10^{-03}$ & $3.99 \times 10^{-04}$ & $3.90 \times 10^{-04}$ & 0.995 & $1.03 \times 10^{-04}$ $^{\dagger\ast}$ & 0.0372 \\
249816490.01 & 20.9789593 & $1.79 \times 10^{-10}$ & $3.29 \times 10^{-14}$ & $3.89 \times 10^{-04}$ & $3.04 \times 10^{-07}$ & $1.30 \times 10^{-04}$ & $6.63 \times 10^{-45}$ & 0.999 & $5.19 \times 10^{-04}$ $^{\dagger}$ & 0.6675 \\
249924395.01 & 1.90808365 & $1.47 \times 10^{-05}$ & $3.56 \times 10^{-08}$ & $1.77 \times 10^{-05}$ & $1.89 \times 10^{-13}$ & $6.74 \times 10^{-06}$ & $0.00 \times 10^{+00}$ & 1.000 & $3.92 \times 10^{-05}$ $^{\dagger}$ & 0.4360 \\
211732116.01 & 4.52195338 & $6.31 \times 10^{-29}$ & $5.21 \times 10^{-09}$ & $9.19 \times 10^{-04}$ & $2.49 \times 10^{-05}$ & $0.00 \times 10^{+00}$ & $2.03 \times 10^{-87}$ & 0.999 & $6.80 \times 10^{-05}$ $^{\dagger\ast}$ & 0.5403 \\
211732116.02 & 16.4344499 & $7.11 \times 10^{-06}$ & $1.1 \times 10^{-05}$ & $1.52 \times 10^{-02}$ & $1.74 \times 10^{-02}$ & $1.04 \times 10^{-07}$ & $3.68 \times 10^{-07}$ & 0.967 & $2.42 \times 10^{-03}$ $^{\dagger\ast}$ & 0.5749 \\
212222383.01 & 5.77647524 & $7.28 \times 10^{-05}$ & $7.41 \times 10^{-08}$ & $1.55 \times 10^{-03}$ & $7.53 \times 10^{-05}$ & $9.60 \times 10^{-29}$ & $2.30 \times 10^{-20}$ & 0.998 & $1.70 \times 10^{-03}$ $^{\dagger}$ & 0.6183 \\
\hline 
\emph{Not yet validated} & & & & & & & & & & \\
205944181.01 & 2.47558277 & $1.47 \times 10^{-04}$ & $4.03 \times 10^{-04}$ & $2.63 \times 10^{-01}$ & $4.13 \times 10^{-02}$ & $0.00 \times 10^{+00}$ & $0.00 \times 10^{+00}$ & 0.695 & $3.05 \times 10^{-01}$ $^{\dagger}$ & 0.3263 \\
205957328.01 & 14.3523671 & $2.83 \times 10^{-04}$ & $1.3 \times 10^{-04}$ & $1.13 \times 10^{-02}$ & $7.23 \times 10^{-03}$ & $9.53 \times 10^{-04}$ & $2.02 \times 10^{-03}$ & 0.978 & $2.2 \times 10^{-02}$ & 0.6039 \\
205999468.01 & 12.2634336 & $2.58 \times 10^{-03}$ & $1.6 \times 10^{-03}$ & $3.67 \times 10^{-02}$ & $3.79 \times 10^{-02}$ & $4.25 \times 10^{-08}$ & $2.05 \times 10^{-05}$ & 0.921 & $7.9 \times 10^{-02}$ $^{\dagger}$ & 0.6805 \\
211428897.01 & 1.61088909 & $0.0 \times 10^{+00}$ & $2.55 \times 10^{-04}$ & $9.24 \times 10^{-03}$ & $1.87 \times 10^{-02}$ & $0.00 \times 10^{+00}$ & $0.00 \times 10^{+00}$ & 0.972 & $2.8 \times 10^{-02}$ & 0.5961 \\
211428897.02 & 2.17805786 & - & - & - & - & - & - & - & error & 0.3319 \\
211428897.03 & 4.96824247 & $0.0 \times 10^{+00}$ & $1.14 \times 10^{-89}$ & $2.27 \times 10^{-03}$ & $3.54 \times 10^{-02}$ & $0.00 \times 10^{+00}$ & $0.00 \times 10^{+00}$ & 0.962 & $3.8 \times 10^{-02}$ & 0.2088 \\
211428897.04 & 6.26462338 & $3.08 \times 10^{-238}$ & $6.04 \times 10^{-03}$ & $2.71 \times 10^{-01}$ & $1.52 \times 10^{-01}$ & $0.00 \times 10^{+00}$ & $0.00 \times 10^{+00}$ & 0.571 & $ 4.29 \times 10^{-01}$ & 0.3817 \\
211711685.01 & 15.462807 & $1.13 \times 10^{-16}$ & $2.82 \times 10^{-07}$ & $8.47 \times 10^{-06}$ & $1.29 \times 10^{-04}$ & $2.89 \times 10^{-28}$ & $4.46 \times 10^{-08}$ & 1.000 & $1.38 \times 10^{-04}$ $^{\dagger}$ & 0.0088 \\ 
211965883.01 & 10.5551697 & $4.34 \times 10^{-05}$ & $1.02 \times 10^{-03}$ & $2.2 \times 10^{-02}$ & $1.26 \times 10^{-02}$ & $7.07 \times 10^{-04}$ & $4.98 \times 10^{-06}$ & 0.964 & $3.6 \times 10^{-02}$ $^{\dagger}$ & 0.4895 \\
212351868.01 & 2.50028488 & $6.01 \times 10^{-02}$ & $7.09 \times 10^{-06}$ & $7.90 \times 10^{-01}$ & $4.47 \times 10^{-16}$ & $1.45 \times 10^{-31}$ & $0.00 \times 10^{+00}$ & 0.149 & $8.51 \times 10^{-01}$ & 0.3189 \\
212585579.01 & 3.0219382 & $2.58 \times 10^{-02}$ & $4.17 \times 10^{-03}$ & $6.09 \times 10^{-01}$ & $6.51 \times 10^{-02}$ & $3.28 \times 10^{-04}$ & $8.33 \times 10^{-08}$ & 0.296 & $ 7.04 \times 10^{-01}$ $^{\dagger}$ & 0.1374 \\
212730483.01 & 3.49096697 & $1.05 \times 10^{-27}$ & $2.29 \times 10^{-03}$ & $4.31 \times 10^{-02}$ & $7.86 \times 10^{-02}$ & $7.36 \times 10^{-04}$ & $3.21 \times 10^{-03}$ & 0.872 & $1.28 \times 10^{-01}$ $^{\dagger}$ & 0.6583 \\
212797028.01 & 29.9835193 & $5.14 \times 10^{-02}$ & $4.03 \times 10^{-02}$ & $3.93 \times 10^{-01}$ & $4.06 \times 10^{-01}$ & $2.15 \times 10^{-09}$ & $7.17 \times 10^{-13}$ & 0.110 & $8.9 \times 10^{-01} $ & 0.4869 \\
213817056.01 & 13.6118493 & $2.90 \times 10^{-99}$ & $5.38 \times 10^{-39}$ & $1.94 \times 10^{-79}$ & $2.80 \times 10^{-44}$ & $4.99 \times 10^{-02}$ & $4.13 \times 10^{-06}$ & 0.950 & $5.00\times 10^{-02}$ & 0.8036 \\
216892056.01 & 2.78585897 & $7.67 \times 10^{-04}$ & $6.54 \times 10^{-03}$ & $1.58 \times 10^{-02}$ & $6.22 \times 10^{-02}$ & $0.00 \times 10^{+00}$ & $0.00 \times 10^{+00}$ & 0.915 & $ $ $8.5 \times 10^{-02}$ $^{\dagger}$ & 0.2193 \\
217192839.04 & 40.9341414 & $4.2 \times 10^{-03}$ & $6.71 \times 10^{-04}$ & $1.08 \times 10^{-01}$ & $1.48 \times 10^{-02}$ & $3.49 \times 10^{-162}$ & $9.49 \times 10^{-53}$ & 0.872 & $ $ $2.6 \times 10^{-02}$ $^{\dagger\ast}$ & 0.6511 \\
220294712.01 & 23.6110924 & $7.03 \times 10^{-02}$ & $2.03 \times 10^{-04}$ & $1.84 \times 10^{-01}$ & $3.93 \times 10^{-02}$ & $3.68 \times 10^{-01}$ & $3.38 \times 10^{-01}$ & $2.89 \times 10^{-48}$ & $1.00$ & 0.7637 \\
220400100.01 & 10.7947008 & $1.90 \times 10^{-05}$ & $1.46 \times 10^{-08}$ & $3.58 \times 10^{-02}$ & $5.17 \times 10^{-04}$ & $2.26 \times 10^{-05}$ & $6.96 \times 10^{-109}$ & 0.964 & $3.6 \times 10^{-02}$ $^{\dagger}$ & 0.4229 \\
220571481.01 & 8.78648628 & $1.06 \times 10^{-02}$ & $8.95 \times 10^{-03}$ & $2.53 \times 10^{-01}$ & $1.27 \times 10^{-01}$ & $5.67 \times 10^{-04}$ & $3.21 \times 10^{-04}$ & 0.599 & $ $ $4.01 \times 10^{-01}$ $^{\dagger}$ & 0.4585 \\
226042826.01 & 3.12897689 & $2.79 \times 10^{-05}$ & $9.32 \times 10^{-05}$ & $5.0 \times 10^{-04}$ & $8.90 \times 10^{-04}$ & $4.53 \times 10^{-01}$ & $5.46 \times 10^{-01}$ & $1.58 \times 10^{-11}$ & $1.00$ & $-$ $\diamond$ \\
247164043.01 & 5.22835211 & $2.28 \times 10^{-07}$ & $1.61 \times 10^{-08}$ & $2.56 \times 10^{-05}$ & $5.56 \times 10^{-12}$ & $0.00 \times 10^{+00}$ & $0.00 \times 10^{+00}$ & 1.000 & $2.59 \times 10^{-05}$ $^{\dagger}$ & 0.0452 \\
247698108.01 & 20.3794342 & $3.98 \times 10^{-03}$ & $3.27 \times 10^{-04}$ & $2.93 \times 10^{-02}$ & $5.85 \times 10^{-03}$ & $4.57 \times 10^{-01}$ & $5.03 \times 10^{-01}$ & $3.12 \times 10^{-04}$ & $1.00$ & 0.1459 \\
247724061.01 & 1.248905 & $1.46 \times 10^{-02}$ & $1.90 \times 10^{-02}$ & $9.41 \times 10^{-02}$ & $3.22 \times 10^{-01}$ & $1.48 \times 10^{-02}$ & $7.94 \times 10^{-03}$ & 0.528 & $4.72 \times 10^{-01}$ $^{\dagger}$ & 0.0795 \\
248222323.01 & 9.42690097 & $3.88 \times 10^{-02}$ & $2.81 \times 10^{-03}$ & $2.38 \times 10^{-01}$ & $1.25 \times 10^{-02}$ & $6.71 \times 10^{-02}$ & $4.71 \times 10^{-02}$ & 0.594 & $4.06 \times 10^{-01}$ & 0.0134 \\
248639411.01 & 3.57739438 & $1.05 \times 10^{-04}$ & $3.35 \times 10^{-03}$ & $4.64 \times 10^{-08}$ & $7.99 \times 10^{-01}$ & $0.00 \times 10^{+00}$ & $0.00 \times 10^{+00}$ & 0.198 & $ $ $8.02 \times 10^{-01}$ $^{\dagger}$ & 0.2883 \\
248740016.01 & 14.3907212 & $2.16 \times 10^{-02}$ & $1.10 \times 10^{-02}$ & $3.53 \times 10^{-01}$ & $2.53 \times 10^{-01}$ & $6.62 \times 10^{-04}$ & $3.28 \times 10^{-04}$ & 0.360 & $6.40 \times 10^{-01}$ $^{\dagger}$ & 0.0892 \\
248827616.01 & 2.57126699 & $2.03 \times 10^{-03}$ & $2.08 \times 10^{-03}$ & $4.78 \times 10^{-02}$ & $9.36 \times 10^{-02}$ & $0.00 \times 10^{+00}$ & $0.00 \times 10^{+00}$ & 0.855 & $1.45 \times 10^{-01}$ $^{\dagger}$ & 0.4882 \\
249827330.01 & 31.3960013 & $1.03 \times 10^{-02}$ & $5.39 \times 10^{-05}$ & $1.10 \times 10^{-01}$ & $2.89 \times 10^{-04}$ & $4.37 \times 10^{-03}$ & $4.35 \times 10^{-103}$ & 0.875 & $1.25 \times 10^{-01}$ $^{\dagger}$ & 0.1697 \\
249865296.01 & 28.8885889 & $9.35 \times 10^{-09}$ & $1.39 \times 10^{-11}$ & $1.52 \times 10^{-22}$ & $2.80 \times 10^{-18}$ & $3.15 \times 10^{-84}$ & $5.95 \times 10^{-230}$ & 1.00 & $9.37 \times 10^{-09}$ & 0.0418 \\
212705192.01 & 2.26837914 & $1.43 \times 10^{-03}$ & $2.5 \times 10^{-04}$ & $9.52 \times 10^{-02}$ & $1.85 \times 10^{-02}$ & $2.83 \times 10^{-04}$ & $0.00 \times 10^{+00}$ & 0.884 & $ 1.16 \times 10^{-01} $ & 0.3976 \\
211784767.01 & 3.58048139 & $1.21 \times 10^{-01}$ & $8.3 \times 10^{-03}$ & $5.31 \times 10^{-01}$ & $6.08 \times 10^{-02}$ & $4.99 \times 10^{-02}$ & $2.21 \times 10^{-01}$ & 0.00895 & $9.91 \times 10^{-01}$ & 0.4486 \\
\enddata
\tablenotetext{\dagger}{Contrast curve/s used.}
\tablenotetext{\ast}{Multiplicity boost used.}
\tablenotetext{\diamond}{Too crowded for valid \texttt{centroid} test.}
\end{deluxetable*}
\end{longrotatetable}

\begin{deluxetable*}{cc}
\tablecaption{Campaign-specific multiplicity boost values. Campaigns 11 and 17 have no multi-planet systems, either in Paper IV or previously published planets and candidates, with which to calculate a multiplicity boost.\label{tab:multiplicity}}
\tablehead{\colhead{\hspace{0.7cm}Campaign} & \colhead{\hspace{1cm}$X2$}}
\startdata
\hspace{0.5cm}1	&	\hspace{1cm}25.9251\\
\hspace{0.5cm}2	&	\hspace{1cm}8.1286\\
\hspace{0.5cm}3	&	\hspace{1cm}15.9445\\
\hspace{0.5cm}4	&	\hspace{1cm}13.1029\\
\hspace{0.5cm}5	&	\hspace{1cm}9.0566\\
\hspace{0.5cm}6	&	\hspace{1cm}11.8727\\
\hspace{0.5cm}7	&	\hspace{1cm}5.5419\\
\hspace{0.5cm}8	&	\hspace{1cm}13.6017\\
\hspace{0.5cm}10	&	\hspace{1cm}35.6843\\
\hspace{0.5cm}11  &   \hspace{1cm} - \\
\hspace{0.5cm}12	&	\hspace{1cm}70.3960\\
\hspace{0.5cm}13	&	\hspace{1cm}21.1136\\
\hspace{0.5cm}14	&	\hspace{1cm}67.4368\\
\hspace{0.5cm}15	&	\hspace{1cm}44.0871\\
\hspace{0.5cm}16	&	\hspace{1cm}13.9069\\
\hspace{0.5cm}17	&	\hspace{1cm}- \\ 
\hspace{0.5cm}18	&  \hspace{1cm}50.3128\\
\enddata
\end{deluxetable*}

\begin{longrotatetable}
\begin{deluxetable*}{lllllllll}
\tablewidth{700pt}
\tablecaption{Derived planet parameters of the newly validated planetary system. Camp: the \emph{K2} campaign/s in which our pipeline detected the signal.}
\tabletypesize{\scriptsize}
\tablehead{
\colhead{Planet} &
\colhead{Candidate ID} & 
\colhead{Camp} &
\colhead{Period} &
\colhead{$R_p/R_s$} &
\colhead{$R_p$} &
\colhead{$T_{\rm mid}$} &
\colhead{$b$} & 
\colhead{$a/R_s$} \\ 
\colhead{} & 
\colhead{} & 
\colhead{} & 
\colhead{[d]} & 
\colhead{} & 
\colhead{[$R_{\oplus}$]} & 
\colhead{[2454833-JD]} & 
\colhead{} &
\colhead{} 
}
\startdata
\epiczerooneonesixbalias & 204750116.01 & $2$ & $23.448172^{+0.000766}_{-0.000996}$ & $0.0253^{+0.001}_{-0.0}$ & $2.906^{+0.195}_{-0.176}$ & $2065.8326^{+0.0019}_{-0.0013}$ & $0.328^{+0.197}_{-0.215}$ & $30.65^{+1.48}_{-2.96}$ \\
\epicsixsixfourbalias & 205111664.01 & $2$ & $15.937039^{+0.000796}_{-0.001144}$ & $0.0228^{+0.001}_{-0.001}$ & $2.359^{+0.187}_{-0.167}$ & $2073.4351^{+0.0028}_{-0.0032}$ & $0.718^{+0.073}_{-0.093}$ & $27.63^{+3.08}_{-3.23}$ \\
\epicnineeightonebalias & 206055981.01 & $3$ & $20.645099^{+0.000995}_{-0.001375}$ & $0.0353^{+0.002}_{-0.002}$ & $2.413^{+0.225}_{-0.196}$ & $2164.3941^{+0.0020}_{-0.0014}$ & $0.83^{+0.052}_{-0.063}$ & $45.05^{+6.05}_{-6.59}$ \\
\epicsixeighttwobalias & 206135682.01 & $3$ & $5.025598^{+0.000287}_{-0.000342}$ & $0.0184^{+0.001}_{-0.001}$ & $1.332^{+0.107}_{-0.1}$ & $2145.7672^{+0.0025}_{-0.0027}$ & $0.272^{+0.209}_{-0.188}$ & $16.2^{+0.88}_{-1.3}$ \\
\epicsixeighttwocalias & 206135682.02 & $3$ & $9.660186^{+0.001164}_{-0.001169}$ & $0.0188^{+0.001}_{-0.001}$ & $1.361^{+0.114}_{-0.106}$ & $2144.8722^{+0.0037}_{-0.0052}$ & $0.306^{+0.219}_{-0.207}$ & $25.98^{+1.87}_{-2.35}$ \\
\epicsixeighttwodalias & 206135682.03 & $3$ & $20.201085^{+0.002354}_{-0.002736}$ & $0.0269^{+0.001}_{-0.001}$ & $1.948^{+0.152}_{-0.143}$ & $2155.5374^{+0.0026}_{-0.0018}$ & $0.287^{+0.2}_{-0.193}$ & $42.92^{+1.77}_{-3.54}$ \\
\epicninefivesevenbalias & 206146957.01 & $3$ & $5.761612^{+0.000846}_{-0.000714}$ & $0.0137^{+0.001}_{-0.001}$ & $1.313^{+0.101}_{-0.095}$ & $2146.8716^{+0.0052}_{-0.0053}$ & $0.727^{+0.076}_{-0.102}$ & $14.31^{+1.66}_{-1.62}$ \\
\epictwoeightsixcalias & 206317286.01 & $3$ & $17.515472^{+0.002288}_{-0.002965}$ & $0.0253^{+0.001}_{-0.001}$ & $2.094^{+0.176}_{-0.16}$ & $2144.8215^{+0.0071}_{-0.0031}$ & $0.268^{+0.205}_{-0.183}$ & $34.54^{+1.69}_{-2.6}$ \\
\epicfiveeightzerobalias & 210797580.01 & $13$ & $2.14084^{+3.6e-05}_{-3.8e-05}$ & $0.0302^{+0.001}_{-0.005}$ & $3.207^{+0.224}_{-0.548}$ & $2989.1840^{+0.0008}_{-0.0008}$ & $0.941^{+0.008}_{-0.216}$ & $3.66^{+3.57}_{-0.23}$ \\
\epicthreefiveninebalias & 211399359.01 & $5$,$18$ & $3.114905^{+9e-06}_{-9e-06}$ & $0.1619^{+0.0}_{-0.0}$ & $13.024^{+0.921}_{-0.868}$ & $2308.4176^{+0.0001}_{-0.0001}$ & $0.666^{+0.006}_{-0.006}$ & $8.64^{+0.06}_{-0.05}$ \\
\epicninenineninebalias & 211490999.01 & $5$,$16$,$18$ & $9.844077^{+0.000565}_{-0.000575}$ & $0.0292^{+0.001}_{-0.0}$ & $2.945^{+0.077}_{-0.075}$ & $2313.3287^{+0.0019}_{-0.0015}$ & $0.337^{+0.103}_{-0.152}$ & $20.01^{+0.77}_{-0.82}$ \\
\epiczerofivefourbalias & 211539054.01 & $5$ & $11.02066^{+0.001431}_{-0.001368}$ & $0.0127^{+0.001}_{-0.001}$ & $2.061^{+0.147}_{-0.14}$ & $2317.2894^{+0.0042}_{-0.0043}$ & $0.666^{+0.086}_{-0.123}$ & $15.29^{+1.8}_{-1.69}$ \\
\epictwooneonesixbalias & 211732116.01 & $16$ & $4.521953^{+0.00032}_{-0.000409}$ & $0.0161^{+0.001}_{-0.001}$ & $1.593^{+0.061}_{-0.058}$ & $3266.6869^{+0.0040}_{-0.0032}$ & $0.235^{+0.139}_{-0.153}$ & $11.44^{+0.33}_{-0.33}$ \\
\epictwooneonesixcalias & 211732116.02 & $16$ & $16.43445^{+0.002378}_{-0.002702}$ & $0.0223^{+0.001}_{-0.001}$ & $2.213^{+0.091}_{-0.086}$ & $3276.2376^{+0.0065}_{-0.0056}$ & $0.649^{+0.05}_{-0.054}$ & $27.11^{+0.88}_{-0.86}$ \\
\epicthreeoneeightbalias & 212006318.01 & $5$ & $14.453895^{+0.004773}_{-0.004175}$ & $0.0133^{+0.001}_{-0.001}$ & $2.243^{+0.186}_{-0.174}$ & $2314.3436^{+0.0105}_{-0.0111}$ & $0.284^{+0.214}_{-0.19}$ & $16.9^{+1.07}_{-1.37}$ \\
\epicthreeeightthreebalias & 212222383.01 & $16$ & $5.776475^{+0.000261}_{-0.000237}$ & $0.0138^{+0.001}_{-0.001}$ & $1.736^{+0.125}_{-0.119}$ & $3263.7426^{+0.0025}_{-0.0022}$ & $0.854^{+0.039}_{-0.048}$ & $11.58^{+1.48}_{-1.39}$ \\
\epiconeoneeightbalias & 212530118.01 & $6$ & $12.832235^{+0.001554}_{-0.001616}$ & $0.0224^{+0.001}_{-0.001}$ & $1.693^{+0.146}_{-0.132}$ & $2397.1429^{+0.0052}_{-0.0050}$ & $0.262^{+0.209}_{-0.181}$ & $30.04^{+1.77}_{-2.32}$ \\
\epiceighttwoeightbalias & 212575828.01 & $6$ & $2.060438^{+0.000118}_{-7.4e-05}$ & $0.036^{+0.002}_{-0.001}$ & $2.88^{+0.27}_{-0.23}$ & $2384.8493^{+0.0018}_{-0.0034}$ & $0.402^{+0.208}_{-0.247}$ & $9.52^{+0.81}_{-1.17}$ \\
\epiczerosixninebalias & 214173069.01 & $7$ & $8.777247^{+0.000625}_{-0.000611}$ & $0.0266^{+0.004}_{-0.001}$ & $2.181^{+0.363}_{-0.198}$ & $2470.9091^{+0.0039}_{-0.0027}$ & $0.481^{+0.334}_{-0.293}$ & $34.56^{+2.71}_{-7.32}$ \\
\epicfivefourfivebalias & 214419545.01 & $7$ & $9.401312^{+0.000469}_{-0.000506}$ & $0.0189^{+0.001}_{-0.001}$ & $2.729^{+0.186}_{-0.176}$ & $2469.9354^{+0.0020}_{-0.0017}$ & $0.883^{+0.024}_{-0.03}$ & $15.19^{+1.58}_{-1.47}$ \\
\epiceightthreeninebalias & 217192839.02 & $7$ & $7.9389334^{+0.000596}_{-0.000592}$ & $0.0146^{+0.001}_{-0.002}$ & $1.075^{+0.142}_{-0.104}$ & $2471.8440^{+0.0033}_{-0.0063}$ & $0.505^{+0.311}_{-0.301}$ & $31.05^{+4.25}_{-7.49}$ \\
\epiceightthreeninecalias & 217192839.01 & $7$ & $16.034656^{+0.001844}_{-0.001572}$ & $0.0271^{+0.001}_{-0.001}$ & $2.0^{+0.184}_{-0.148}$ & $2471.2916^{+0.0030}_{-0.0027}$ & $0.338^{+0.249}_{-0.221}$ & $39.39^{+2.21}_{-5.33}$ \\
\epiceightthreeninedalias & 217192839.03 & $7$ & $26.803023^{+0.002775}_{-0.003048}$ & $0.0229^{+0.002}_{-0.001}$ & $1.689^{+0.18}_{-0.139}$ & $2474.9152^{+0.0038}_{-0.0043}$ & $0.497^{+0.225}_{-0.286}$ & $59.43^{+7.25}_{-11.72}$ \\
\epiceightninefivebalias & 217977895.01 & $7$ & $21.700156^{+0.002021}_{-0.001915}$ & $0.0234^{+0.001}_{-0.001}$ & $2.072^{+0.152}_{-0.133}$ & $2481.3647^{+0.0045}_{-0.0036}$ & $0.346^{+0.197}_{-0.213}$ & $40.81^{+2.57}_{-4.03}$ \\
\epicsixzerotwobalias & 218668602.01 & $7$ & $1.865962^{+8.5e-05}_{-6.8e-05}$ & $0.0178^{+0.001}_{-0.0}$ & $1.564^{+0.118}_{-0.106}$ & $2469.0928^{+0.0016}_{-0.0023}$ & $0.268^{+0.211}_{-0.183}$ & $7.77^{+0.35}_{-0.64}$ \\
\epictwoseventwobalias & 220221272.05 & $8$ & $2.231527^{+0.000194}_{-0.000325}$ & $0.0284^{+0.003}_{-0.012}$ & $1.076^{+0.232}_{-0.06}$ & $2560.7157^{+0.0045}_{-0.0075}$ & $0.517^{+0.296}_{-0.295}$ & $16.59^{+33.69}_{-3.33}$ \\
\epictwoseventwocalias & 220221272.03 & $8$ & $4.194766^{+0.000309}_{-0.000189}$ & $0.0314^{+0.003}_{-0.002}$ & $1.191^{+0.053}_{-0.075}$ & $2561.4158^{+0.0023}_{-0.0035}$ & $0.446^{+0.345}_{-0.273}$ & $24.63^{+3.74}_{-5.04}$ \\
\epictwoseventwodalias & 220221272.02 & $8$ & $6.679582^{+0.000719}_{-0.000498}$ & $0.0367^{+0.003}_{-0.002}$ & $1.392^{+0.118}_{-0.075}$ & $2559.4243^{+0.0042}_{-0.0039}$ & $0.367^{+0.385}_{-0.265}$ & $27.93^{+2.83}_{-5.65}$ \\
\epictwoseventwoealias & 220221272.04 & $8$ & $9.715043^{+0.001003}_{-0.000739}$ & $0.0354^{+0.004}_{-0.002}$ & $1.345^{+0.139}_{-0.082}$ & $2565.9666^{+0.0026}_{-0.0049}$ & $0.471^{+0.33}_{-0.325}$ & $38.39^{+5.44}_{-11.22}$ \\
\epictwoseventwofalias & 220221272.01 & $8$ & $13.62749^{+0.000342}_{-0.000574}$ & $0.0585^{+0.002}_{-0.001}$ & $2.222^{+0.091}_{-0.082}$ & $2559.6002^{+0.0009}_{-0.0012}$ & $0.467^{+0.079}_{-0.072}$ & $47.7^{+1.46}_{-1.42}$ \\
\epicfoursevensevenbalias & 220459477.01 & $8$ & $2.380867^{+0.000161}_{-0.000216}$ & $0.0185^{+0.001}_{-0.001}$ & $1.539^{+0.145}_{-0.126}$ & $2560.4544^{+0.0049}_{-0.0038}$ & $0.361^{+0.213}_{-0.236}$ & $10.07^{+0.72}_{-1.05}$ \\
\epiceightsevenfourbalias & 220510874.01 & $8$ & $7.473223^{+0.000374}_{-0.000312}$ & $0.0219^{+0.0}_{-0.0}$ & $2.32^{+0.071}_{-0.067}$ & $2562.9056^{+0.0024}_{-0.0025}$ & $0.246^{+0.124}_{-0.154}$ & $16.23^{+0.46}_{-0.58}$ \\
\epictwothreethreebalias & 220696233.01 & $8$ & $28.735327^{+0.000992}_{-0.000994}$ & $0.1146^{+0.002}_{-0.002}$ & $7.322^{+0.268}_{-0.264}$ & $2568.9787^{+0.0011}_{-0.0011}$ & $0.711^{+0.03}_{-0.034}$ & $55.42^{+2.09}_{-2.1}$ \\
\epicfourfivefivebalias & 245943455.01 & $12$ & $6.339069^{+0.000225}_{-0.000186}$ & $0.0374^{+0.002}_{-0.001}$ & $3.66^{+0.33}_{-0.229}$ & $2908.7944^{+0.0017}_{-0.0019}$ & $0.388^{+0.301}_{-0.248}$ & $14.71^{+1.04}_{-3.11}$ \\
\epiczerofoureightbalias & 245991048.01 & $12$ & $8.584043^{+0.000888}_{-0.002026}$ & $0.0185^{+0.001}_{-0.0}$ & $2.184^{+0.118}_{-0.114}$ & $2909.7820^{+0.0007}_{-0.0004}$ & $0.32^{+0.25}_{-0.204}$ & $20.68^{+1.38}_{-2.14}$ \\
\epiczerofoureightcalias & 245991048.02 & $12$ & $20.851337^{+0.002804}_{-0.002981}$ & $0.0172^{+0.001}_{-0.001}$ & $2.04^{+0.151}_{-0.138}$ & $2914.1012^{+0.0050}_{-0.0055}$ & $0.399^{+0.178}_{-0.243}$ & $28.9^{+2.04}_{-3.02}$ \\
\epicninesevensevenbalias & 245995977.01 & $12$ & $3.312792^{+0.0001}_{-0.000104}$ & $0.0442^{+0.001}_{-0.002}$ & $4.662^{+0.313}_{-0.343}$ & $2906.9029^{+0.0013}_{-0.0012}$ & $0.893^{+0.018}_{-0.045}$ & $5.51^{+0.93}_{-0.44}$ \\
\epicthreeonefourbalias & 246074314.01 & $12$ & $4.622654^{+0.000223}_{-0.000252}$ & $0.0222^{+0.001}_{-0.001}$ & $1.371^{+0.284}_{-0.188}$ & $2908.5165^{+0.0017}_{-0.0017}$ & $0.434^{+0.285}_{-0.297}$ & $11.92^{+1.2}_{-2.64}$ \\
\epicthreenineeightbalias & 246084398.01 & $12$ & $15.398723^{+0.002422}_{-0.00193}$ & $0.016^{+0.001}_{-0.001}$ & $1.85^{+0.161}_{-0.154}$ & $2906.6737^{+0.0037}_{-0.0091}$ & $0.711^{+0.078}_{-0.116}$ & $24.32^{+2.8}_{-2.86}$ \\
\epiczerofourninebalias & 246429049.01 & $12$ & $10.413181^{+0.000932}_{-0.000697}$ & $0.0201^{+0.001}_{-0.001}$ & $2.352^{+0.175}_{-0.167}$ & $2906.3926^{+0.0037}_{-0.0053}$ & $0.803^{+0.045}_{-0.061}$ & $18.38^{+2.1}_{-1.98}$ \\
\epiczerofourzerobalias & 246876040.01 & $13$ & $5.095717^{+0.000115}_{-0.000115}$ & $0.0245^{+0.001}_{-0.001}$ & $1.692^{+0.136}_{-0.126}$ & $2992.2859^{+0.0010}_{-0.0012}$ & $0.636^{+0.107}_{-0.171}$ & $15.96^{+2.29}_{-2.08}$ \\
\epiceightoneninebalias & 246891819.01 & $13$ & $4.803352^{+0.000357}_{-0.000466}$ & $0.0304^{+0.001}_{-0.001}$ & $2.597^{+0.192}_{-0.178}$ & $2992.0089^{+0.0042}_{-0.0030}$ & $0.25^{+0.206}_{-0.176}$ & $14.44^{+0.51}_{-1.08}$ \\
\epiceightoneninecalias & 246891819.02 & $13$ & $8.491282^{+0.000335}_{-0.000304}$ & $0.0397^{+0.002}_{-0.002}$ & $3.395^{+0.28}_{-0.257}$ & $2990.5786^{+0.0016}_{-0.0016}$ & $0.677^{+0.095}_{-0.119}$ & $22.43^{+2.81}_{-3.05}$ \\
\epicthreeninetwobalias & 246953392.01 & $13$ & $0.673862^{+1.6e-05}_{-1.9e-05}$ & $0.0157^{+0.0}_{-0.0}$ & $1.579^{+0.114}_{-0.103}$ & $2987.9636^{+0.0016}_{-0.0014}$ & $0.266^{+0.206}_{-0.177}$ & $3.29^{+0.14}_{-0.24}$ \\
\epicthreeninetwocalias & 246953392.02 & $13$ & $25.76051^{+0.001999}_{-0.001887}$ & $0.0311^{+0.003}_{-0.001}$ & $3.124^{+0.178}_{-0.112}$ & $2996.4180^{+0.0018}_{-0.0027}$ & $0.475^{+0.32}_{-0.301}$ & $39.31^{+4.12}_{-12.04}$ \\
\epiczerozerothreebalias & 247383003.01 & $13$ & $3.572326^{+0.00018}_{-0.000155}$ & $0.0229^{+0.001}_{-0.001}$ & $2.417^{+0.187}_{-0.183}$ & $2990.5751^{+0.0020}_{-0.0022}$ & $0.892^{+0.029}_{-0.036}$ & $10.11^{+1.3}_{-1.24}$ \\
\epicthreefivezerobalias  & 248463350.01 & $14$ & $6.393025^{+0.001139}_{-0.00075}$ & $0.0168^{+0.001}_{-0.001}$ & $2.274^{+0.113}_{-0.111}$ & $3076.0449^{+0.0042}_{-0.0048}$ & $0.523^{+0.073}_{-0.086}$ & $11.59^{+0.43}_{-0.46}$ \\
\epicthreefivezerocalias  & 248463350.02 & $14$ & $18.787839^{+0.000877}_{-0.000993}$ & $0.0377^{+0.001}_{-0.001}$ & $5.098^{+0.154}_{-0.146}$ & $3078.6845^{+0.0015}_{-0.0016}$ & $0.701^{+0.028}_{-0.031}$ & $23.55^{+0.92}_{-0.89}$ \\
\epiconefourzerobalias & 248472140.01 & $14$ & $0.759978^{+1e-05}_{-1e-05}$ & $0.0359^{+0.006}_{-0.003}$ & $6.049^{+1.047}_{-0.481}$ & $3073.2191^{+0.0006}_{-0.0006}$ & $0.974^{+0.012}_{-0.008}$ & $1.89^{+0.09}_{-0.09}$ \\
\epicthreezerosevenbalias & 248518307.01 & $14$ & $3.865053^{+0.000186}_{-0.000182}$ & $0.0279^{+0.001}_{-0.001}$ & $1.163^{+0.051}_{-0.049}$ & $3075.6234^{+0.0025}_{-0.0026}$ & $0.208^{+0.165}_{-0.14}$ & $19.12^{+0.79}_{-0.85}$ \\
\epicfiveonefourbalias & 248527514.01 & $14$ & $6.293099^{+0.000571}_{-0.000707}$ & $0.0306^{+0.002}_{-0.002}$ & $2.262^{+0.209}_{-0.192}$ & $3073.9531^{+0.0067}_{-0.0048}$ & $0.621^{+0.115}_{-0.223}$ & $20.97^{+3.0}_{-2.66}$ \\
\epicfiveninesevenbalias & 248621597.01 & $14$ & $17.27474^{+0.002593}_{-0.003175}$ & $0.0192^{+0.001}_{-0.001}$ & $2.678^{+0.213}_{-0.195}$ & $3087.4286^{+0.0036}_{-0.0044}$ & $0.468^{+0.165}_{-0.249}$ & $23.11^{+2.15}_{-2.66}$ \\
\epicthreefivethreebalias & 248758353.01 & $14$ & $33.589979^{+0.000768}_{-0.000768}$ & $0.0551^{+0.001}_{-0.002}$ & $5.333^{+0.357}_{-0.387}$ & $3075.9613^{+0.0014}_{-0.0011}$ & $0.723^{+0.05}_{-0.147}$ & $39.87^{+7.11}_{-3.12}$ \\
\epictwosevenninebalias & 248861279.01 & $14$ & $13.115365^{+0.000443}_{-0.000609}$ & $0.042^{+0.001}_{-0.001}$ & $2.53^{+0.084}_{-0.083}$ & $3084.8369^{+0.0010}_{-0.0008}$ & $0.481^{+0.05}_{-0.06}$ & $34.74^{+1.15}_{-1.08}$ \\
\epicninetwoeightbalias & 248874928.01 & $14$ & $3.435471^{+3.6e-05}_{-3.1e-05}$ & $0.0548^{+0.001}_{-0.004}$ & $4.569^{+0.085}_{-0.123}$ & $3075.9959^{+0.0004}_{-0.0004}$ & $0.871^{+0.02}_{-0.097}$ & $6.33^{+4.28}_{-0.57}$ \\
\epicfoursevenonebalias & 249223471.01 & $15$ & $22.549406^{+0.001377}_{-0.001433}$ & $0.0439^{+0.001}_{-0.001}$ & $4.602^{+0.311}_{-0.292}$ & $3177.3895^{+0.0025}_{-0.0023}$ & $0.544^{+0.121}_{-0.255}$ & $32.15^{+4.42}_{-3.45}$ \\
\epicsixfiveonebalias & 249403651.01 & $15$ & $4.941907^{+0.000661}_{-0.000696}$ & $0.0127^{+0.001}_{-0.001}$ & $1.283^{+0.097}_{-0.089}$ & $3160.5101^{+0.0075}_{-0.0071}$ & $0.309^{+0.209}_{-0.207}$ & $12.77^{+0.86}_{-1.1}$ \\
\epicsixfiveonecalias & 249403651.02 & $15$ & $9.22453^{+0.001322}_{-0.000911}$ & $0.0134^{+0.0}_{-0.0}$ & $1.36^{+0.096}_{-0.088}$ & $3164.6545^{+0.0040}_{-0.0043}$ & $0.259^{+0.207}_{-0.17}$ & $18.76^{+1.03}_{-1.44}$ \\
\epicfourninezerobalias & 249816490.01 & $15$ & $20.978959^{+0.001305}_{-0.001247}$ & $0.0196^{+0.001}_{-0.0}$ & $1.677^{+0.118}_{-0.109}$ & $3158.5363^{+0.0039}_{-0.0042}$ & $0.213^{+0.206}_{-0.154}$ & $36.24^{+1.02}_{-2.4}$ \\
\epicthreeninefivebalias & 249924395.01 & $15$ & $1.908084^{+4.7e-05}_{-5.8e-05}$ & $0.0185^{+0.001}_{-0.0}$ & $2.482^{+0.174}_{-0.158}$ & $3157.2841^{+0.0011}_{-0.0013}$ & $0.418^{+0.167}_{-0.236}$ & $5.32^{+0.41}_{-0.55}$ \\
\enddata
\end{deluxetable*}
\end{longrotatetable}


\section{Validated Planets}
\label{sec:planets}

\begin{figure}
\centering
\includegraphics[width=0.75\textwidth]{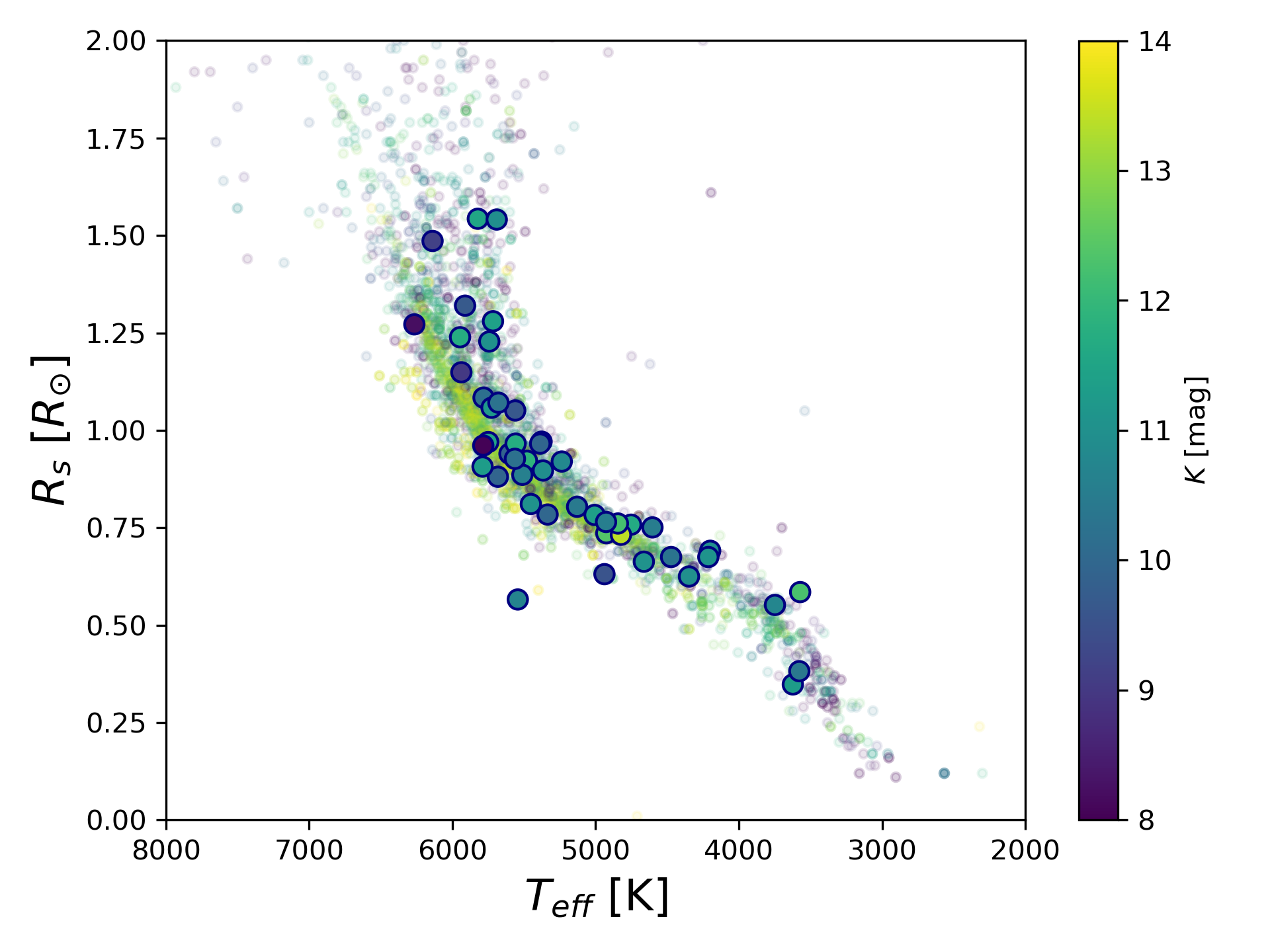}
\includegraphics[width=0.75\textwidth]{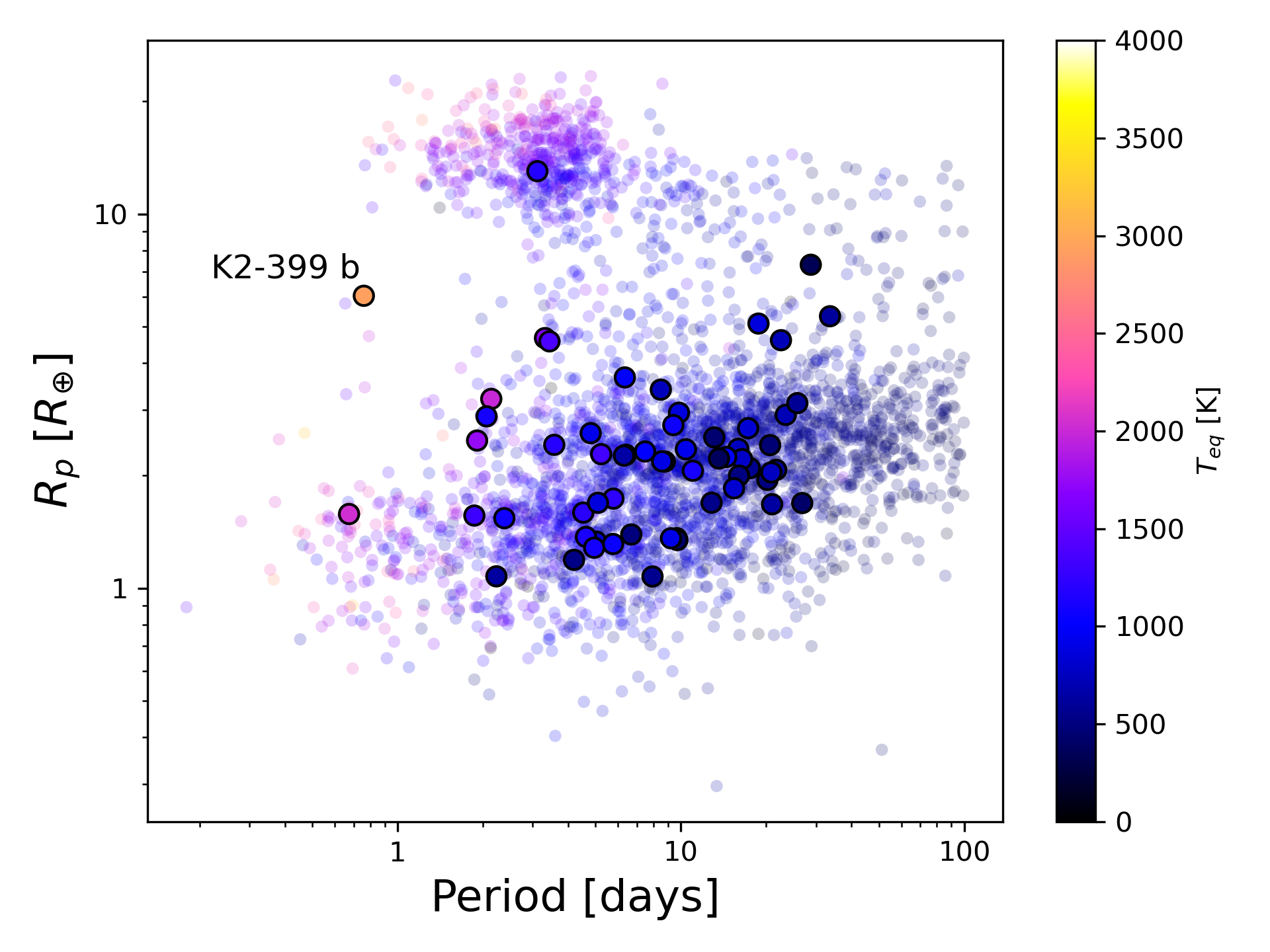}
\caption{Some properties of the systems containing newly validated planets. \emph{Top}: the new host stars (large circles with black edges) compared to the host stars of known planets at the NASA Exoplanet Archive. \emph{Bottom}: the new planets (large circles with black edges) compared to the planets with measured radii and orbital periods shorter than 100 days at the NASA Exoplanet Archive. \epiconefourzerobalias\ is highlighted, which is an ultra-hot Saturn-sized planet in a relatively unpopulated region of parameter space.}
\label{fig:hrdiagram}
\end{figure}

\begin{figure}
\centering
\includegraphics[width=0.325\textwidth]{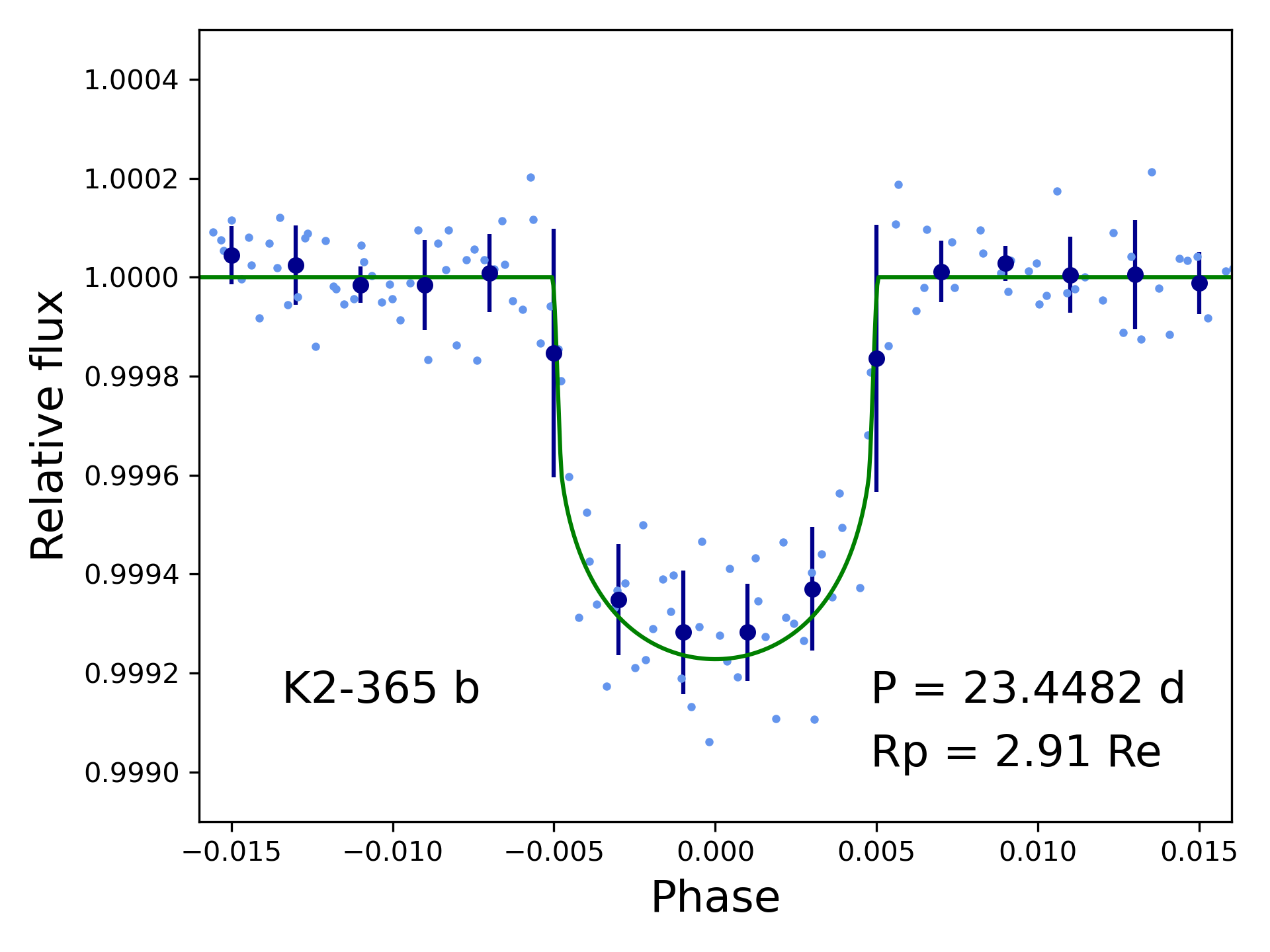}
\includegraphics[width=0.325\textwidth]{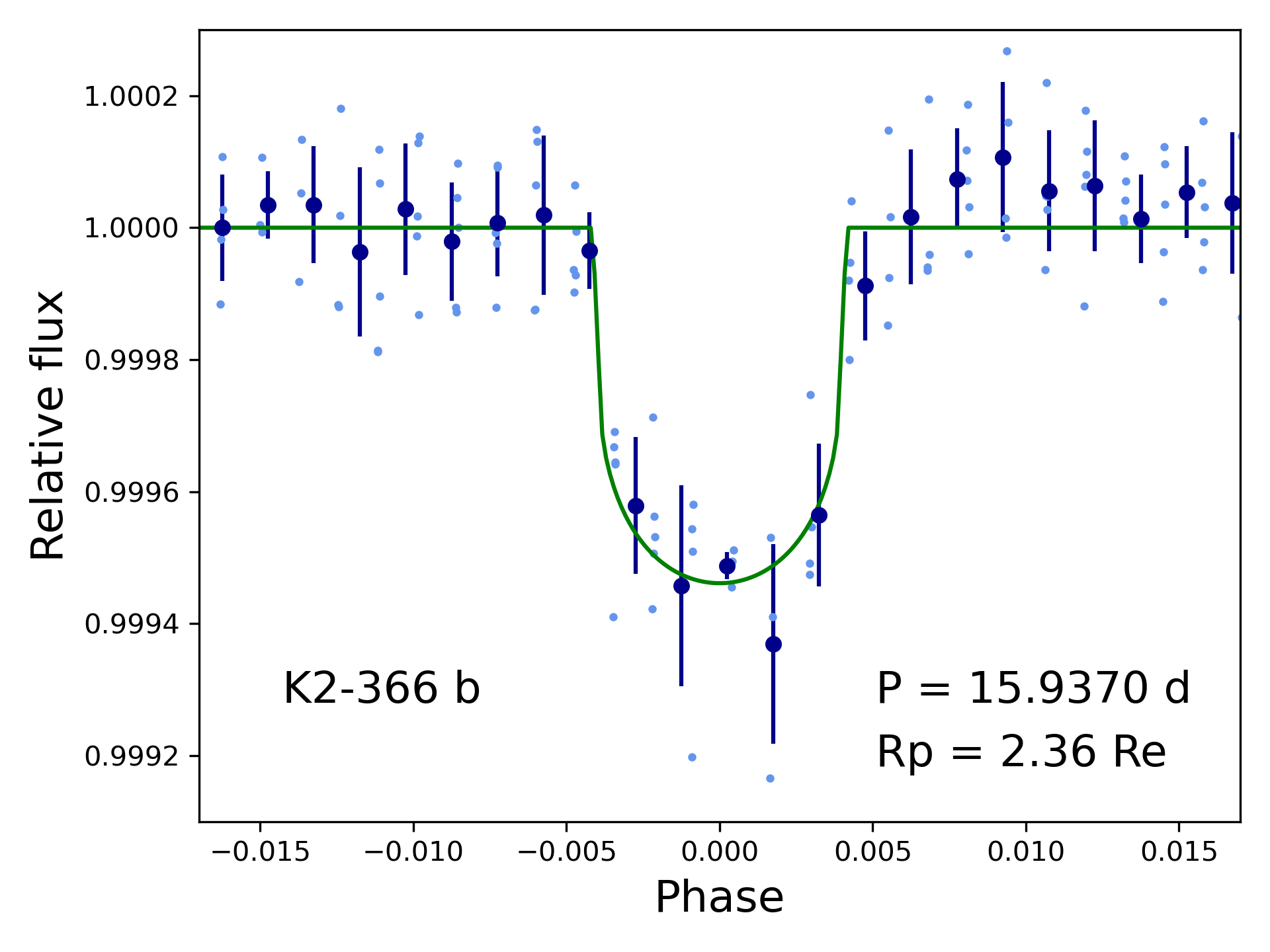}
\includegraphics[width=0.325\textwidth]{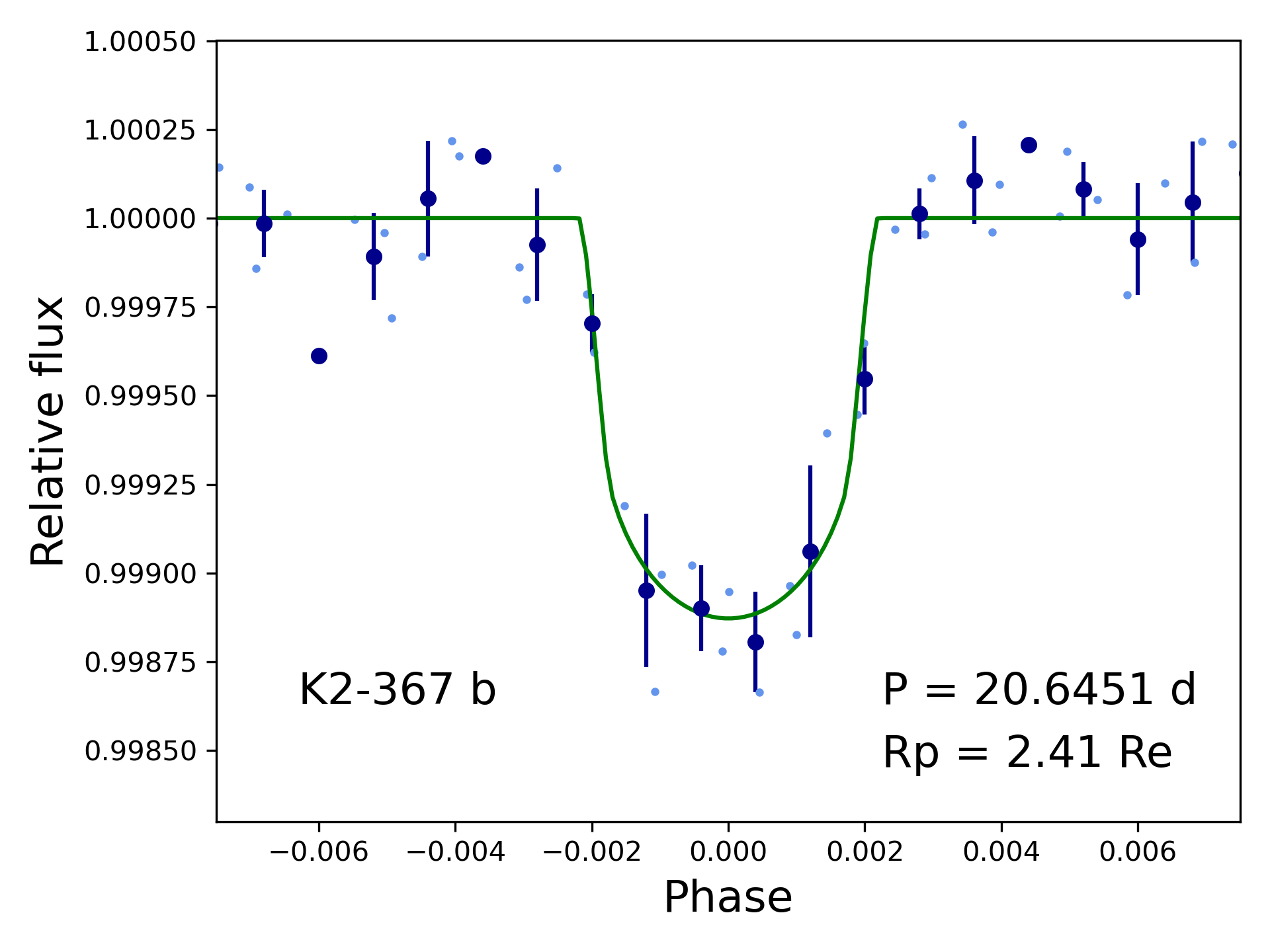}
\includegraphics[width=0.325\textwidth]{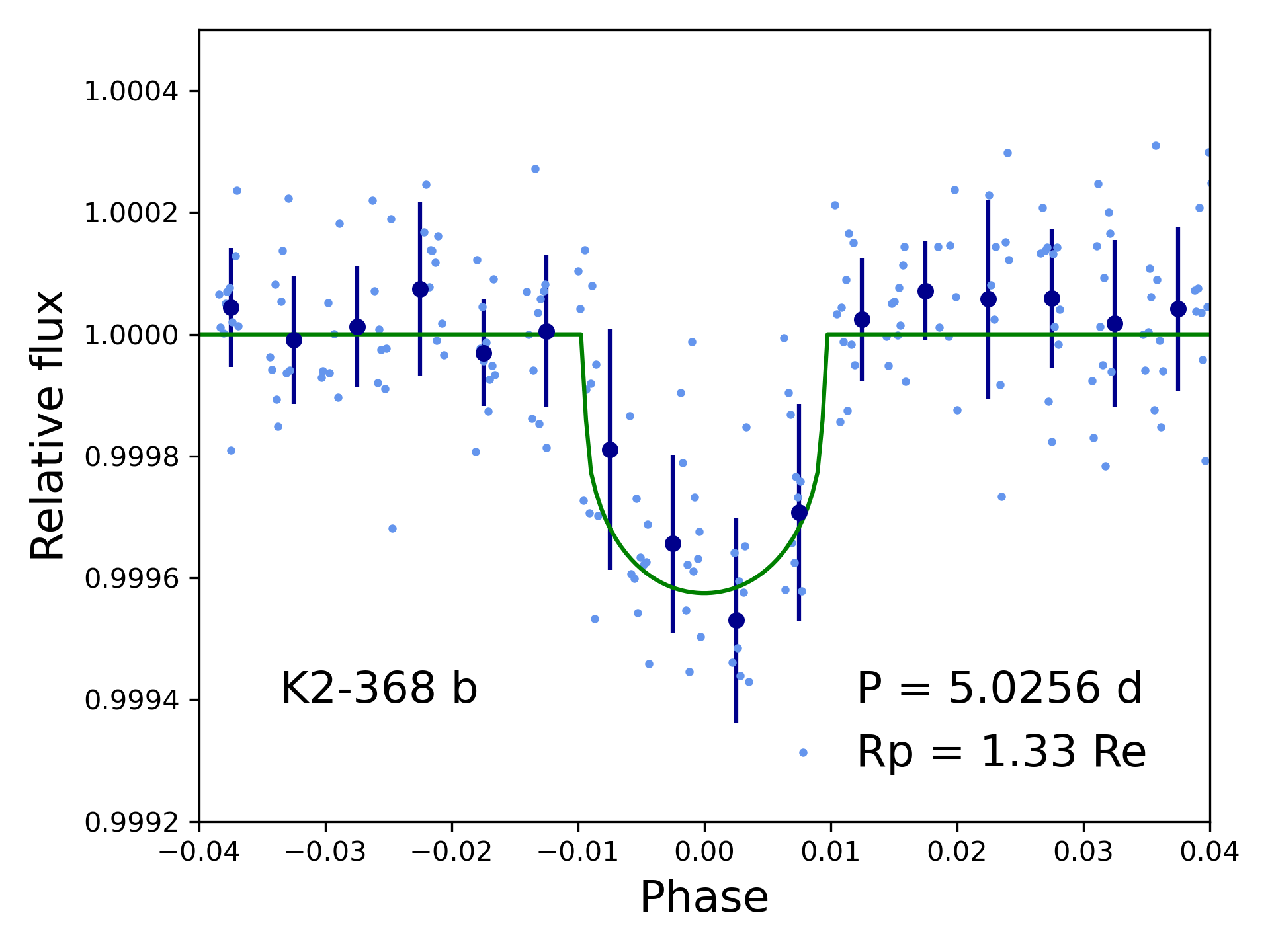}
\includegraphics[width=0.325\textwidth]{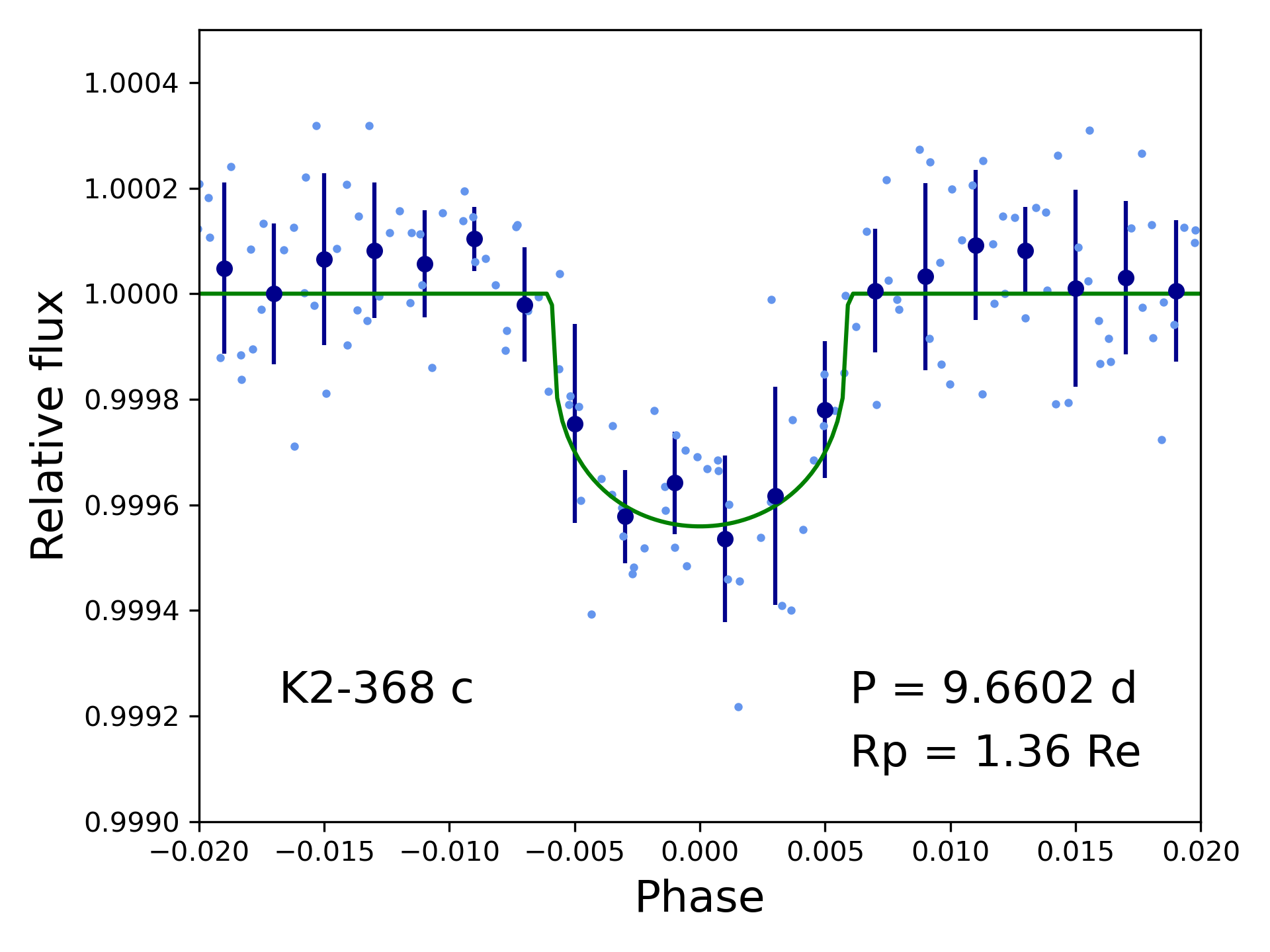}
\includegraphics[width=0.325\textwidth]{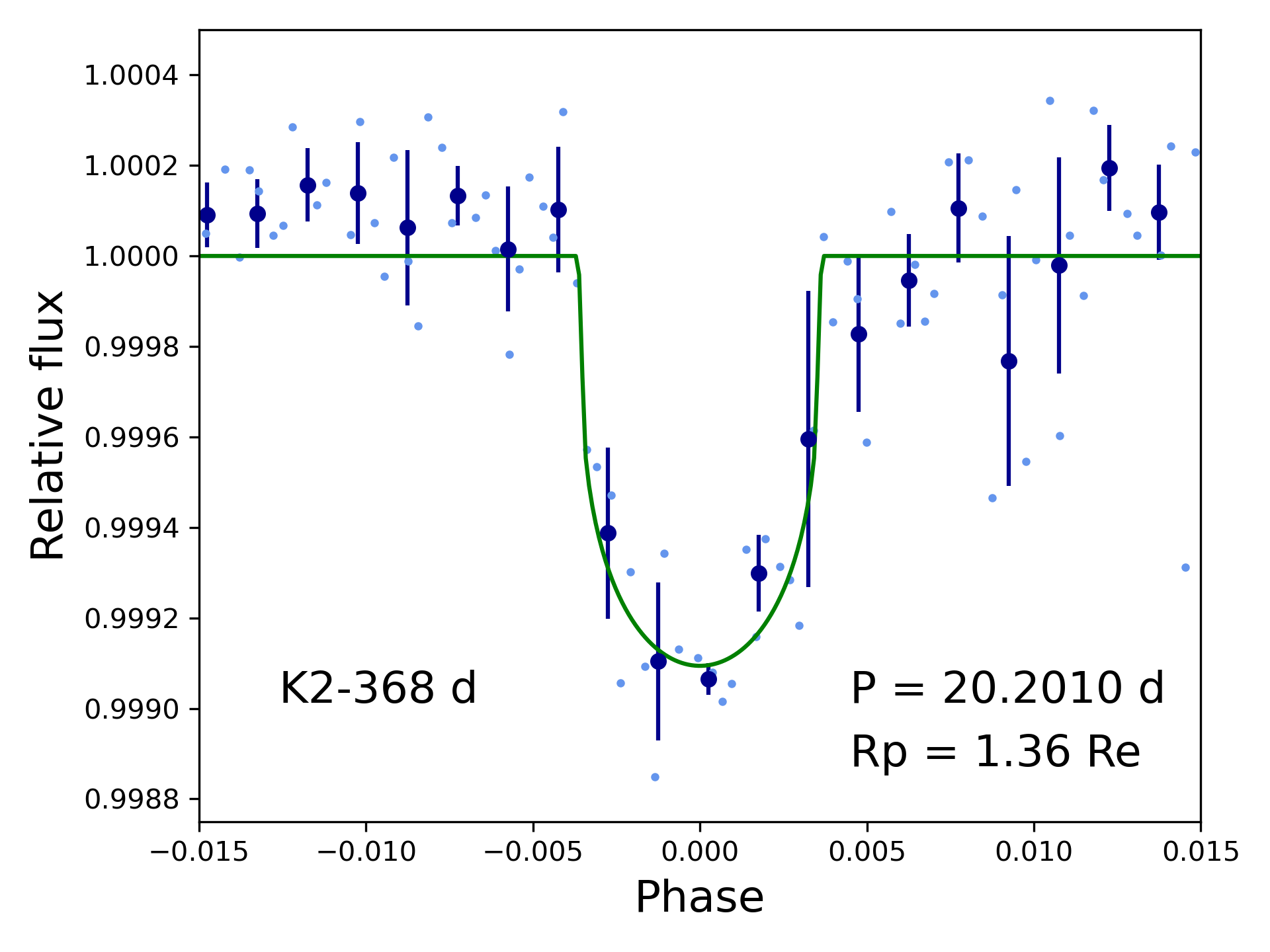}
\includegraphics[width=0.325\textwidth]{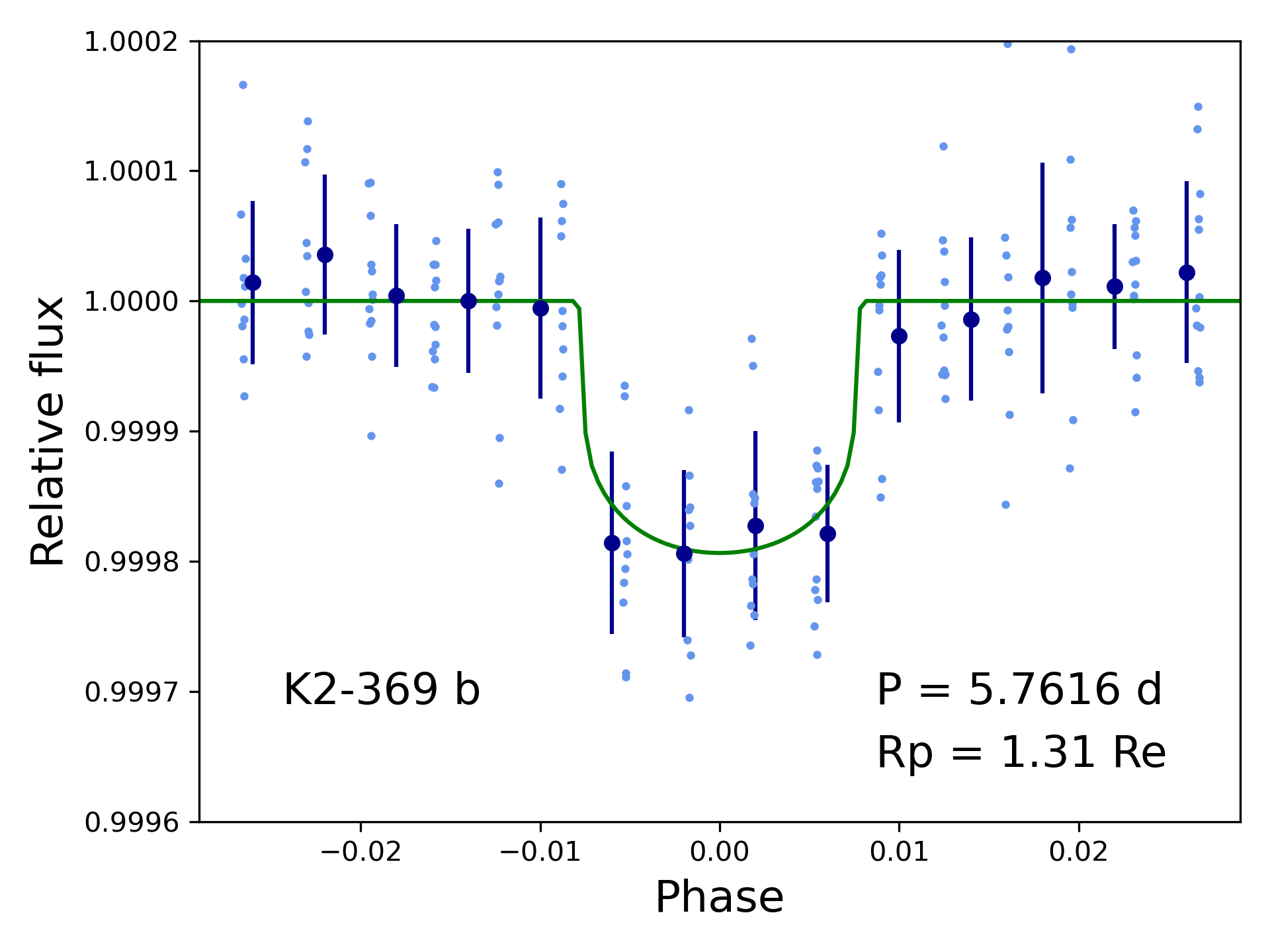}
\includegraphics[width=0.325\textwidth]{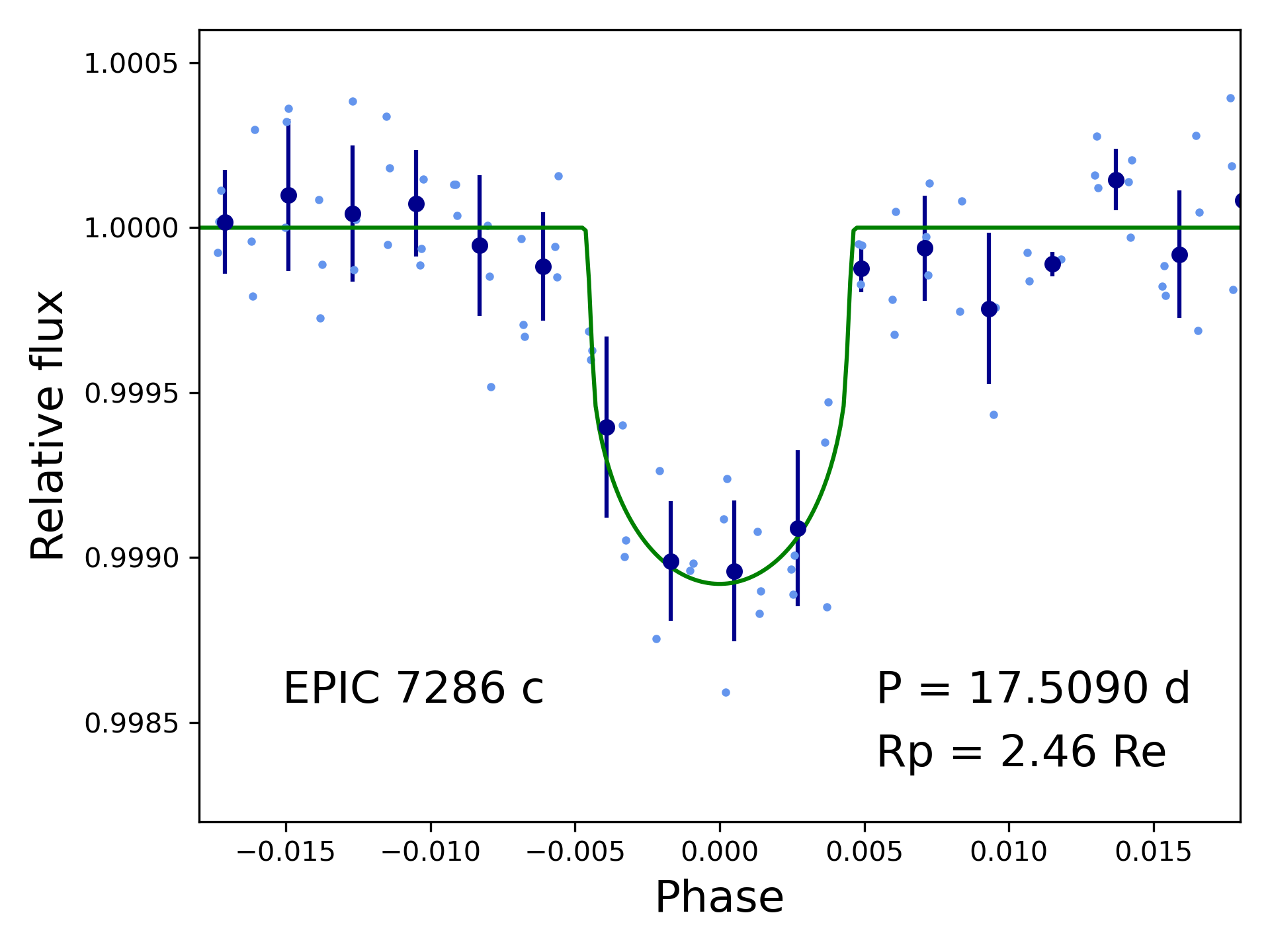}
\includegraphics[width=0.325\textwidth]{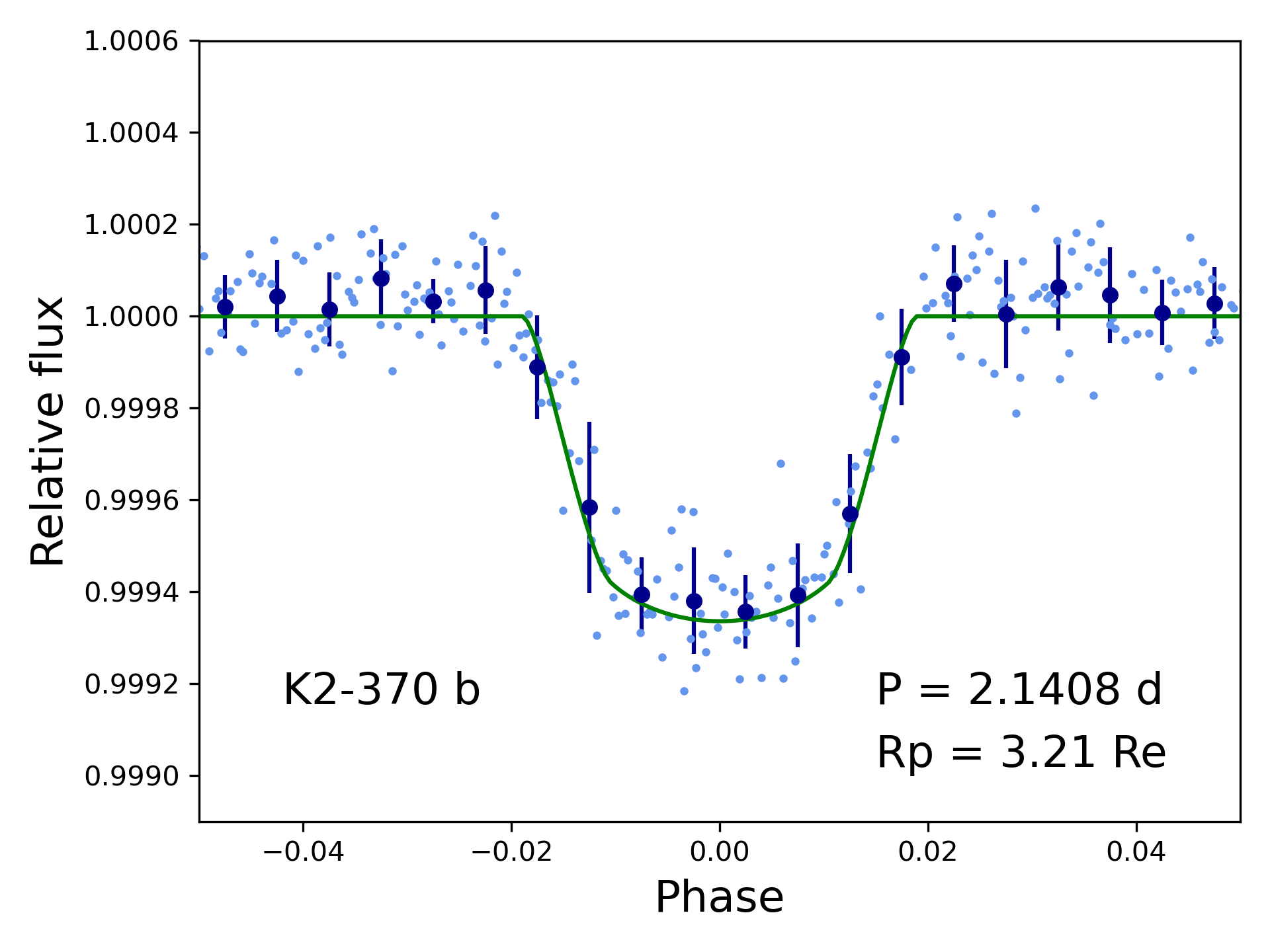}
\includegraphics[width=0.325\textwidth]{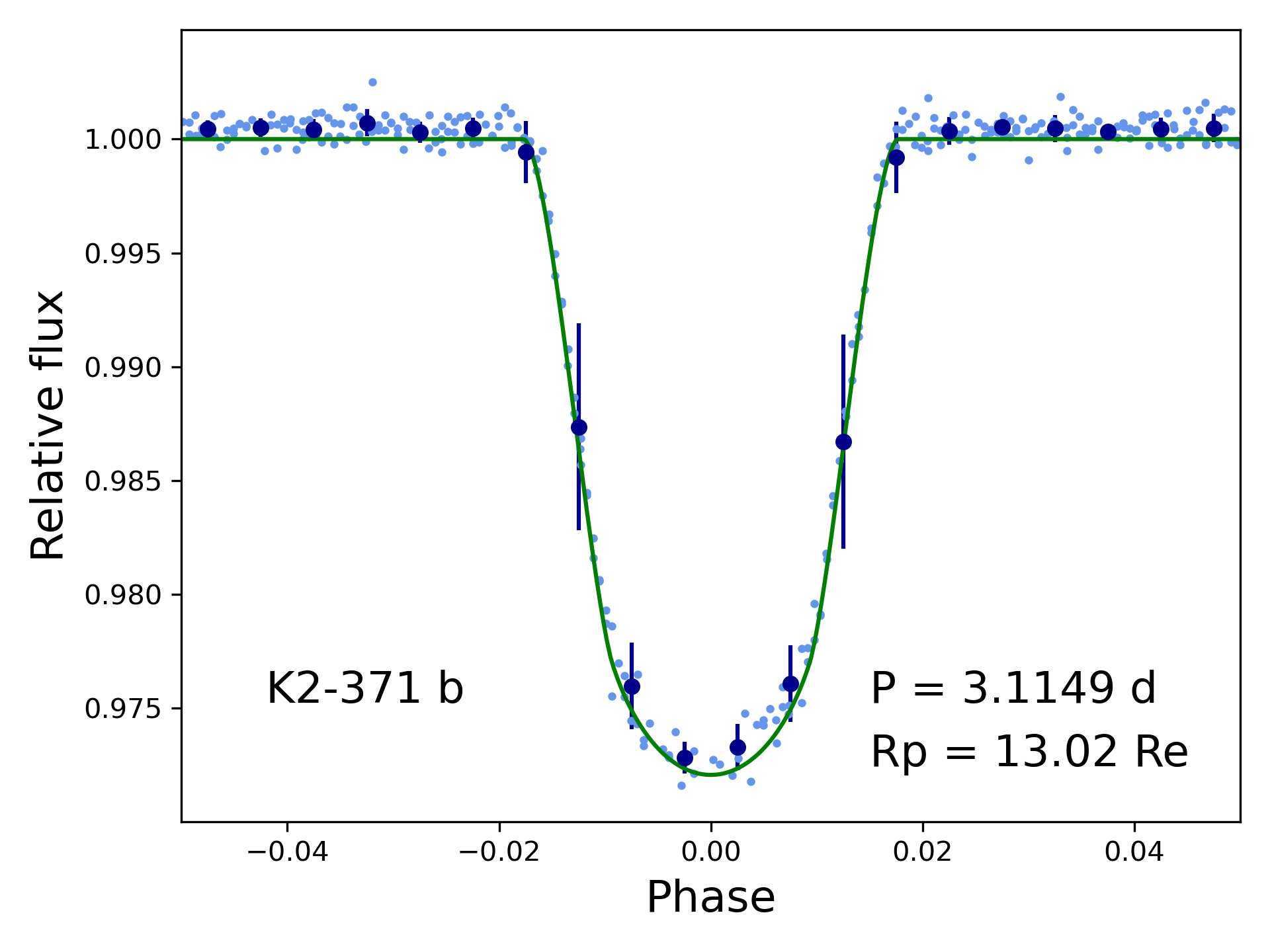}
\includegraphics[width=0.325\textwidth]{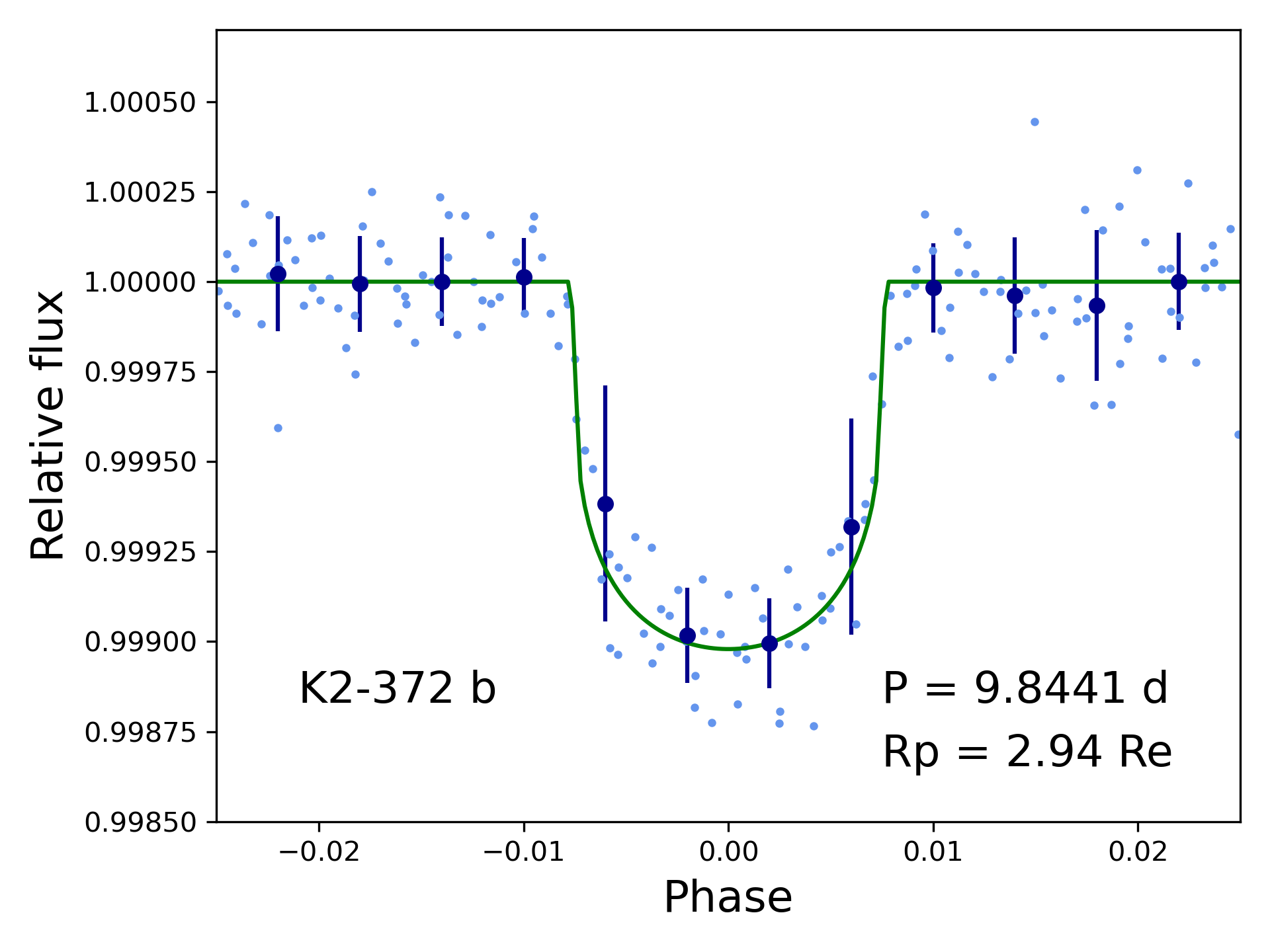}
\includegraphics[width=0.325\textwidth]{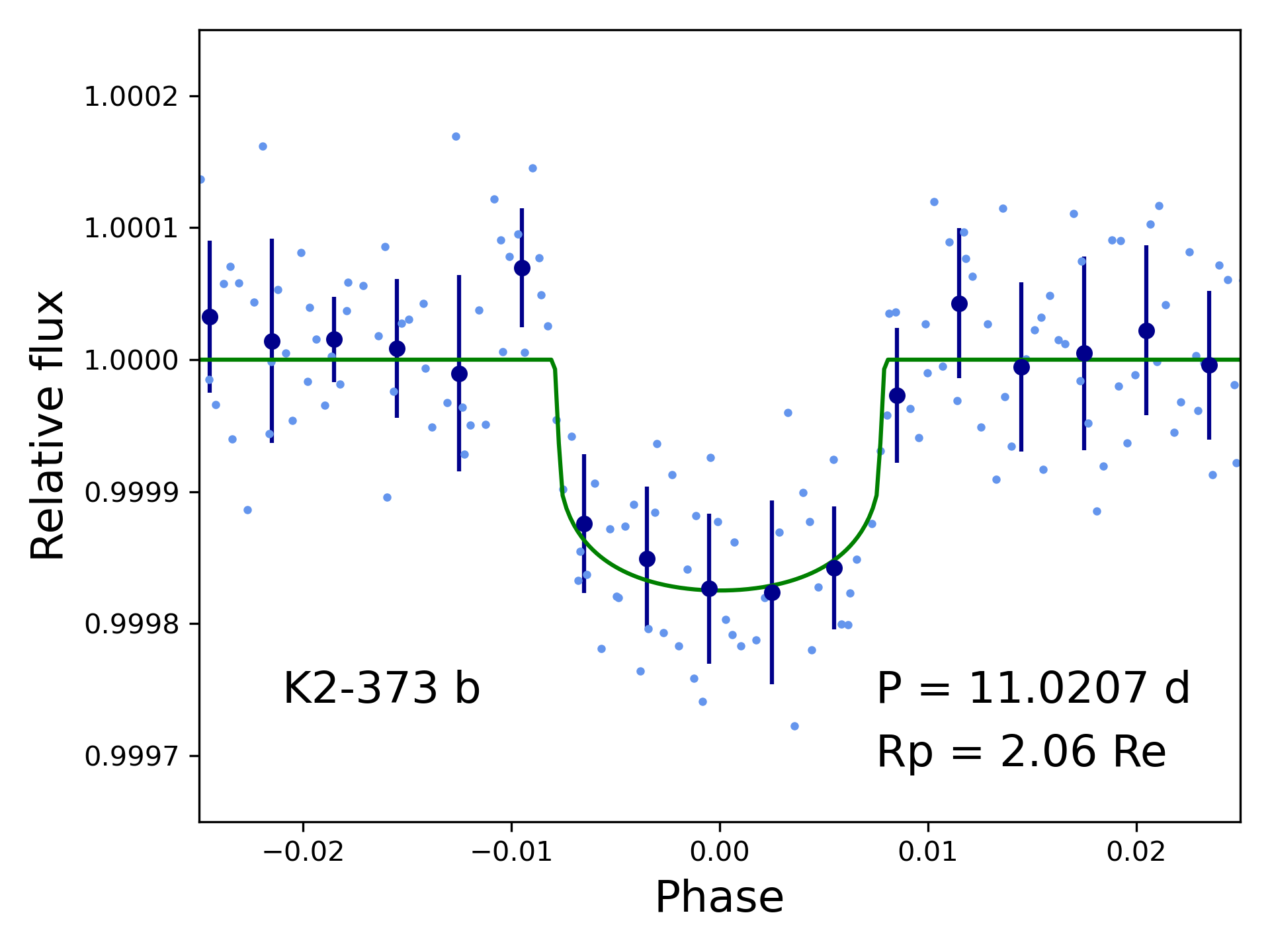}
\includegraphics[width=0.325\textwidth]{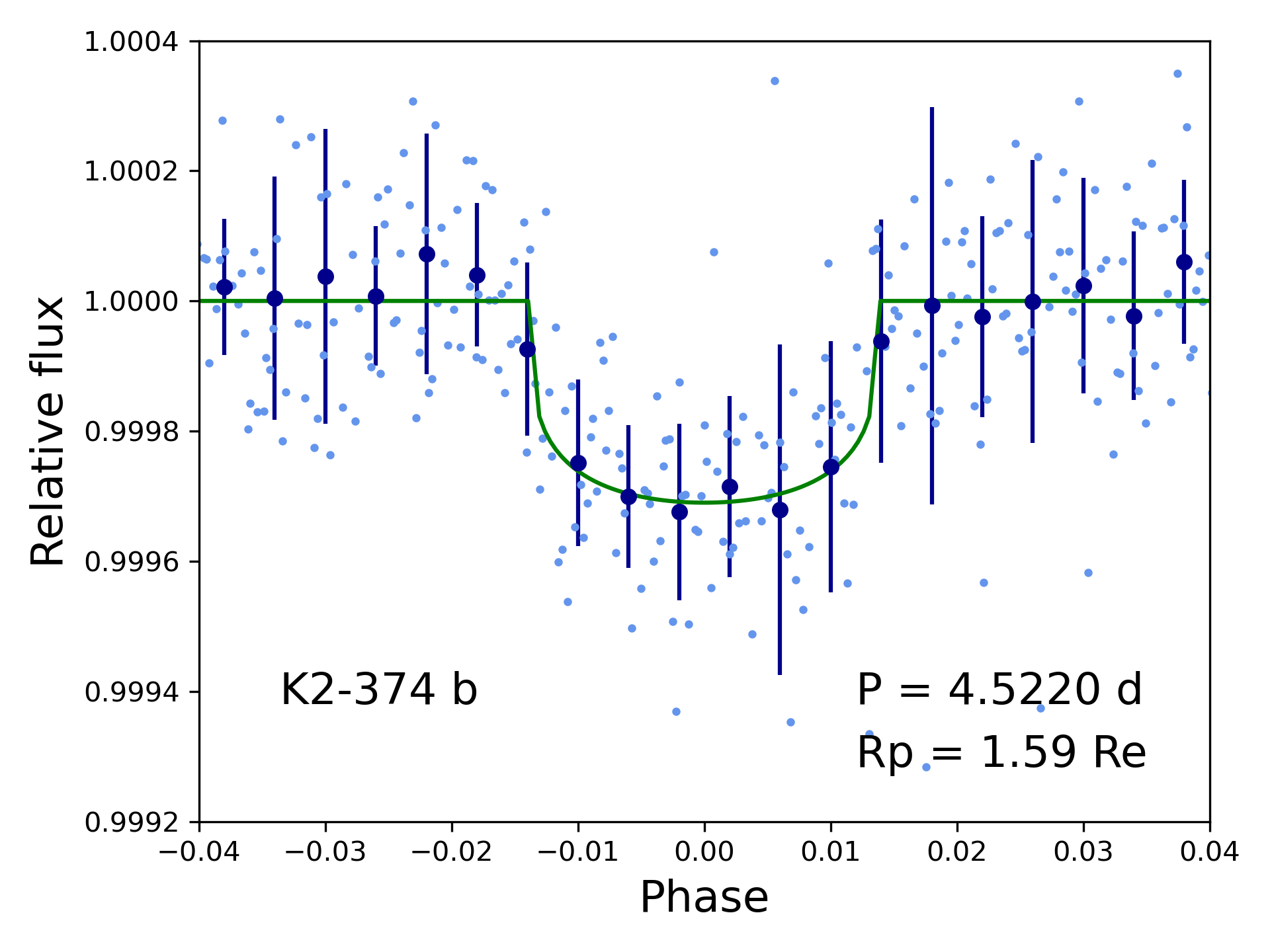}
\includegraphics[width=0.325\textwidth]{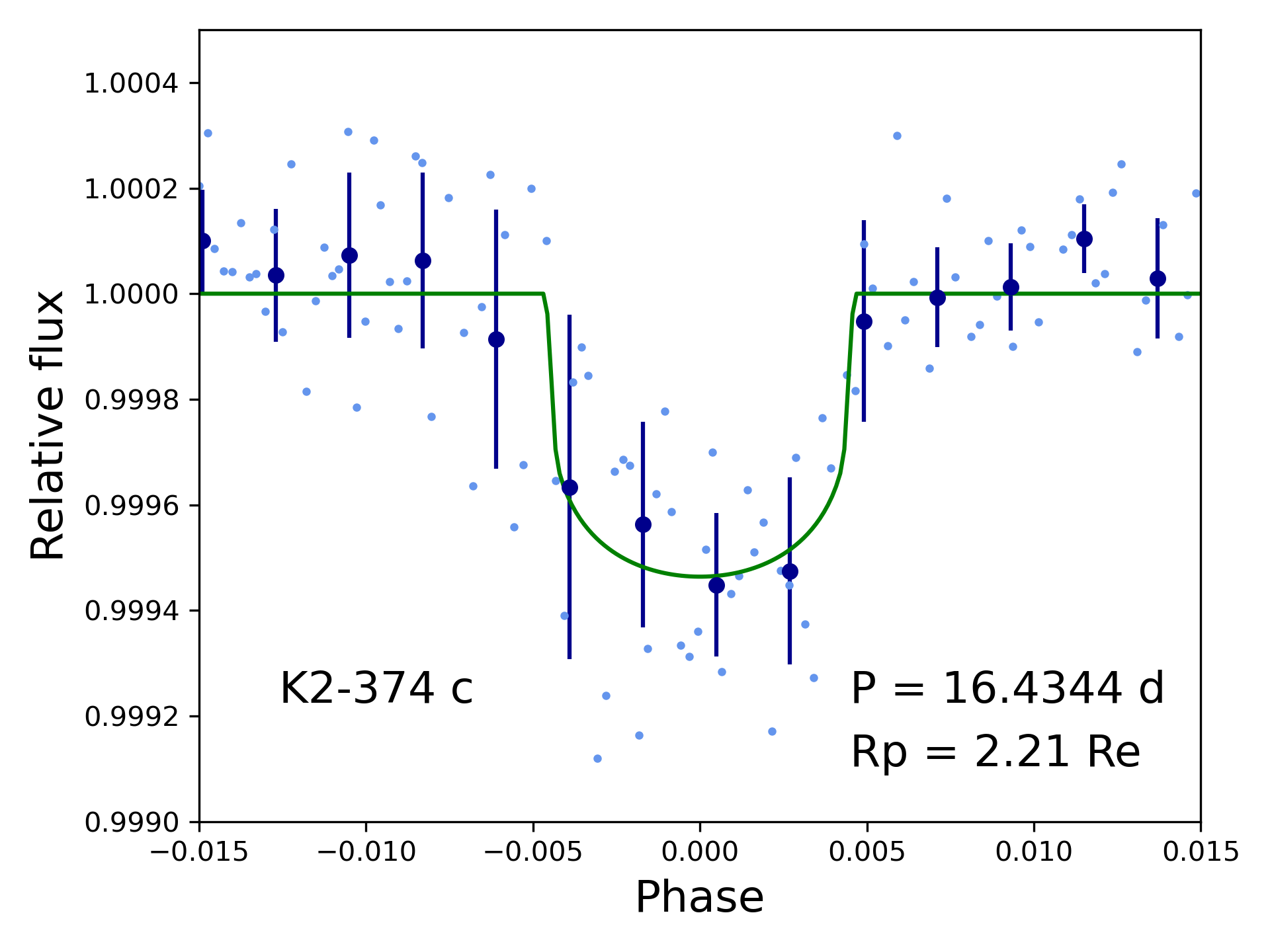}
\includegraphics[width=0.325\textwidth]{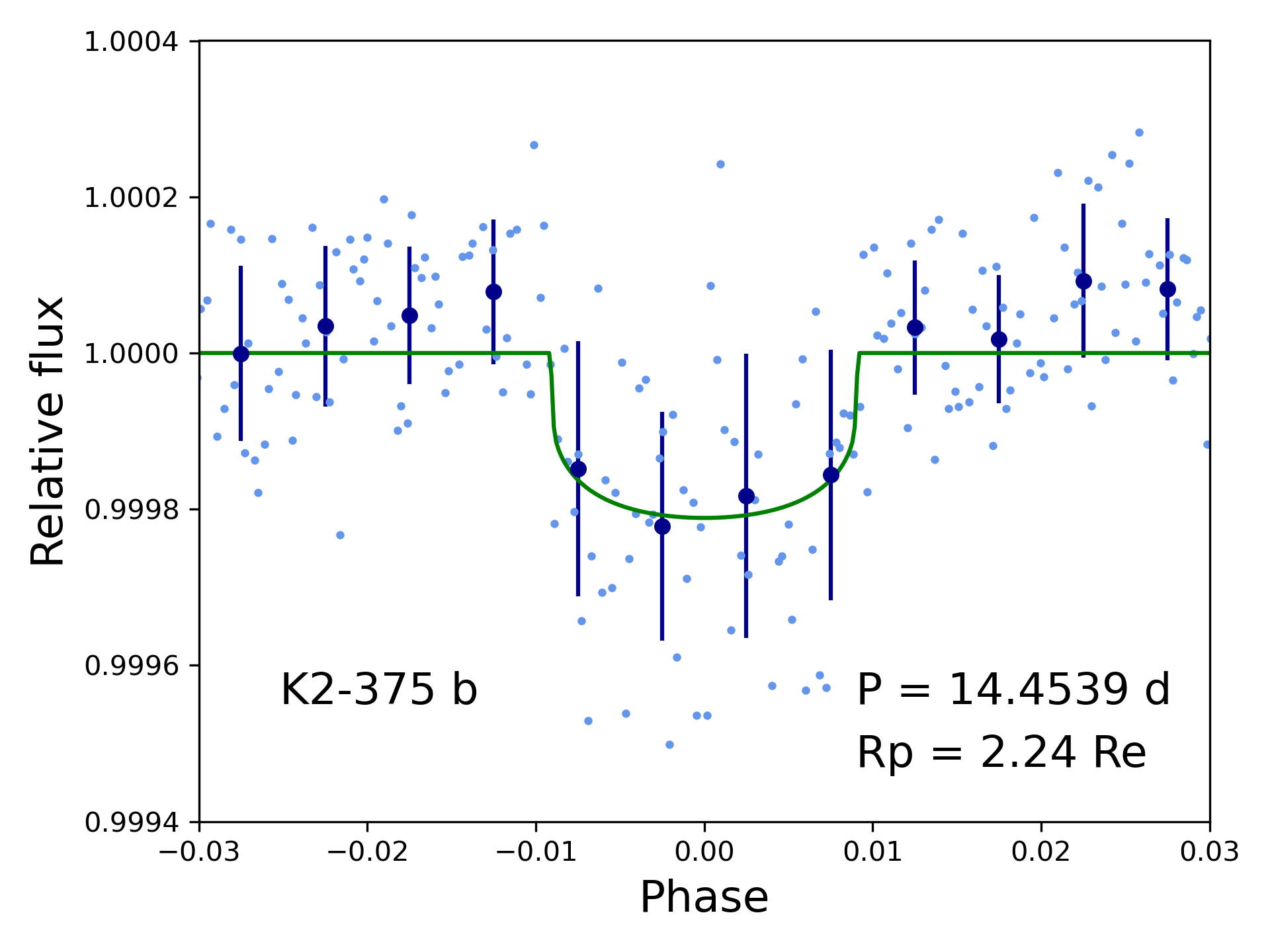}
\caption{Validated planet folded light curves. The solid green line is the best-fit transit model; the large blue points are binned observations.}
\label{fig:phasedlcs1}
\end{figure}

\begin{figure}
\centering
\includegraphics[width=0.325\textwidth]{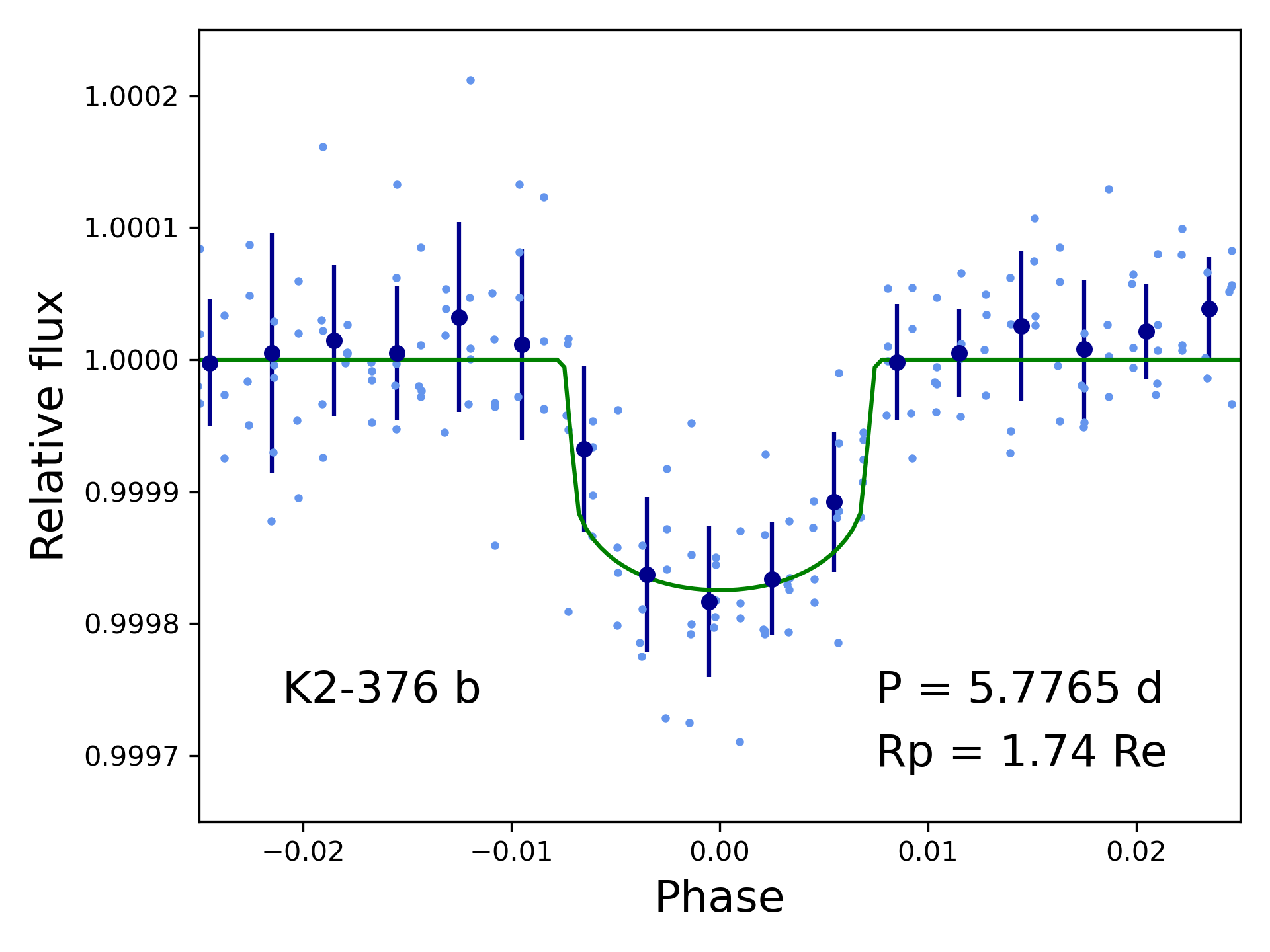}
\includegraphics[width=0.325\textwidth]{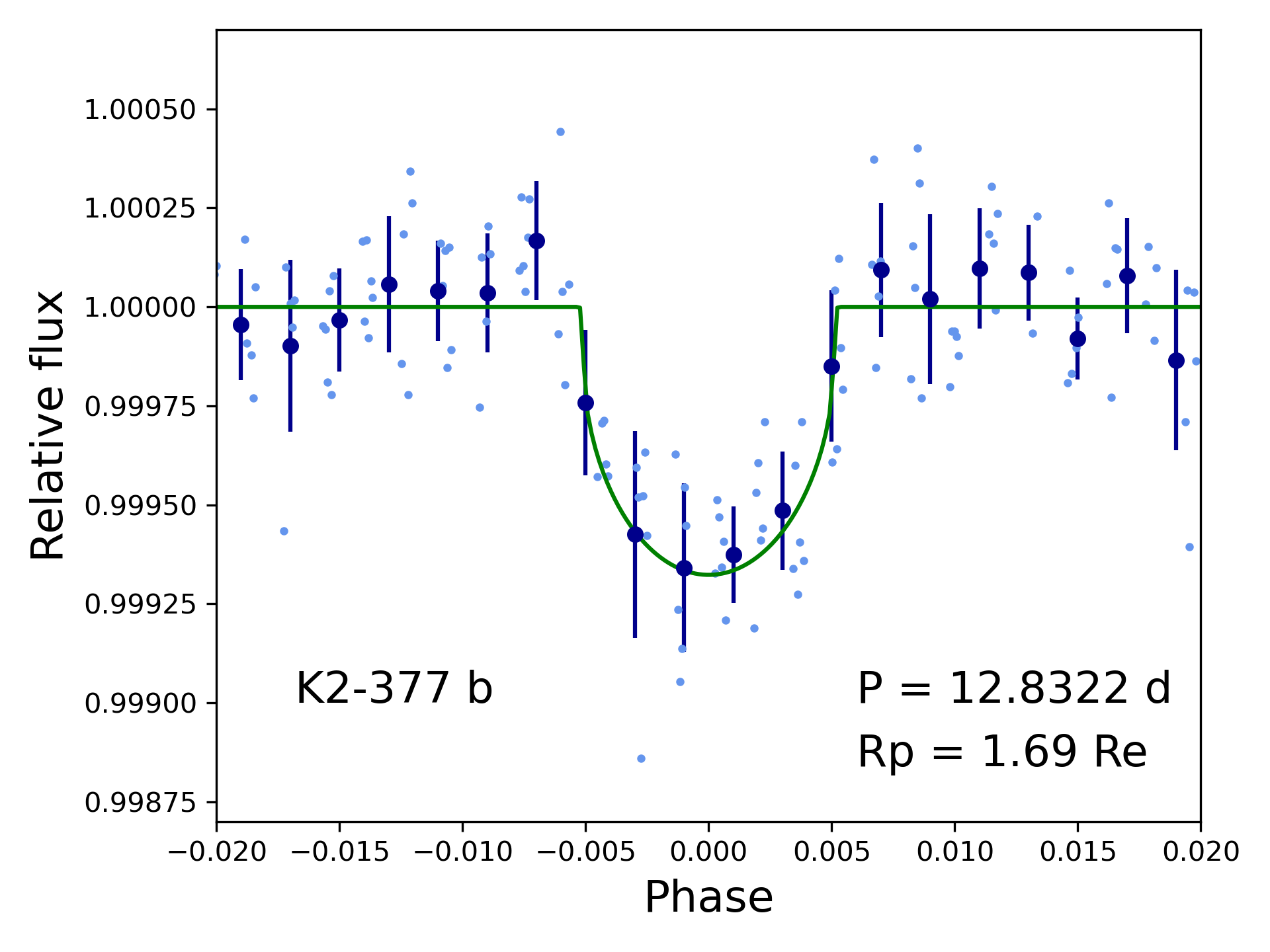}
\includegraphics[width=0.325\textwidth]{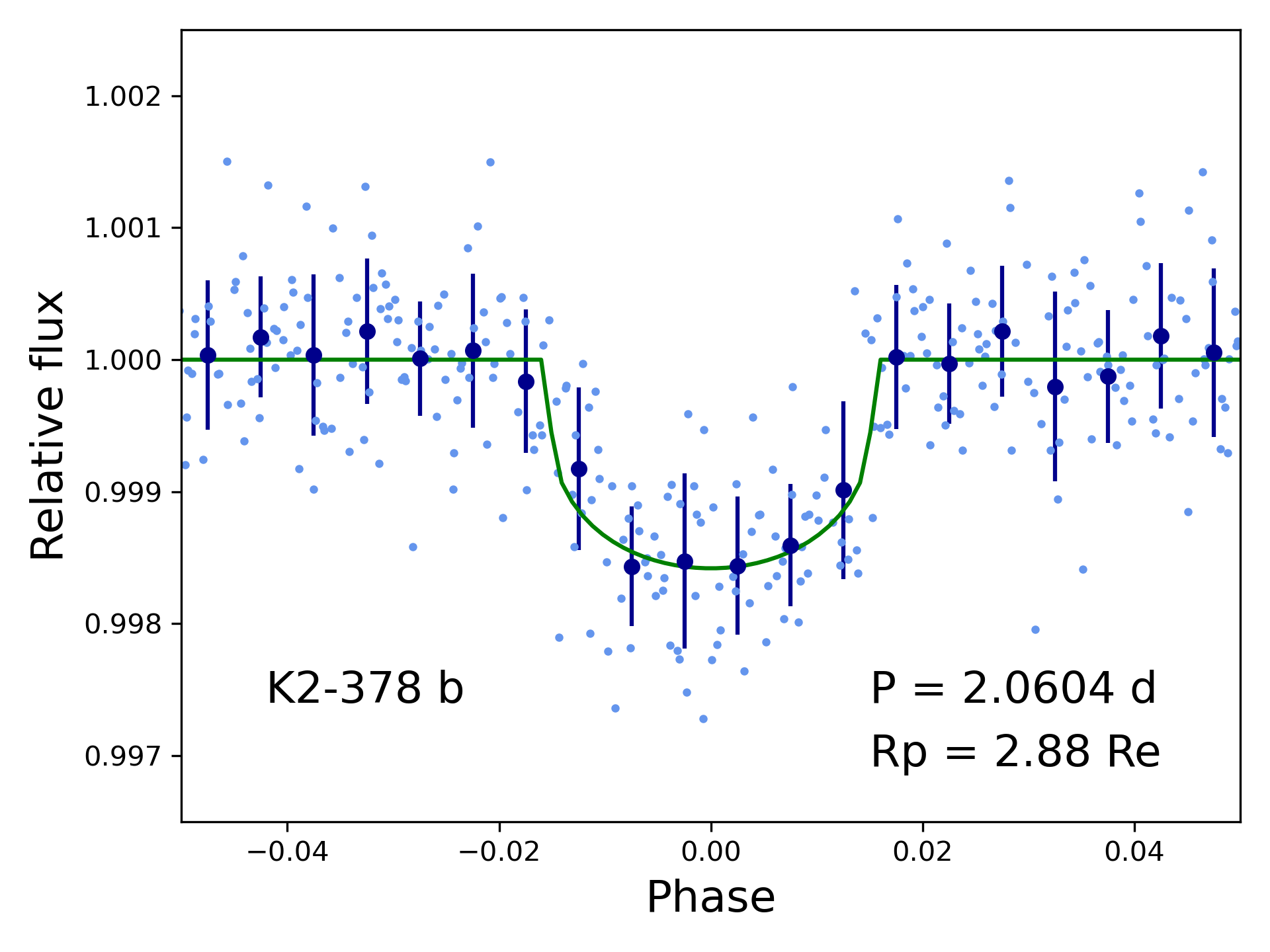}
\includegraphics[width=0.325\textwidth]{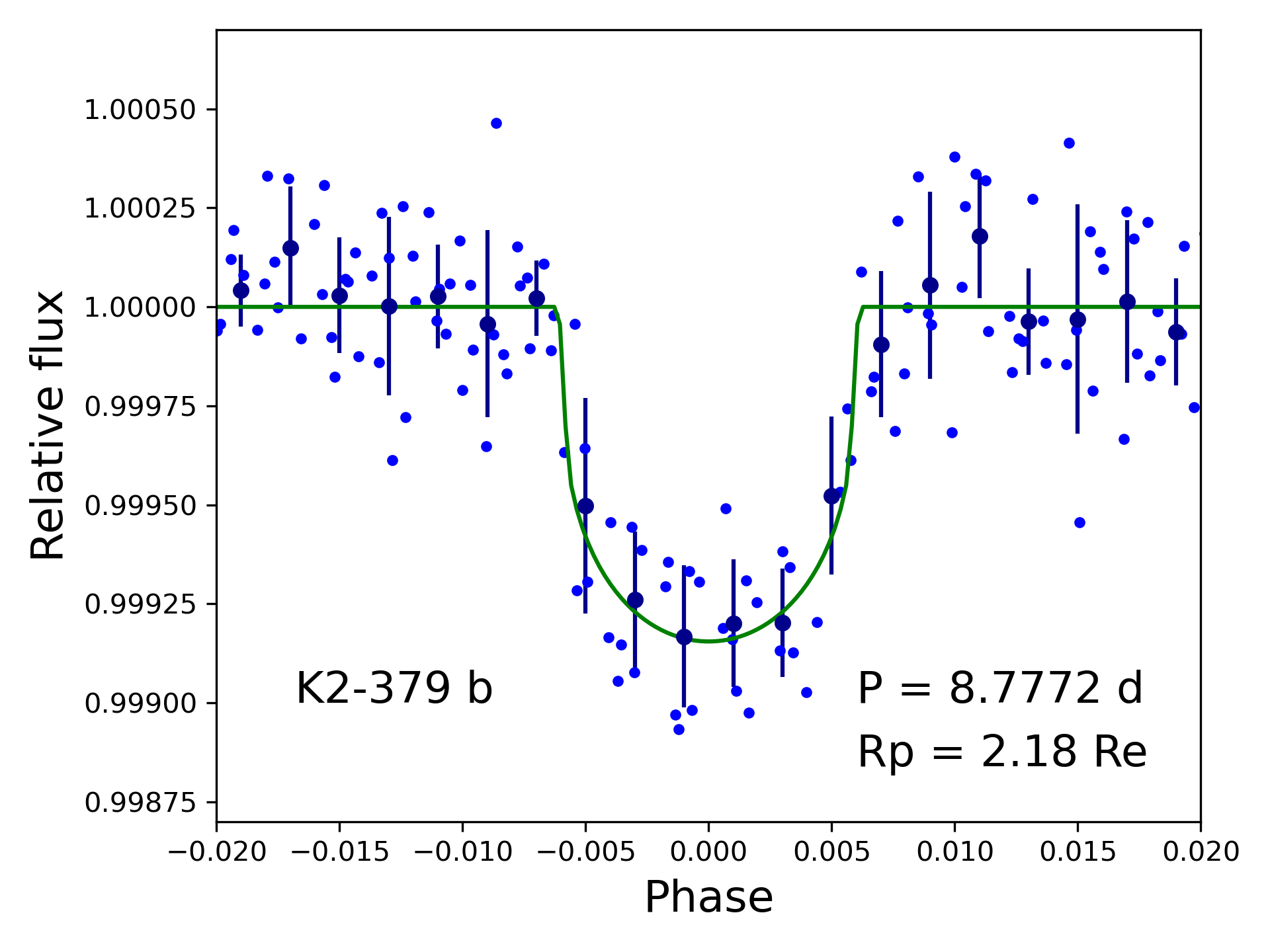}
\includegraphics[width=0.325\textwidth]{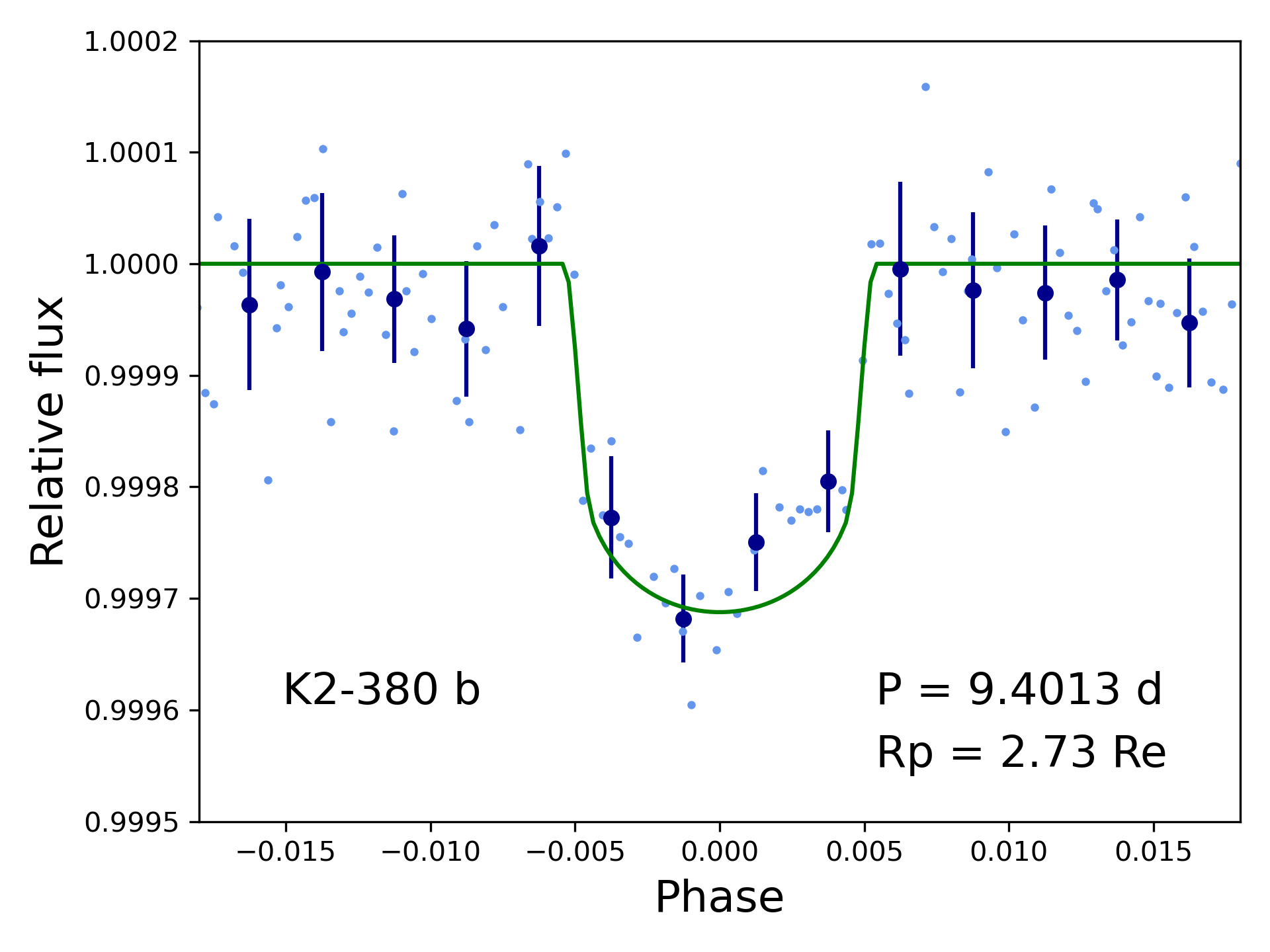}
\includegraphics[width=0.325\textwidth]{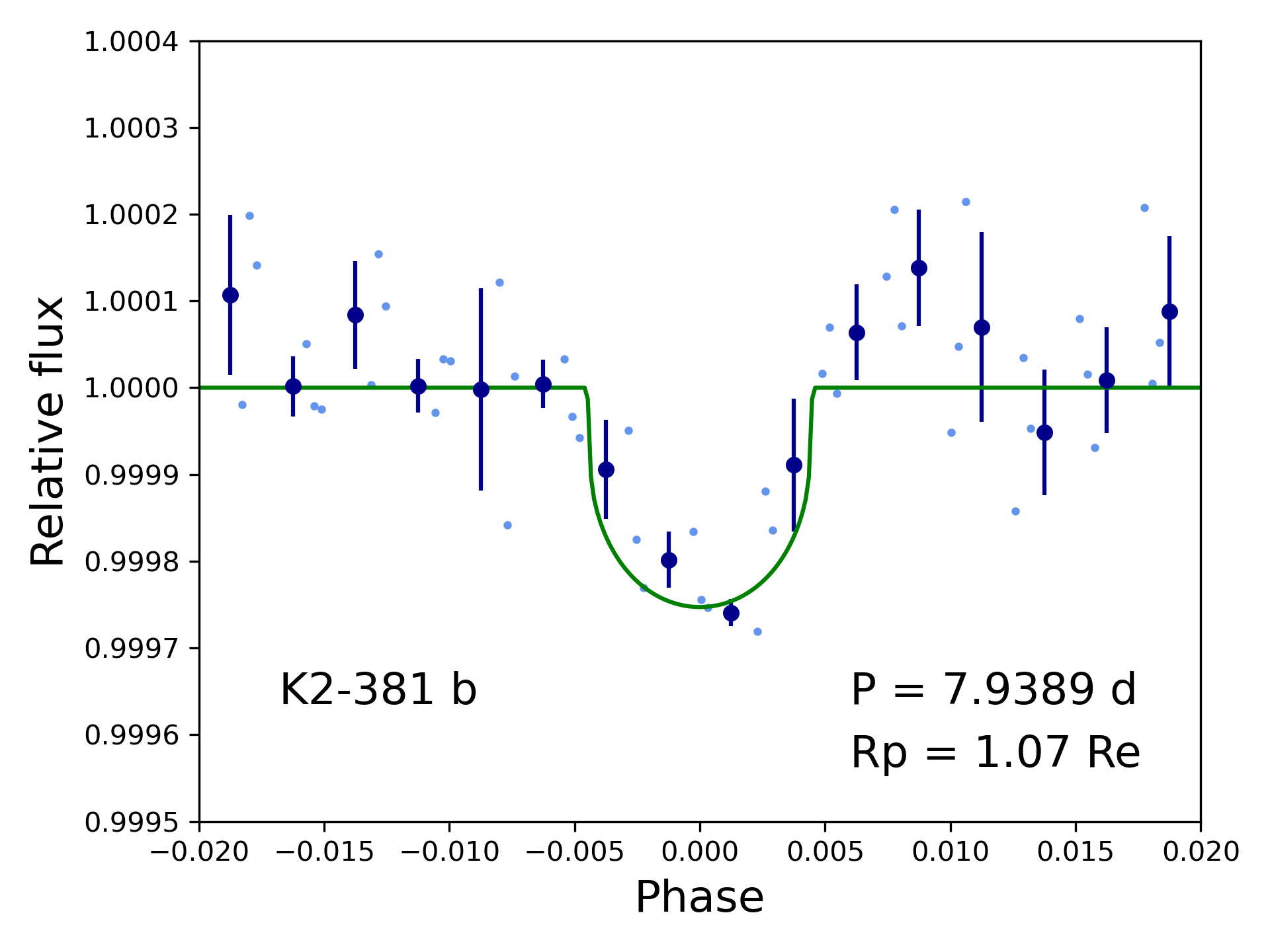}
\includegraphics[width=0.325\textwidth]{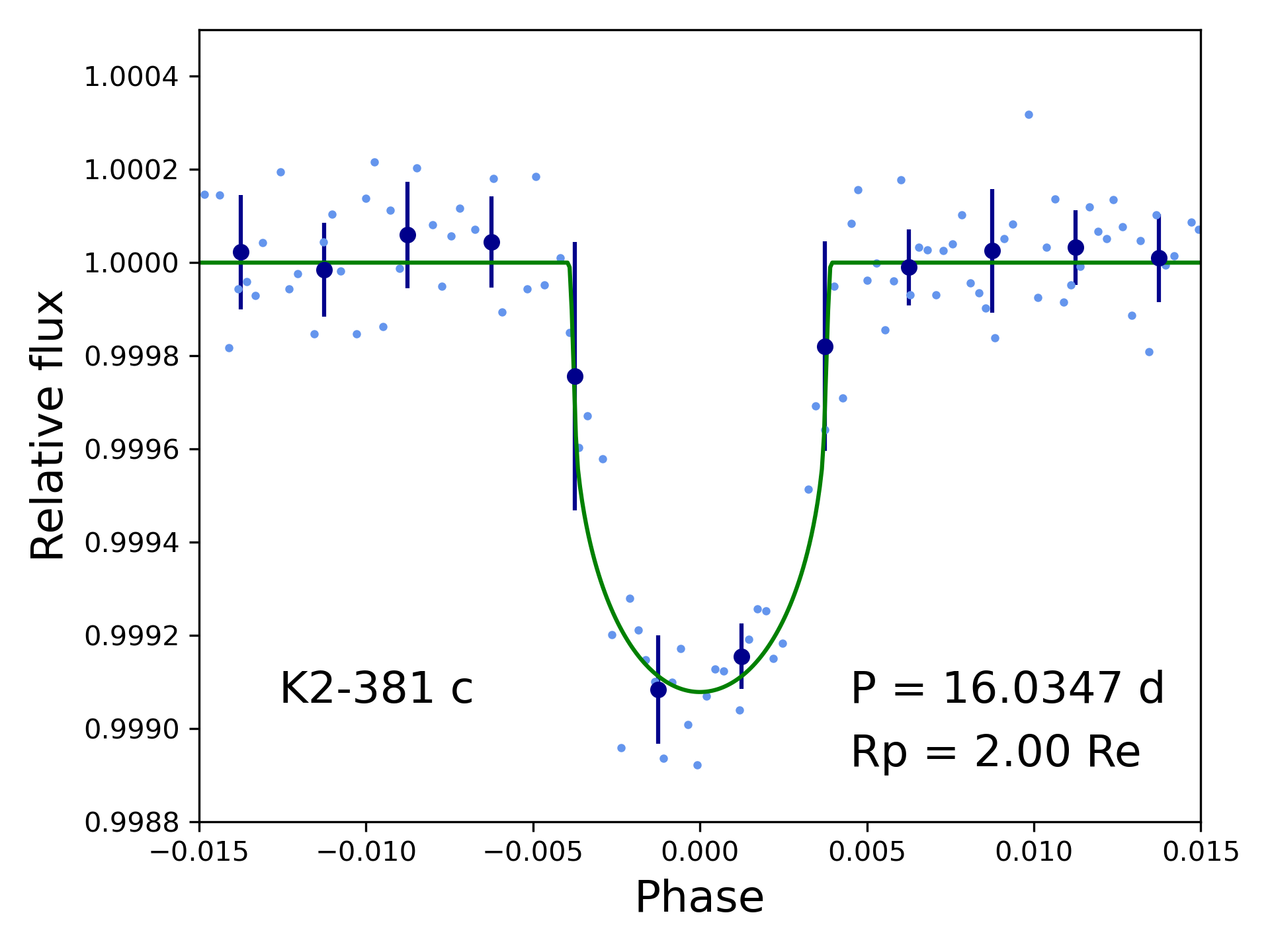}
\includegraphics[width=0.325\textwidth]{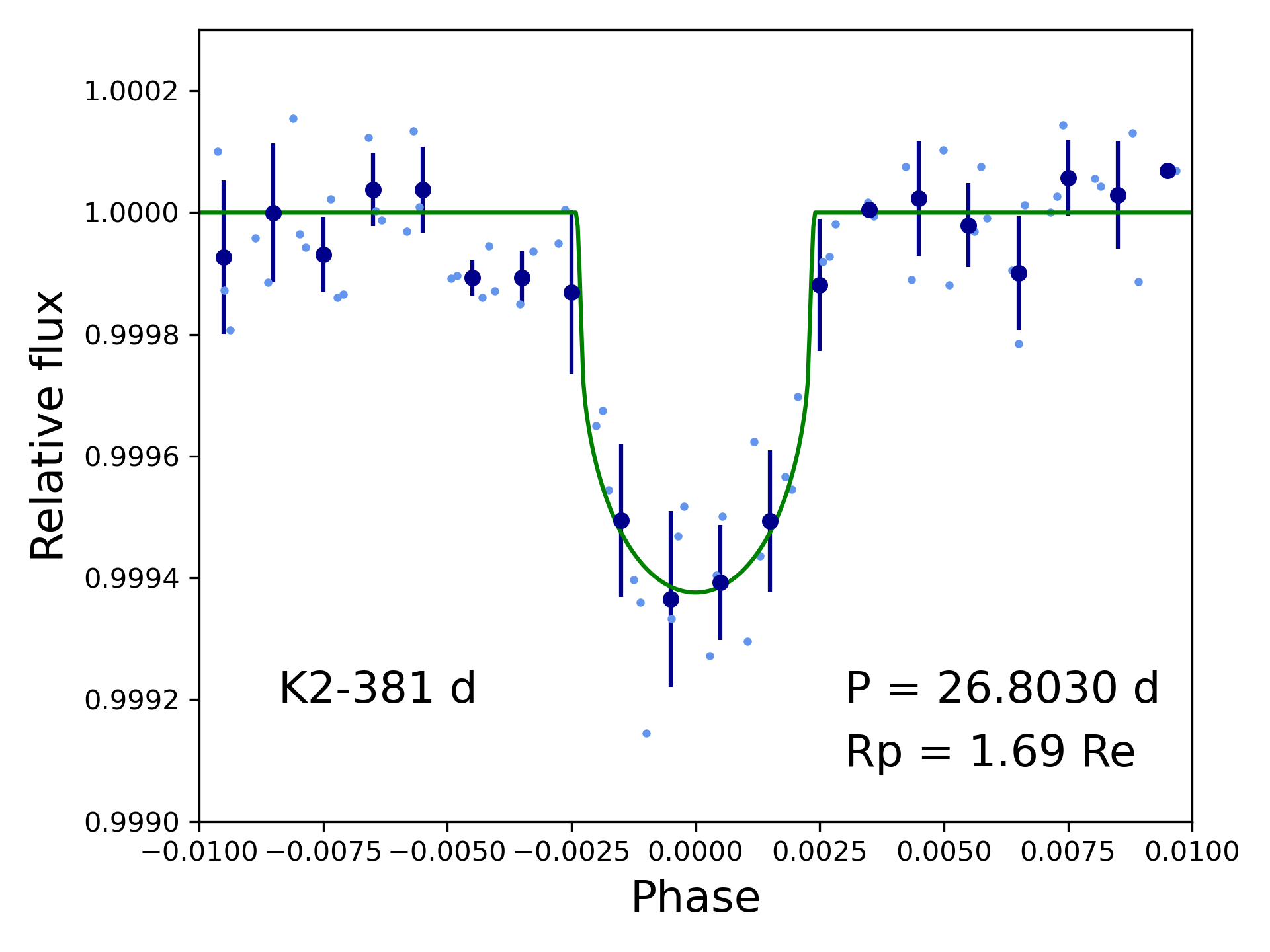}
\includegraphics[width=0.325\textwidth]{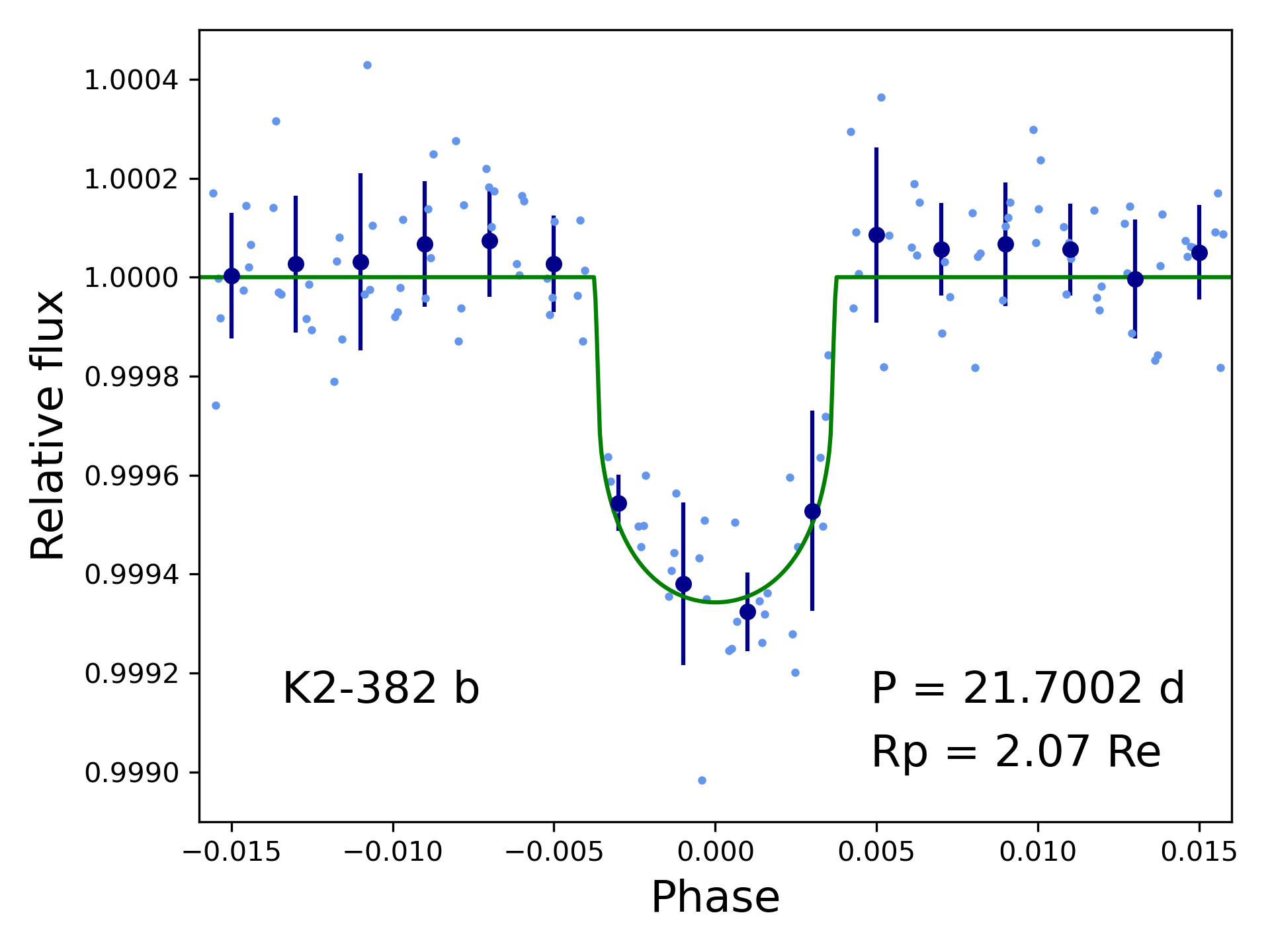}
\includegraphics[width=0.325\textwidth]{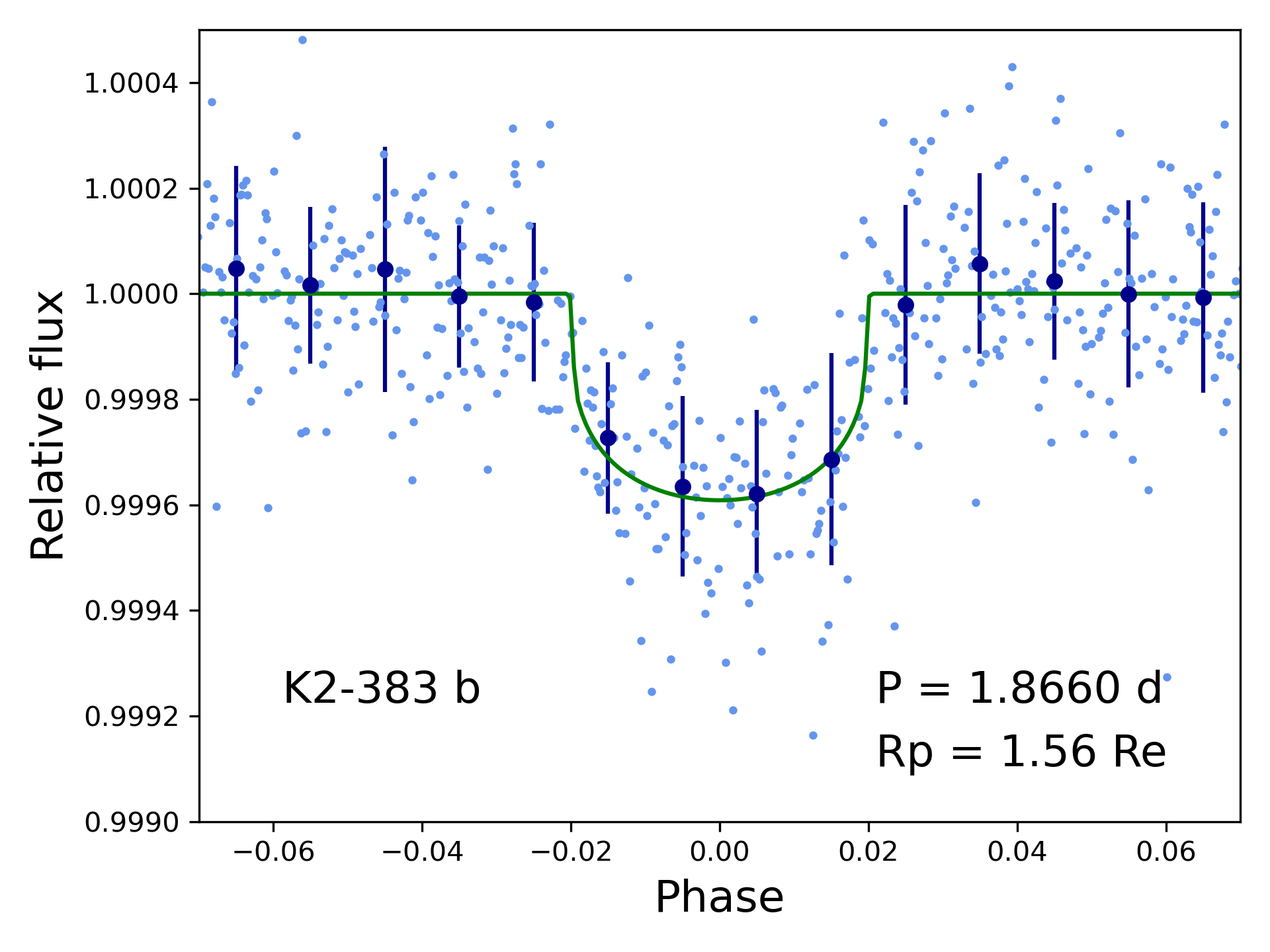}
\includegraphics[width=0.325\textwidth]{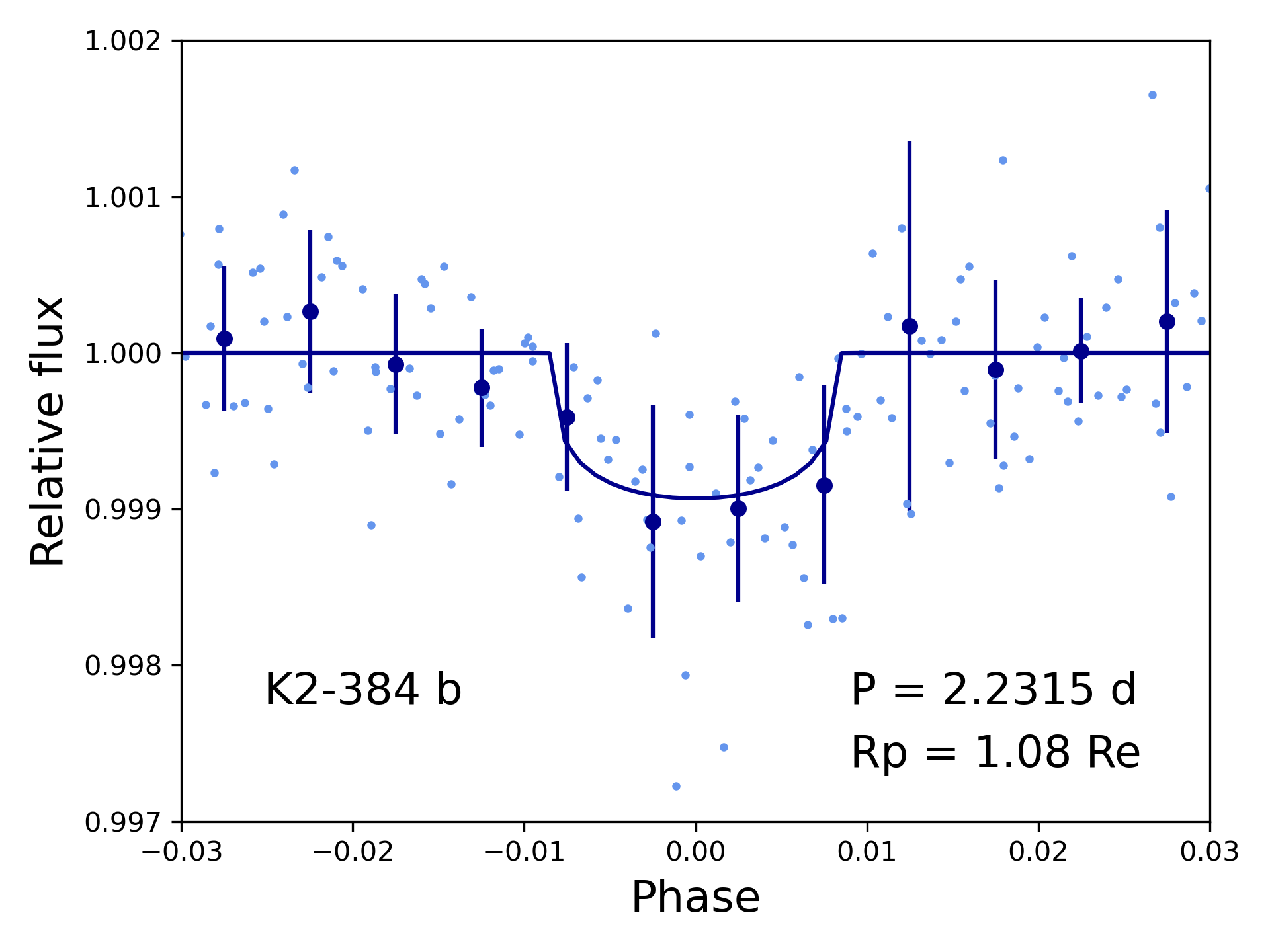}
\includegraphics[width=0.325\textwidth]{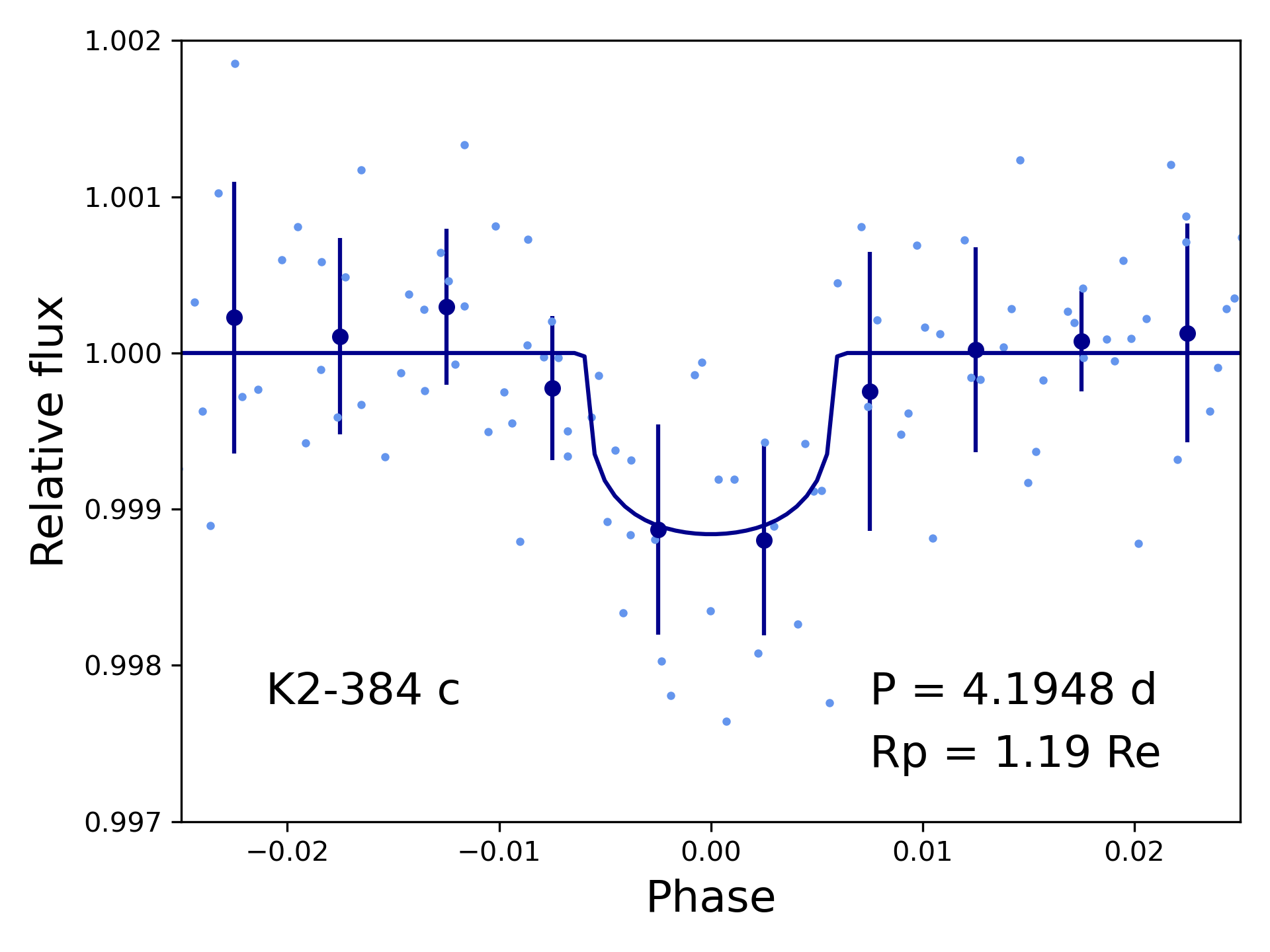}
\includegraphics[width=0.325\textwidth]{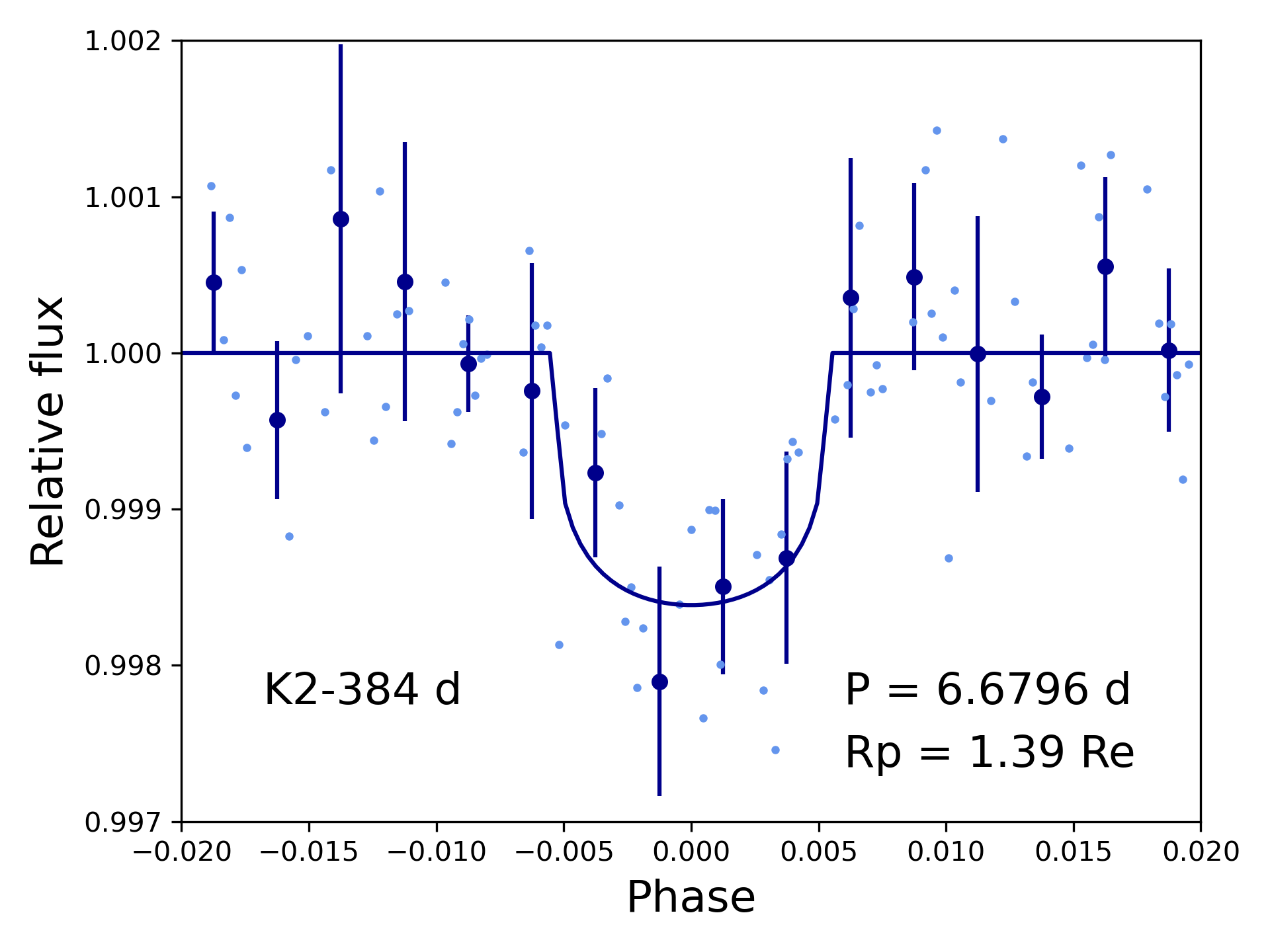}
\includegraphics[width=0.325\textwidth]{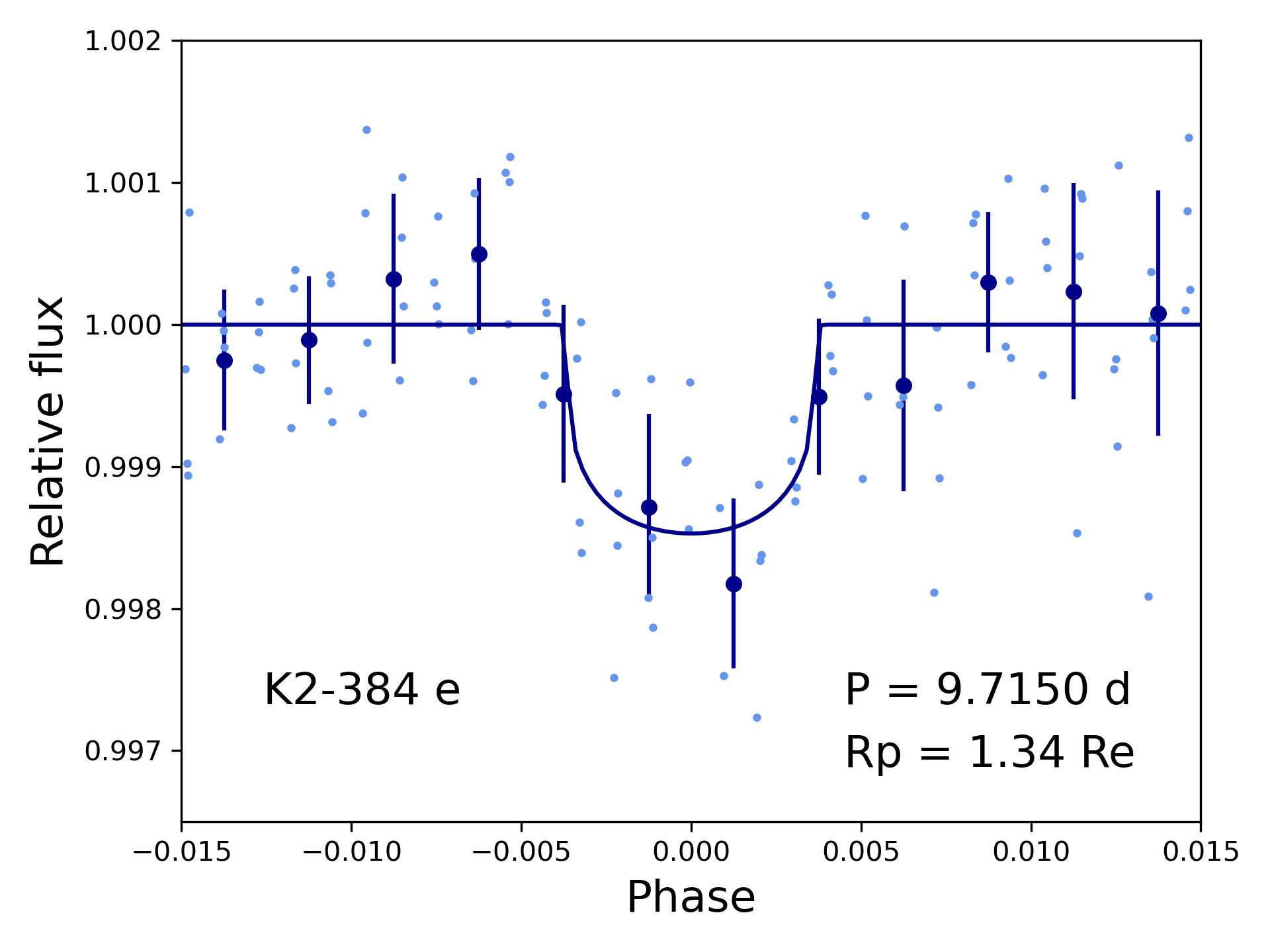}
\includegraphics[width=0.325\textwidth]{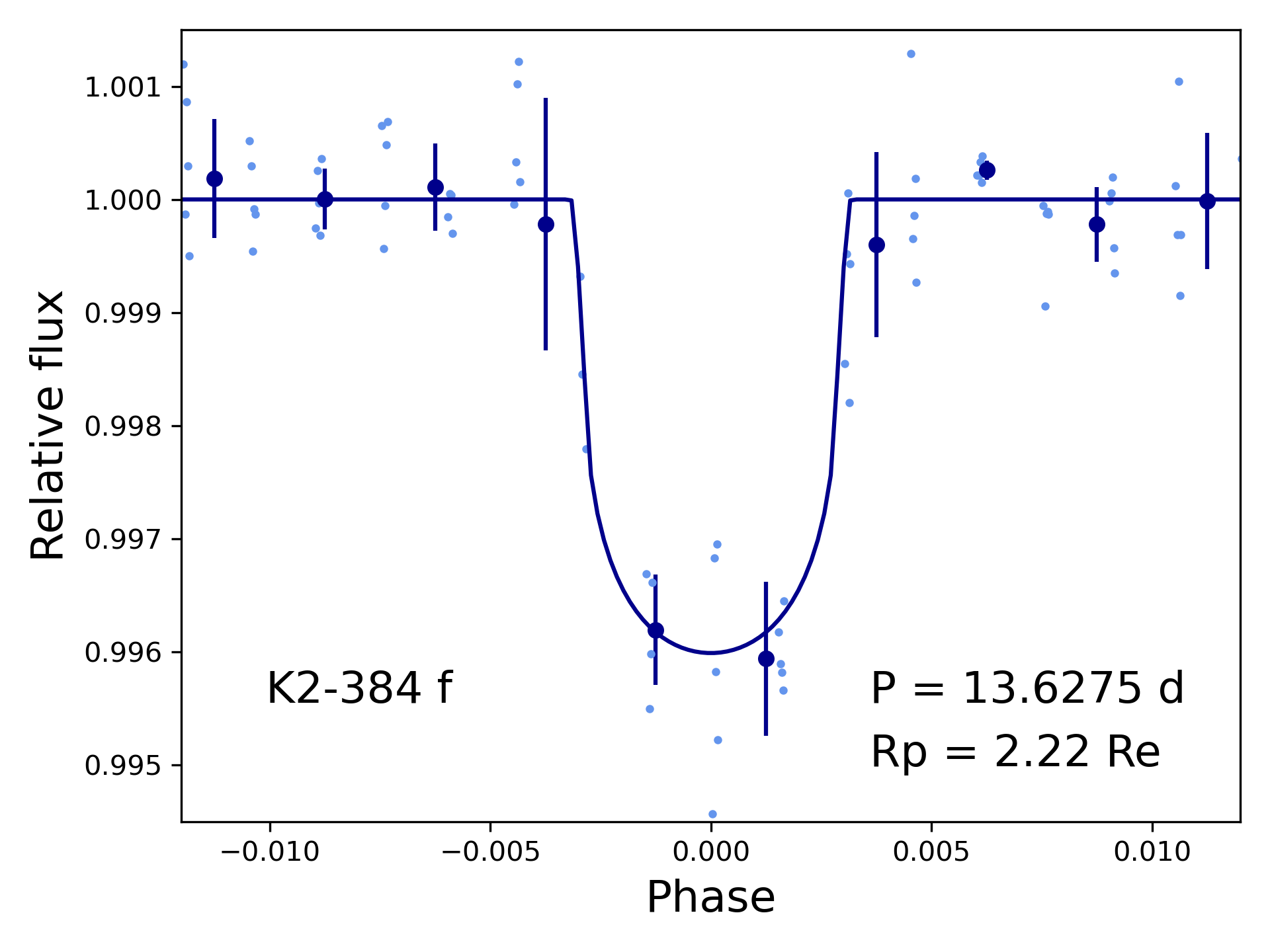}
\caption{Validated planet folded light curves. Description as for Figure \ref{fig:phasedlcs1}.}
\label{fig:phasedlcs2}
\end{figure}

\begin{figure}
\centering
\includegraphics[width=0.325\textwidth]{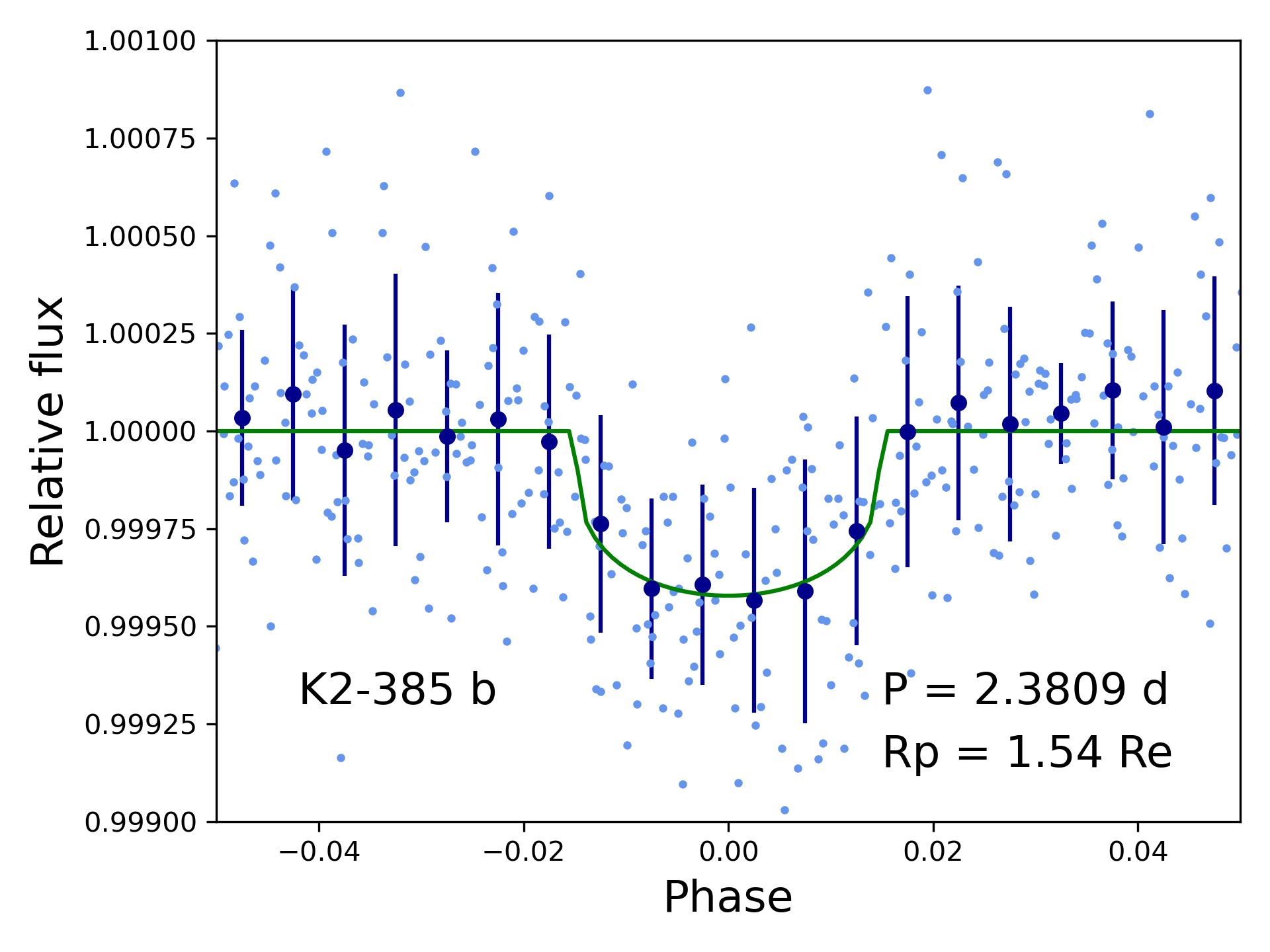}
\includegraphics[width=0.325\textwidth]{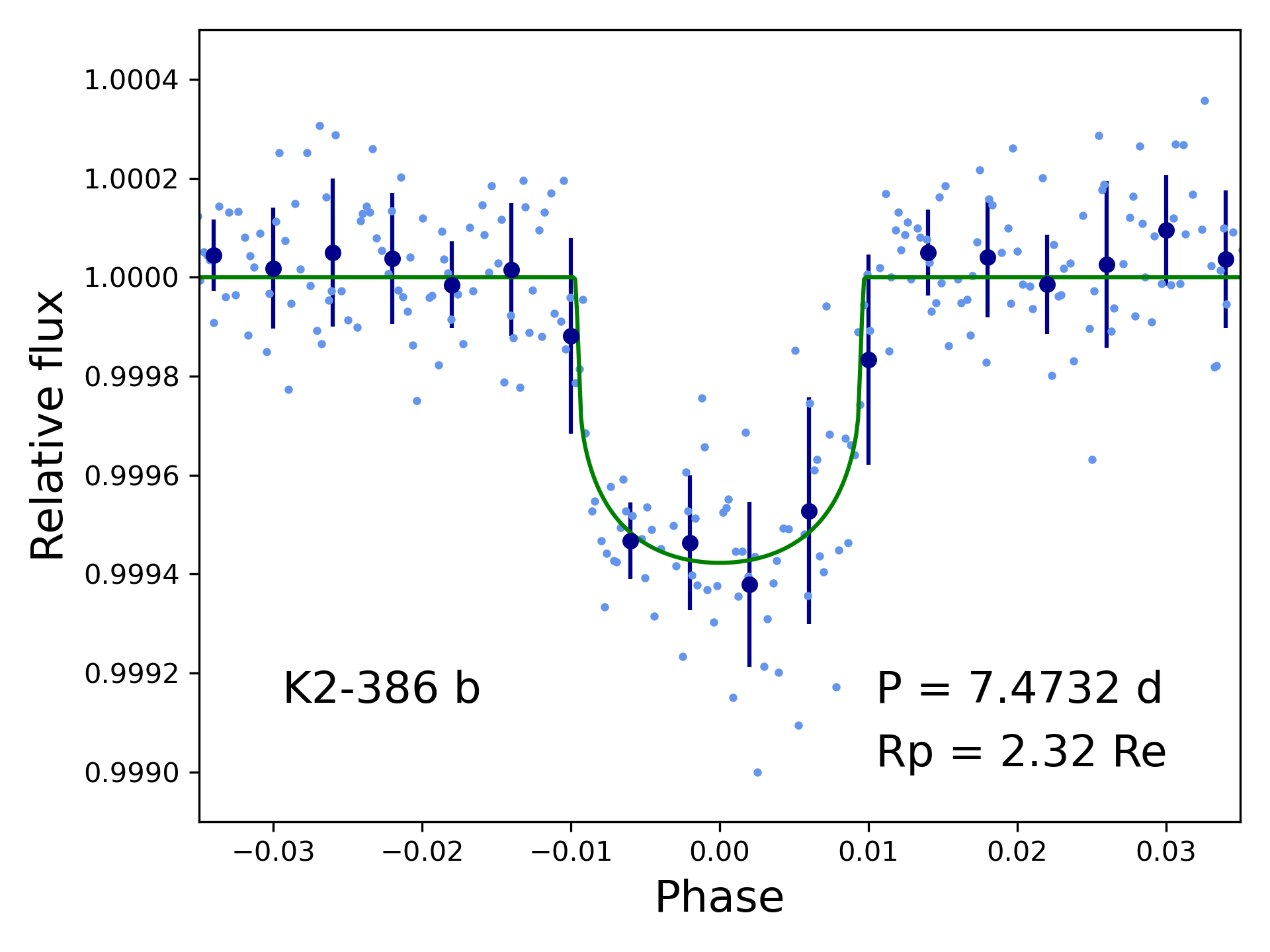}
\includegraphics[width=0.325\textwidth]{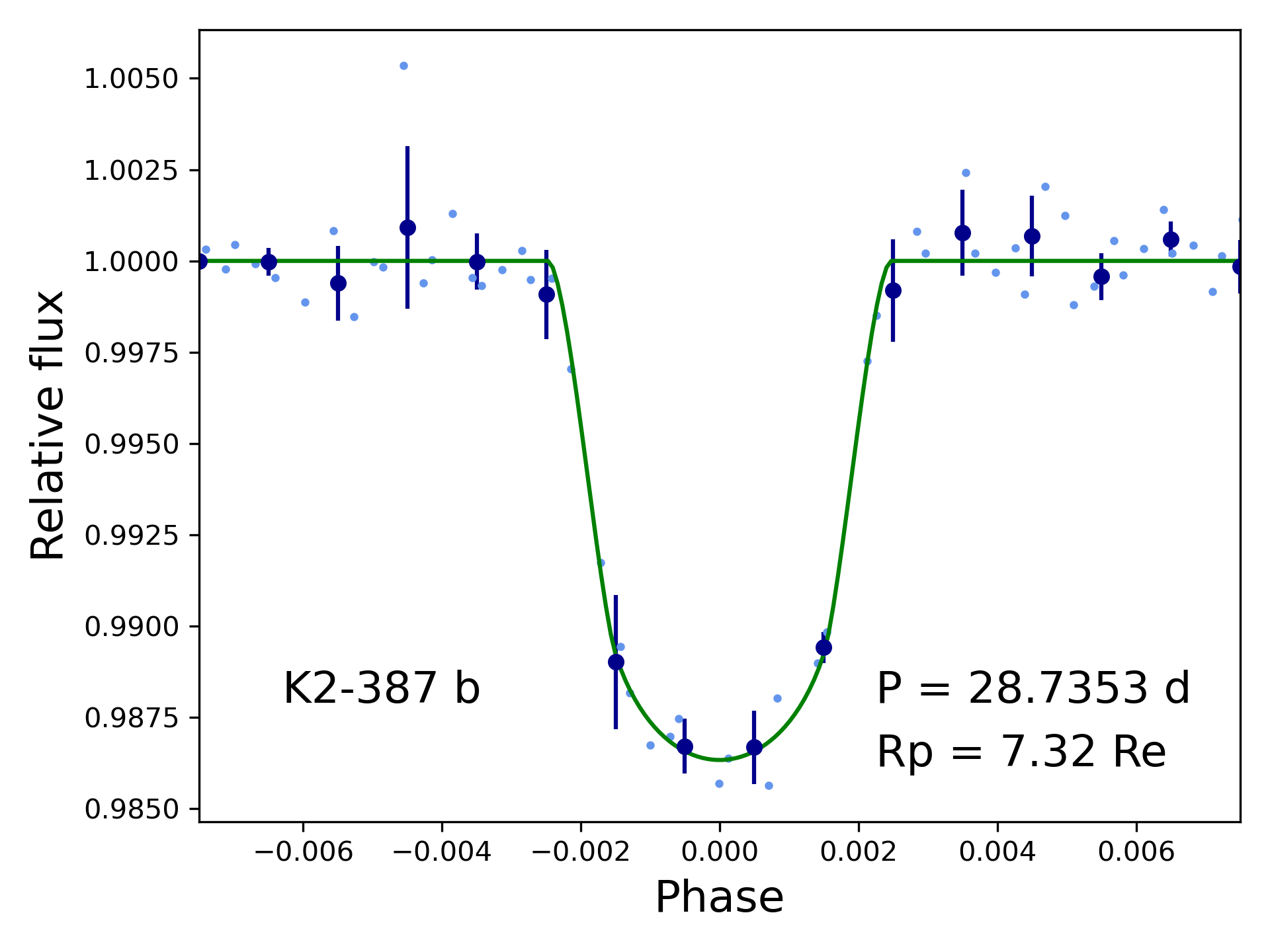}
\includegraphics[width=0.325\textwidth]{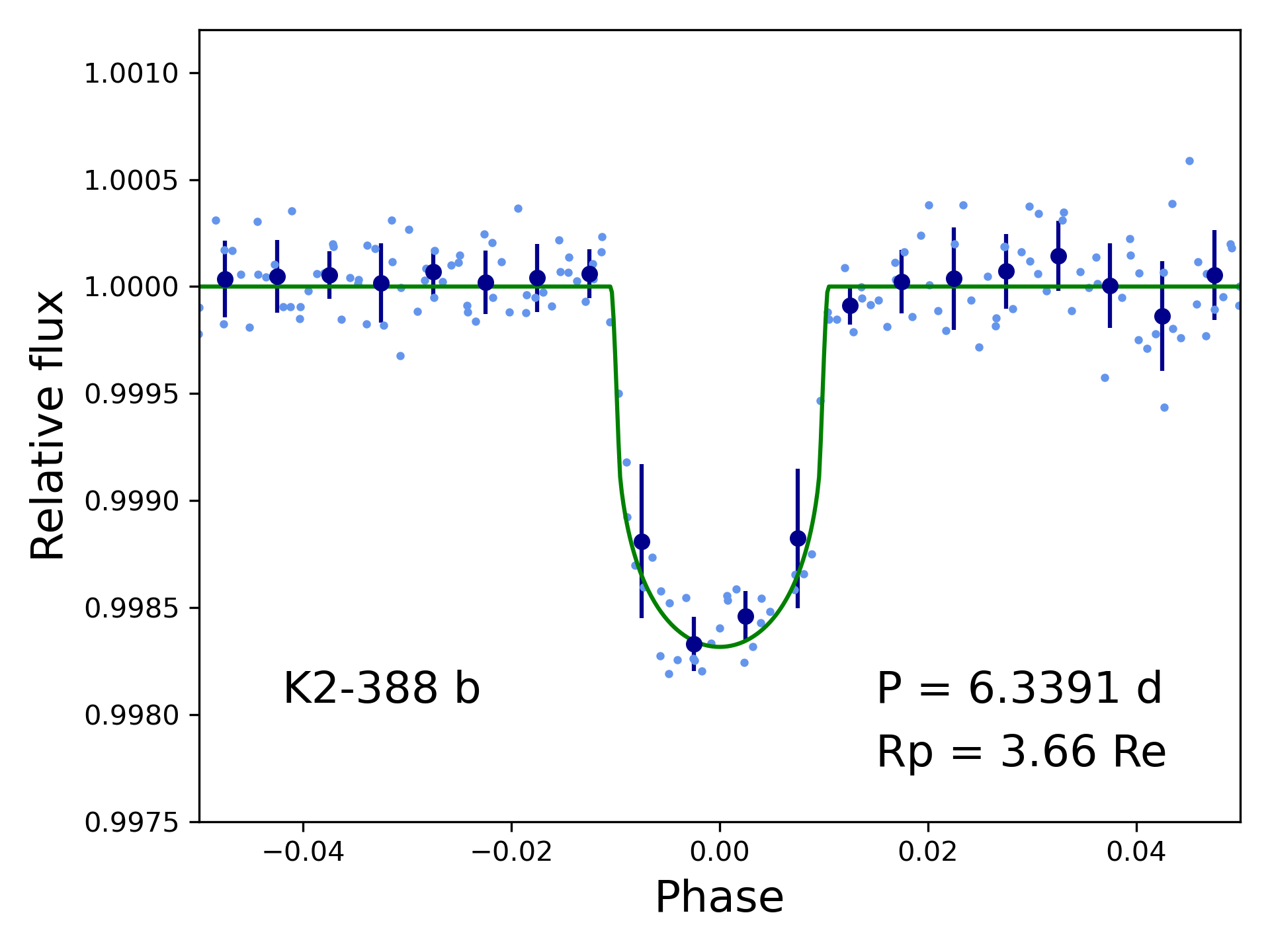}
\includegraphics[width=0.325\textwidth]{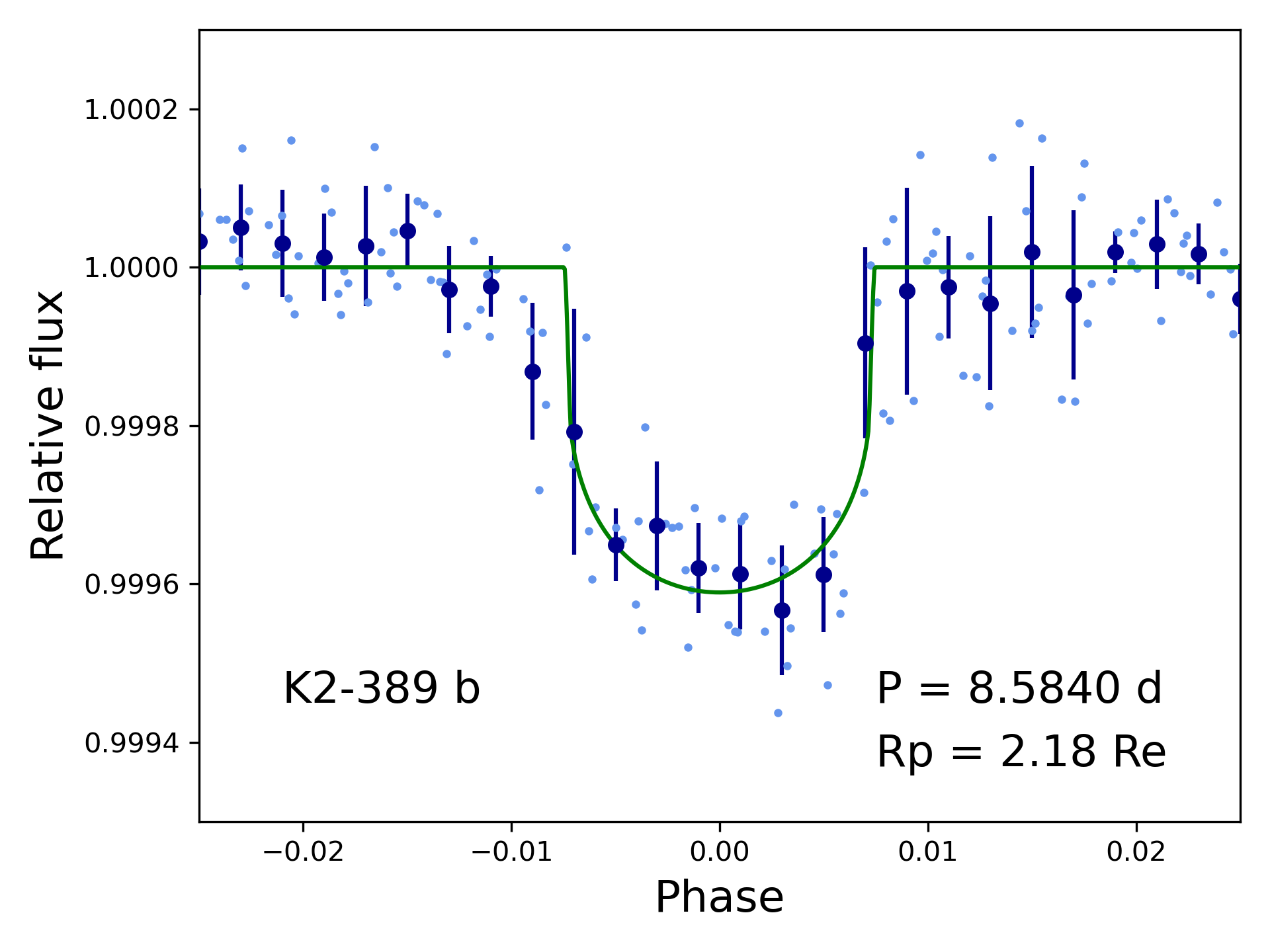}
\includegraphics[width=0.325\textwidth]{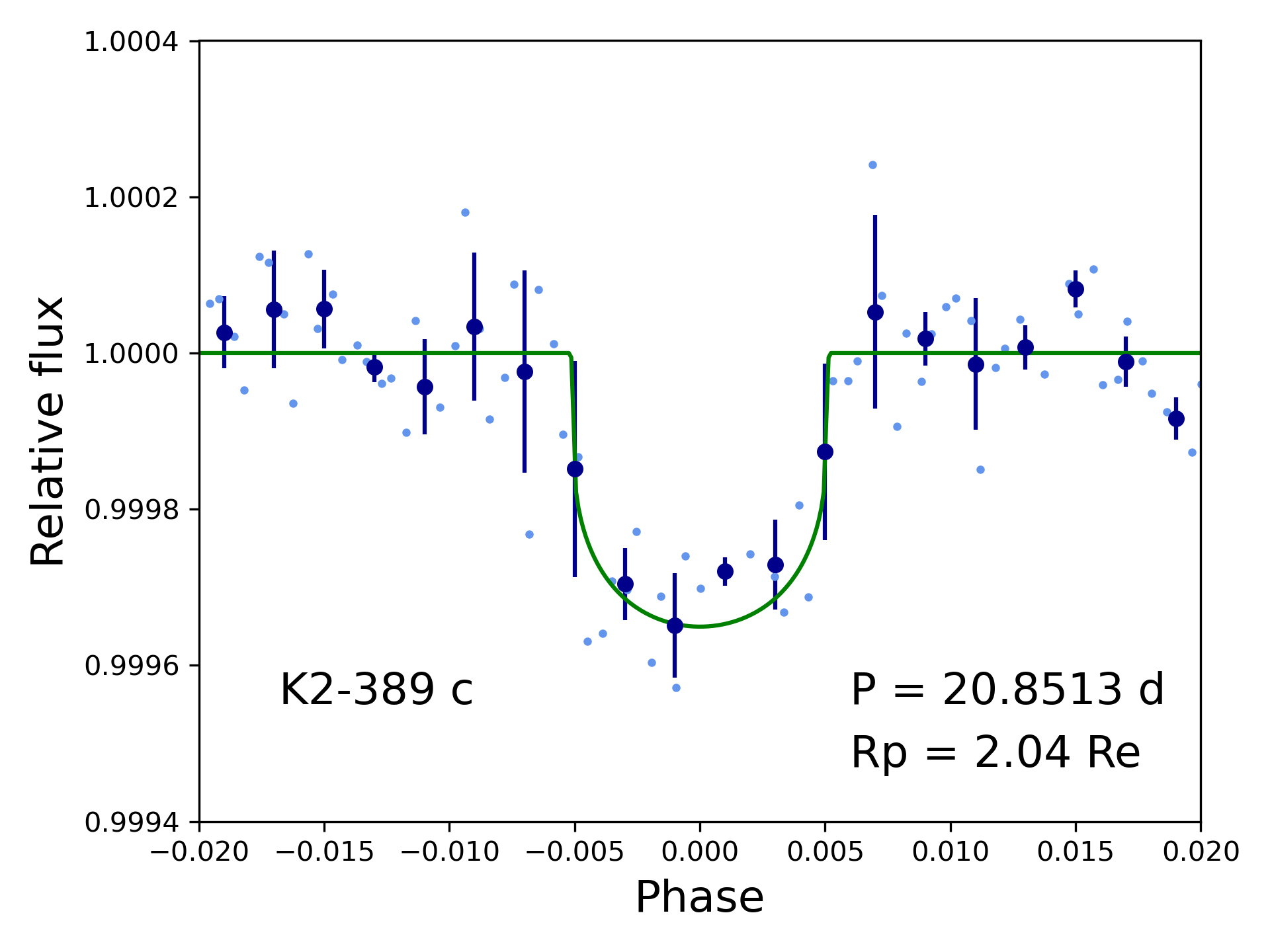}
\includegraphics[width=0.325\textwidth]{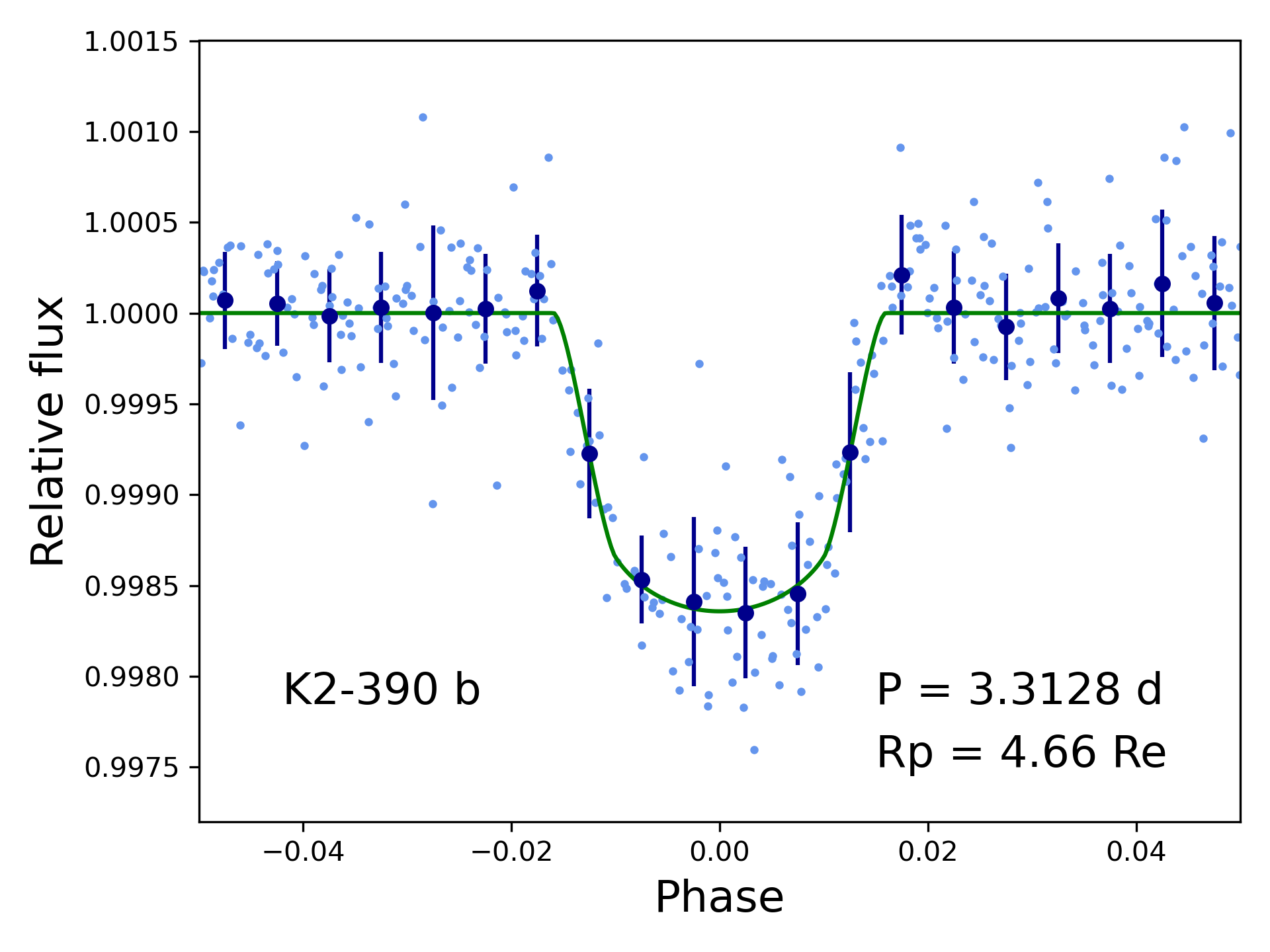}
\includegraphics[width=0.325\textwidth]{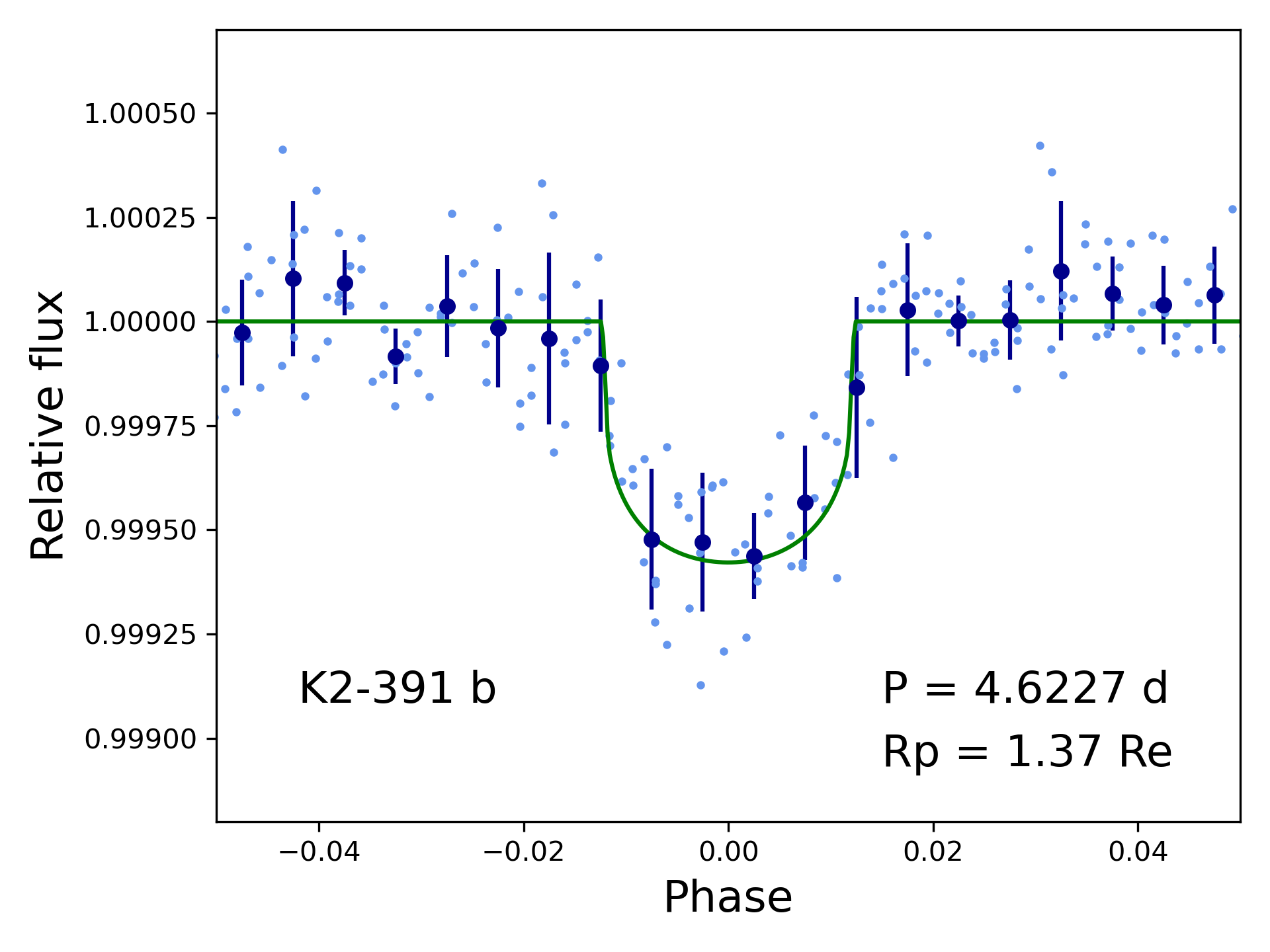}
\includegraphics[width=0.325\textwidth]{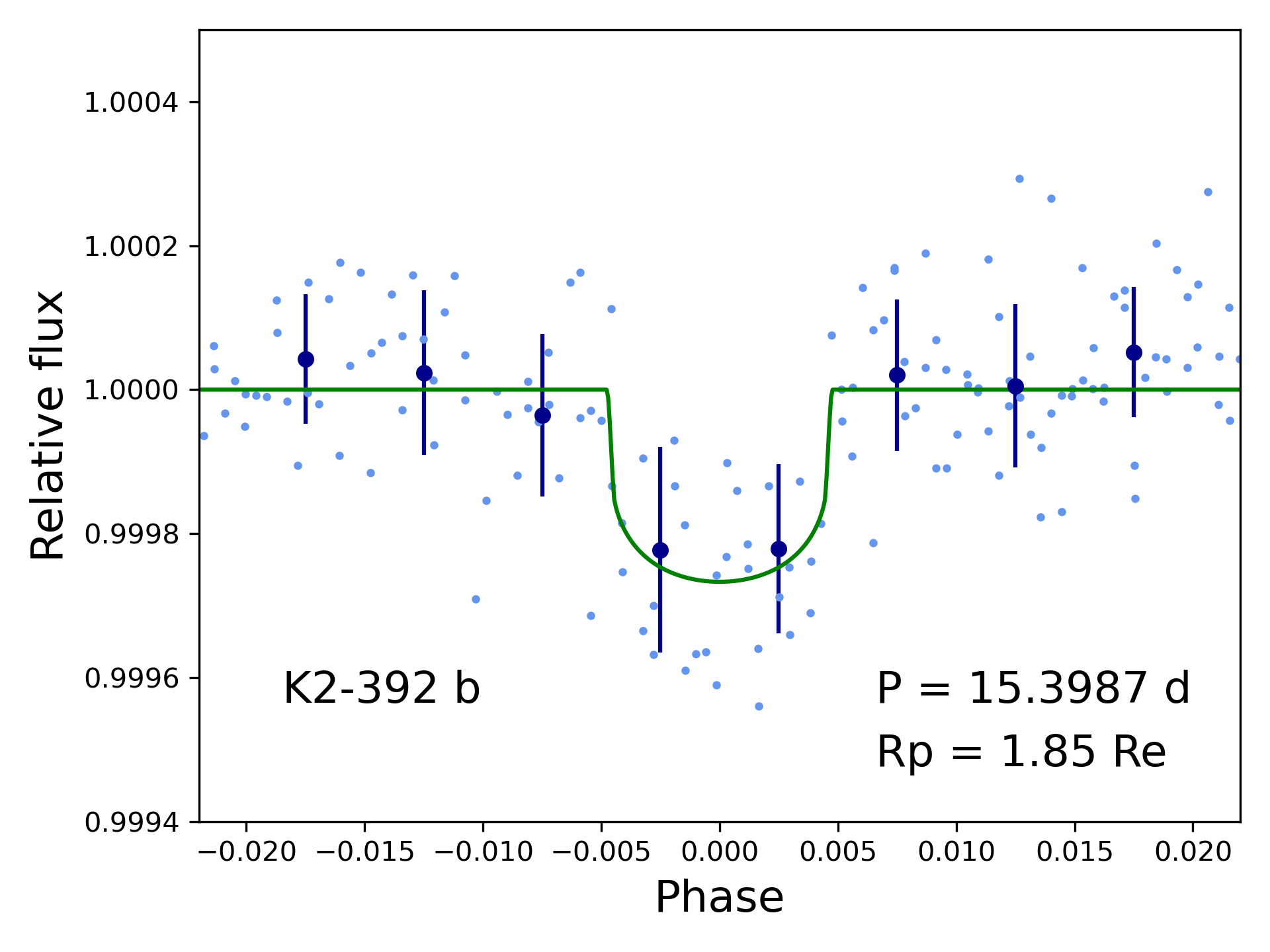}
\includegraphics[width=0.325\textwidth]{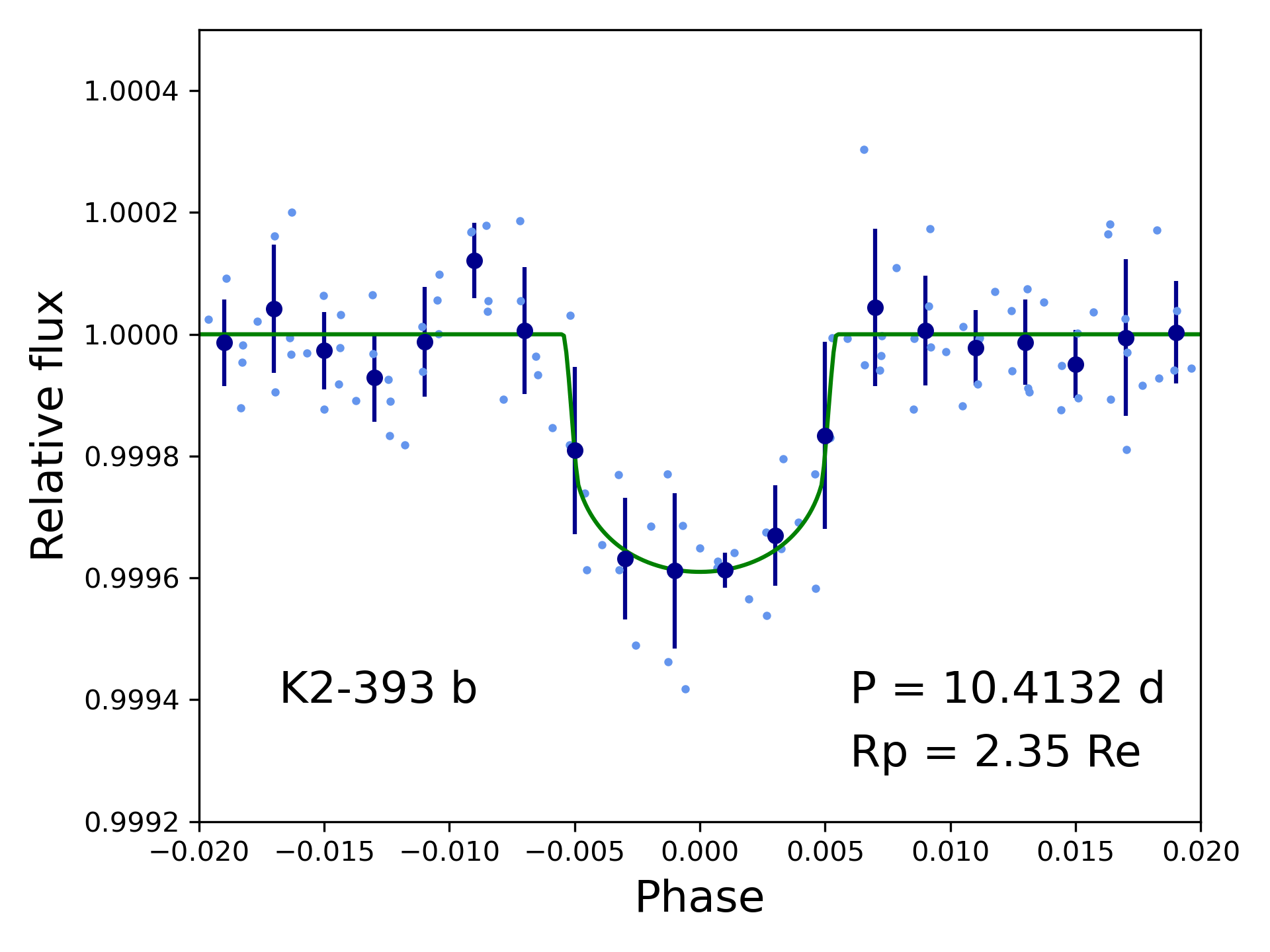}
\includegraphics[width=0.325\textwidth]{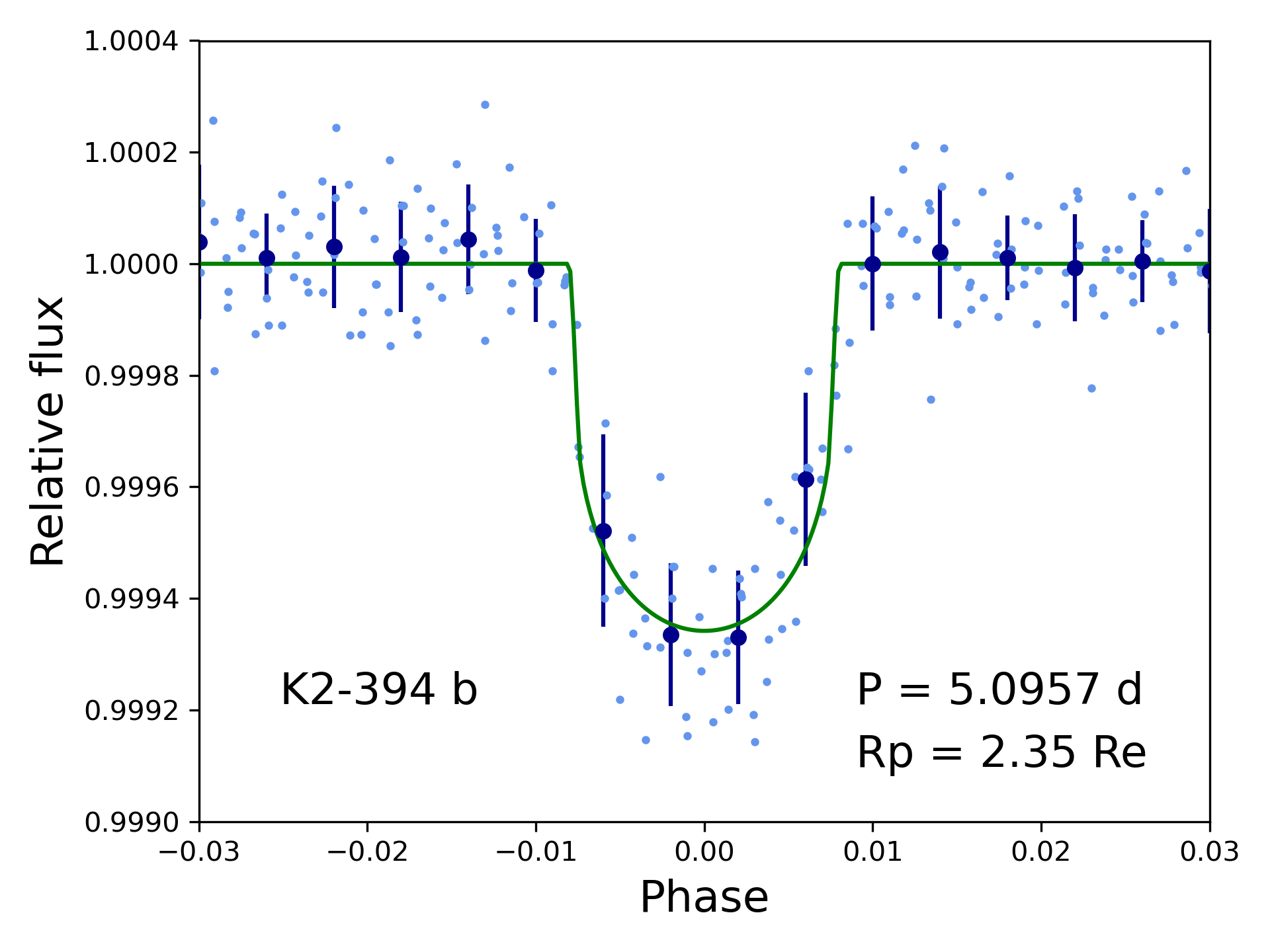}
\includegraphics[width=0.325\textwidth]{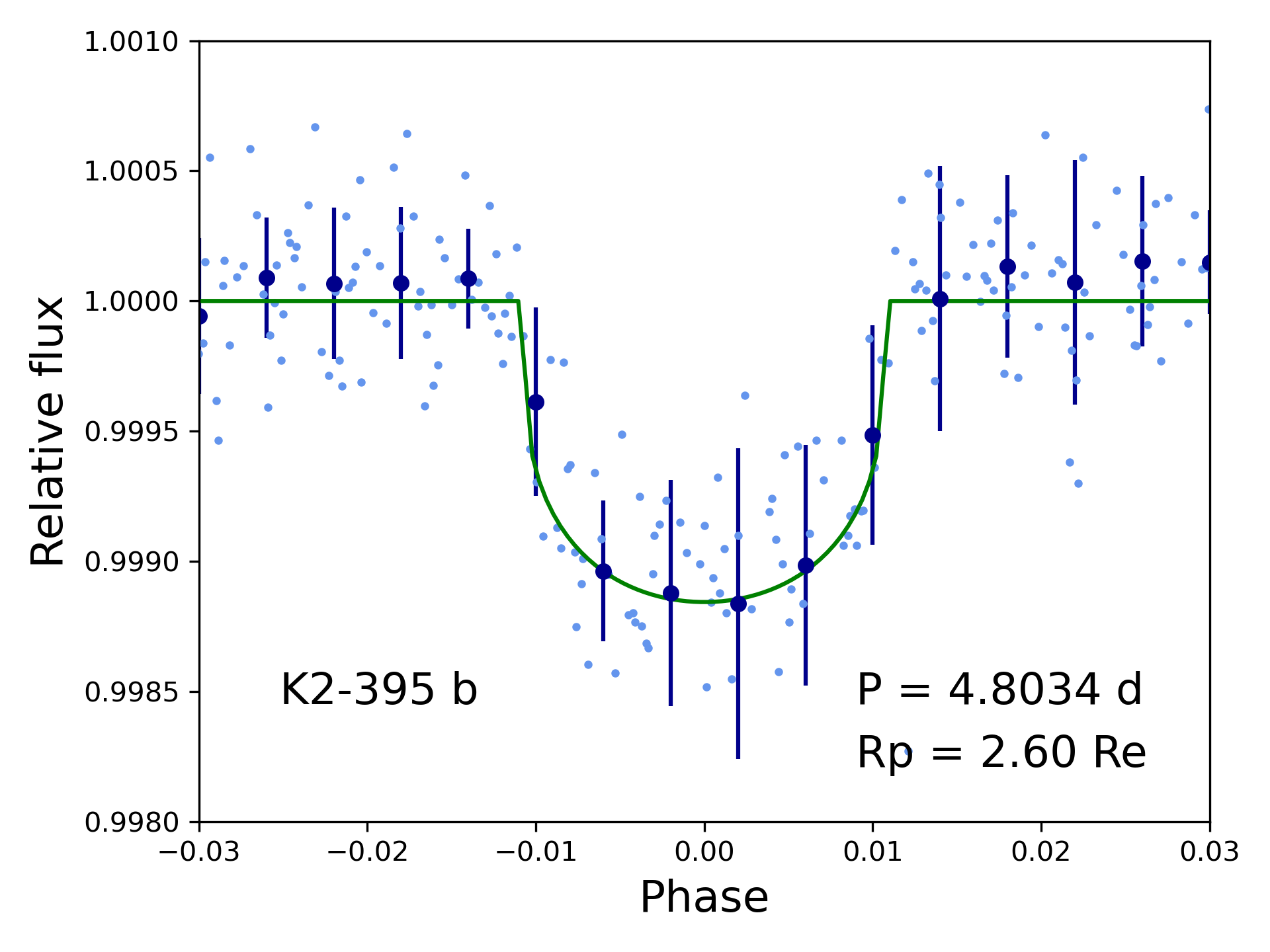}
\includegraphics[width=0.325\textwidth]{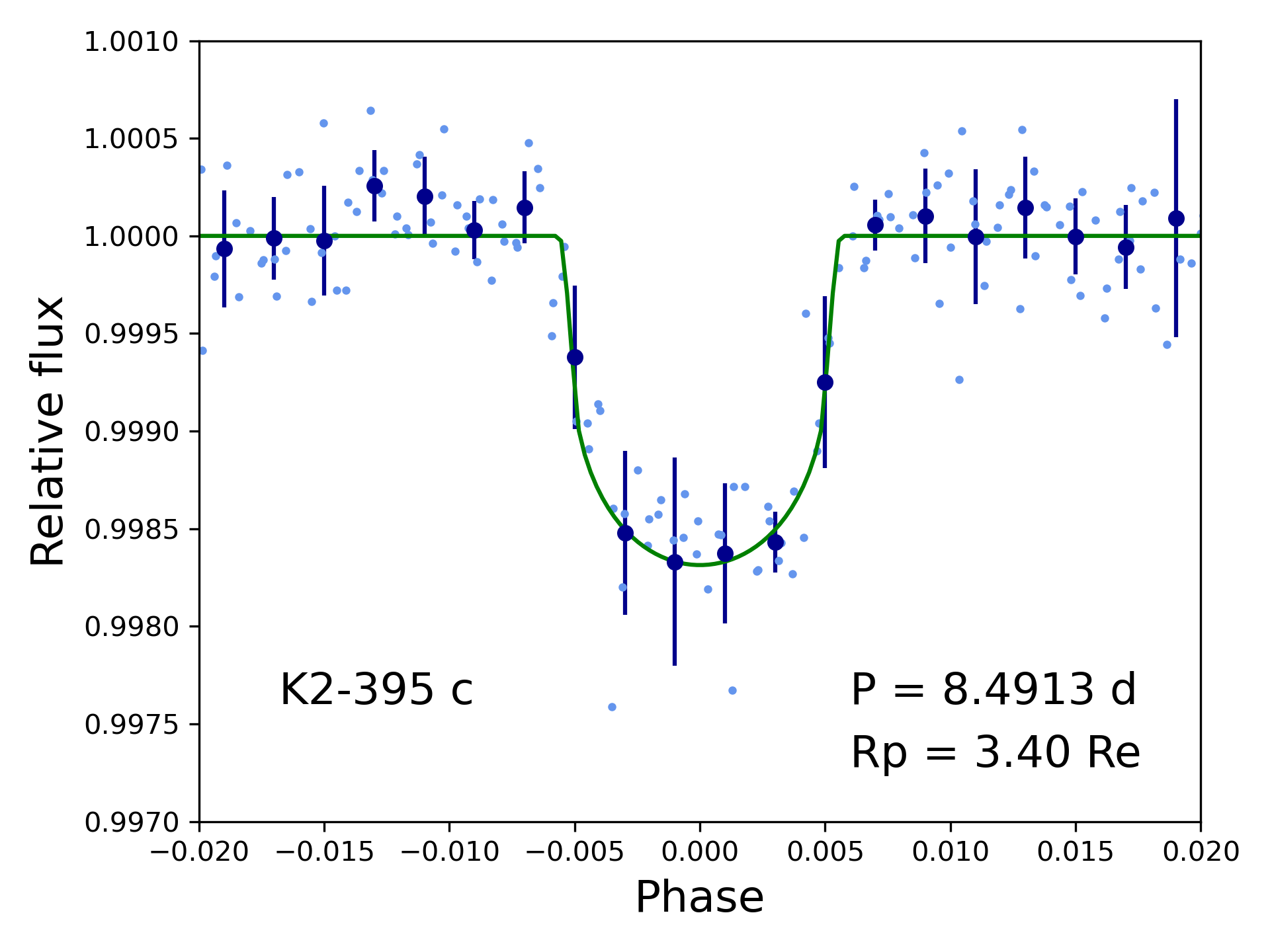}
\includegraphics[width=0.325\textwidth]{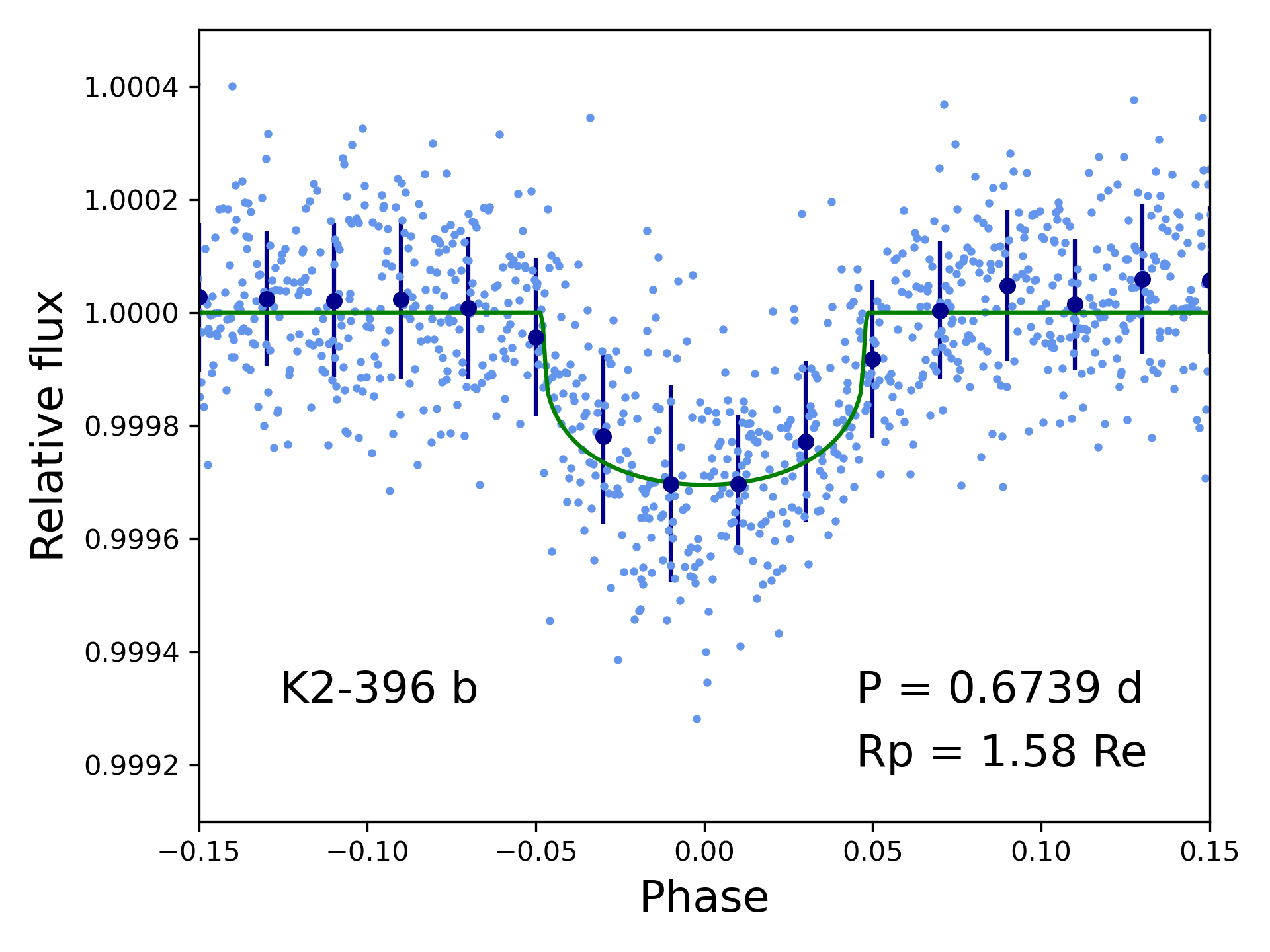}
\includegraphics[width=0.325\textwidth]{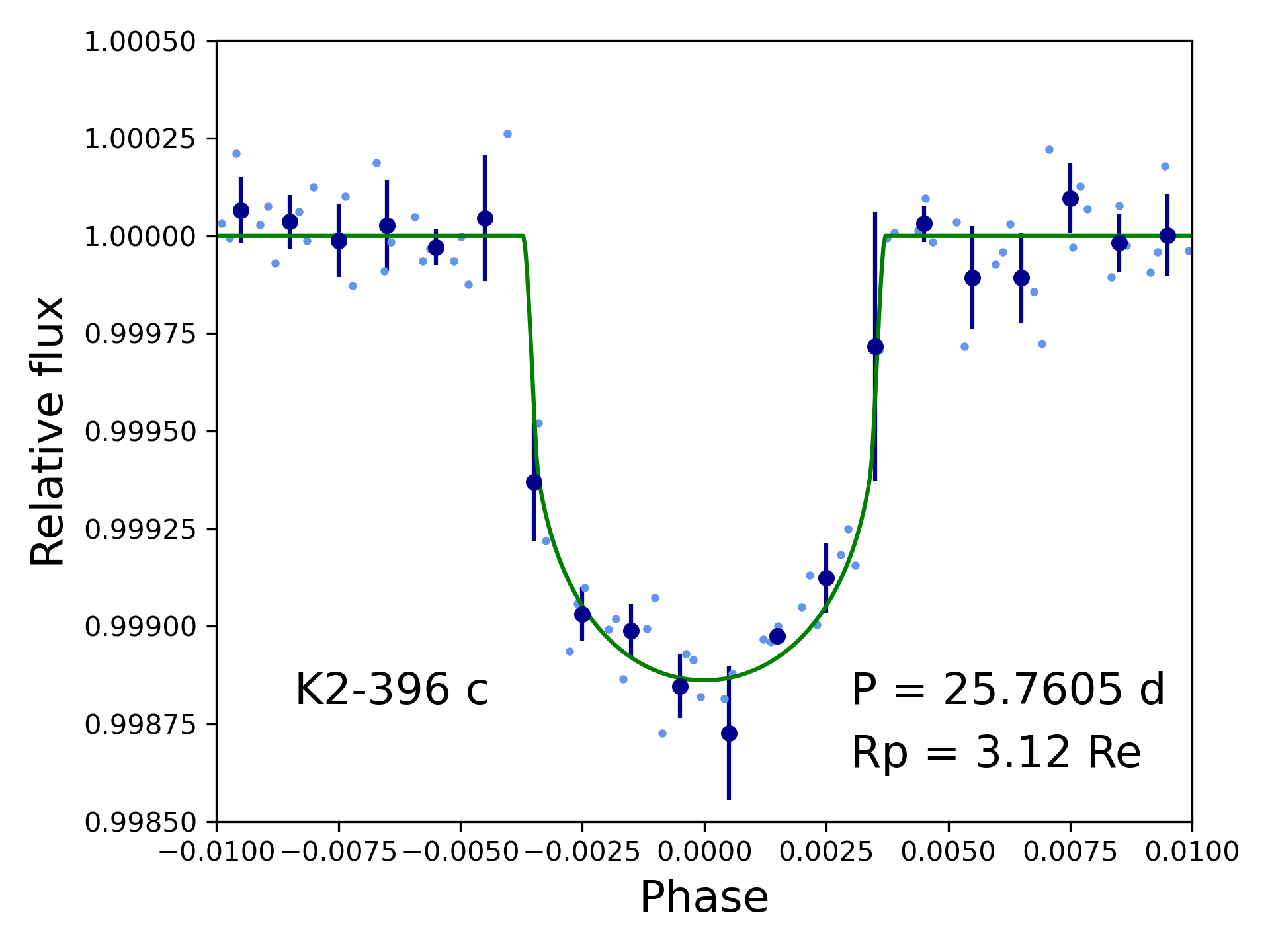}
\caption{Validated planet folded light curves. Description as for Figure \ref{fig:phasedlcs1}.}
\label{fig:phasedlcs3}
\end{figure}

\begin{figure}
\centering
\includegraphics[width=0.325\textwidth]{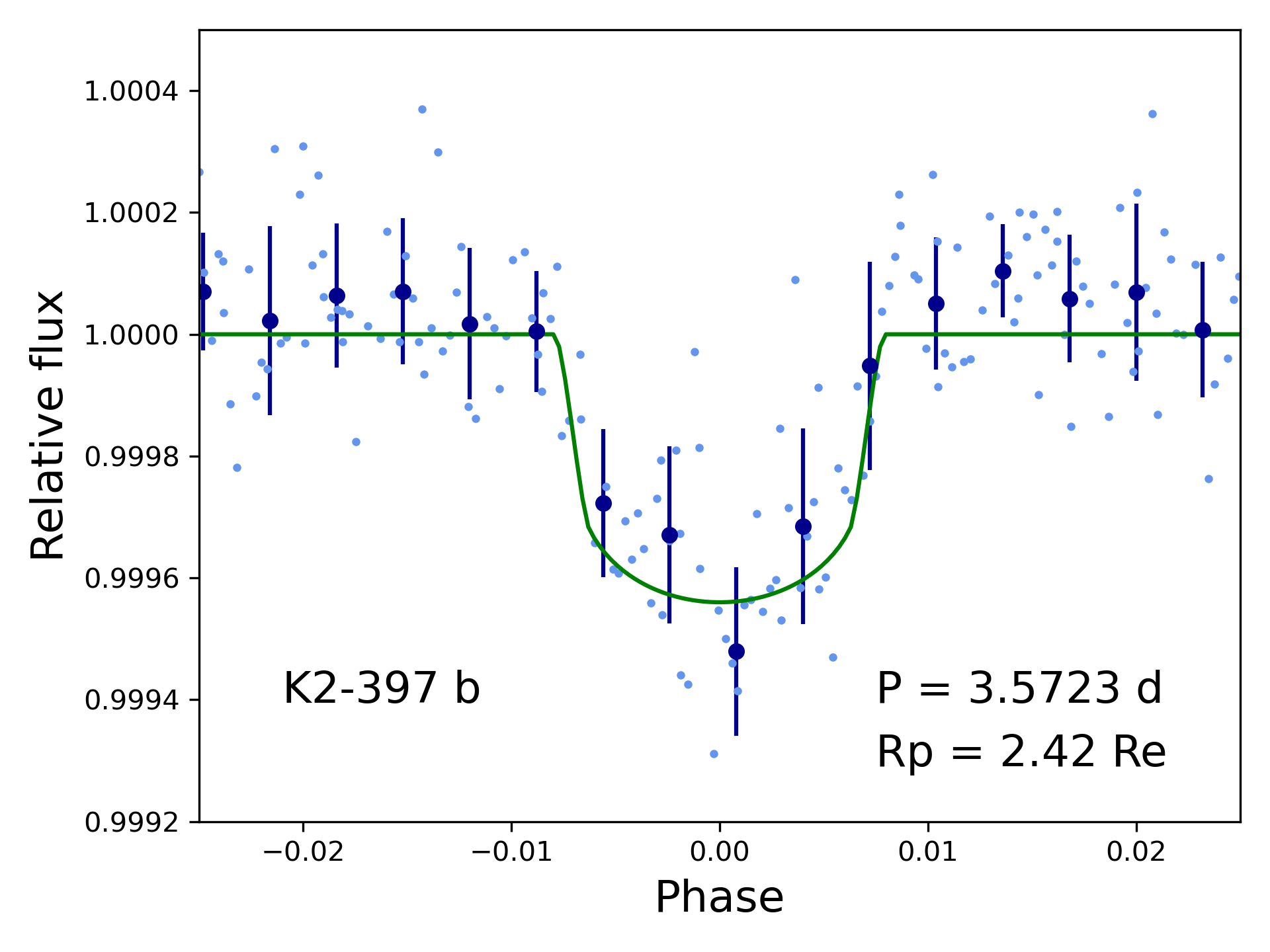}
\includegraphics[width=0.325\textwidth]{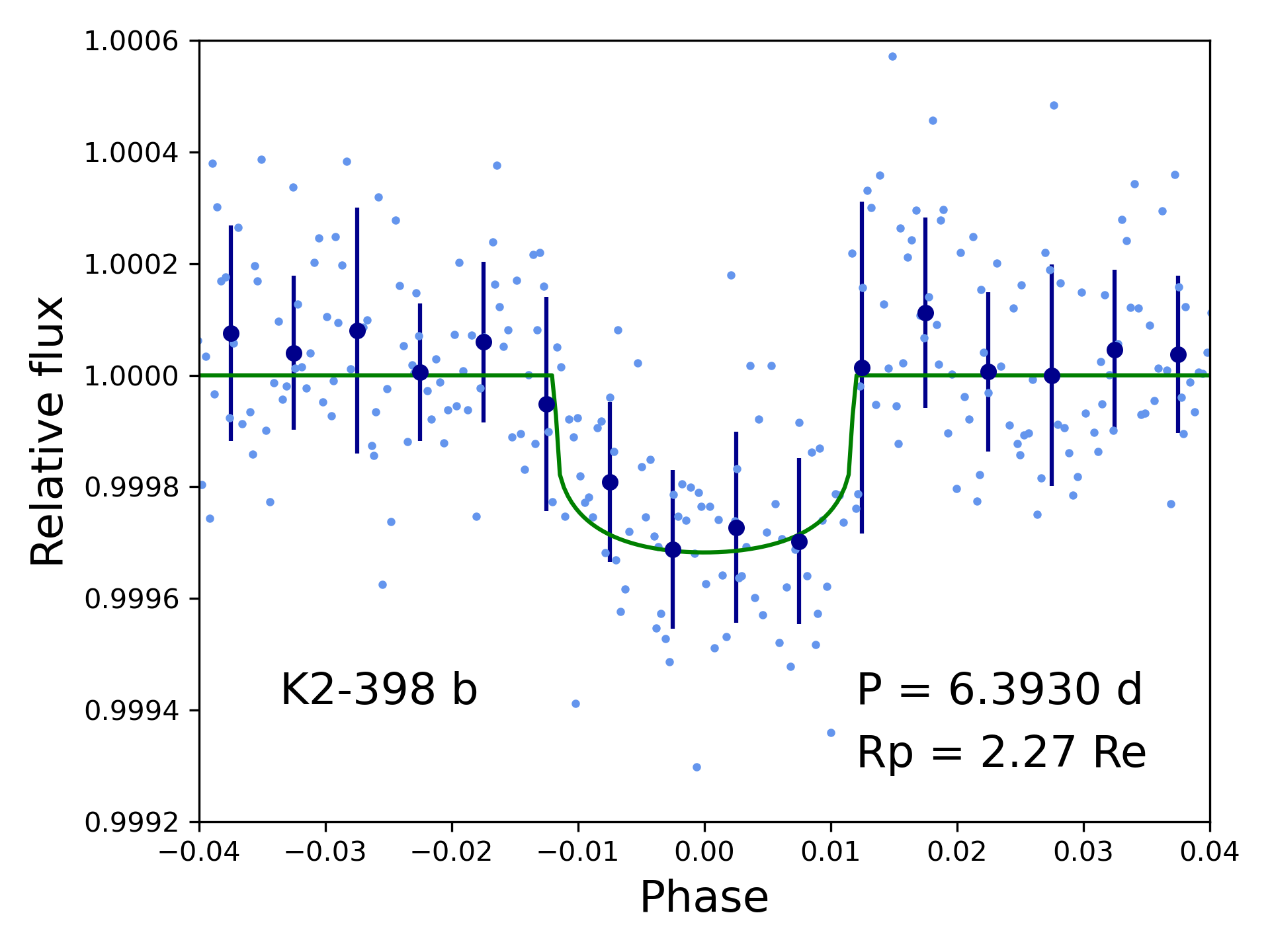}
\includegraphics[width=0.325\textwidth]{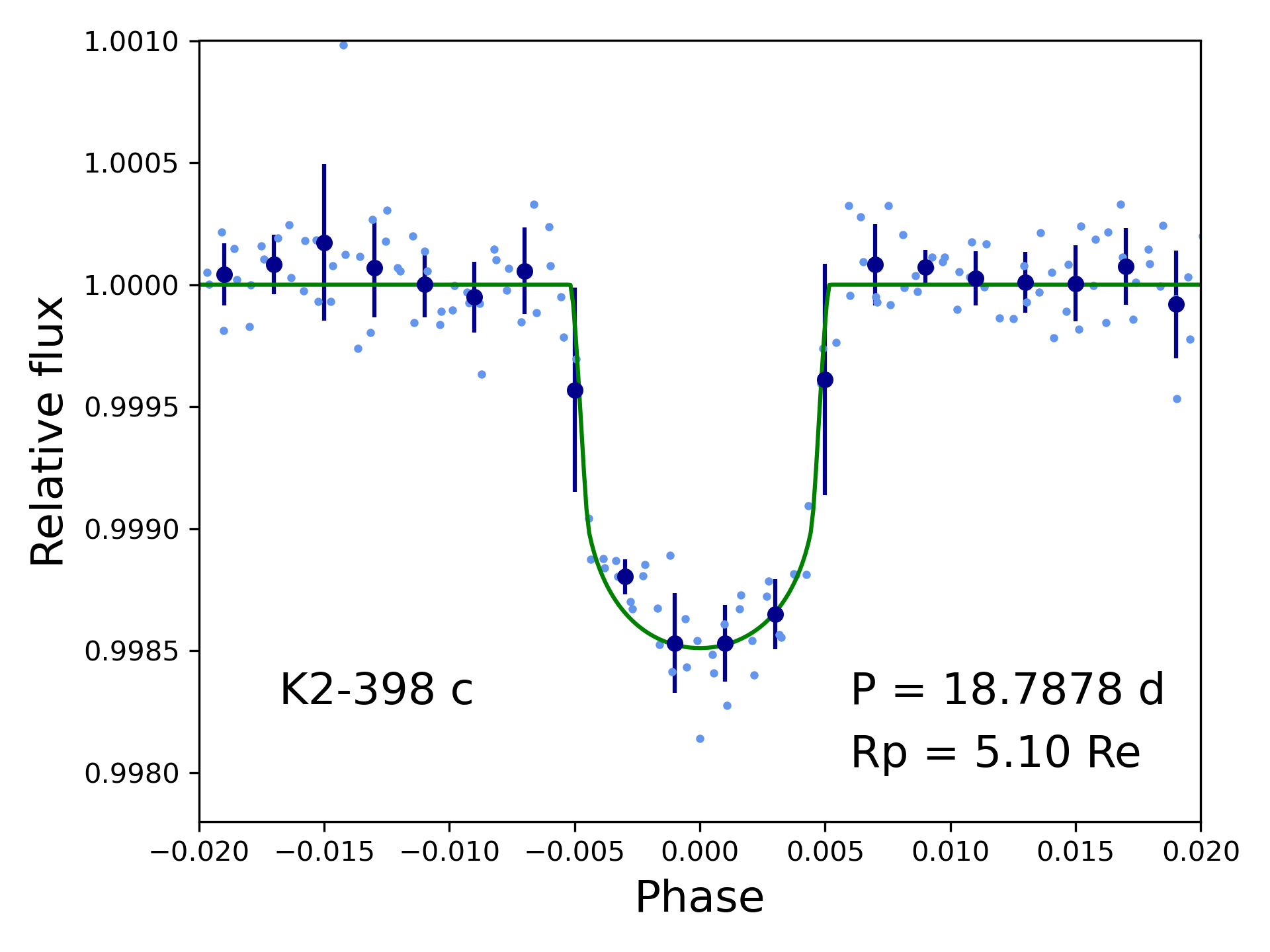}
\includegraphics[width=0.325\textwidth]{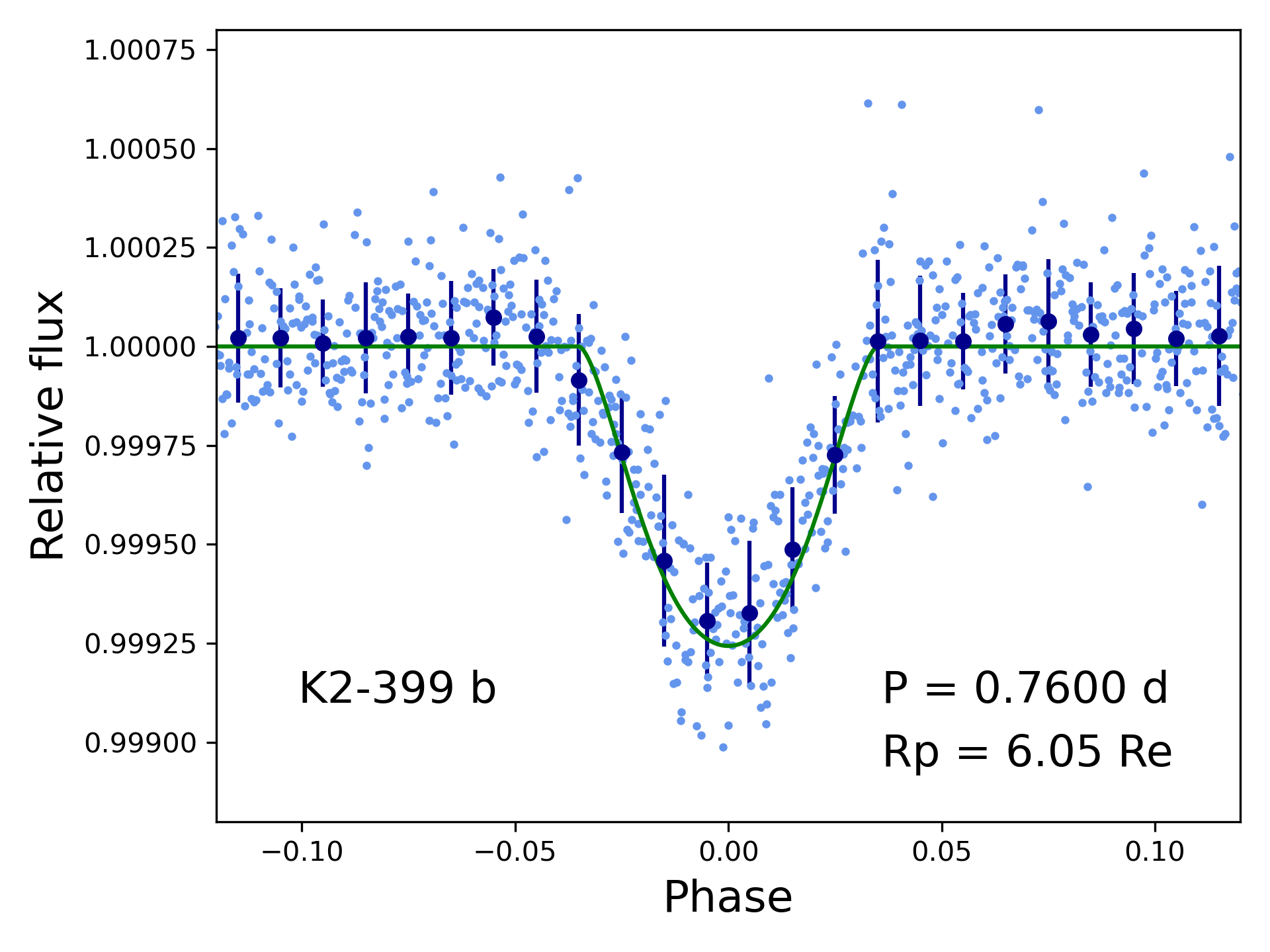}
\includegraphics[width=0.325\textwidth]{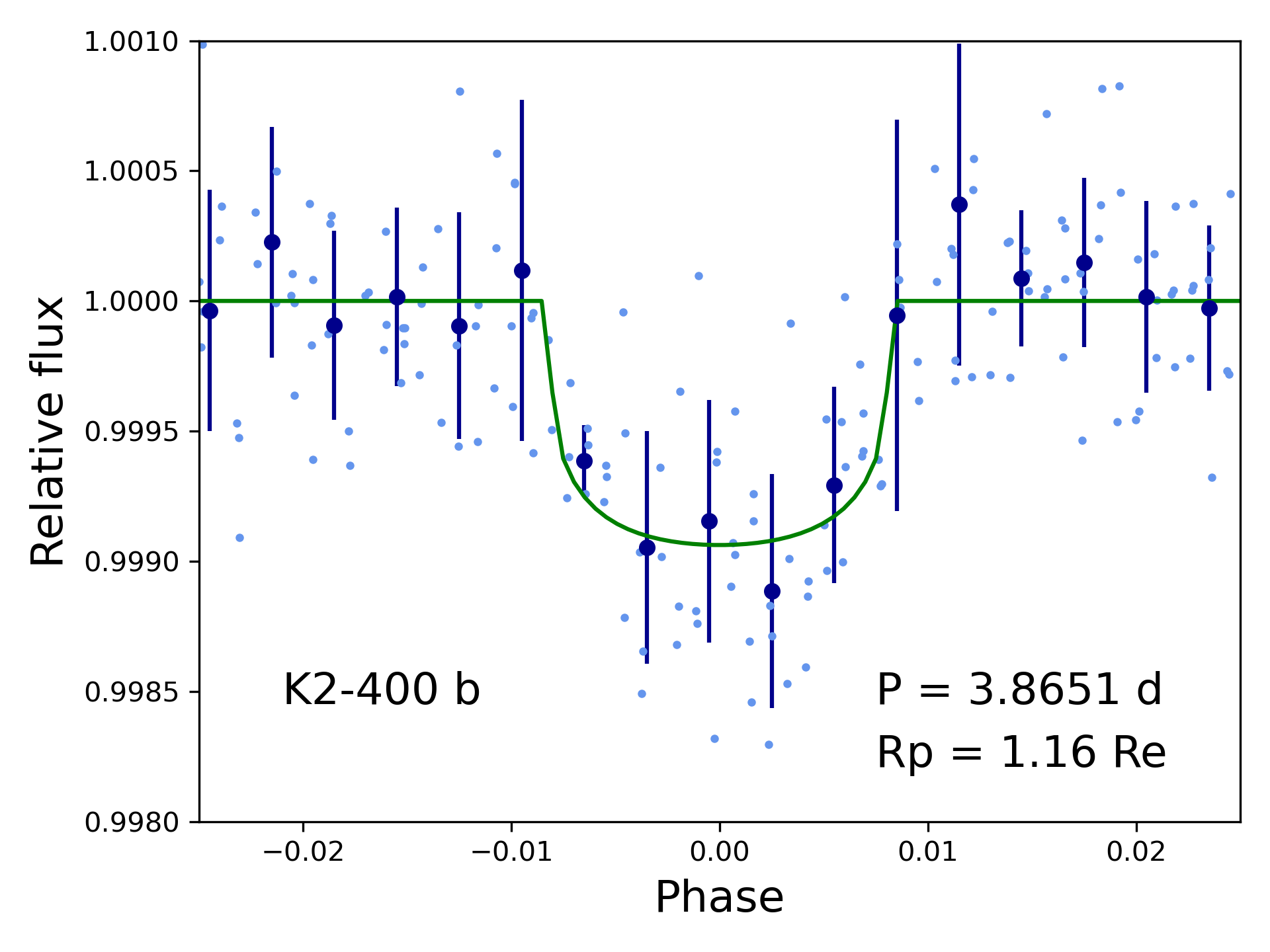}
\includegraphics[width=0.325\textwidth]{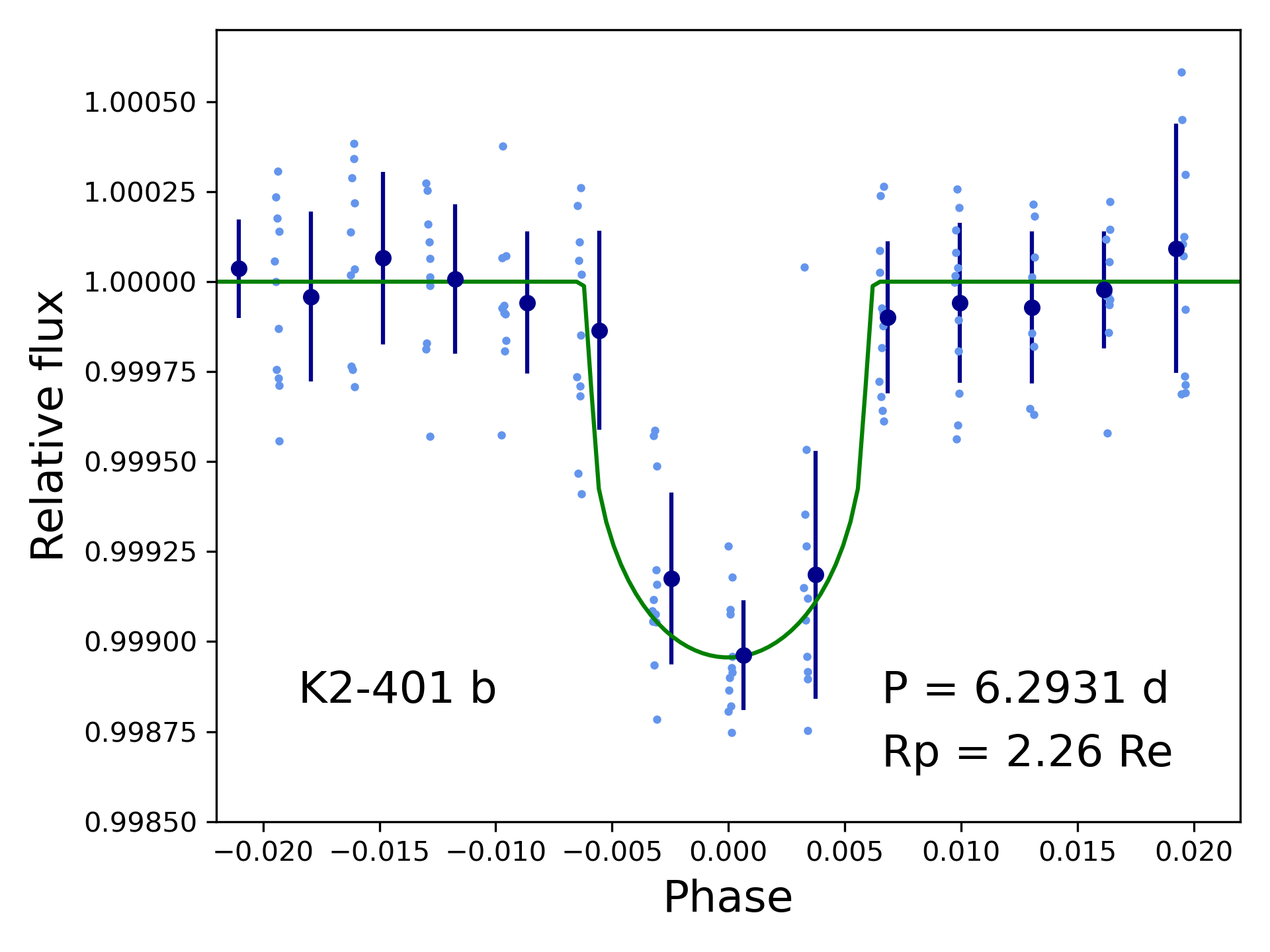}
\includegraphics[width=0.325\textwidth]{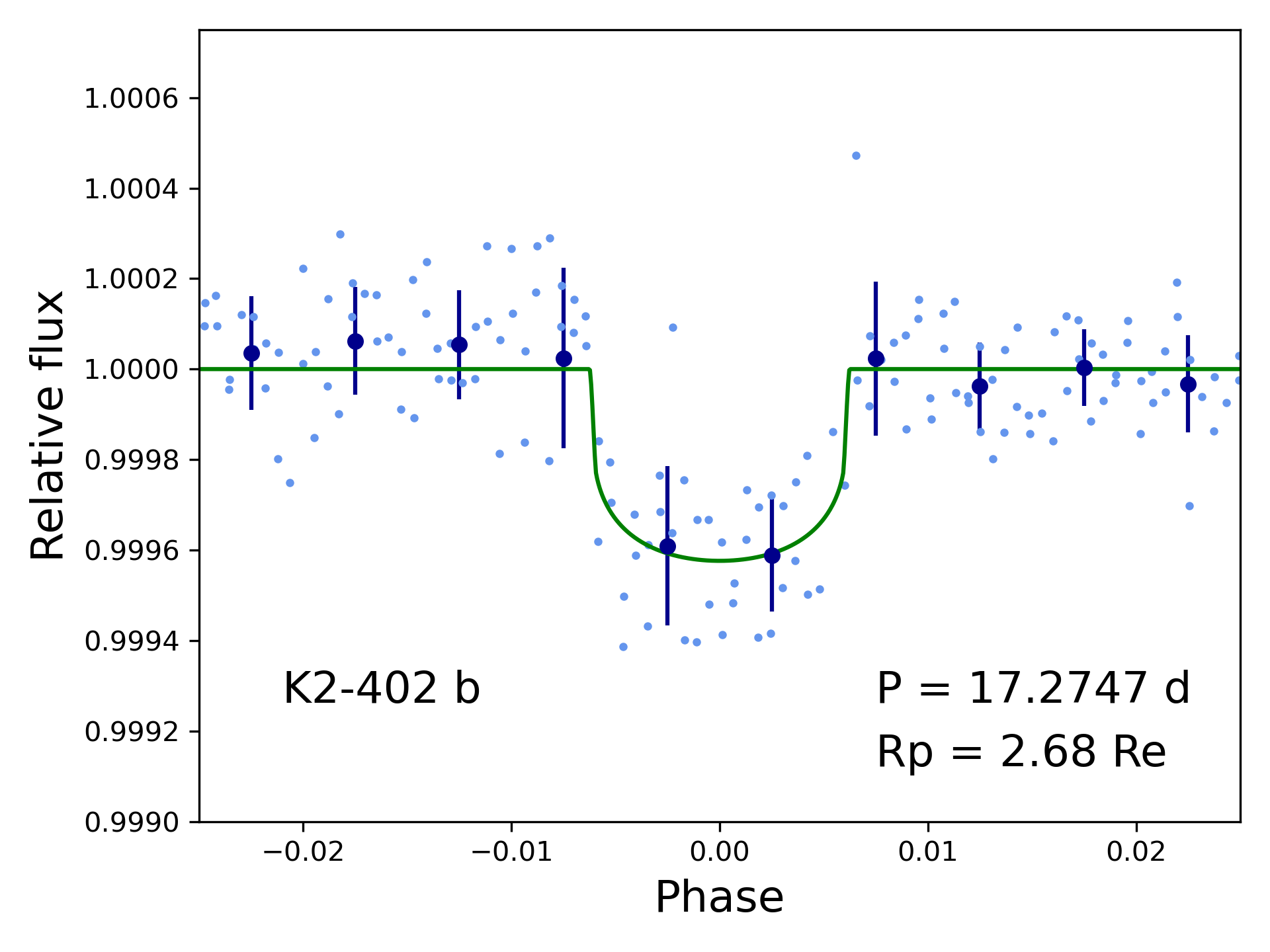}
\includegraphics[width=0.325\textwidth]{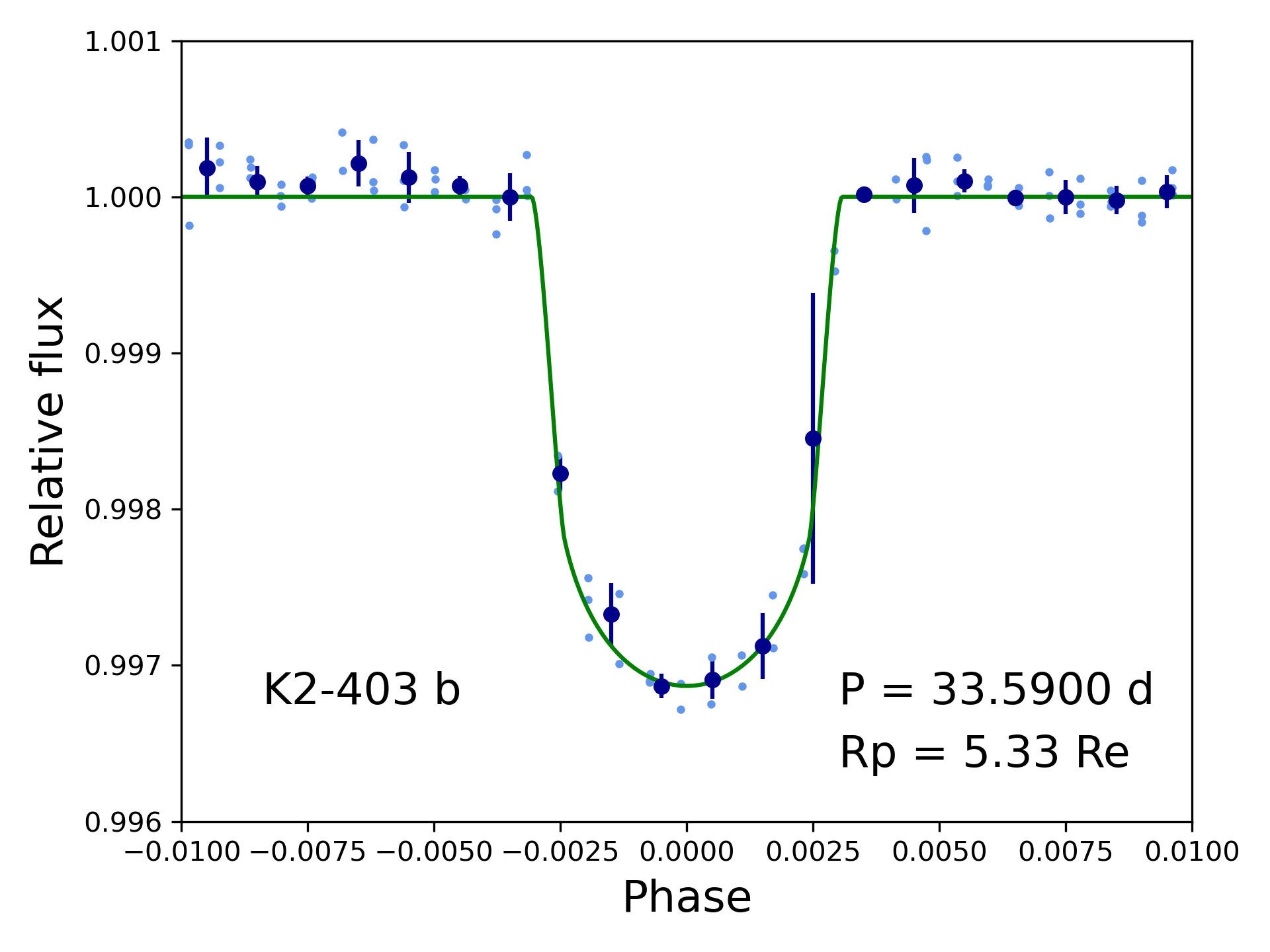}
\includegraphics[width=0.325\textwidth]{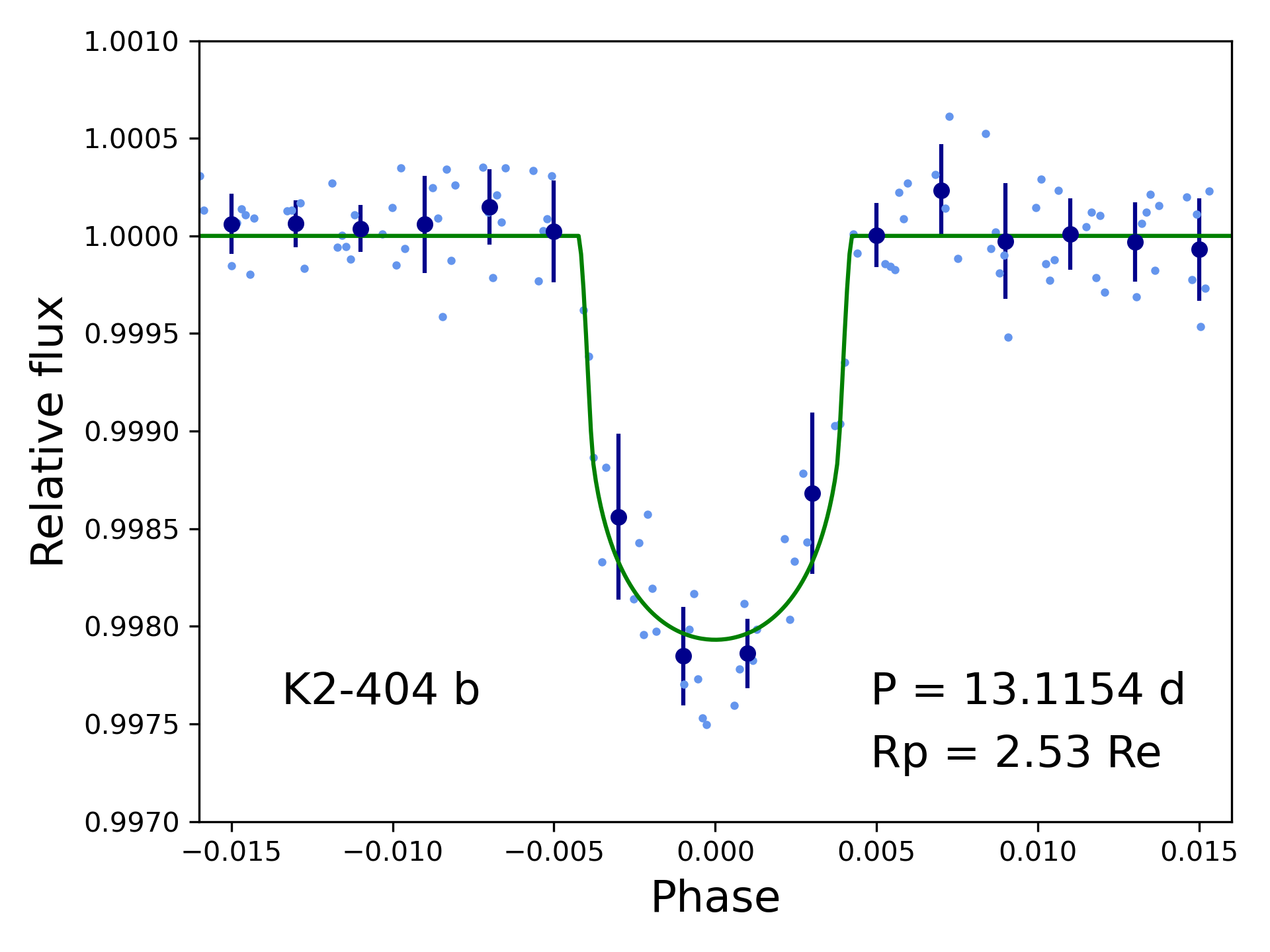}
\includegraphics[width=0.325\textwidth]{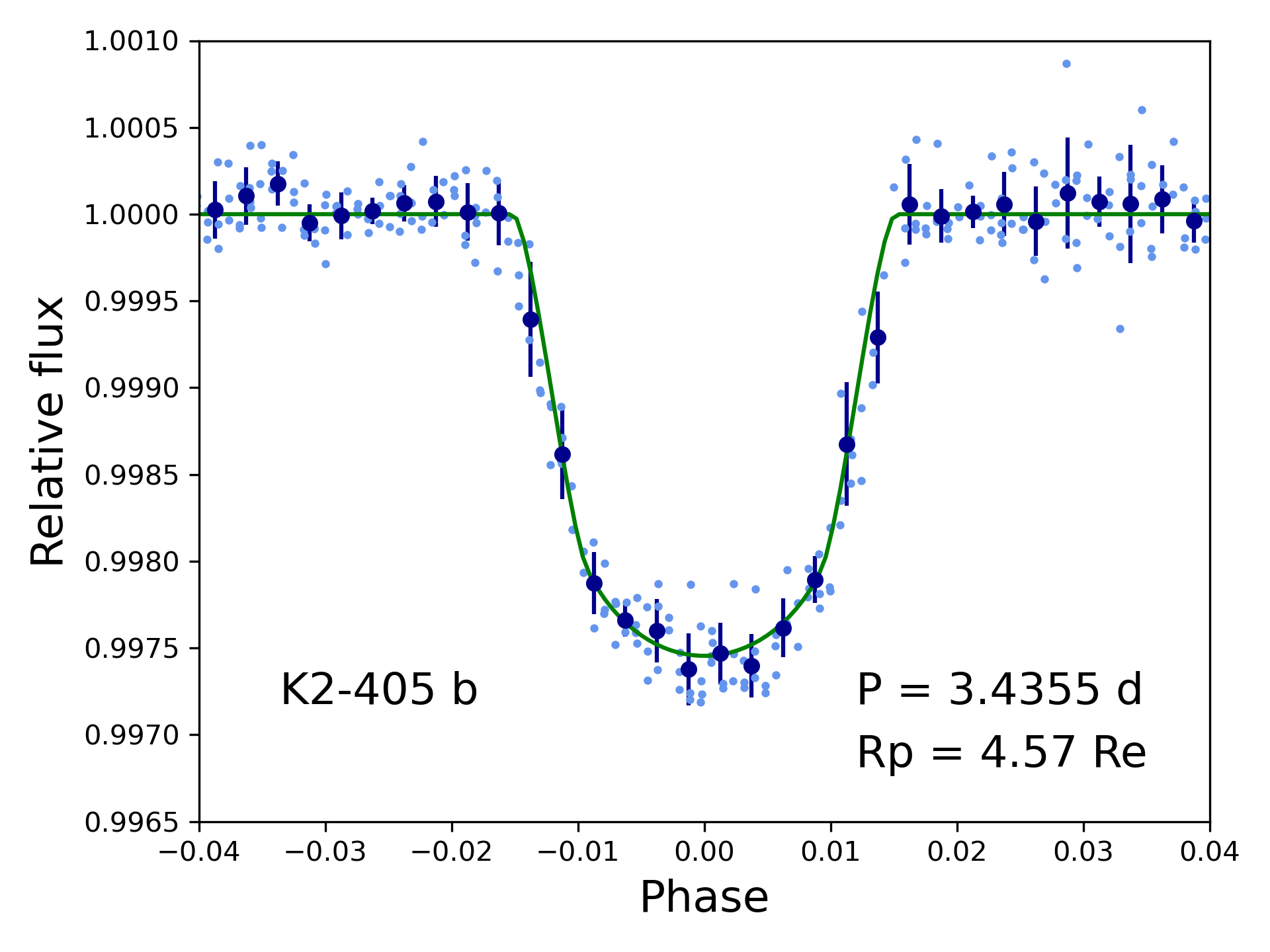}`
\includegraphics[width=0.325\textwidth]{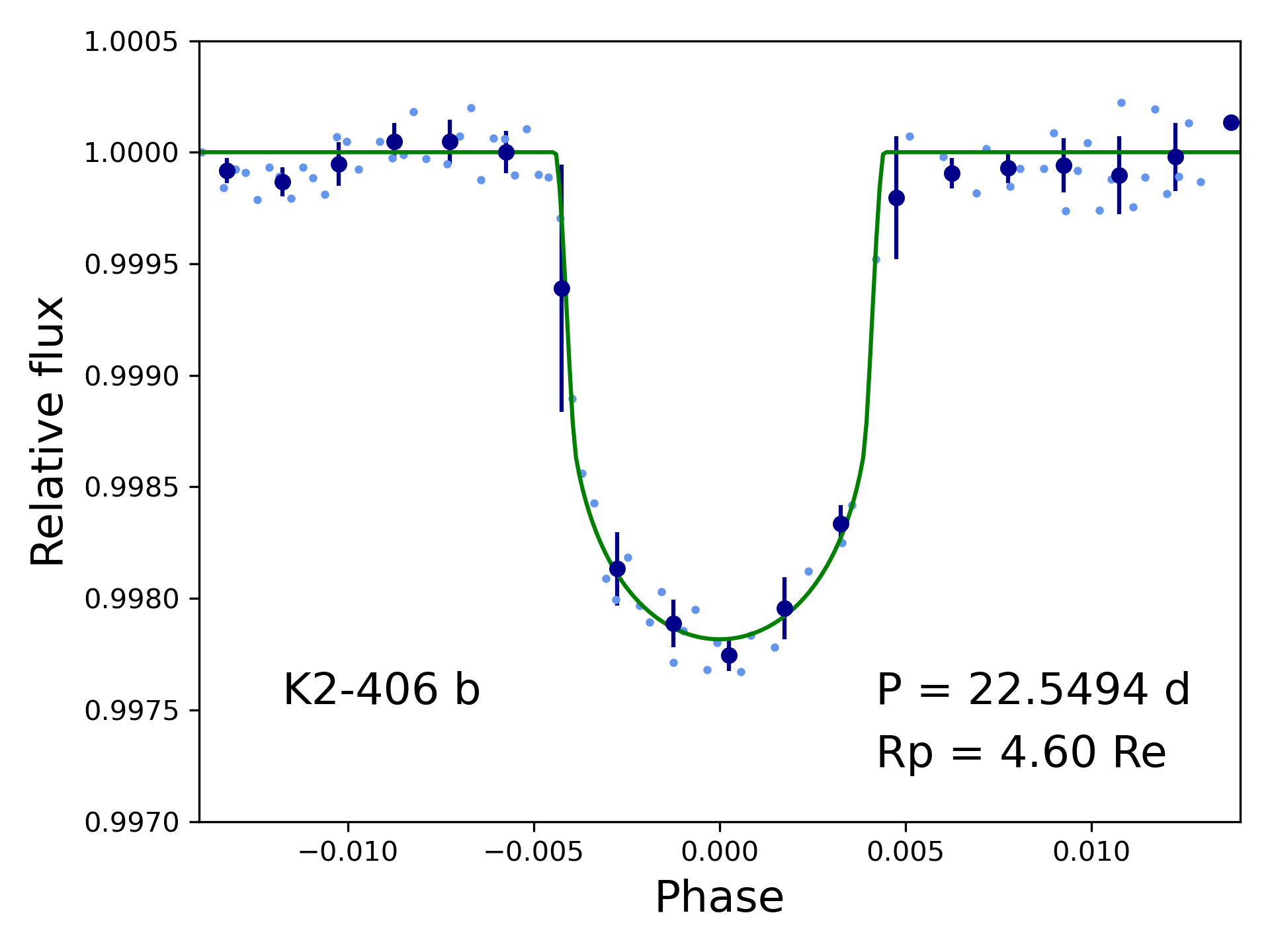}
\includegraphics[width=0.325\textwidth]{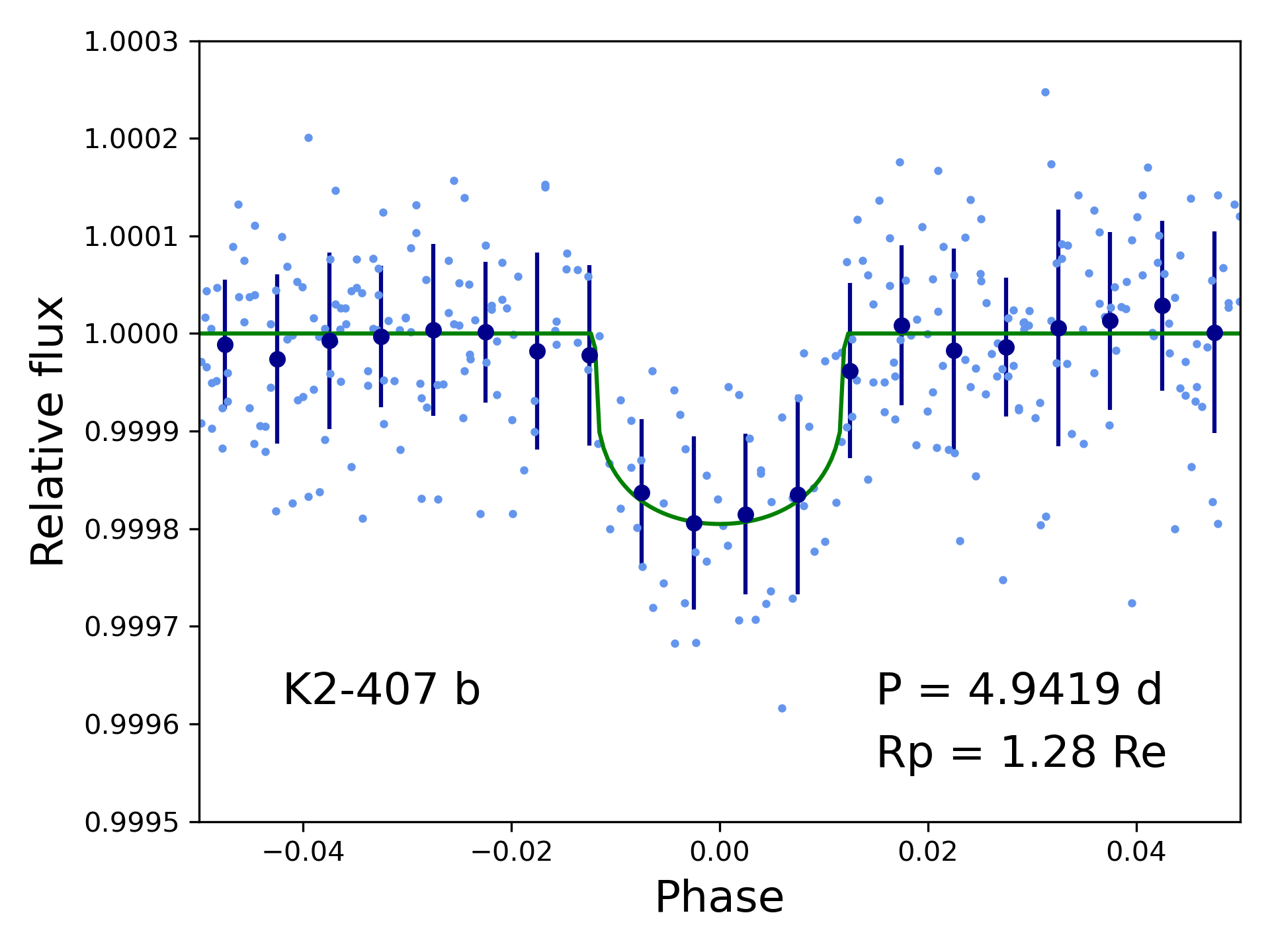}
\includegraphics[width=0.325\textwidth]{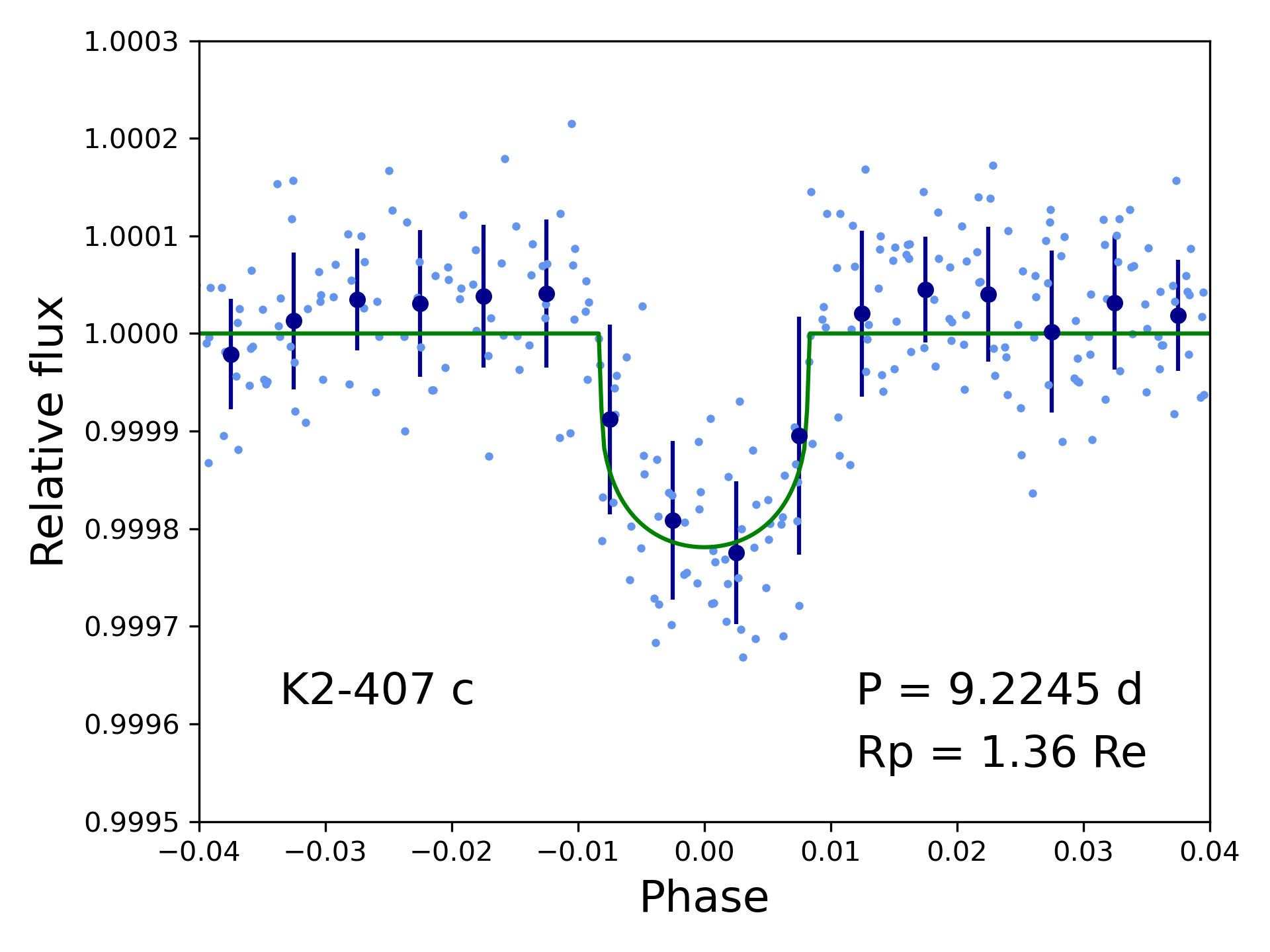}
\includegraphics[width=0.325\textwidth]{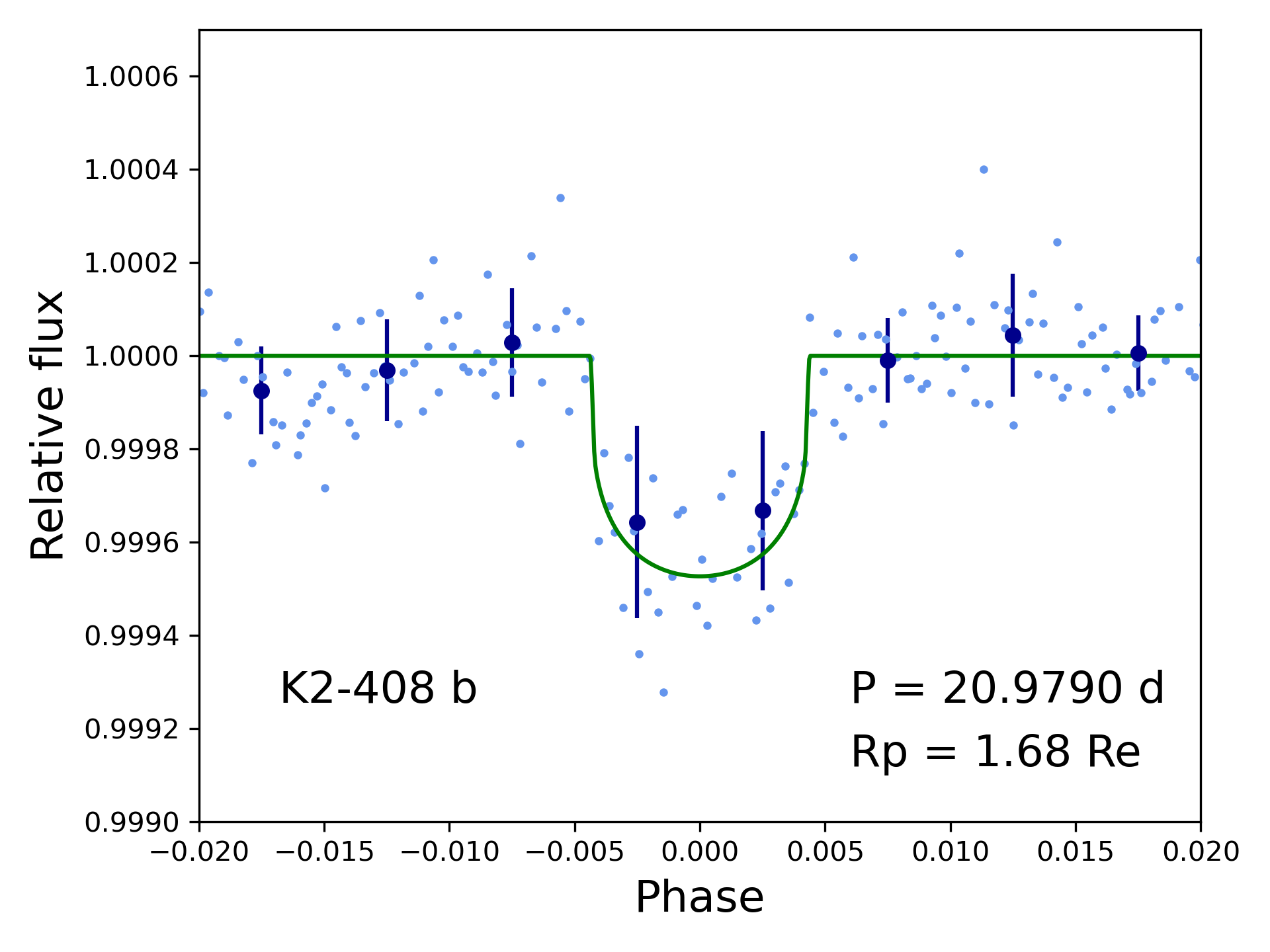}
\includegraphics[width=0.325\textwidth]{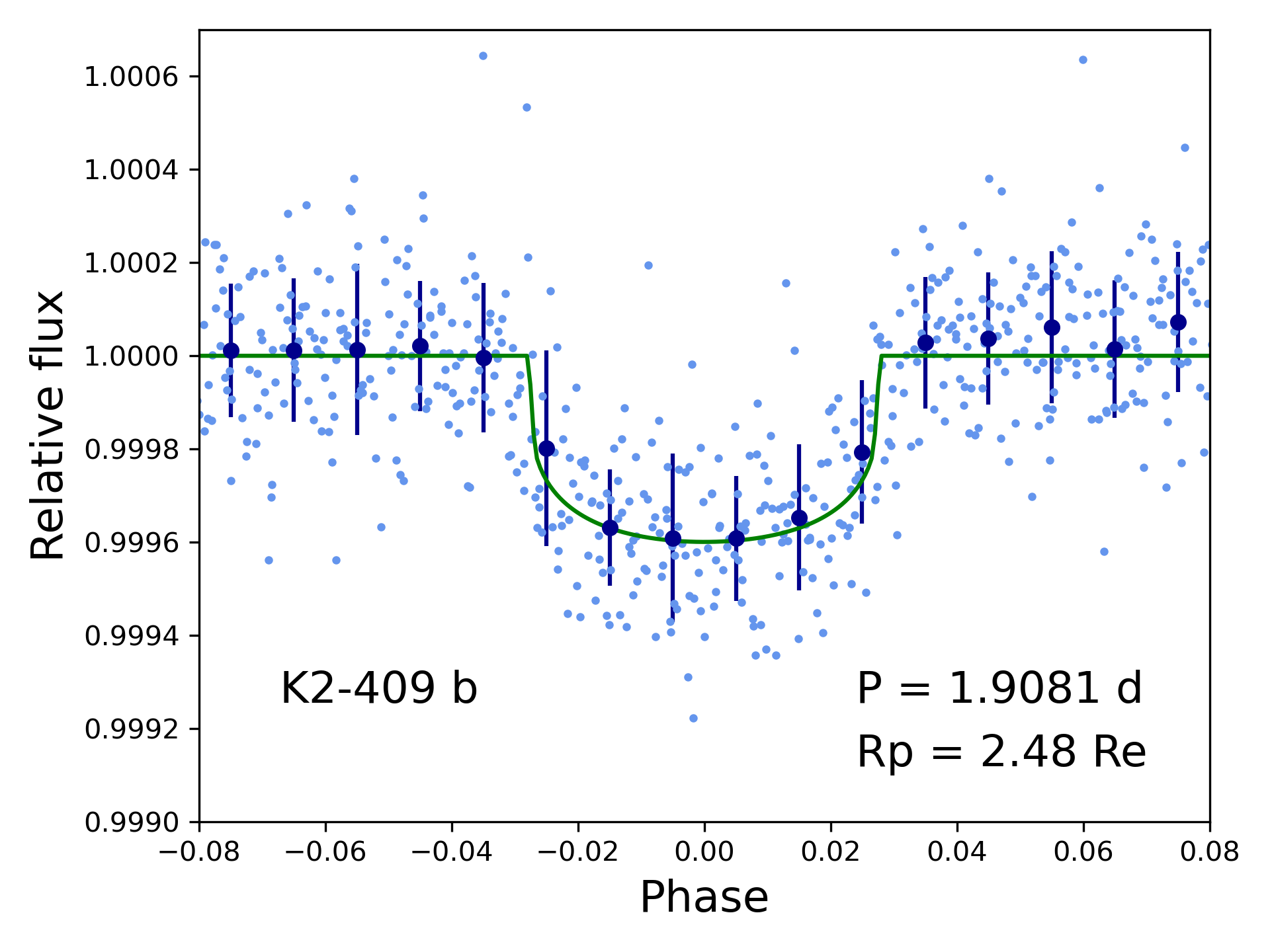}
\caption{Validated planet folded light curves. Description as for Figure \ref{fig:phasedlcs1}.}
\label{fig:phasedlcs4}
\end{figure}

\begin{figure}
\centering
\includegraphics[width=0.245\textwidth]{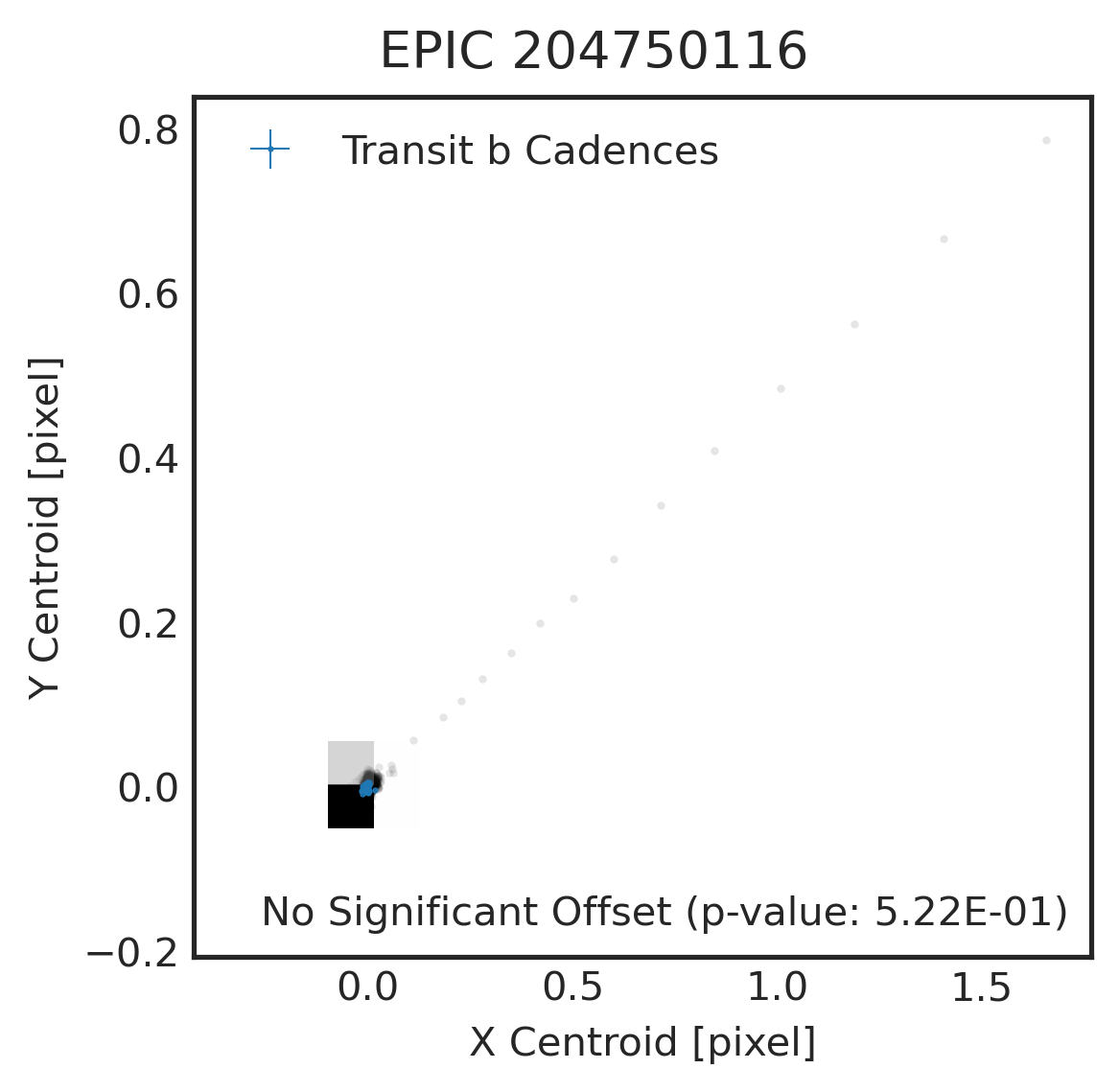}
\includegraphics[width=0.245\textwidth]{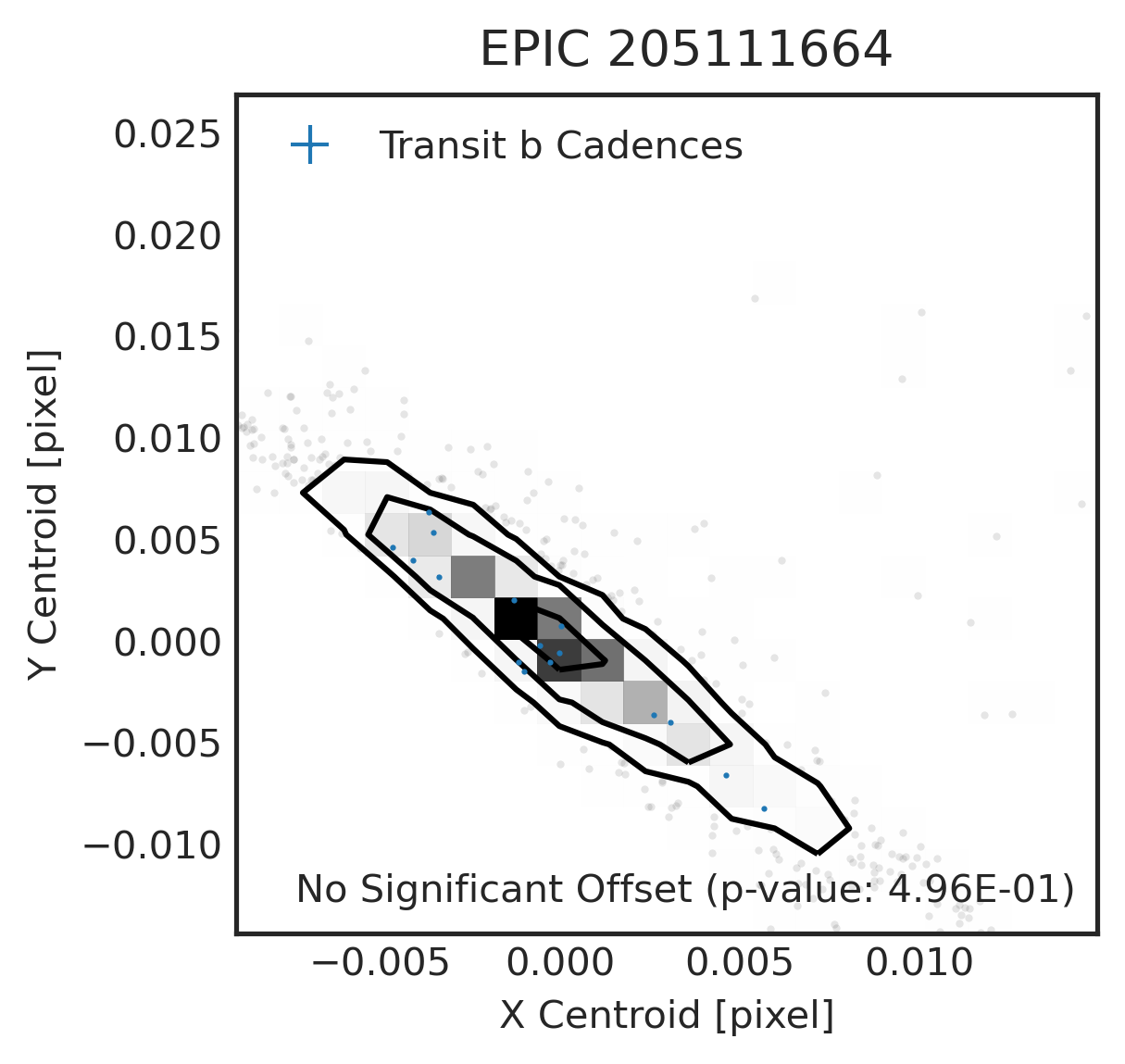}
\includegraphics[width=0.245\textwidth]{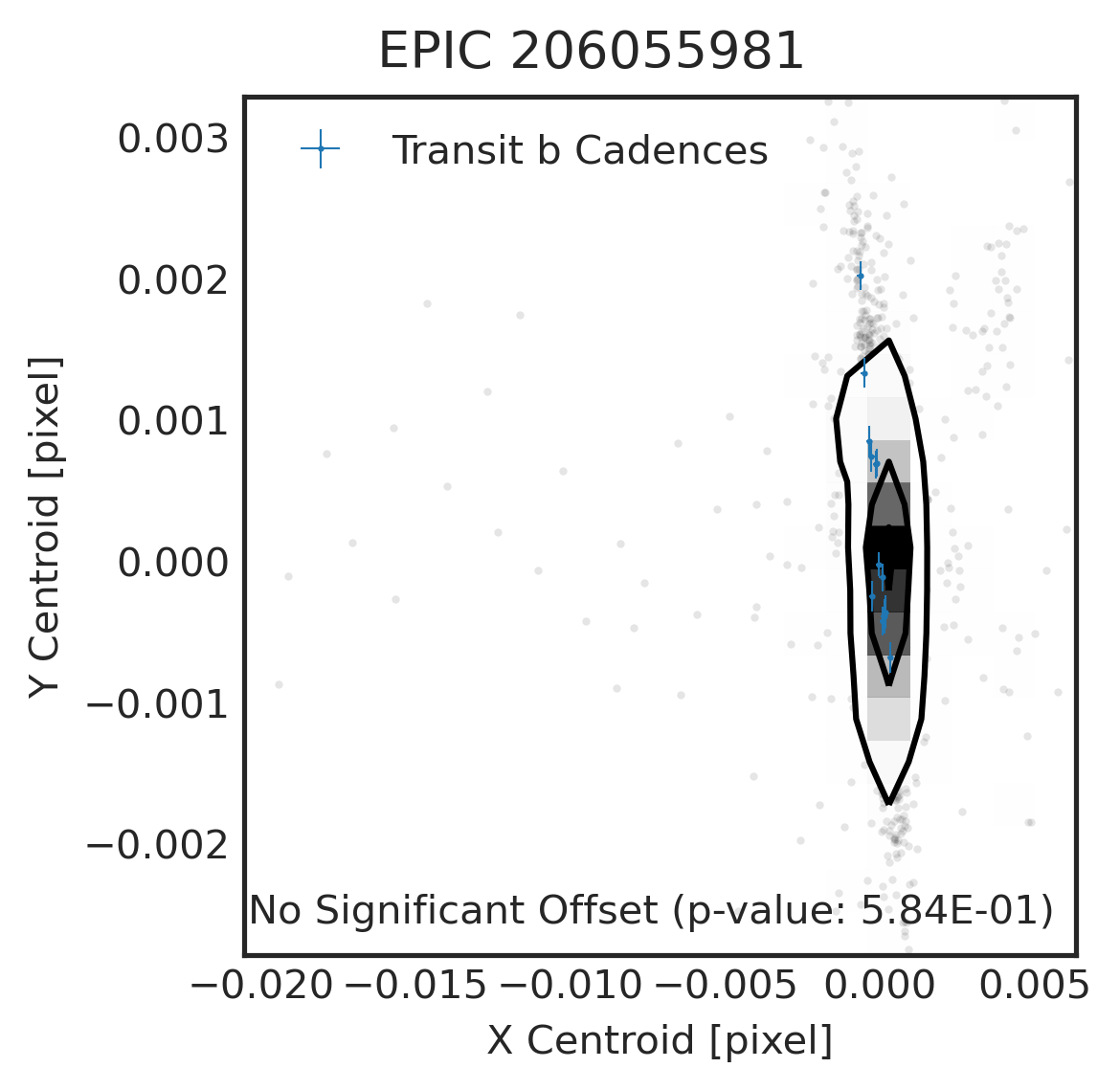}
\includegraphics[width=0.245\textwidth]{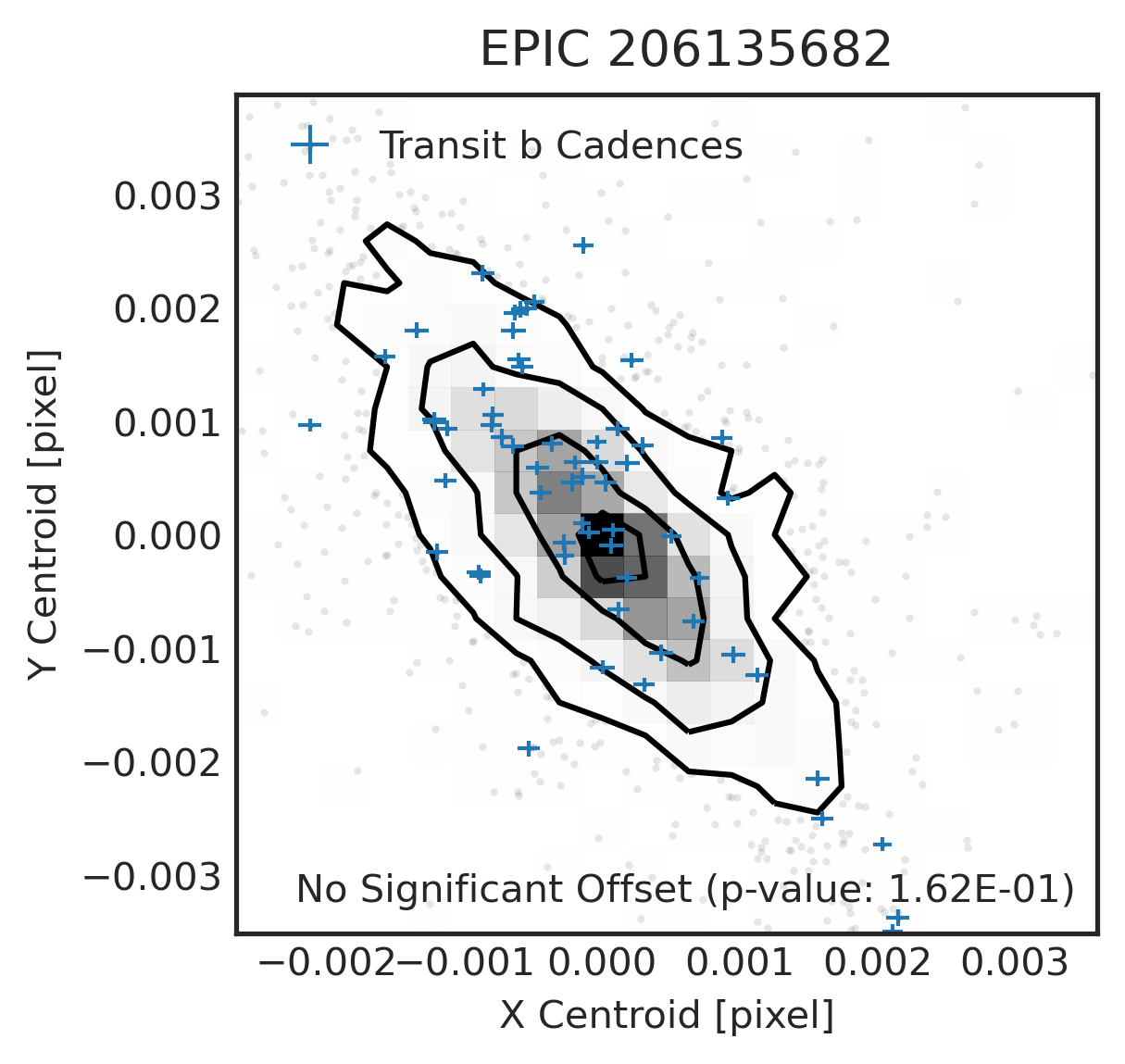}
\includegraphics[width=0.245\textwidth]{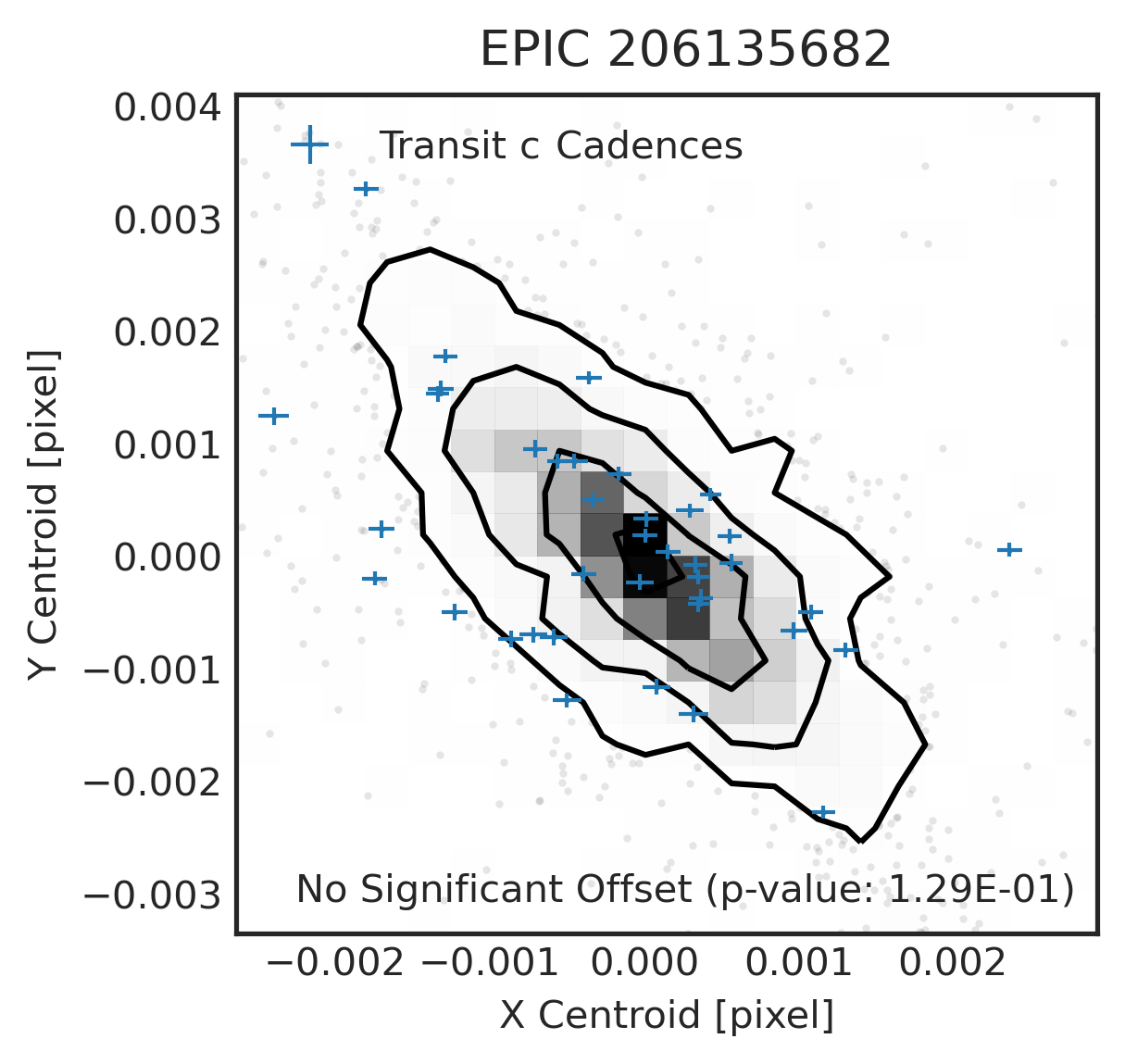}
\includegraphics[width=0.245\textwidth]{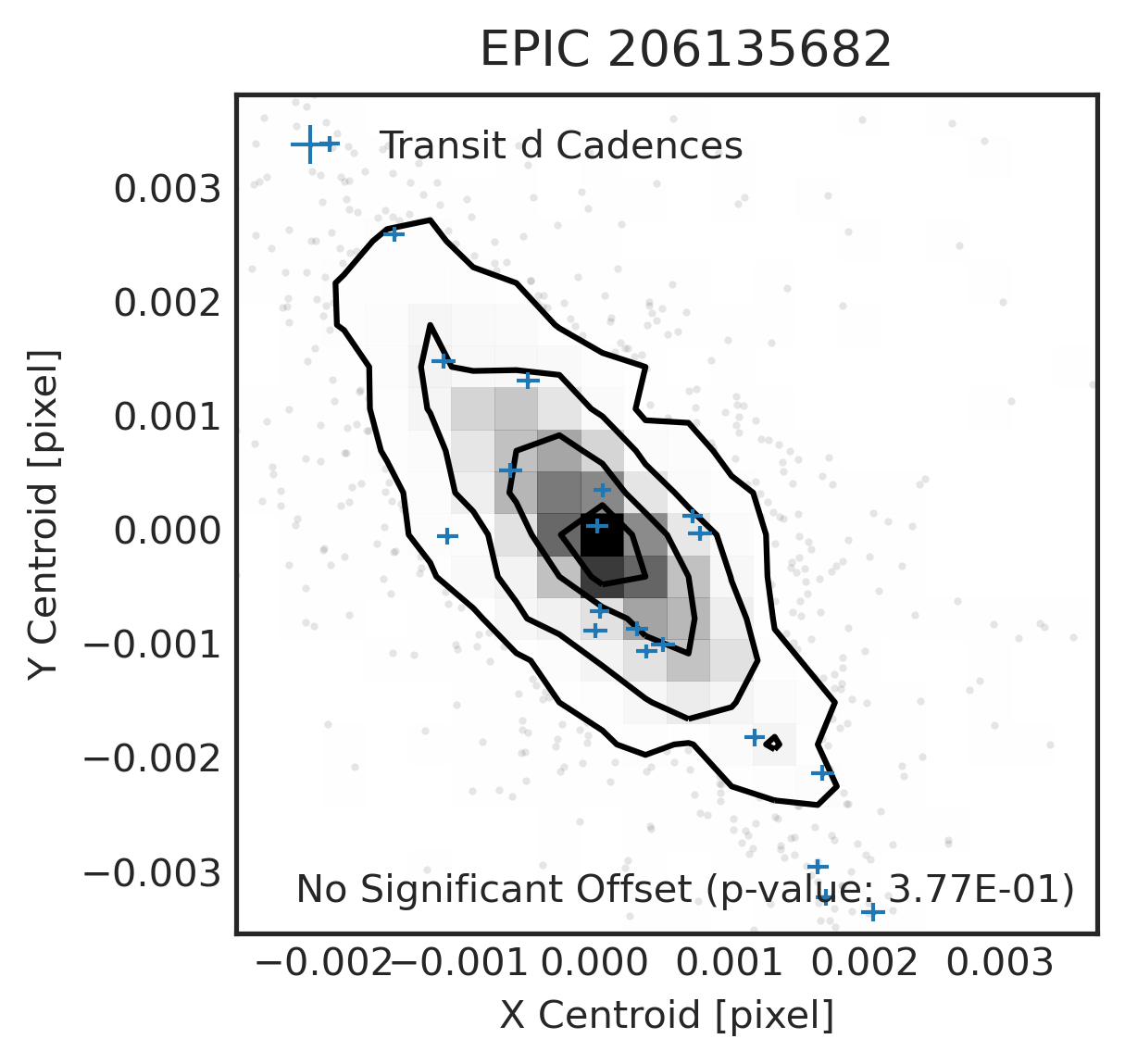}
\includegraphics[width=0.245\textwidth]{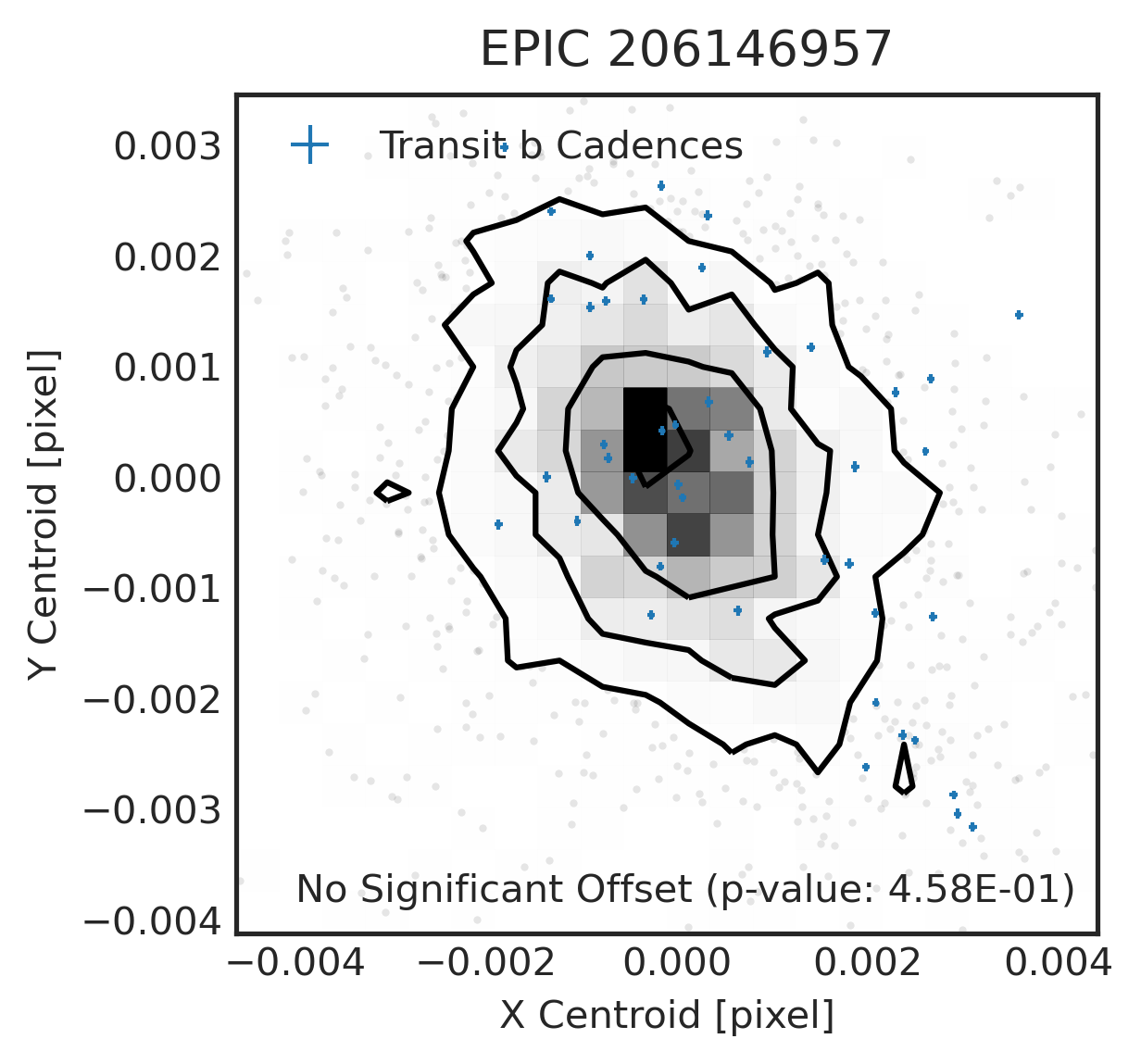}
\includegraphics[width=0.245\textwidth]{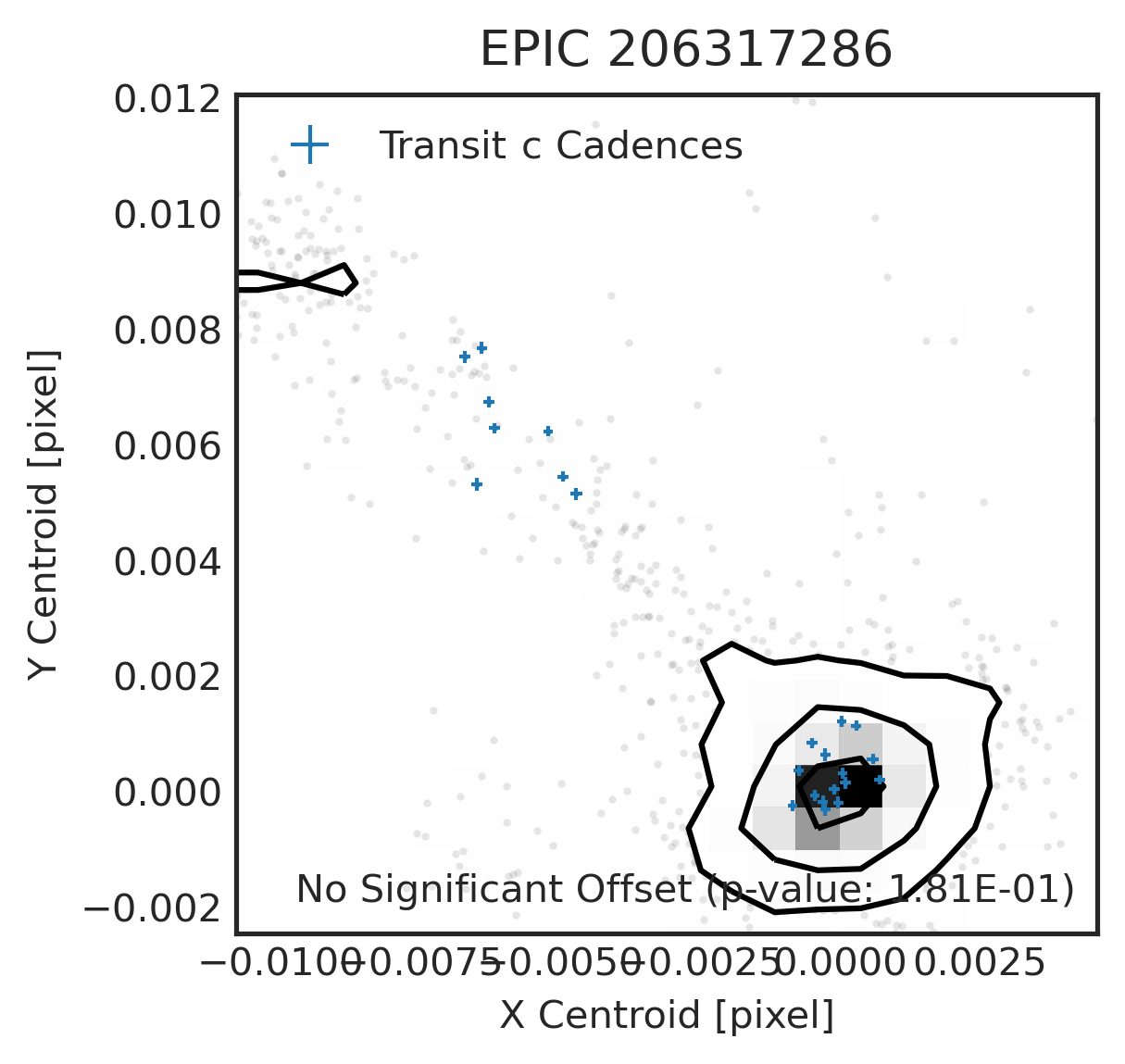} 
\includegraphics[width=0.245\textwidth]{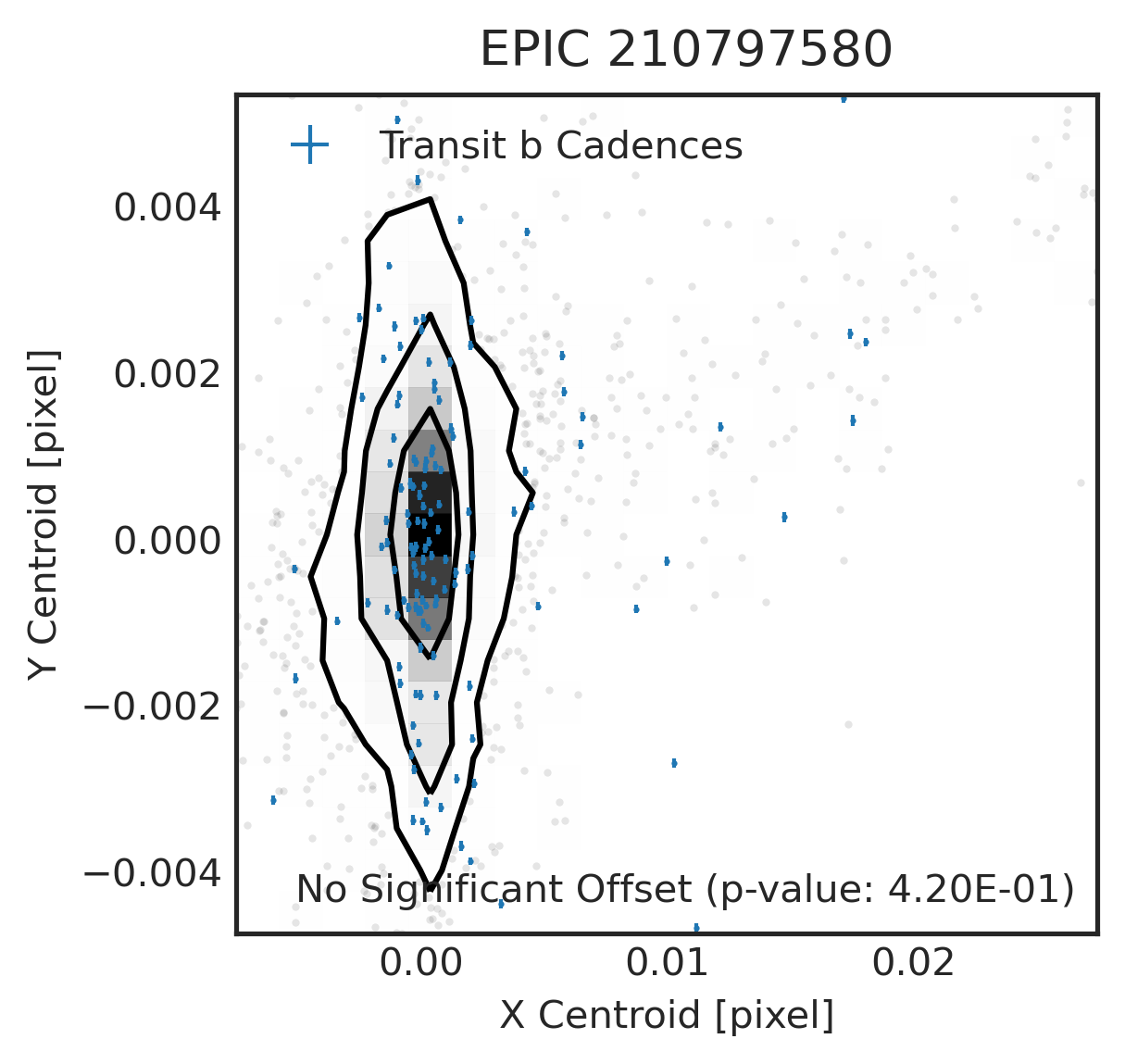}
\includegraphics[width=0.245\textwidth]{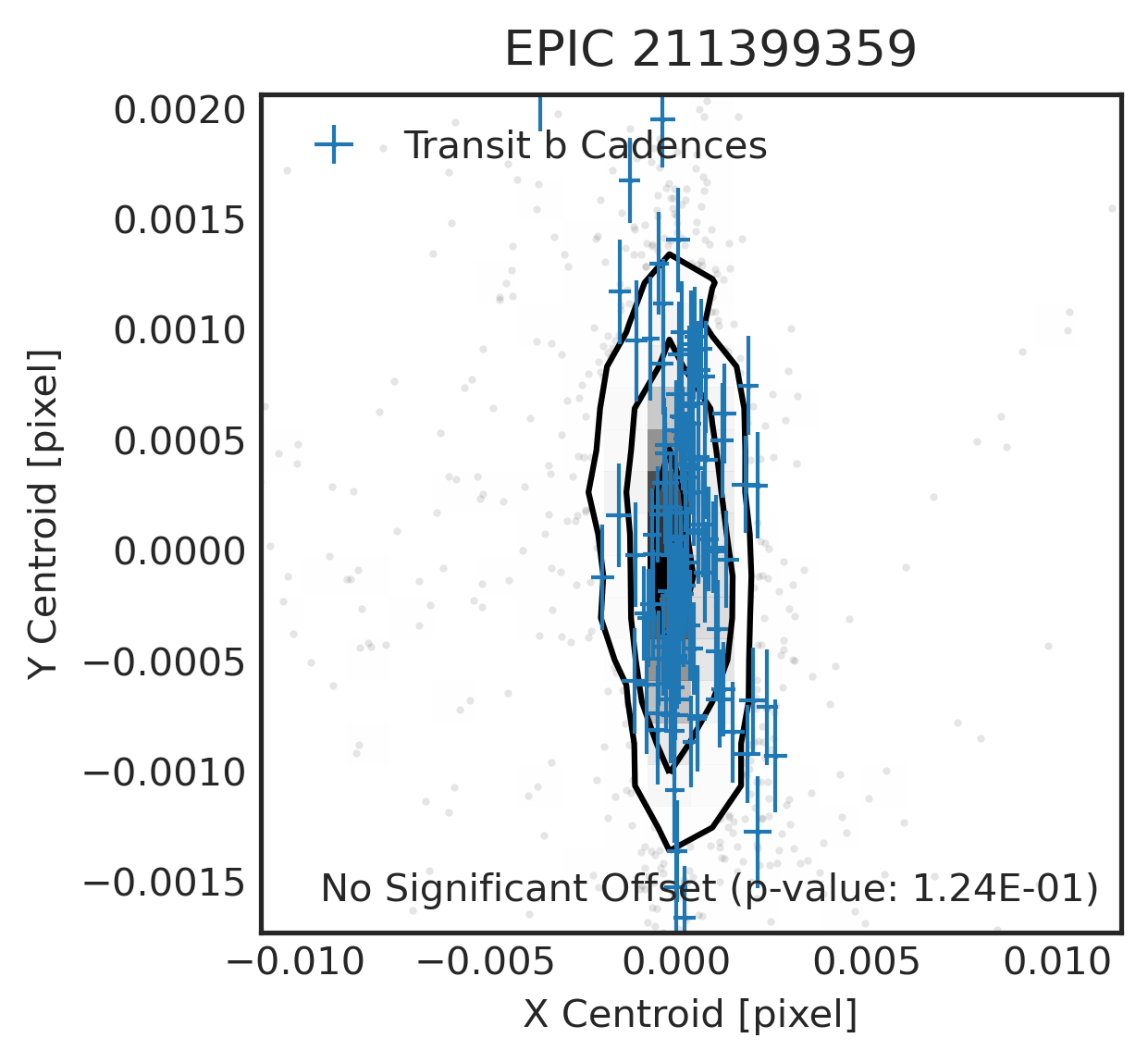}
\includegraphics[width=0.245\textwidth]{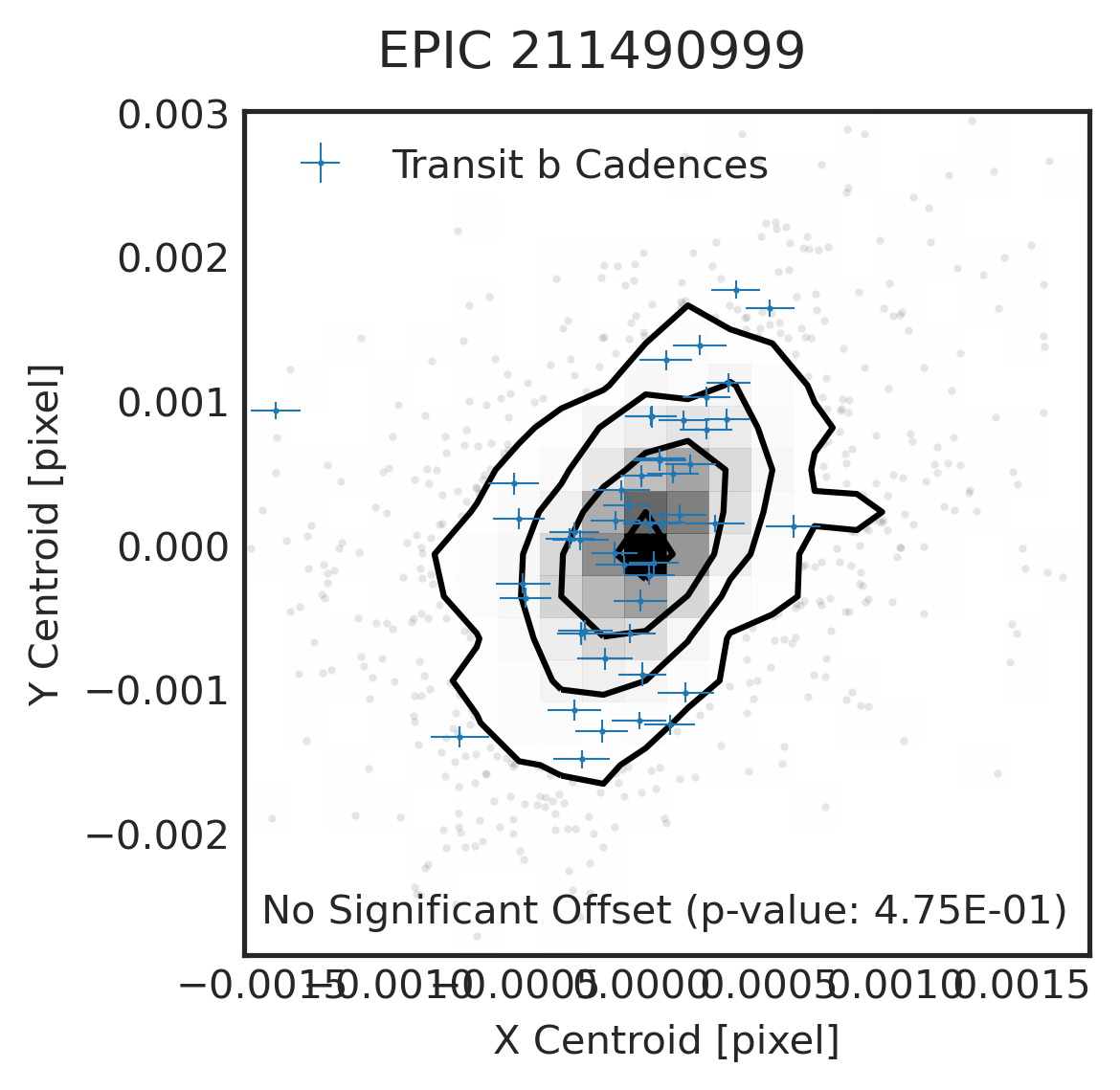}
\includegraphics[width=0.245\textwidth]{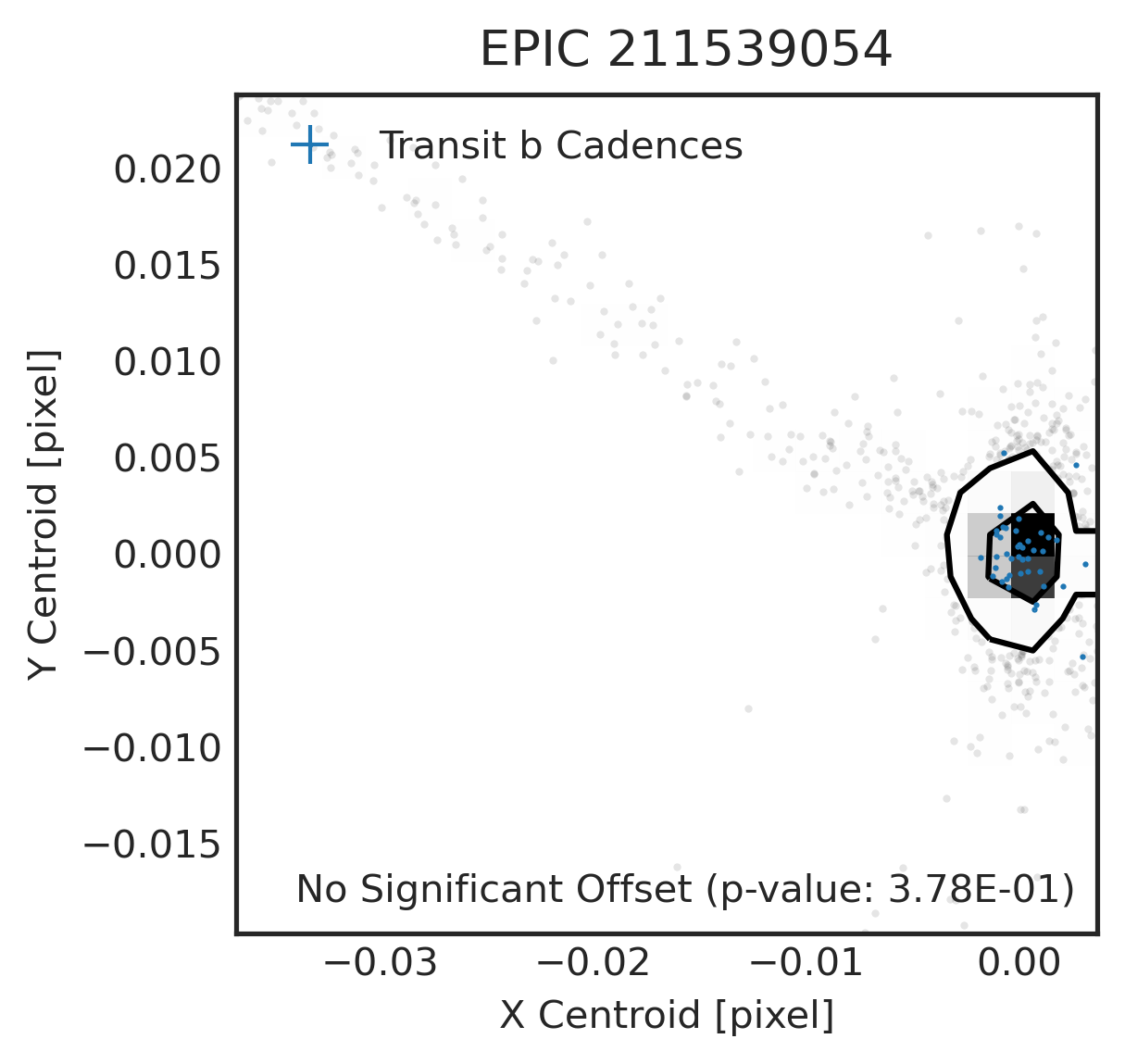}
\includegraphics[width=0.245\textwidth]{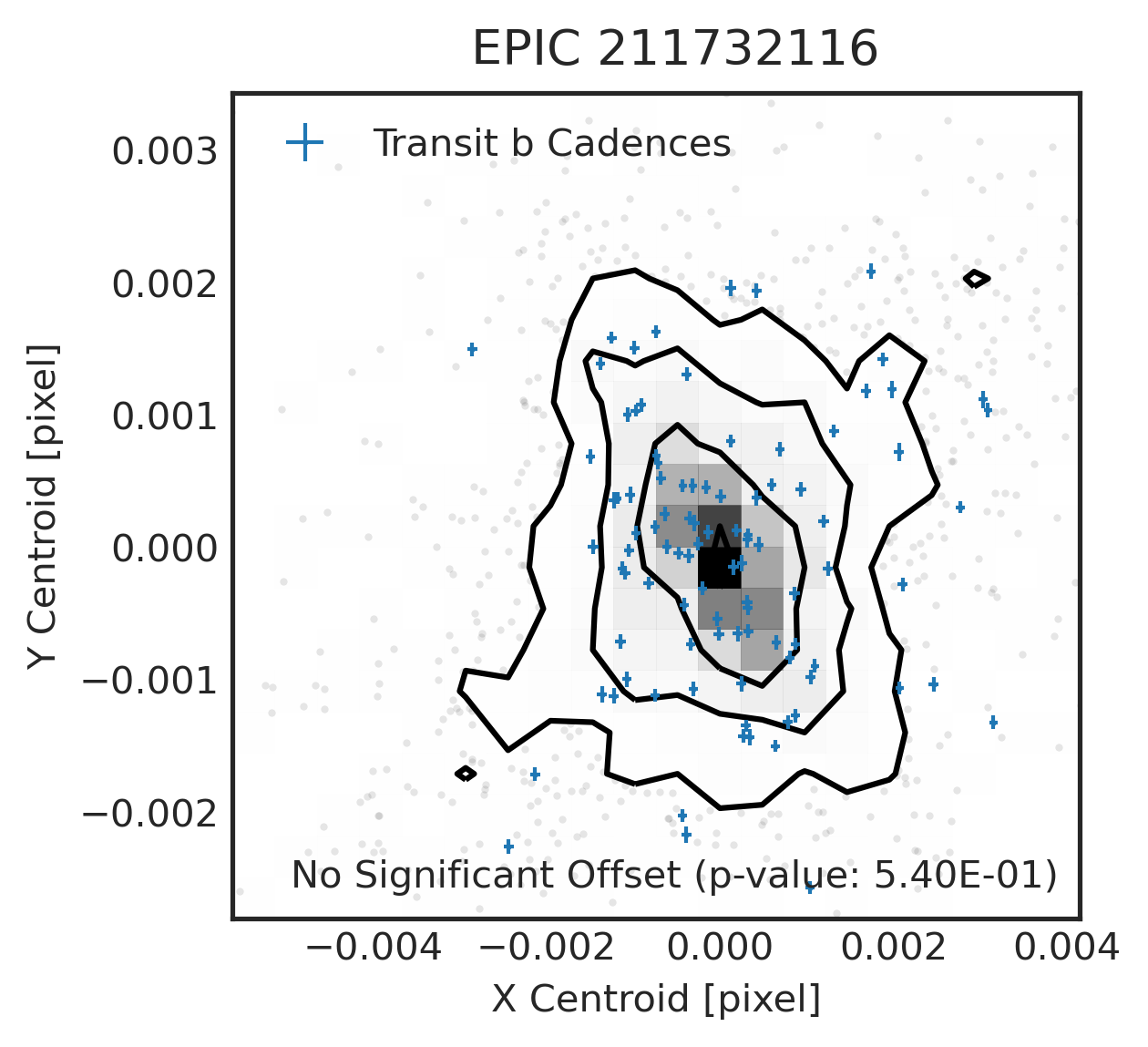}
\includegraphics[width=0.245\textwidth]{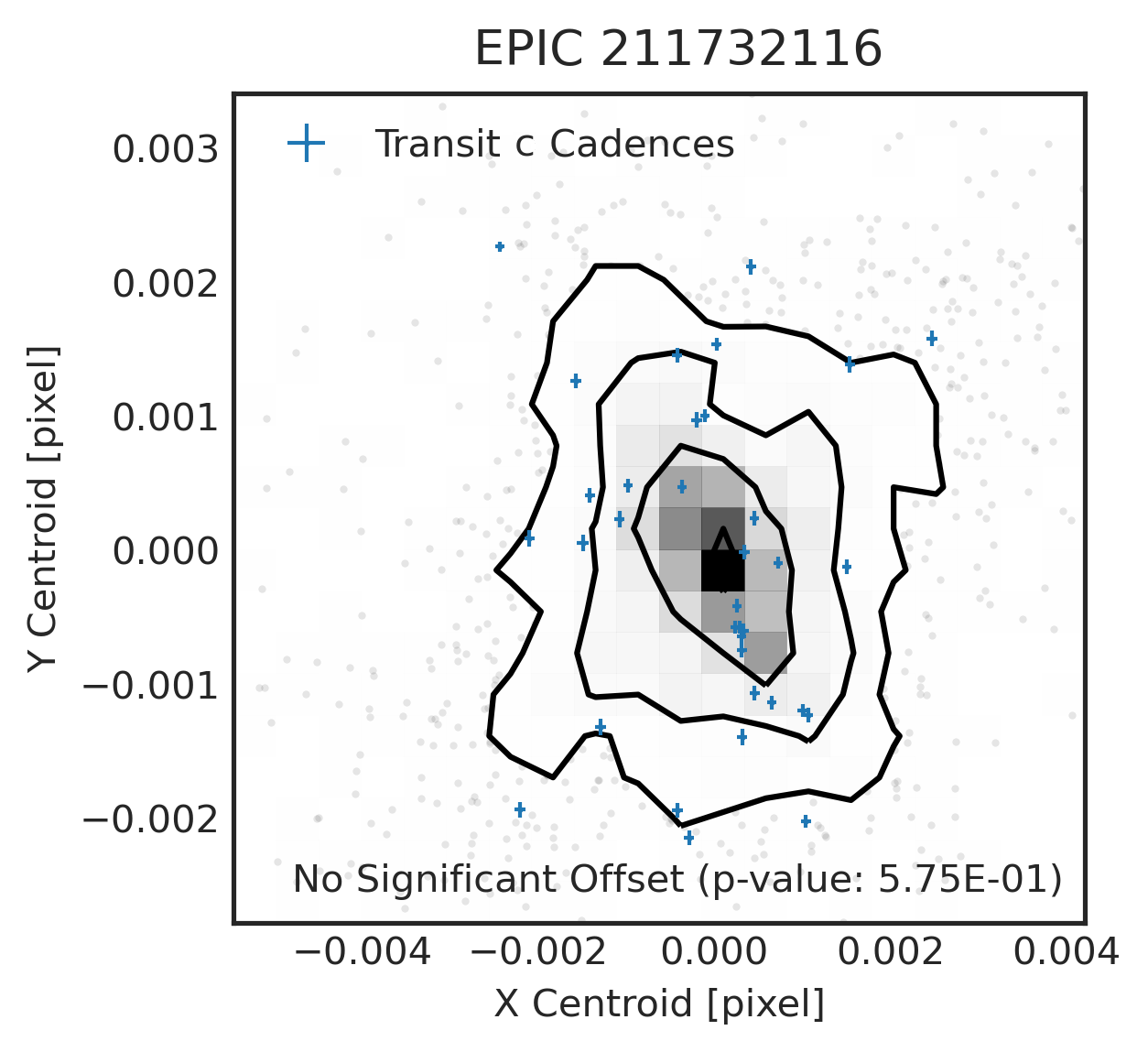}
\includegraphics[width=0.245\textwidth]{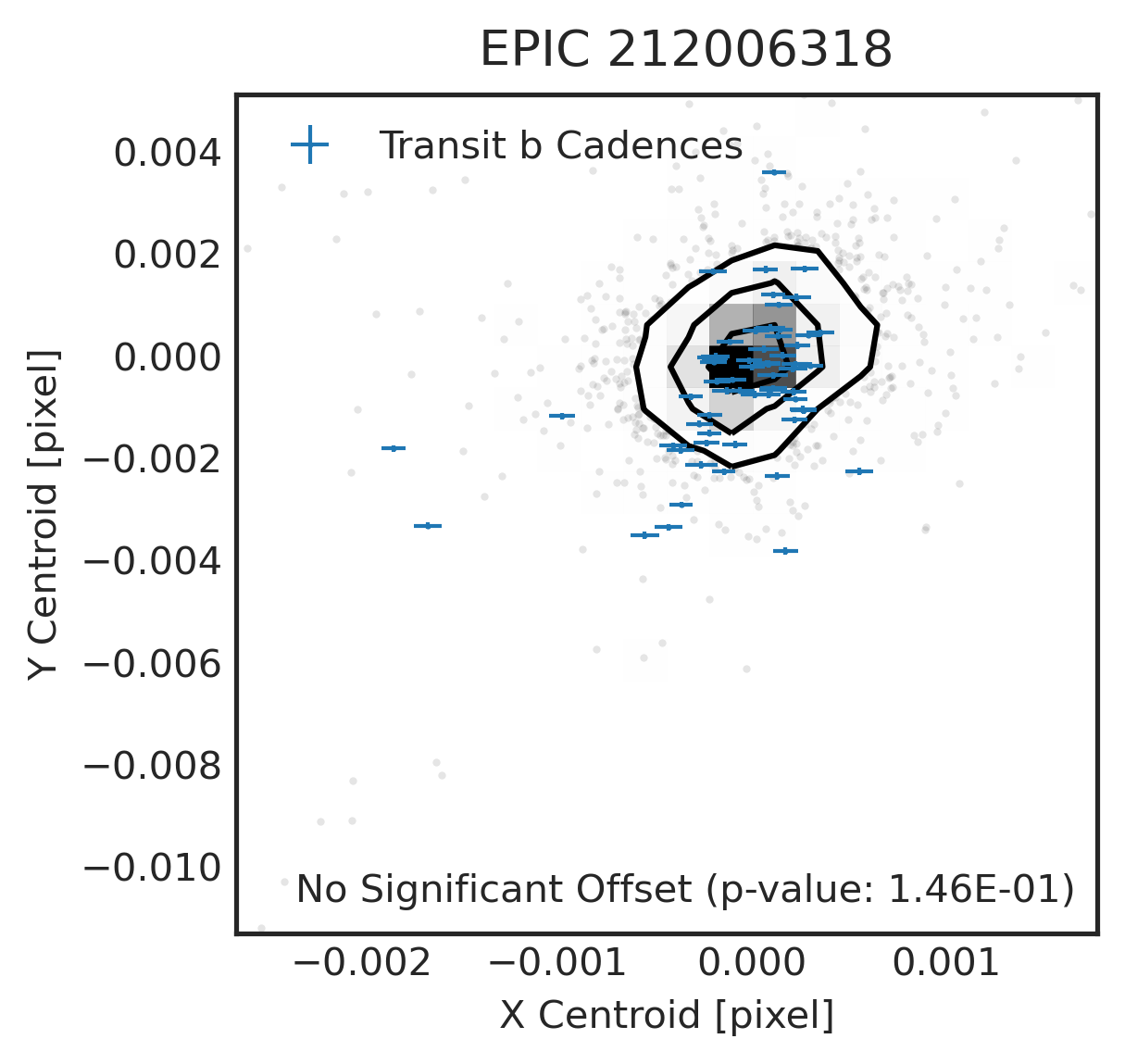}
\includegraphics[width=0.245\textwidth]{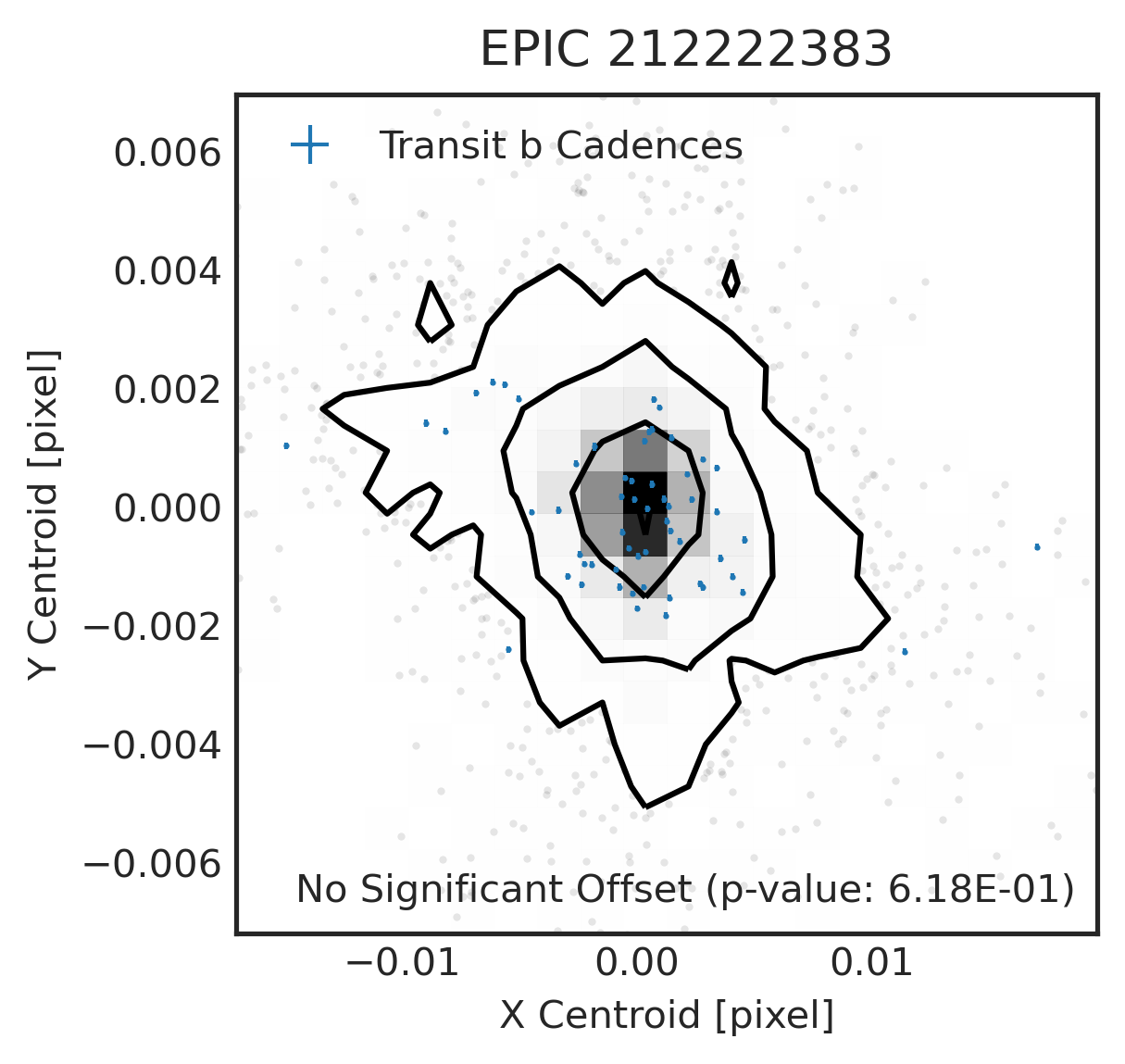}
\includegraphics[width=0.245\textwidth]{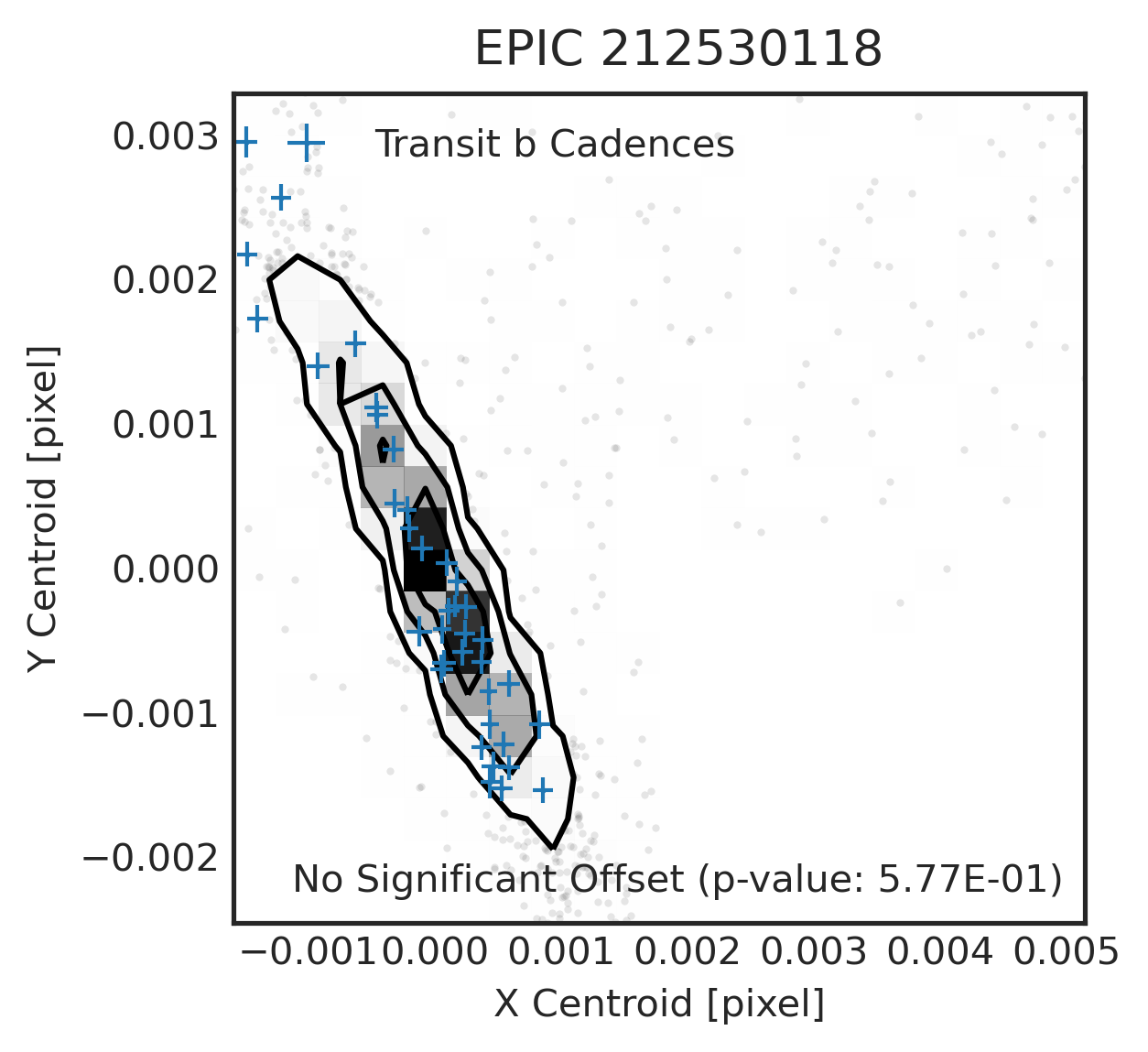}
\includegraphics[width=0.245\textwidth]{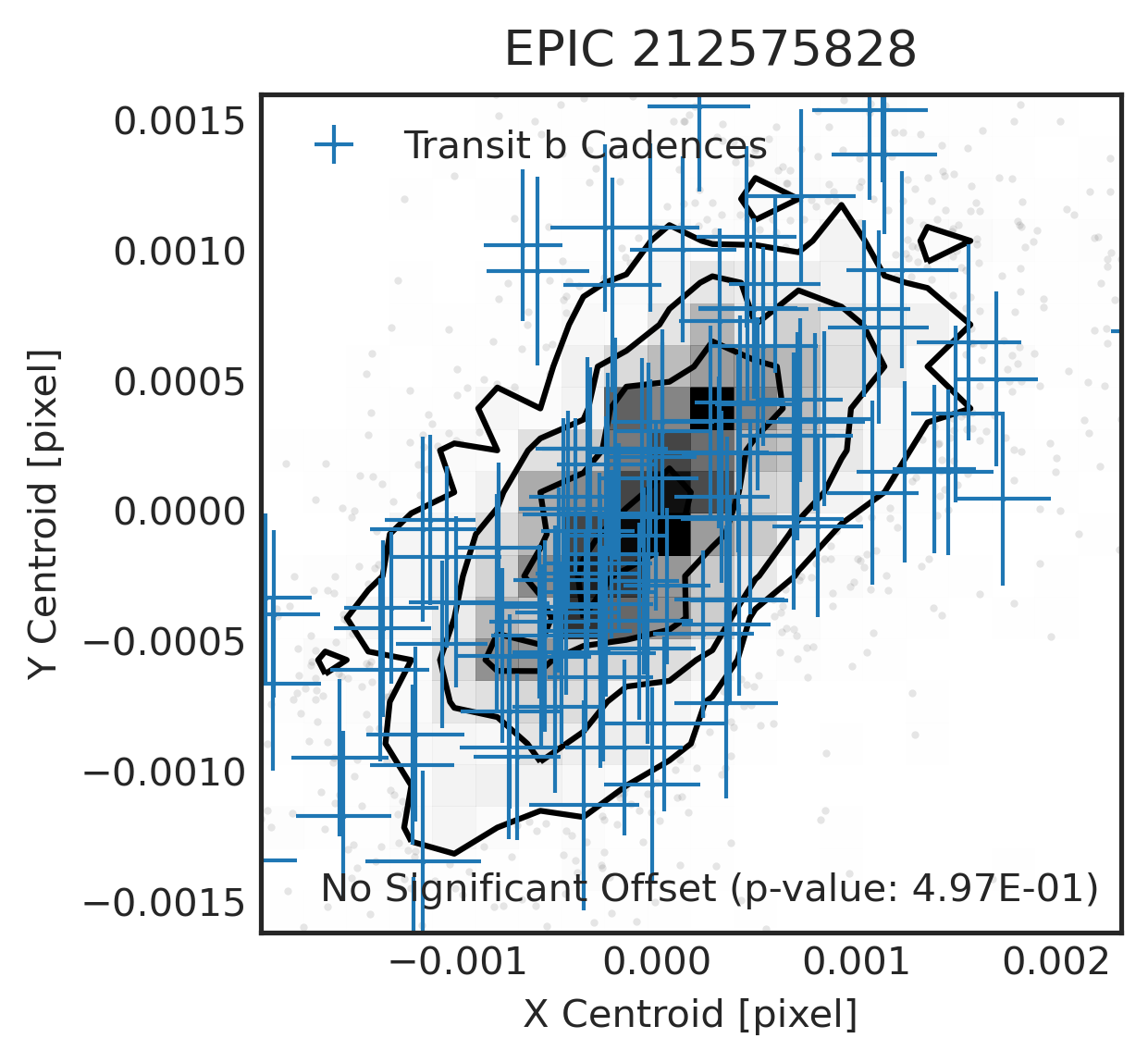}
\includegraphics[width=0.245\textwidth]{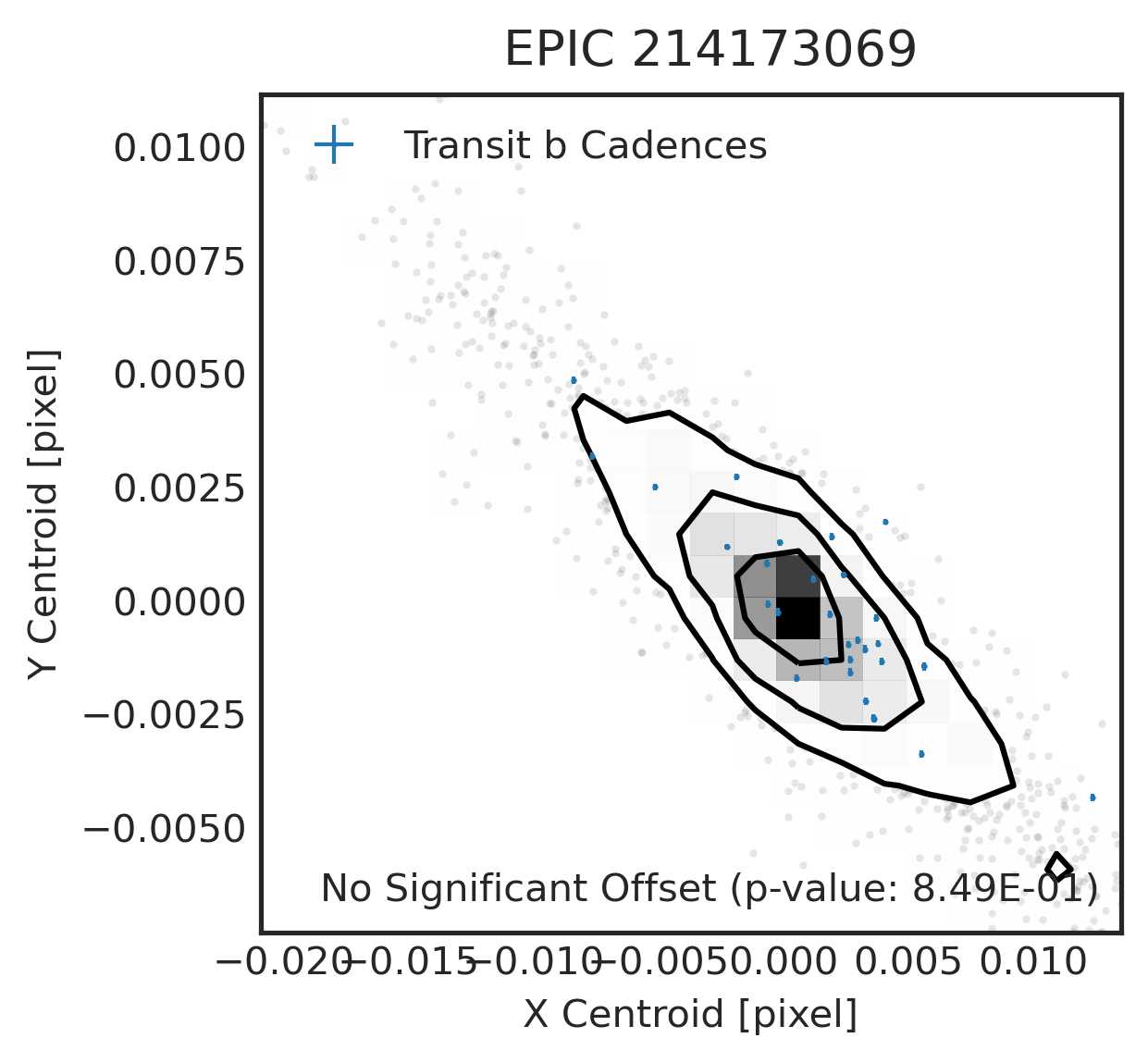}
\includegraphics[width=0.245\textwidth]{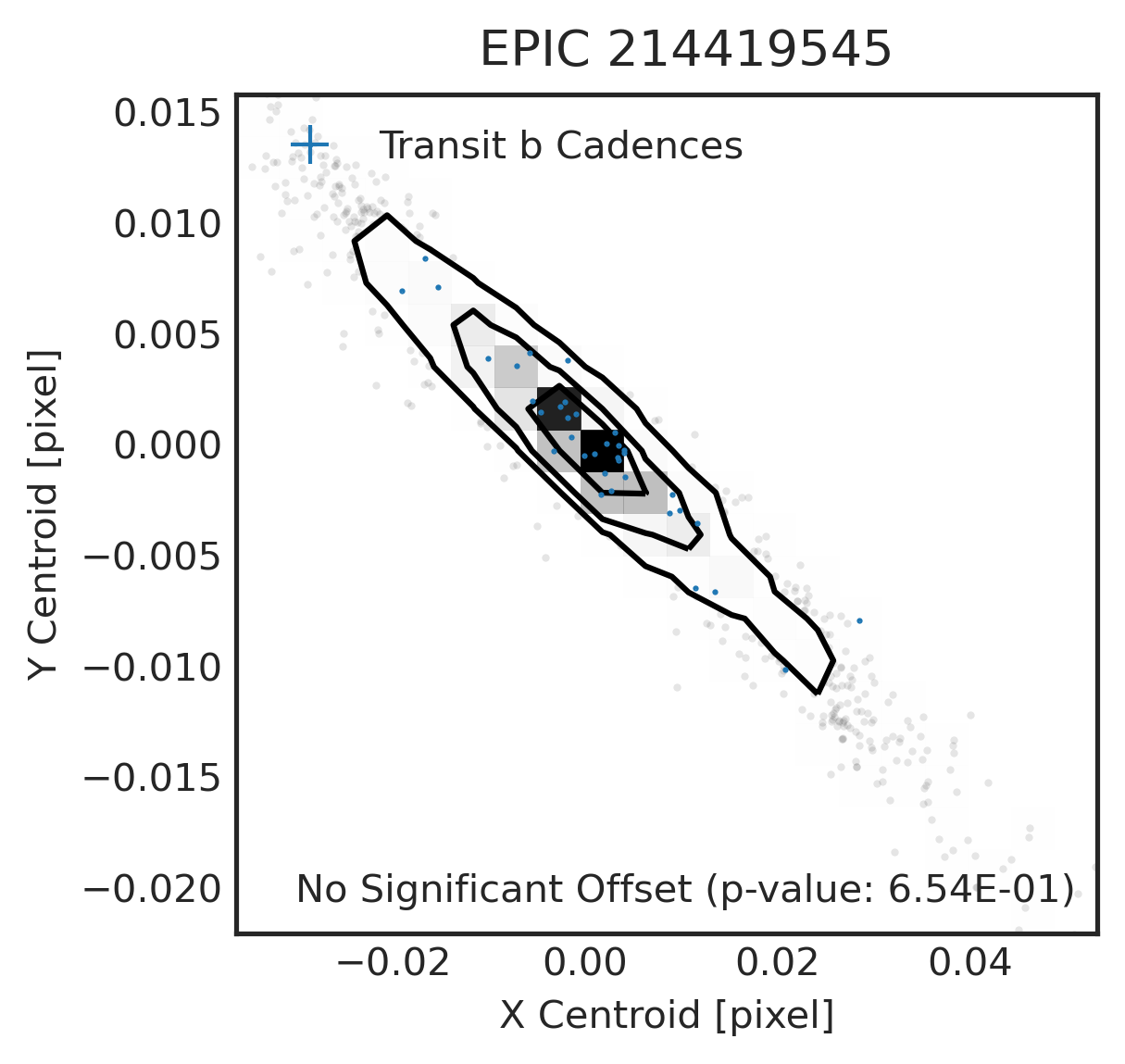}
\caption{Validated planet centroid plots. The contours show the $1\sigma$, $2\sigma$, and $3\sigma$ contours for the centroid locations for out-of-transit cadences; the grey points show the remaining out-of-transit centroid locations that fall outside those contours. The blue points are the centroid locations of the in-transit cadences.}
\label{fig:centroidplots1}
\end{figure}

\begin{figure}
\centering

\includegraphics[width=0.245\textwidth]{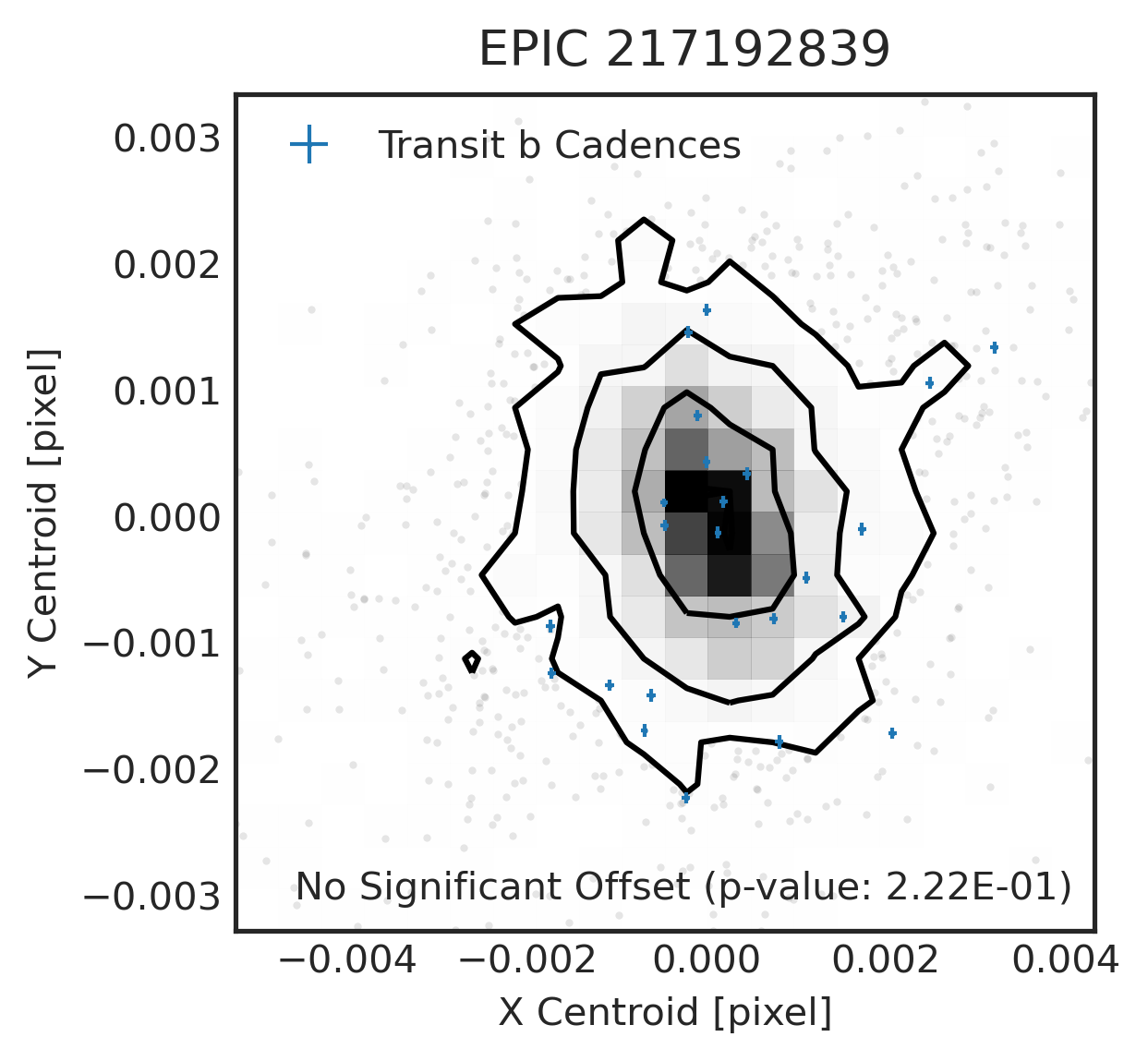}
\includegraphics[width=0.245\textwidth]{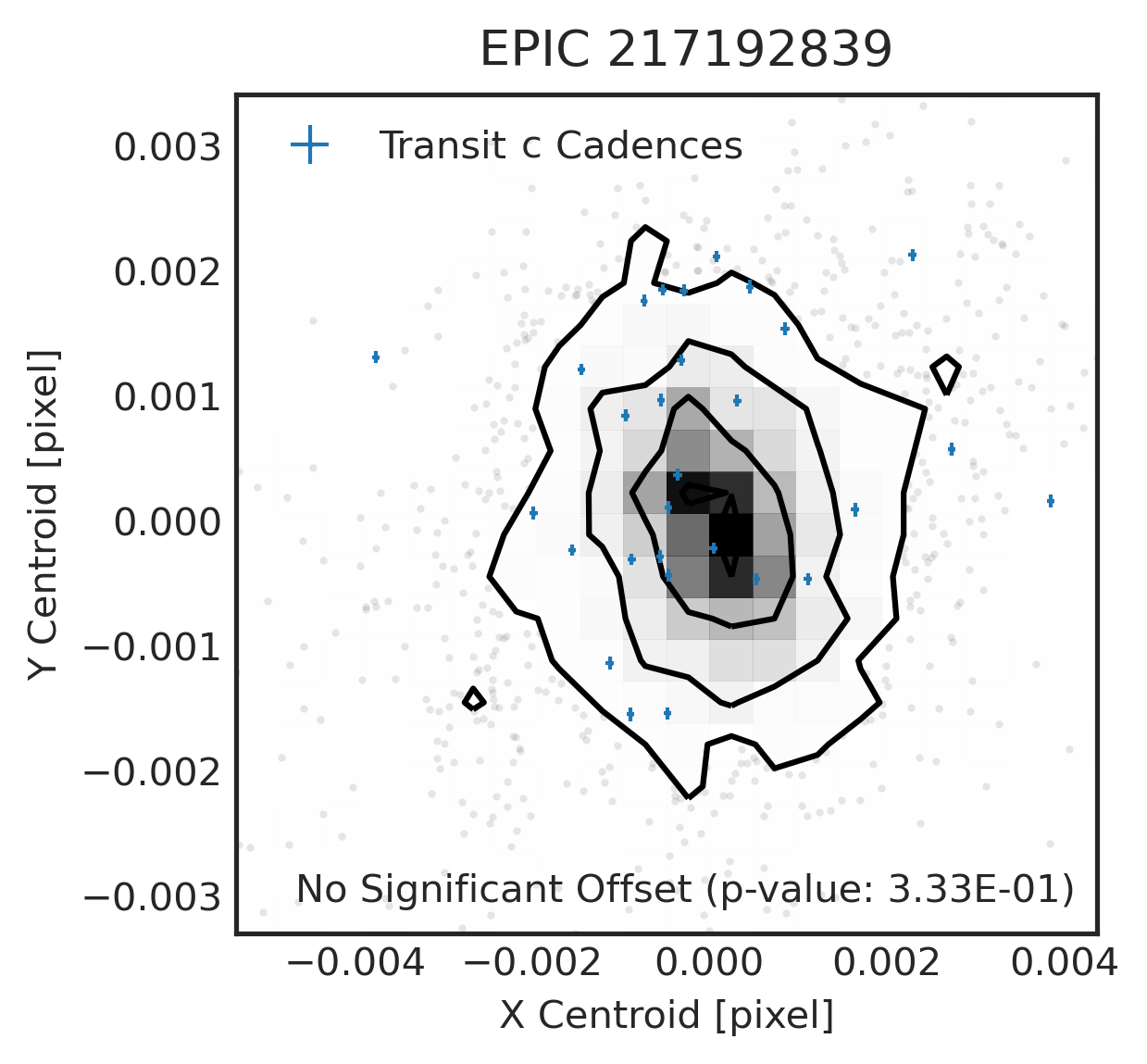}
\includegraphics[width=0.245\textwidth]{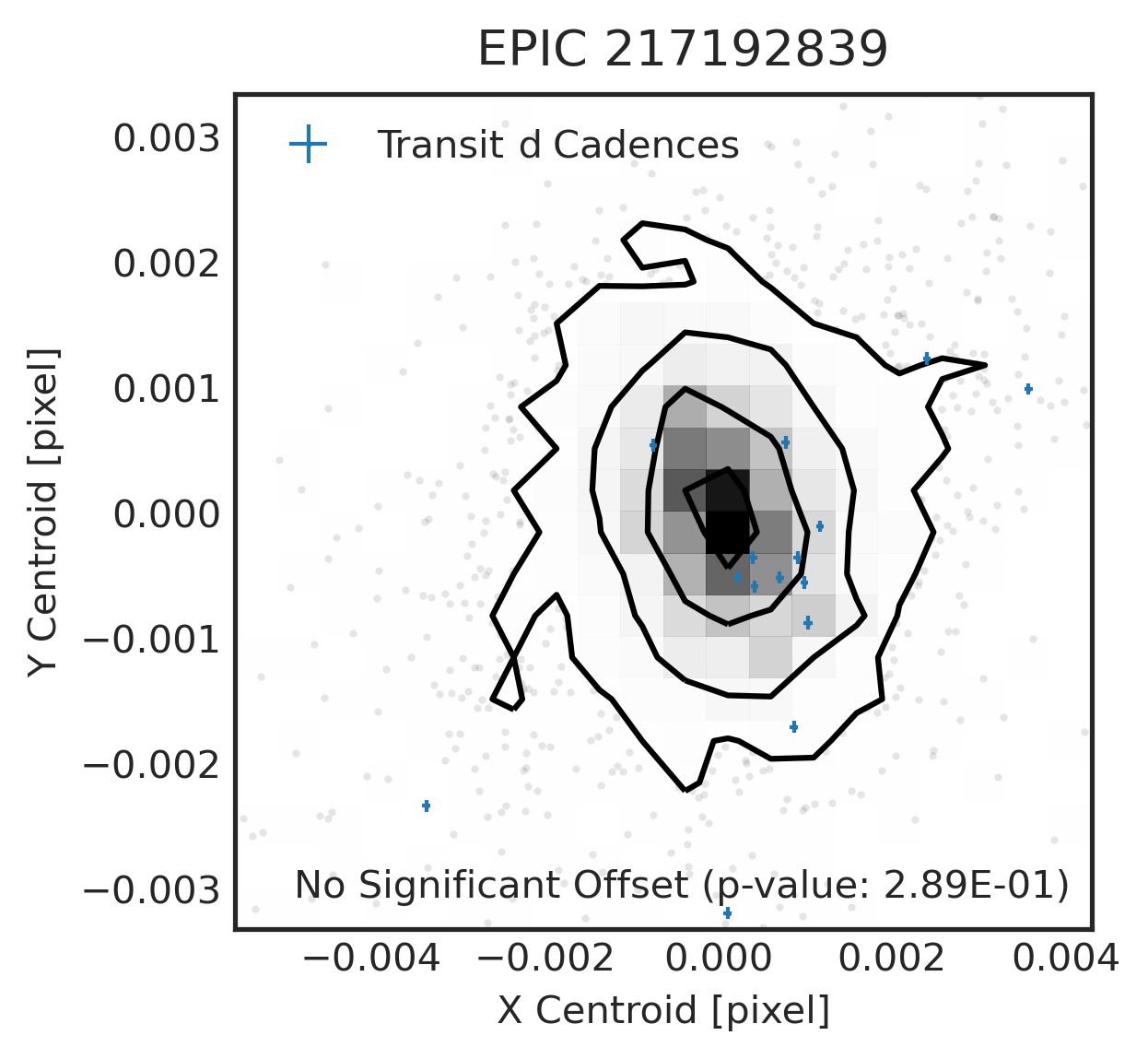}
\includegraphics[width=0.245\textwidth]{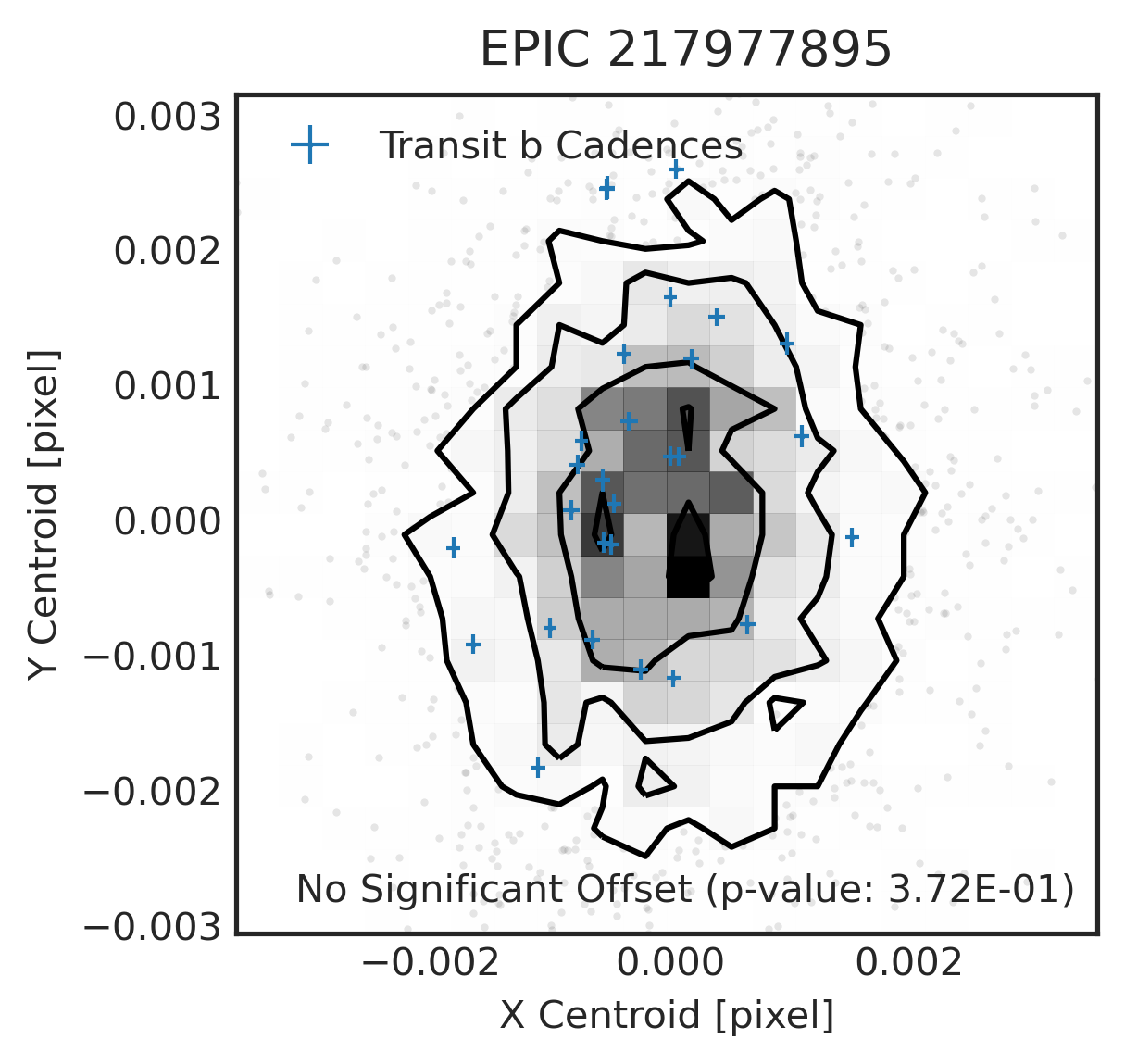}
\includegraphics[width=0.245\textwidth]{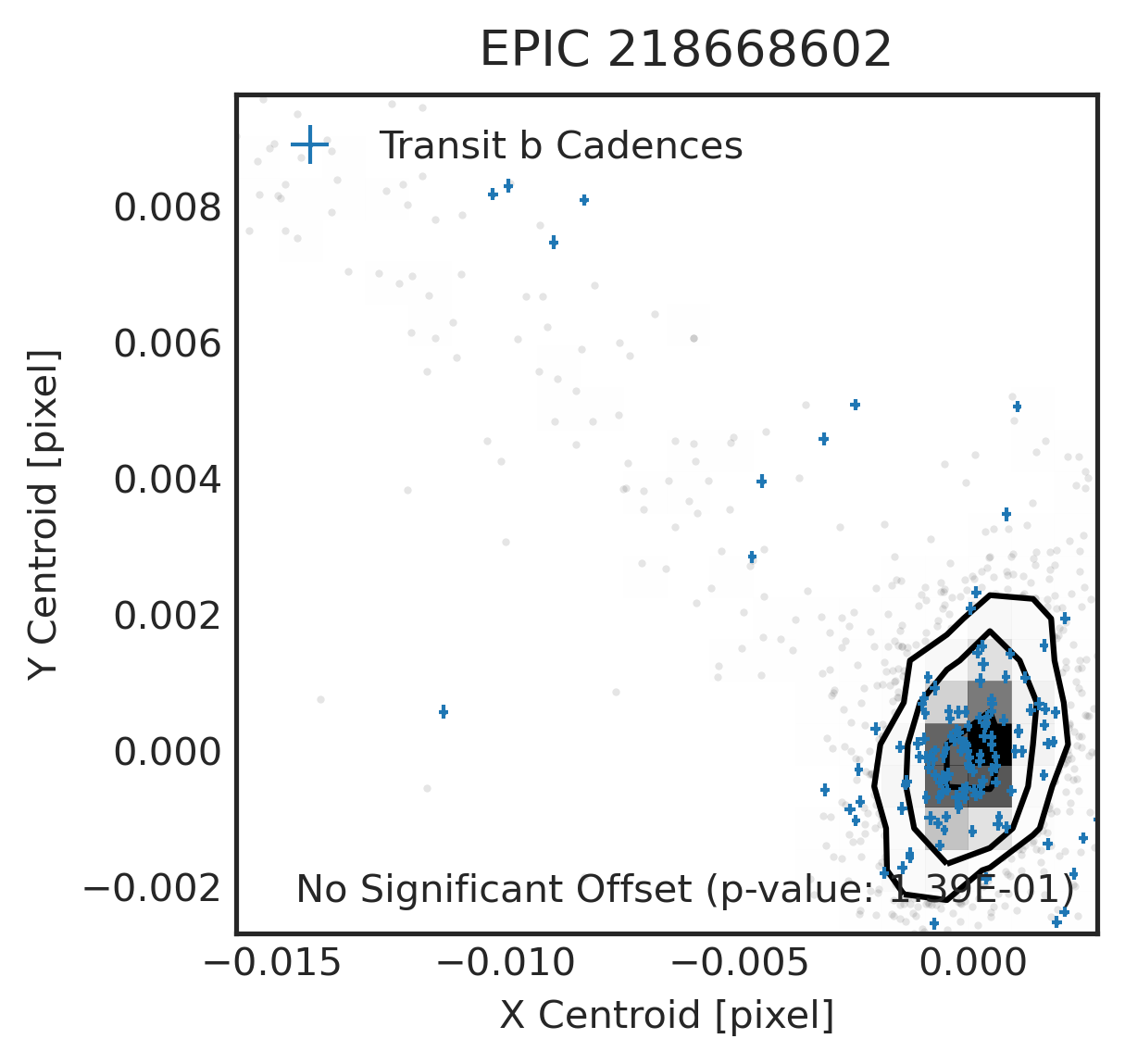}
\includegraphics[width=0.245\textwidth]{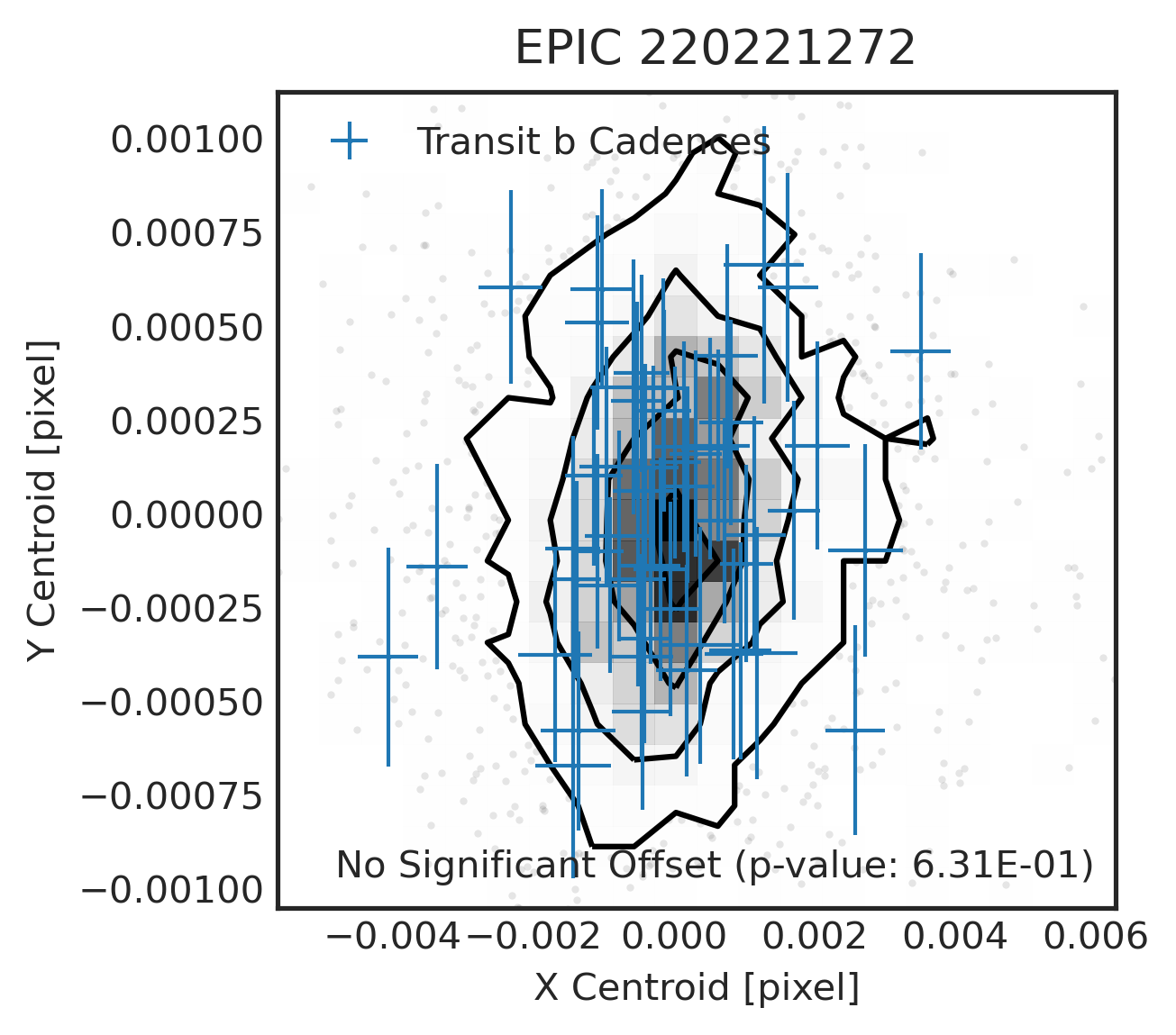}
\includegraphics[width=0.245\textwidth]{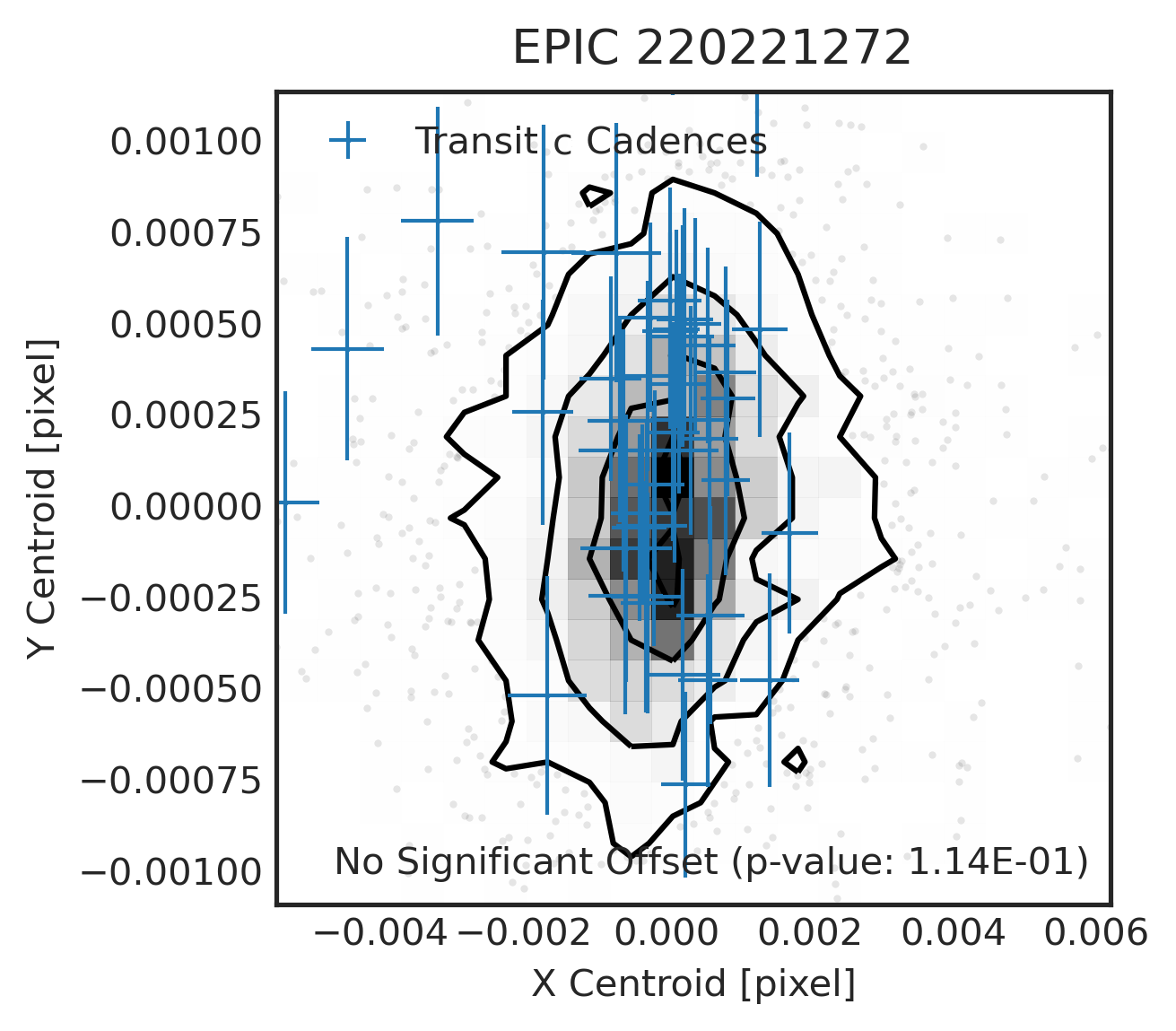}
\includegraphics[width=0.245\textwidth]{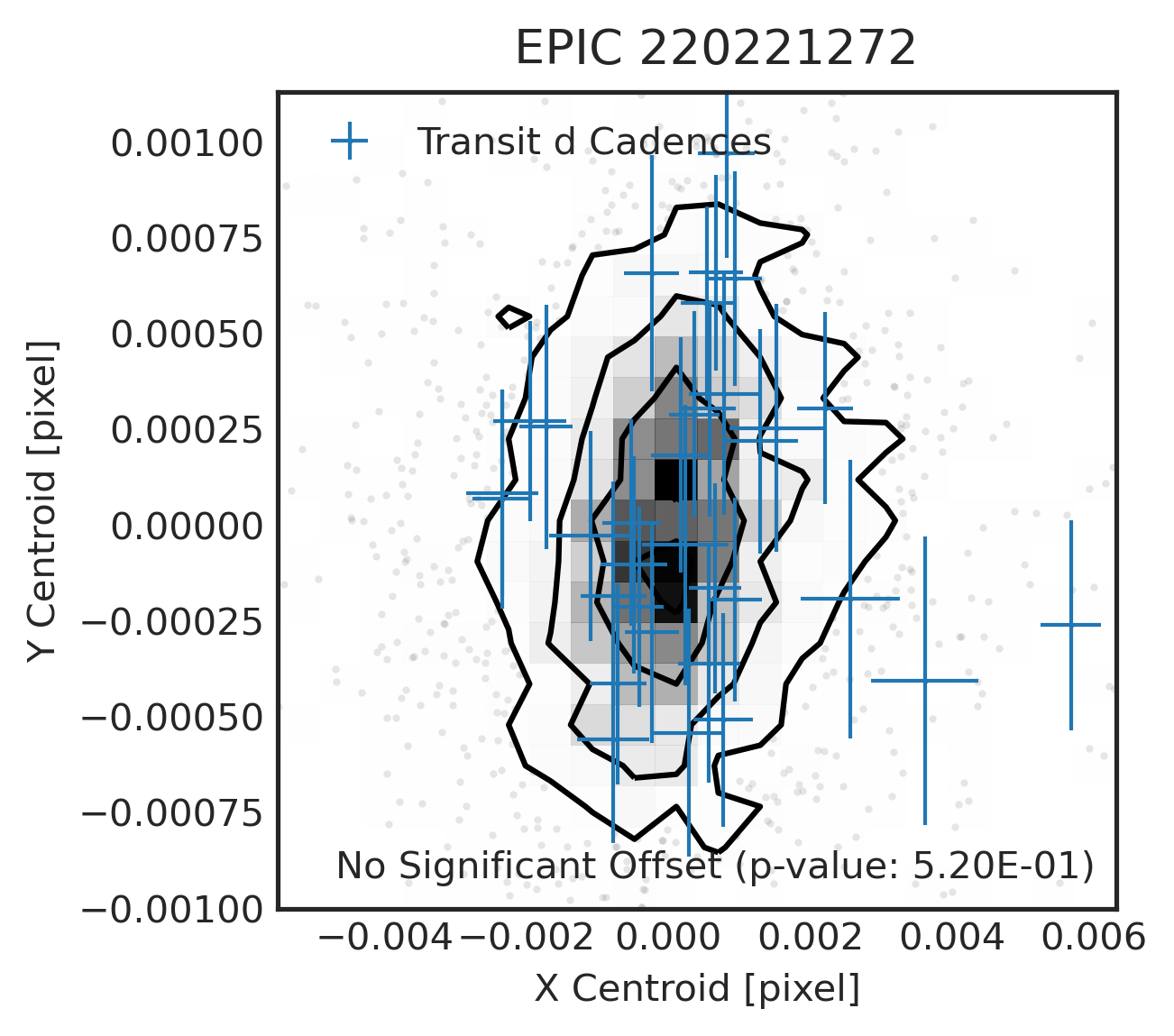}
\includegraphics[width=0.245\textwidth]{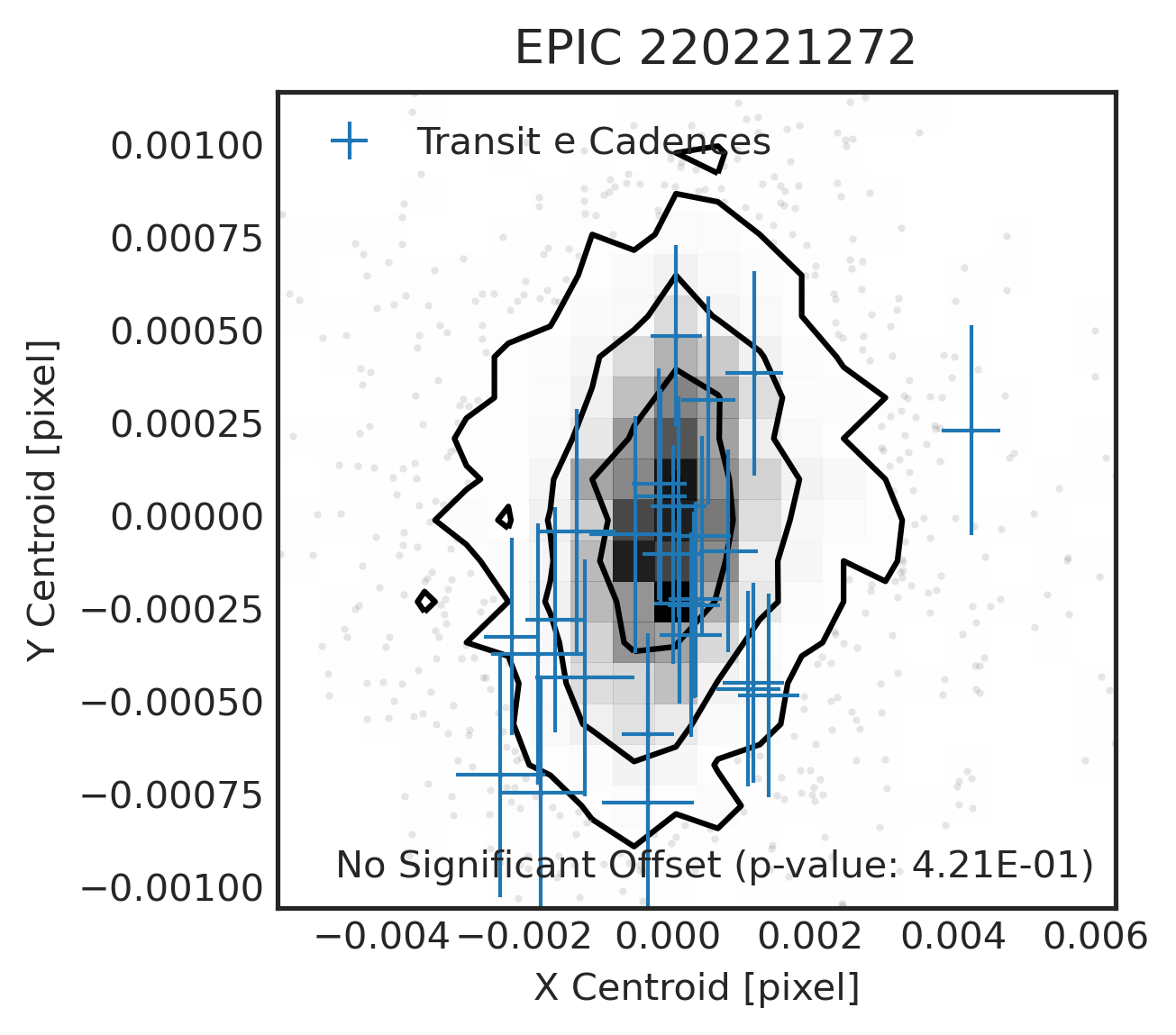}
\includegraphics[width=0.245\textwidth]{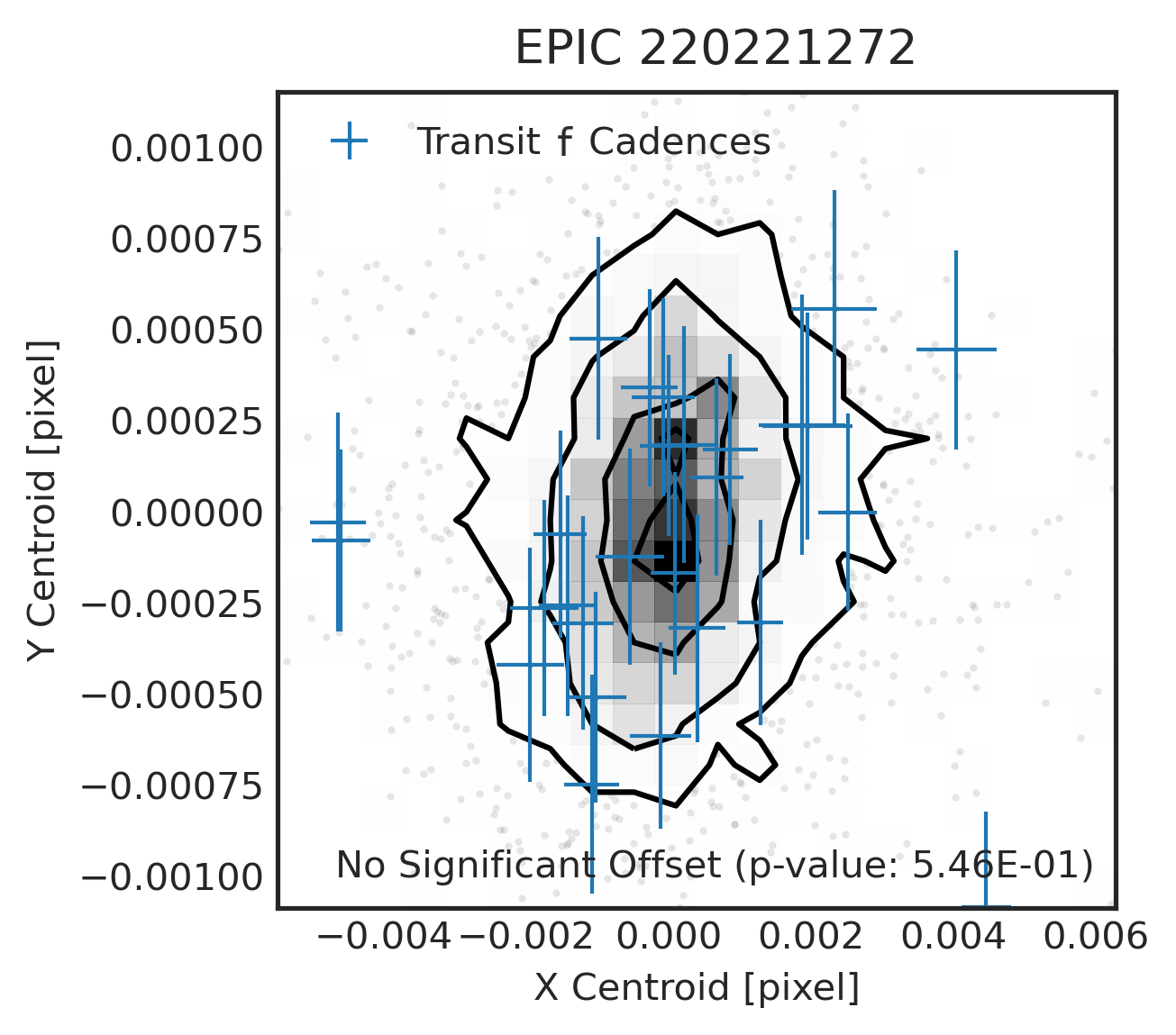}
\includegraphics[width=0.245\textwidth]{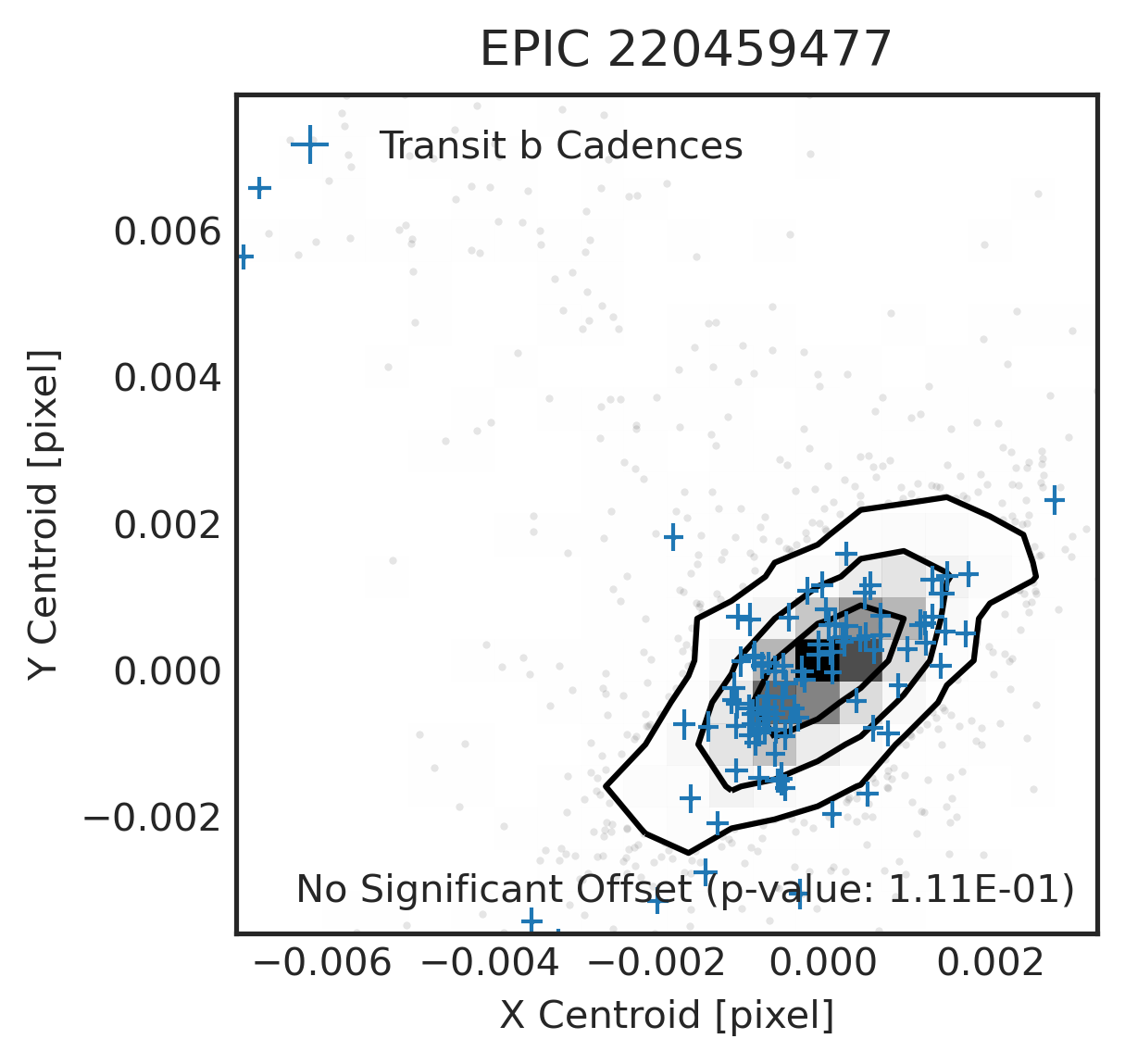}
\includegraphics[width=0.245\textwidth]{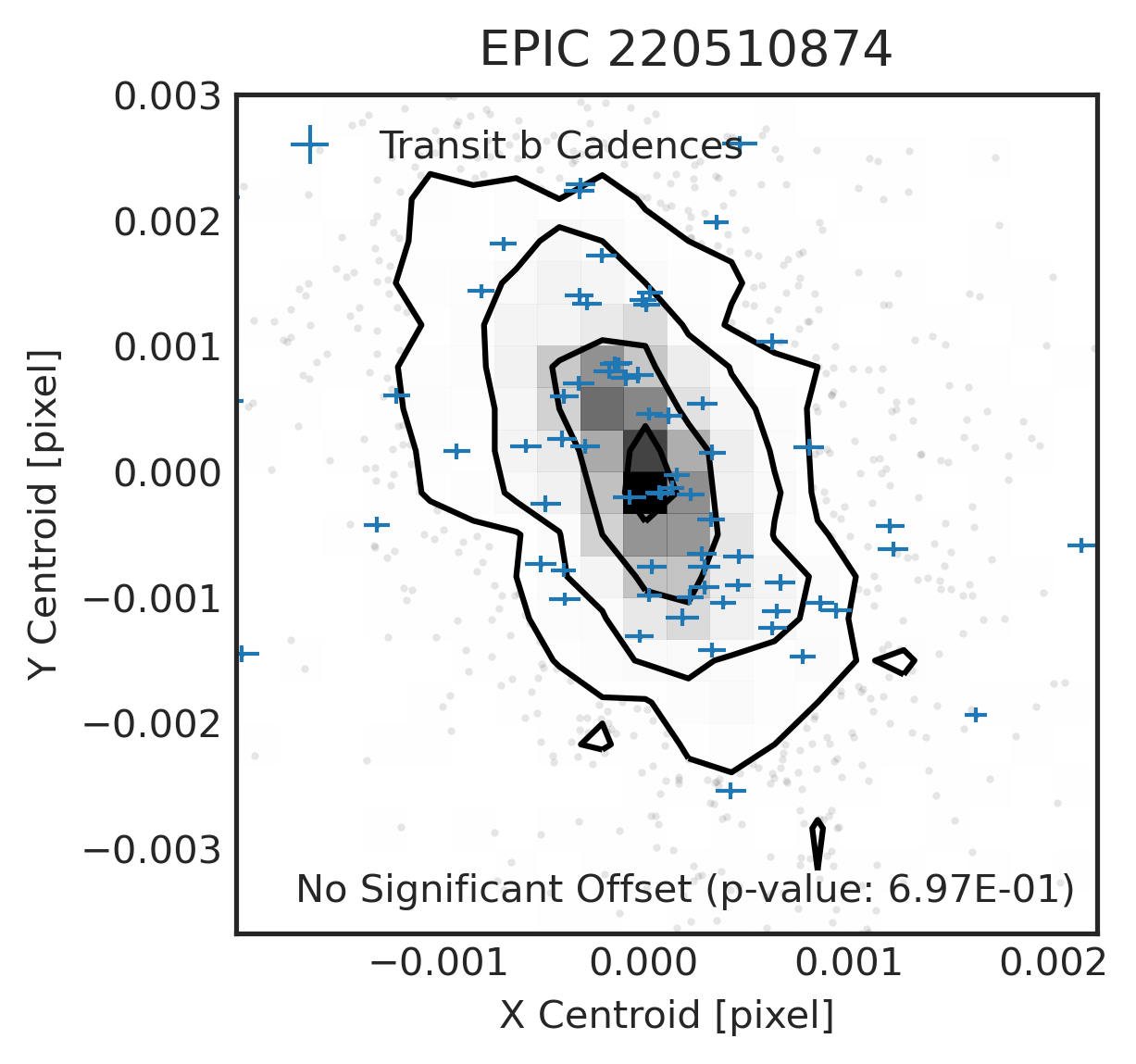} 
\includegraphics[width=0.245\textwidth]{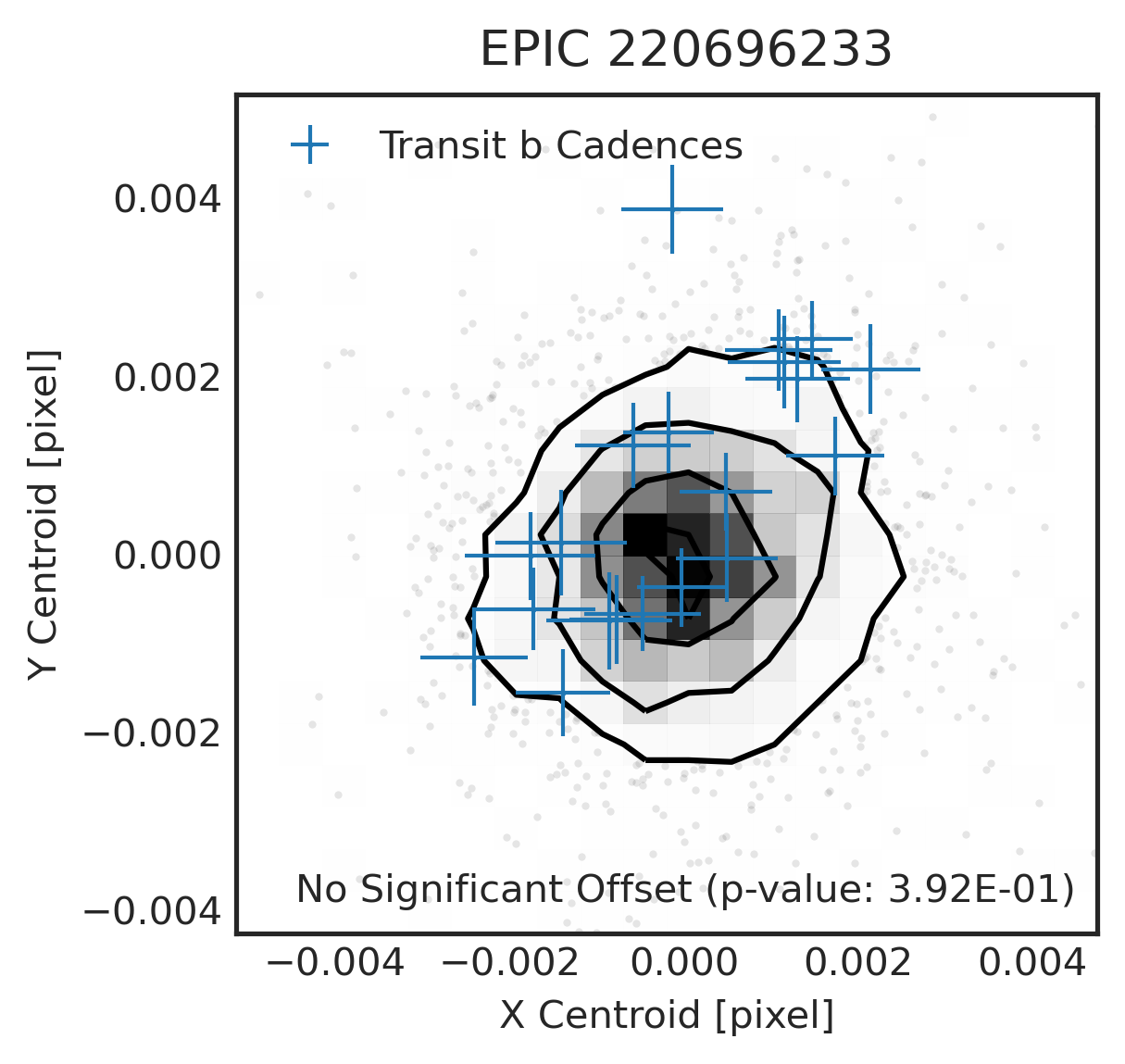}
\includegraphics[width=0.245\textwidth]{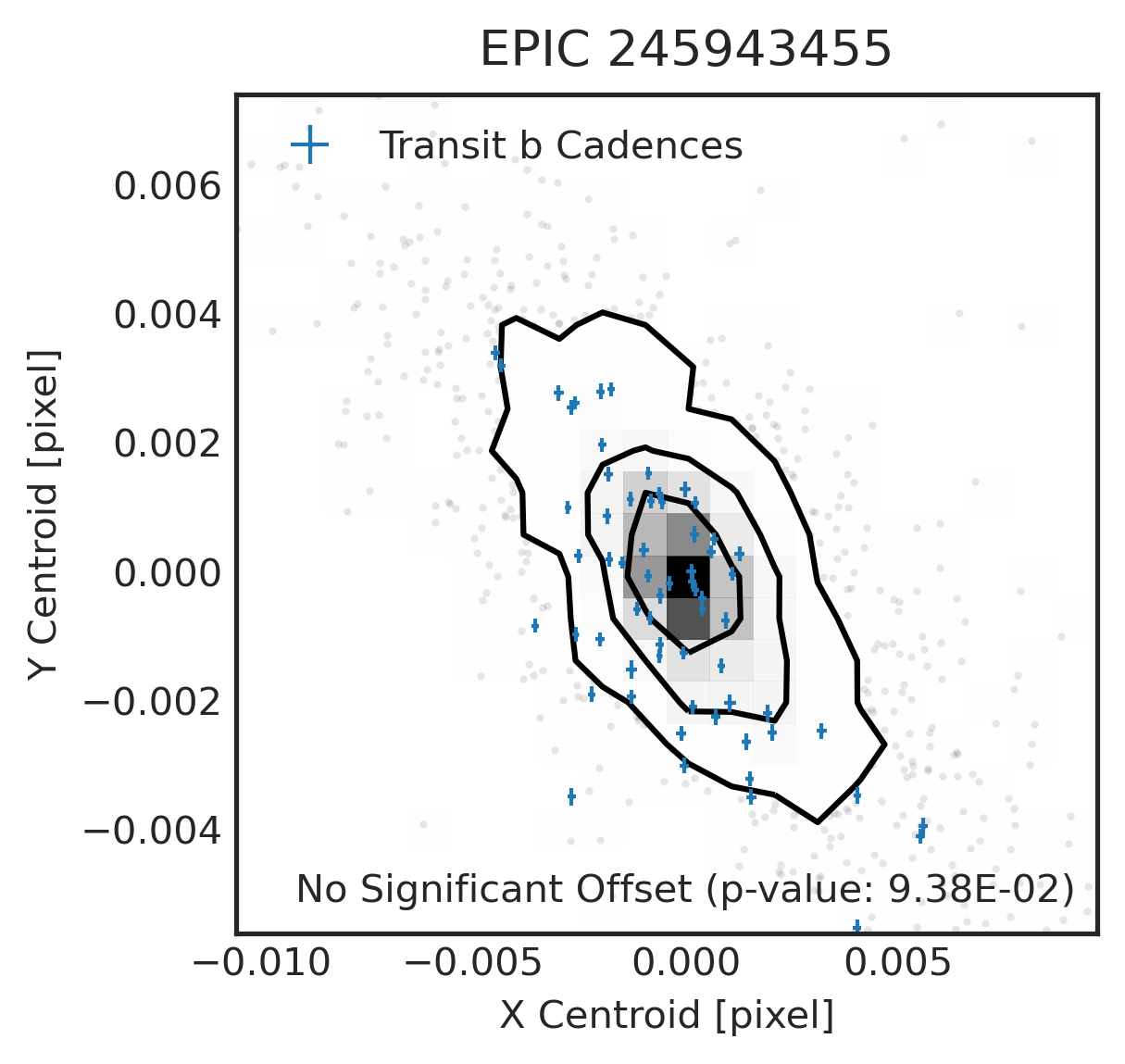}
\includegraphics[width=0.245\textwidth]{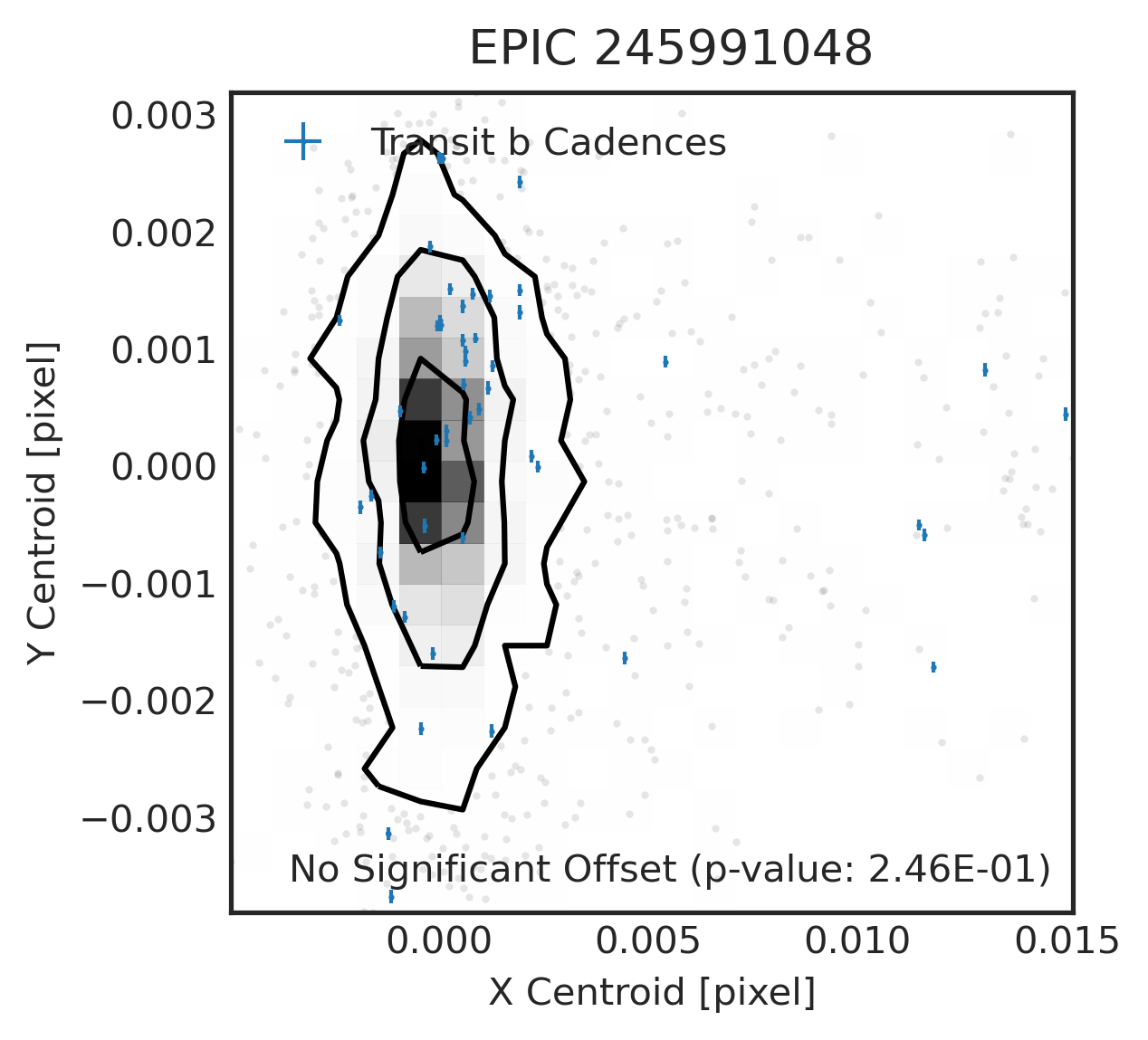}
\includegraphics[width=0.245\textwidth]{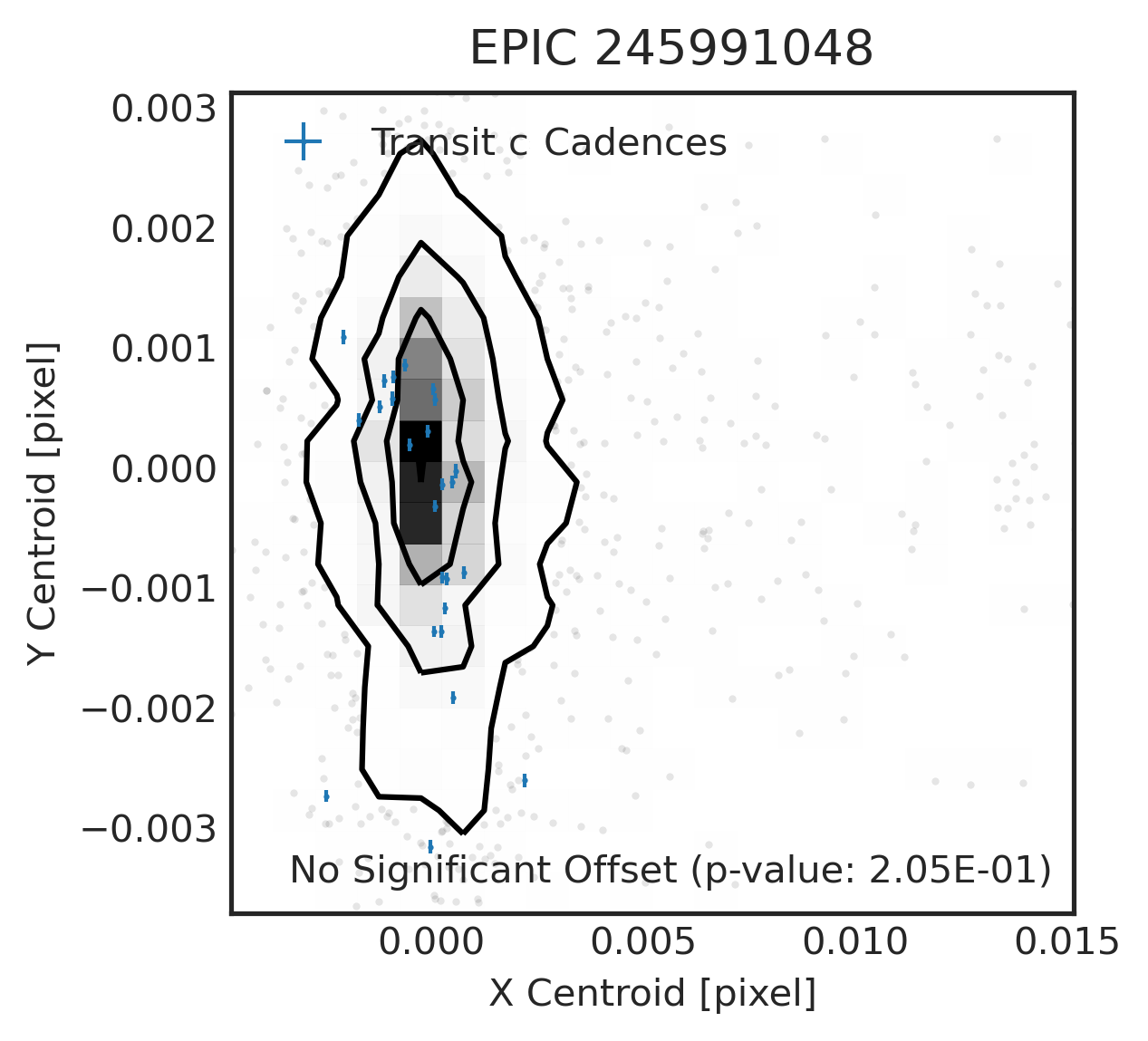}
\includegraphics[width=0.245\textwidth]{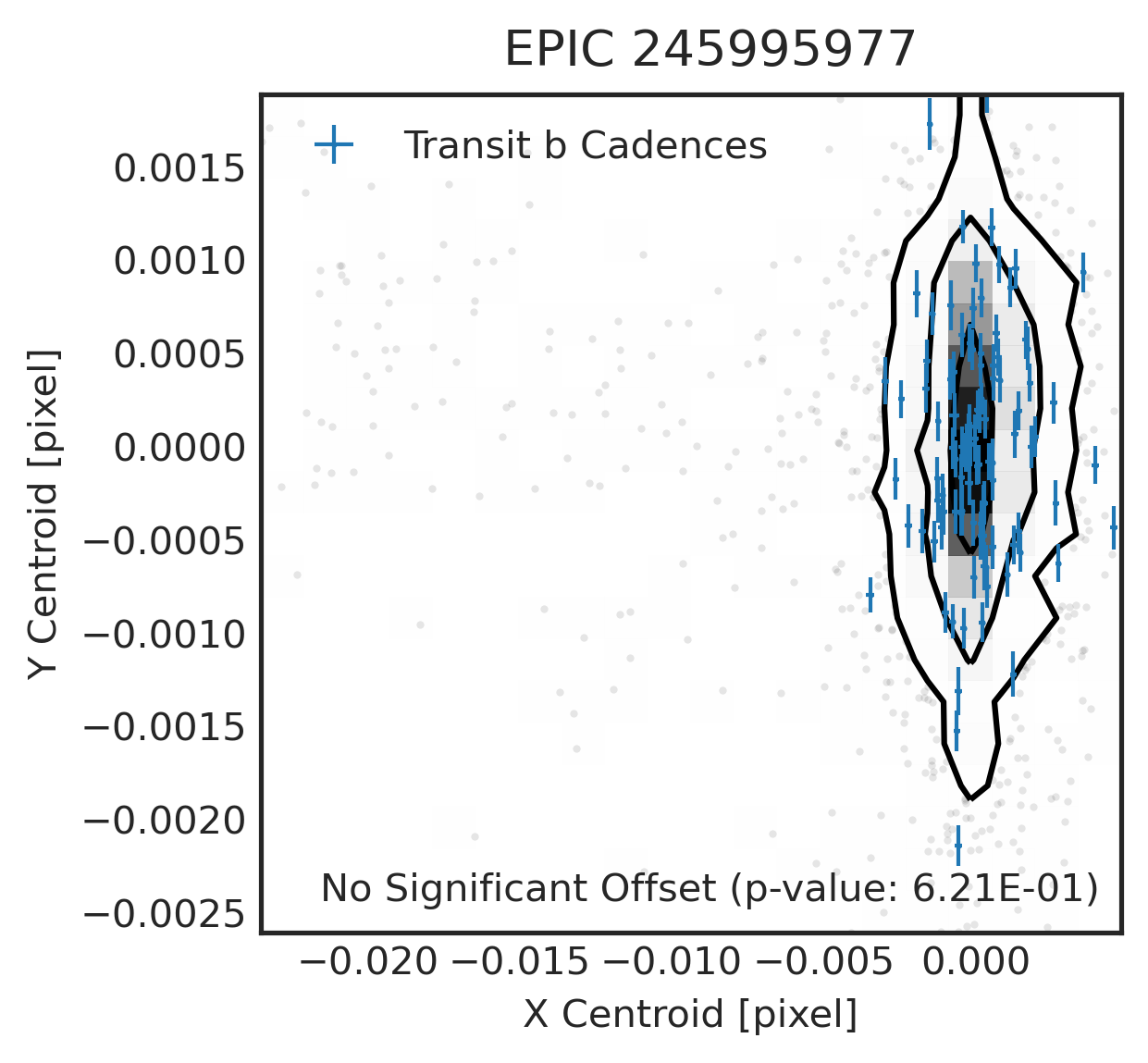}
\includegraphics[width=0.245\textwidth]{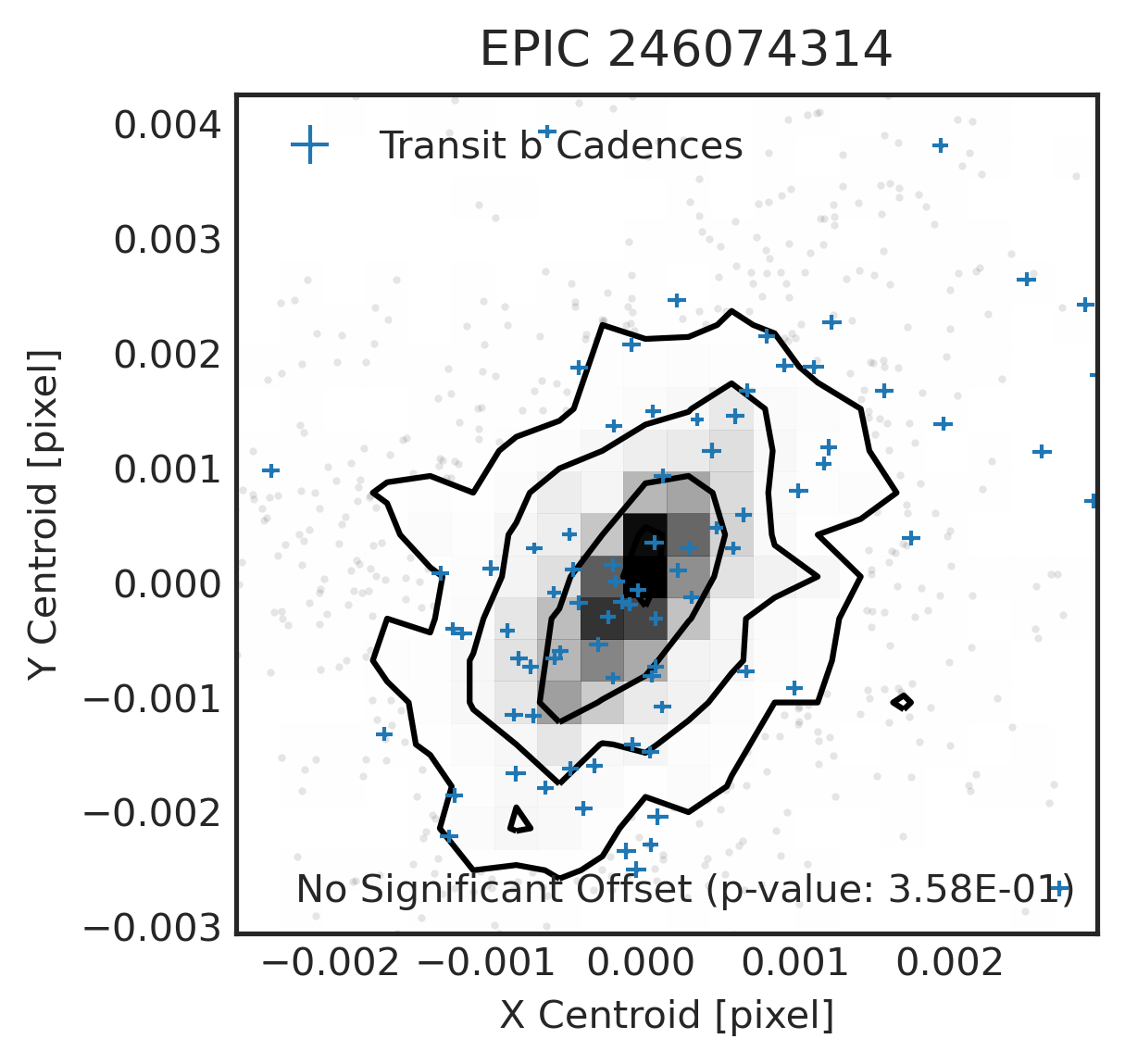}
\includegraphics[width=0.245\textwidth]{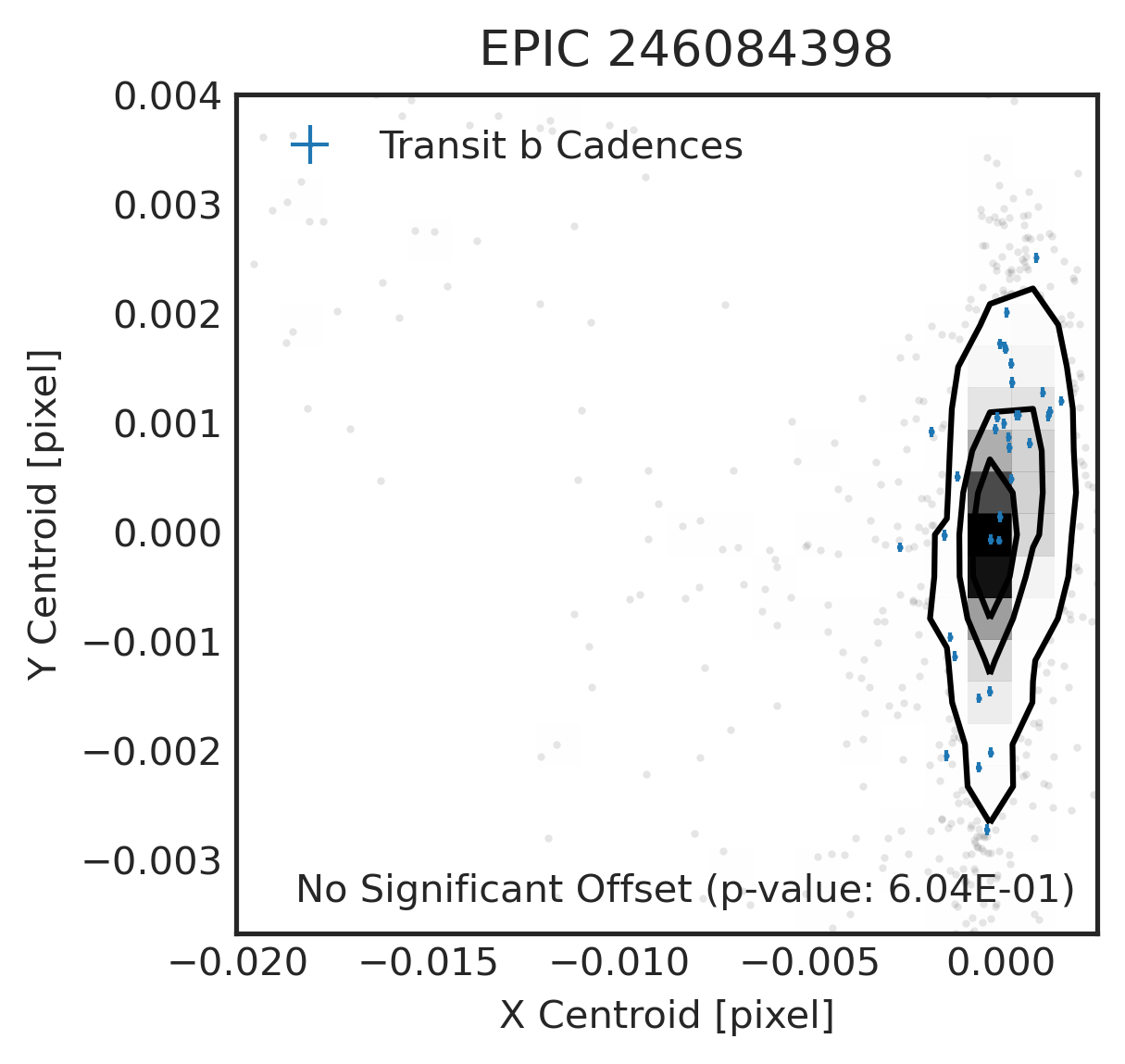}
\includegraphics[width=0.245\textwidth]{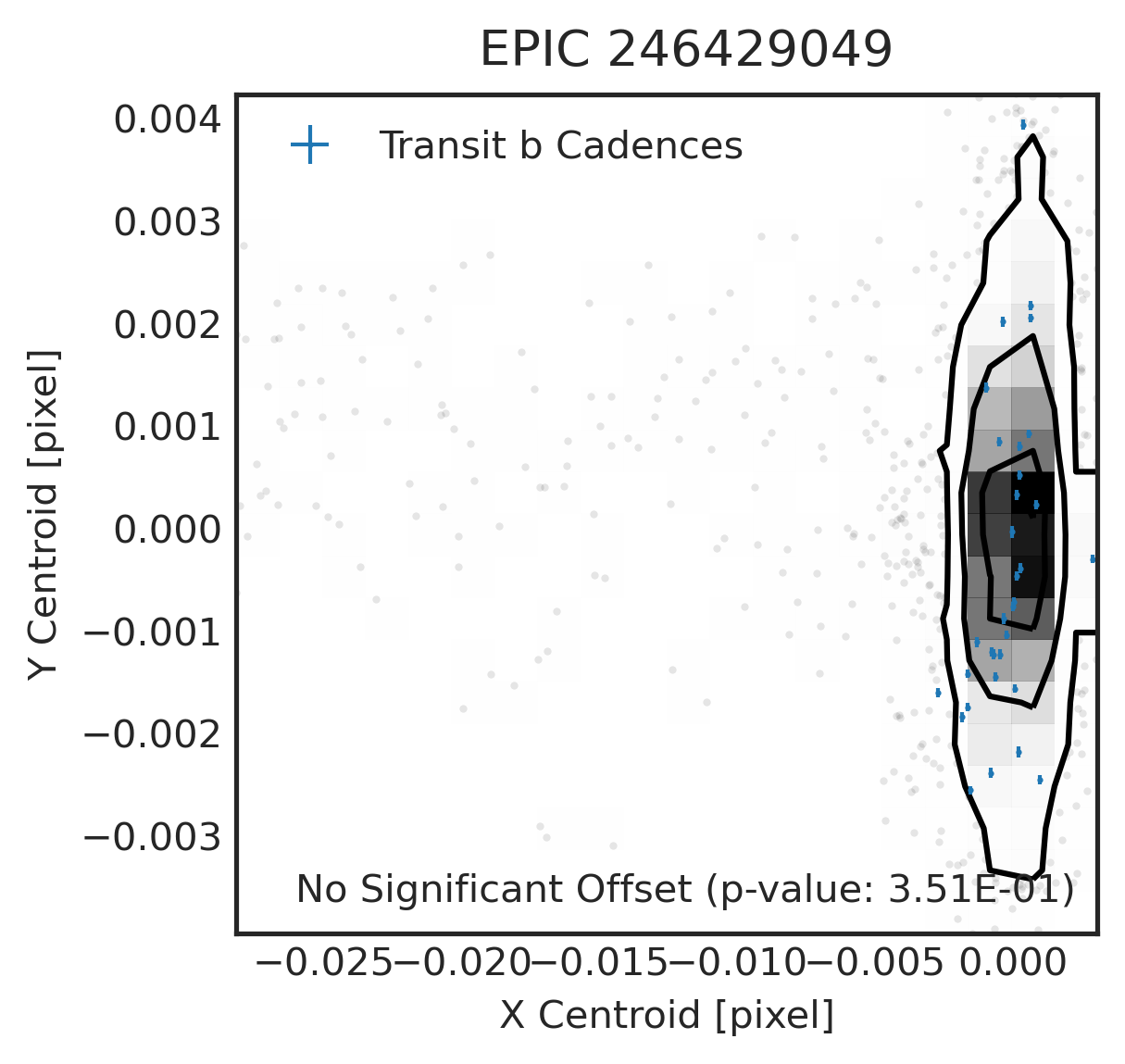}
\caption{Validated planet centroid plots. Description as for Figure \ref{fig:centroidplots1}.}
\label{fig:centroidplots2}
\end{figure}

\begin{figure}
\centering
\includegraphics[width=0.245\textwidth]{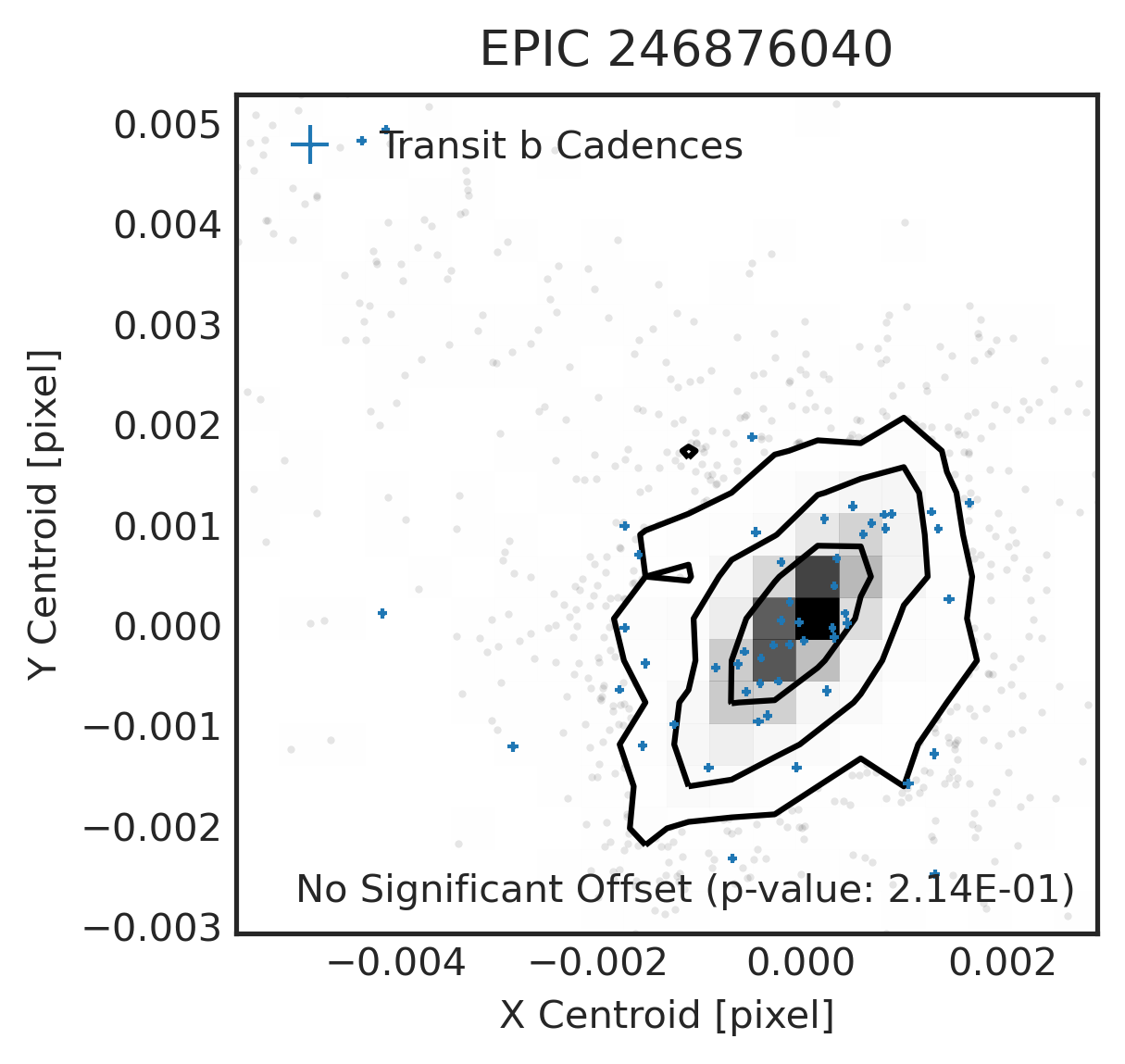}
\includegraphics[width=0.245\textwidth]{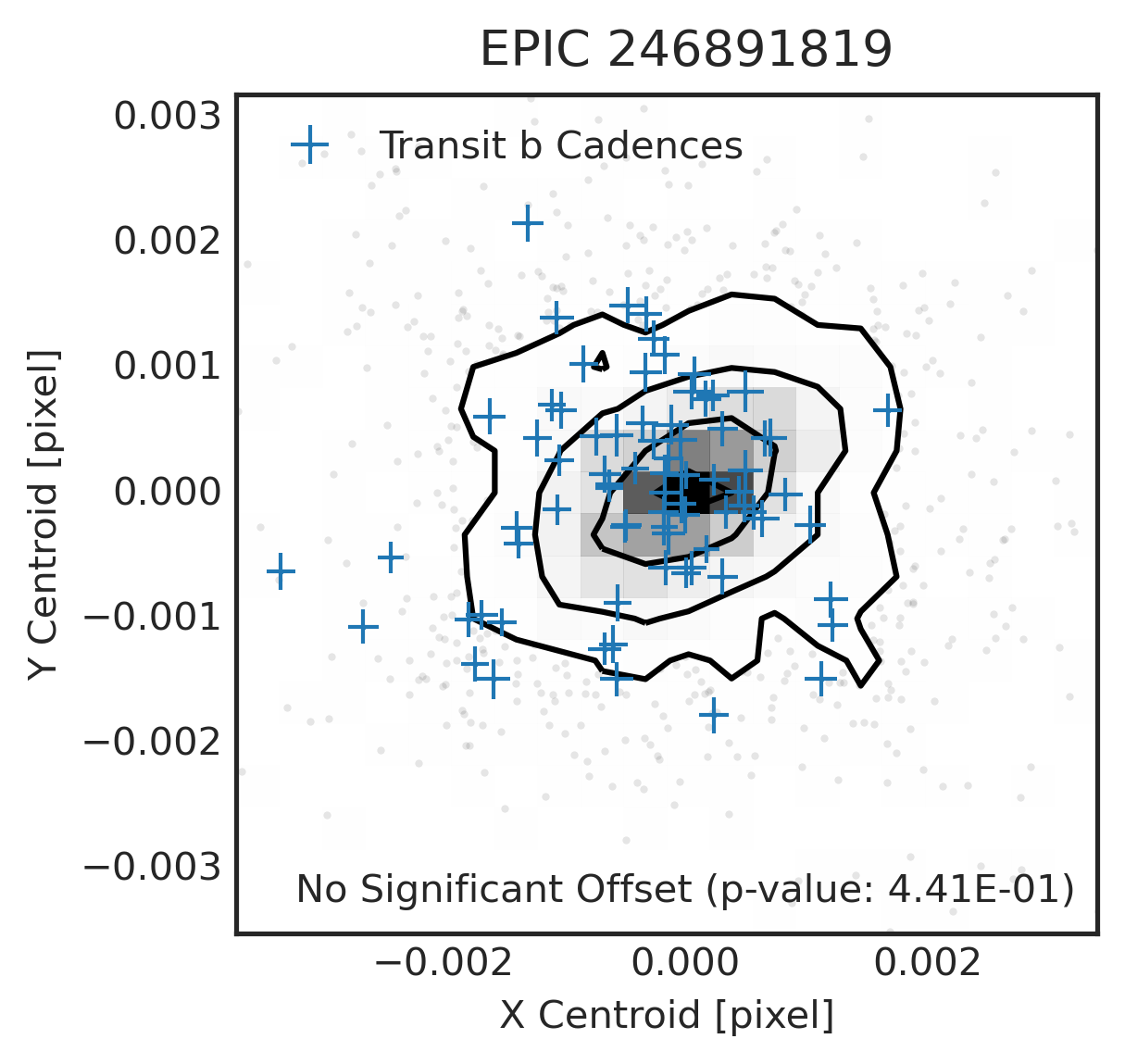}	
\includegraphics[width=0.245\textwidth]{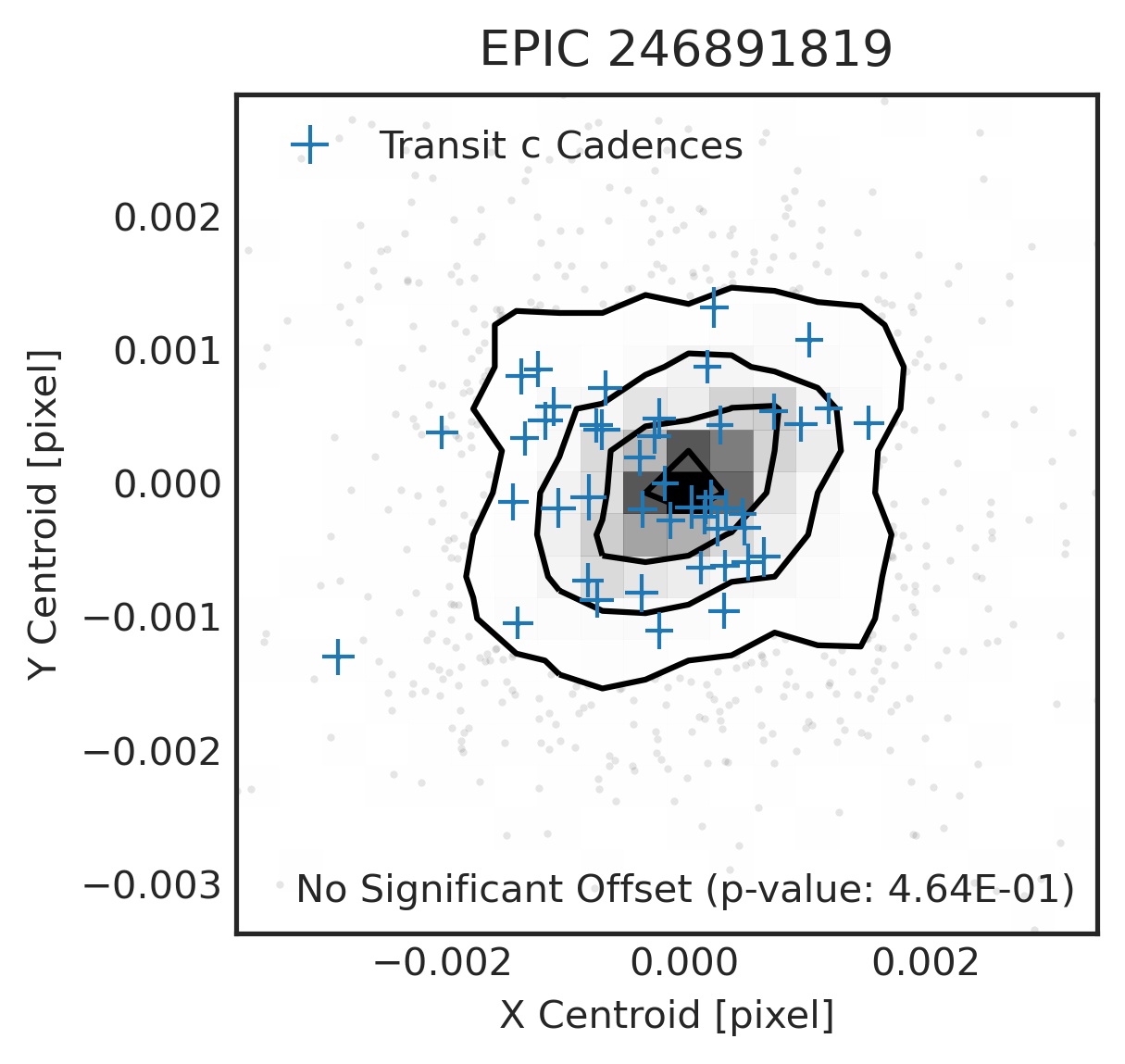}
\includegraphics[width=0.245\textwidth]{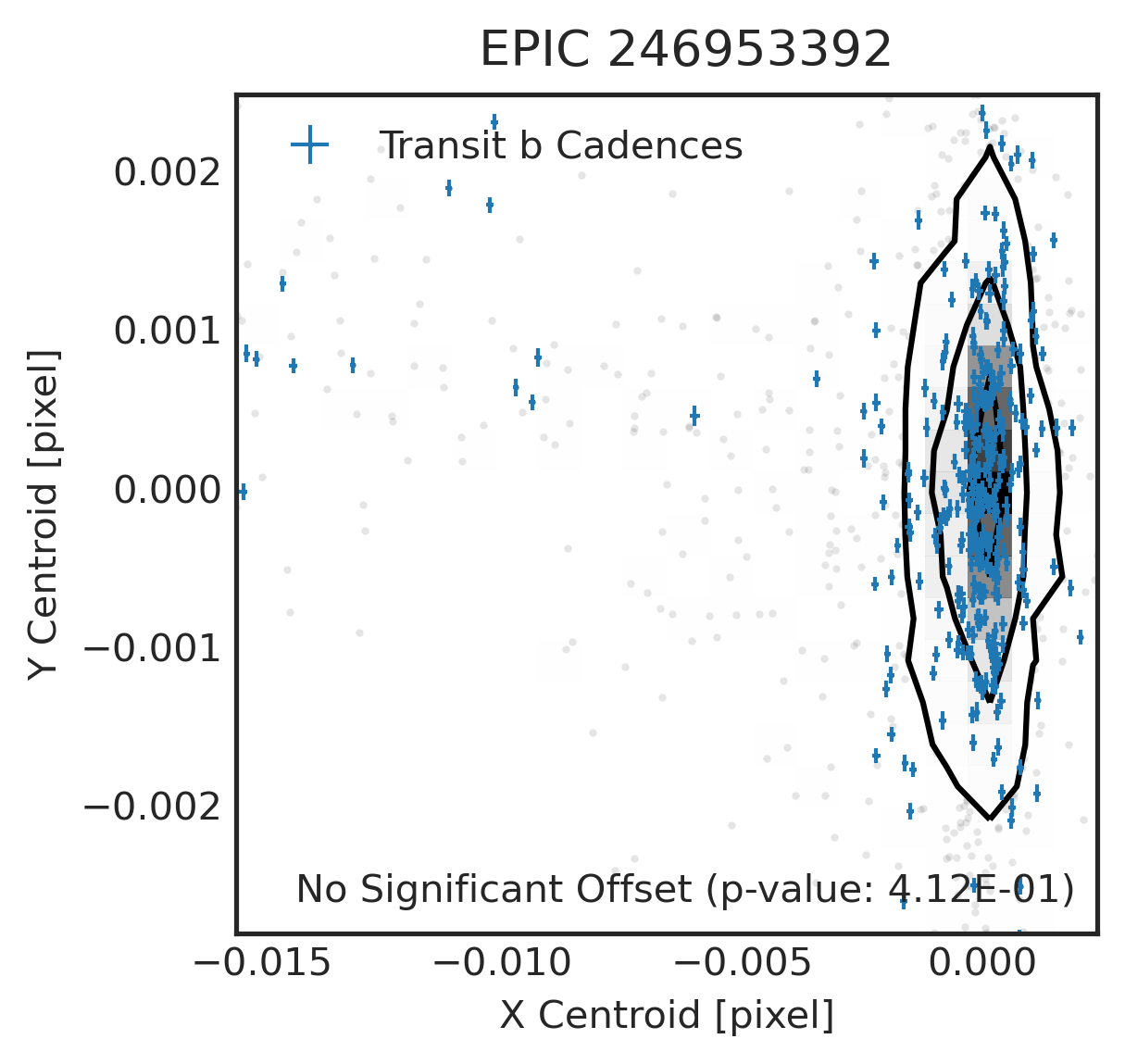}
\includegraphics[width=0.245\textwidth]{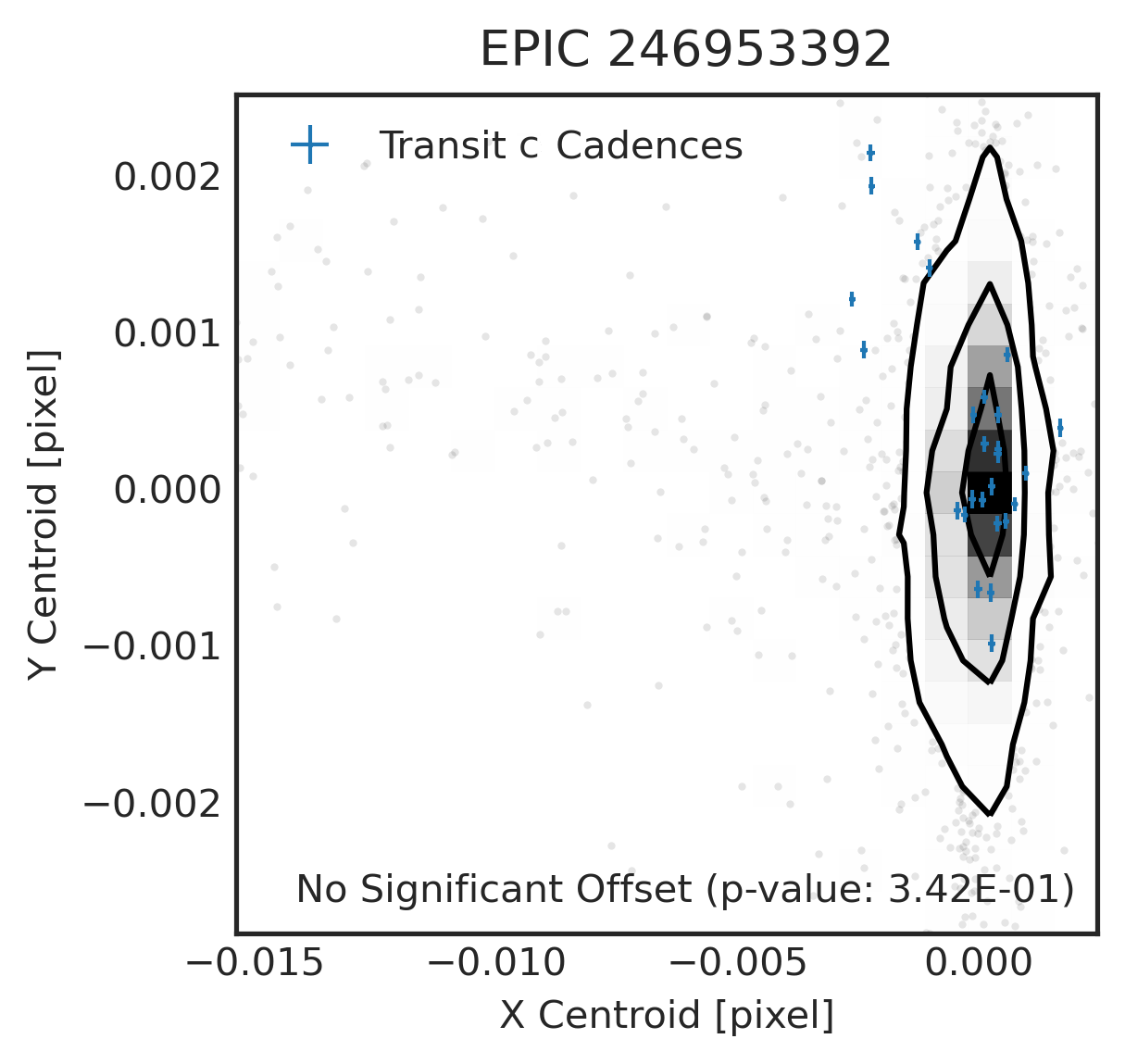}
\includegraphics[width=0.245\textwidth]{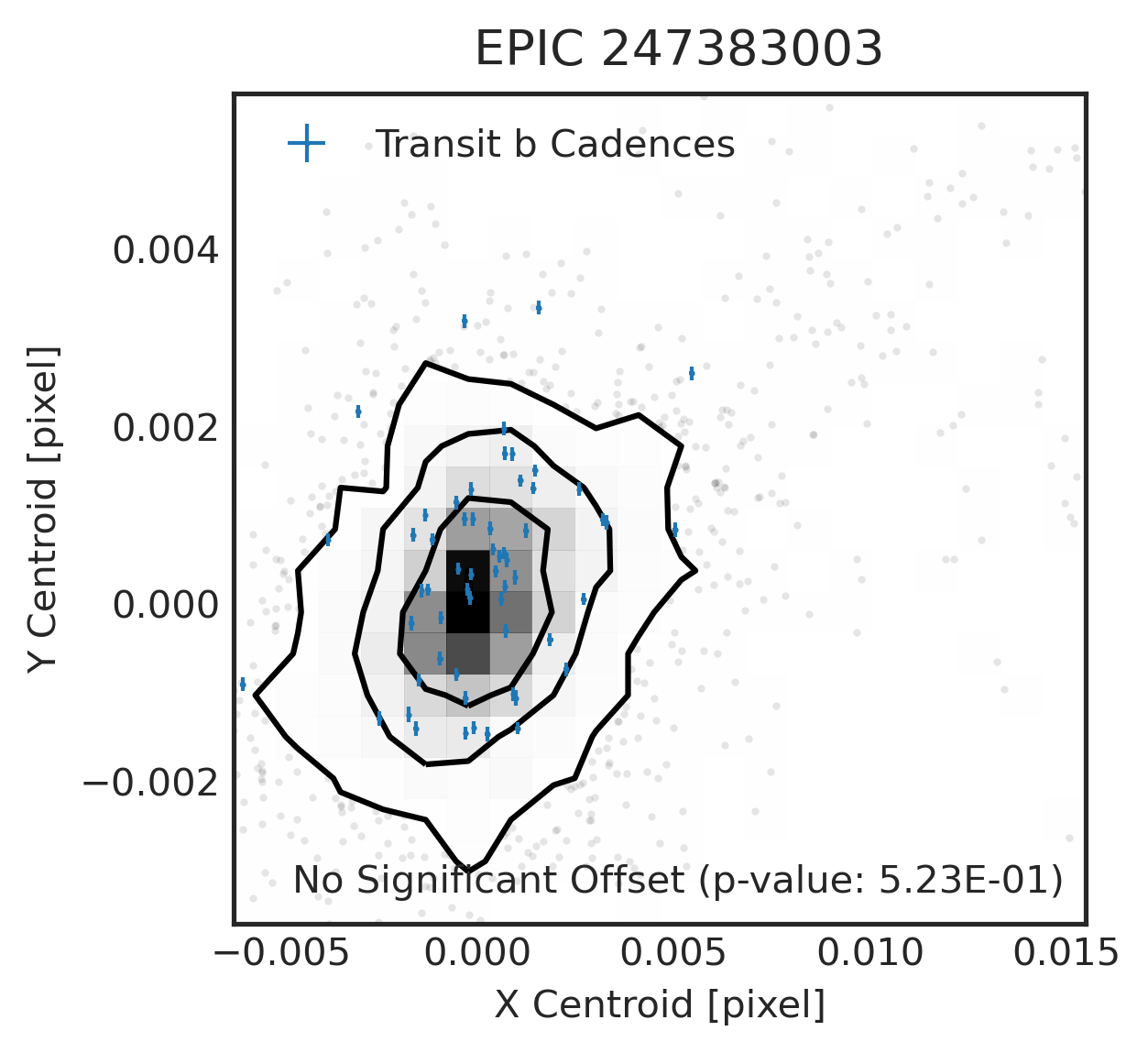}
\includegraphics[width=0.245\textwidth]{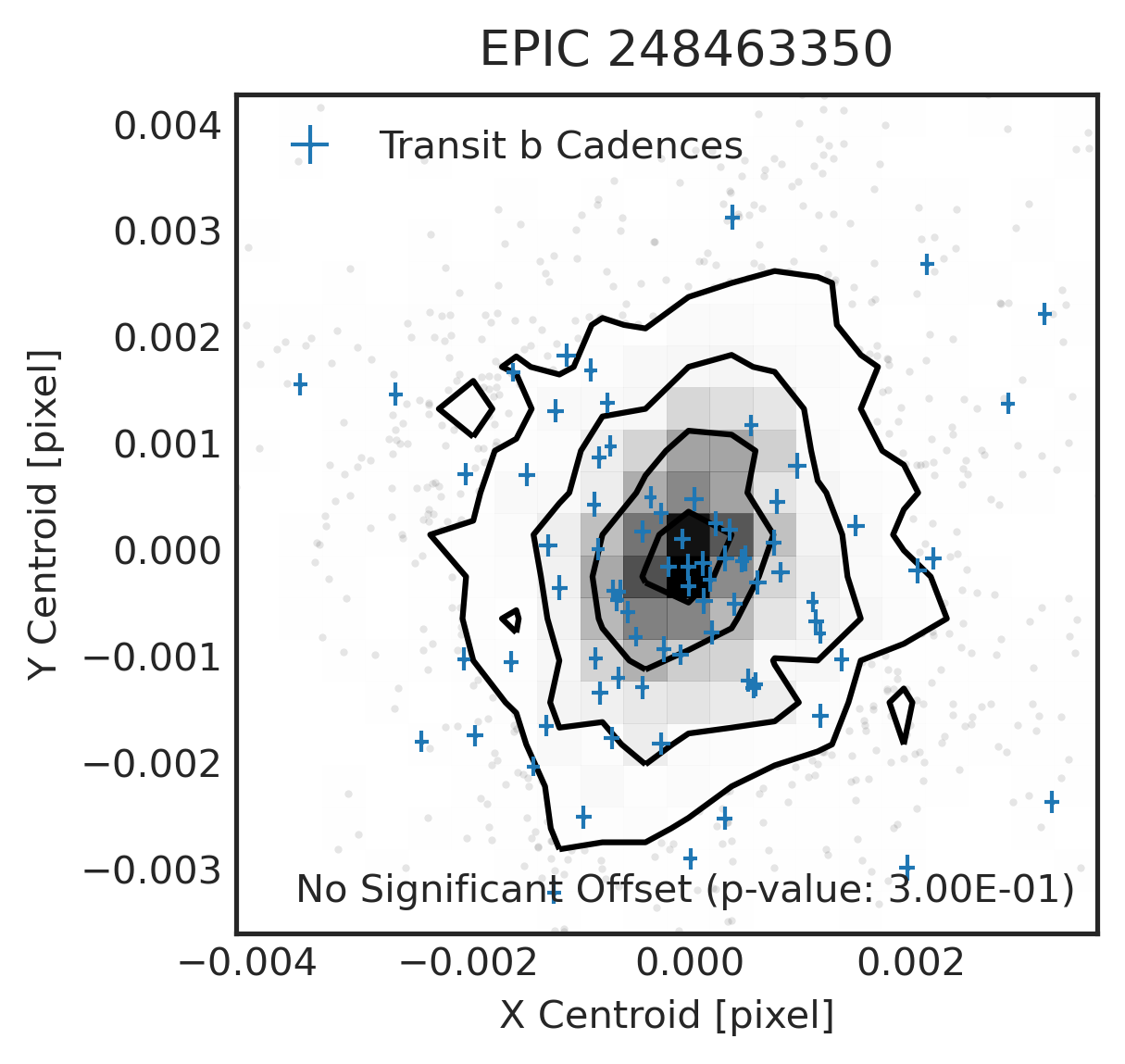}
\includegraphics[width=0.245\textwidth]{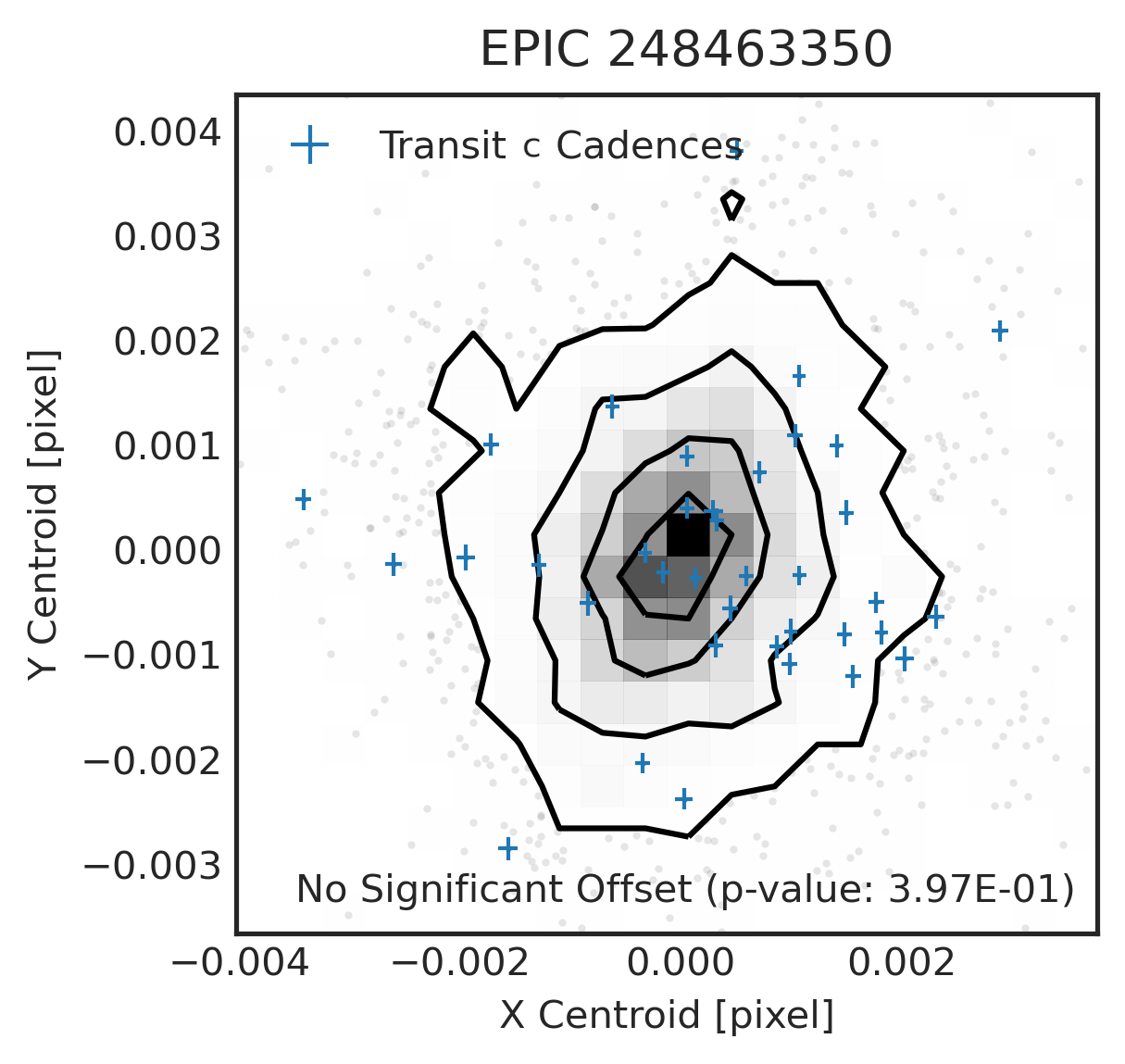}
\includegraphics[width=0.245\textwidth]{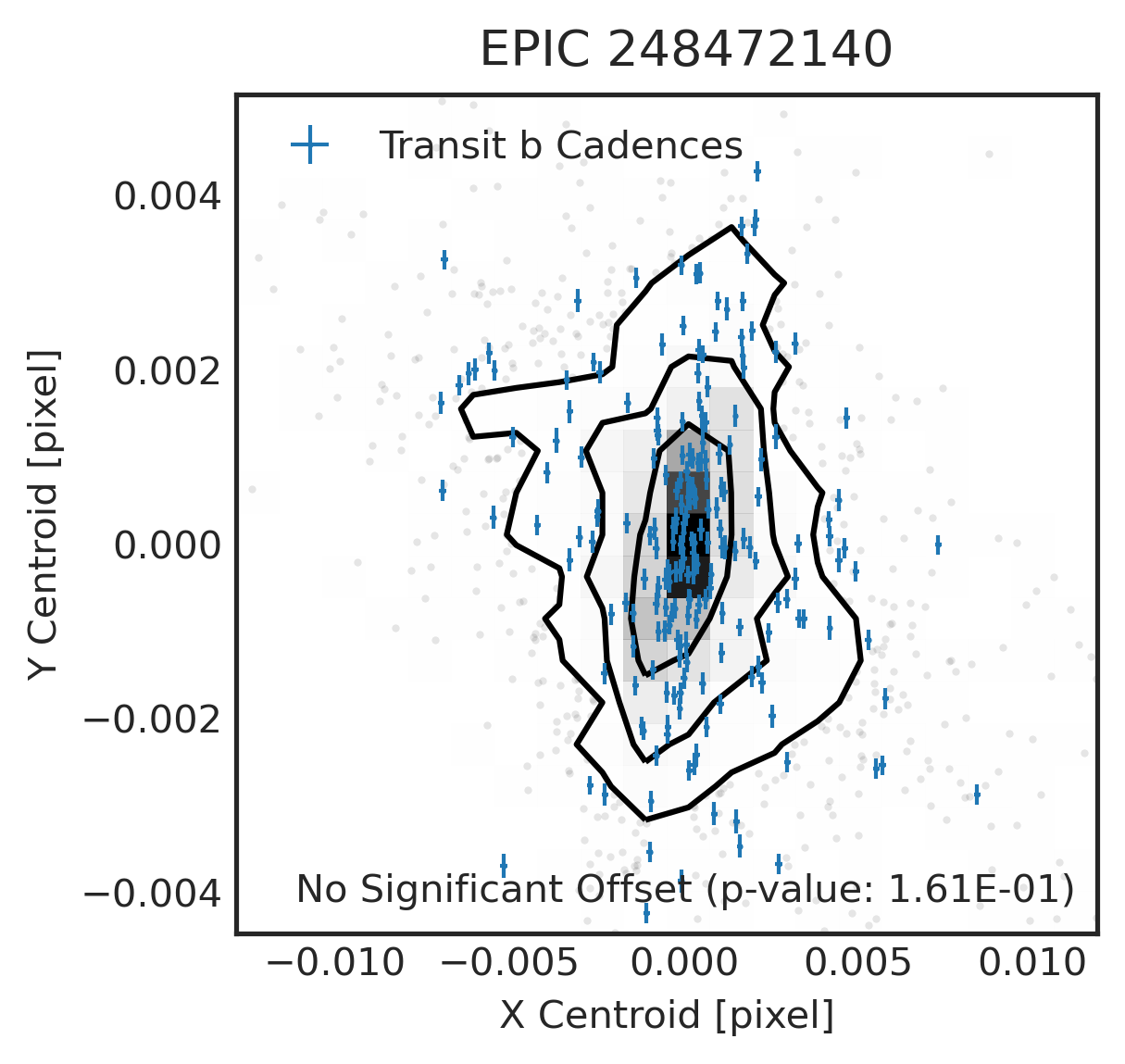}
\includegraphics[width=0.245\textwidth]{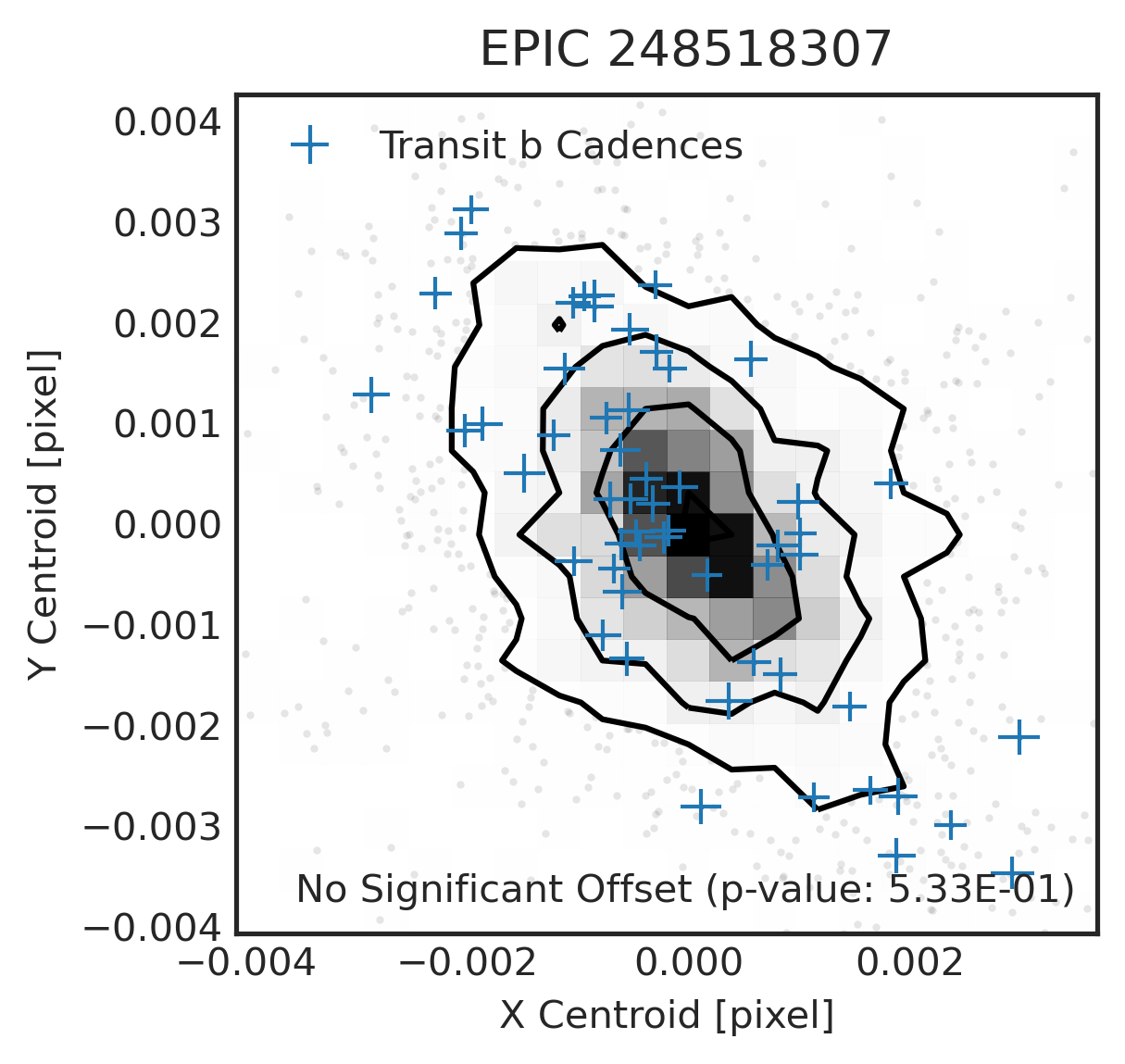}
\includegraphics[width=0.245\textwidth]{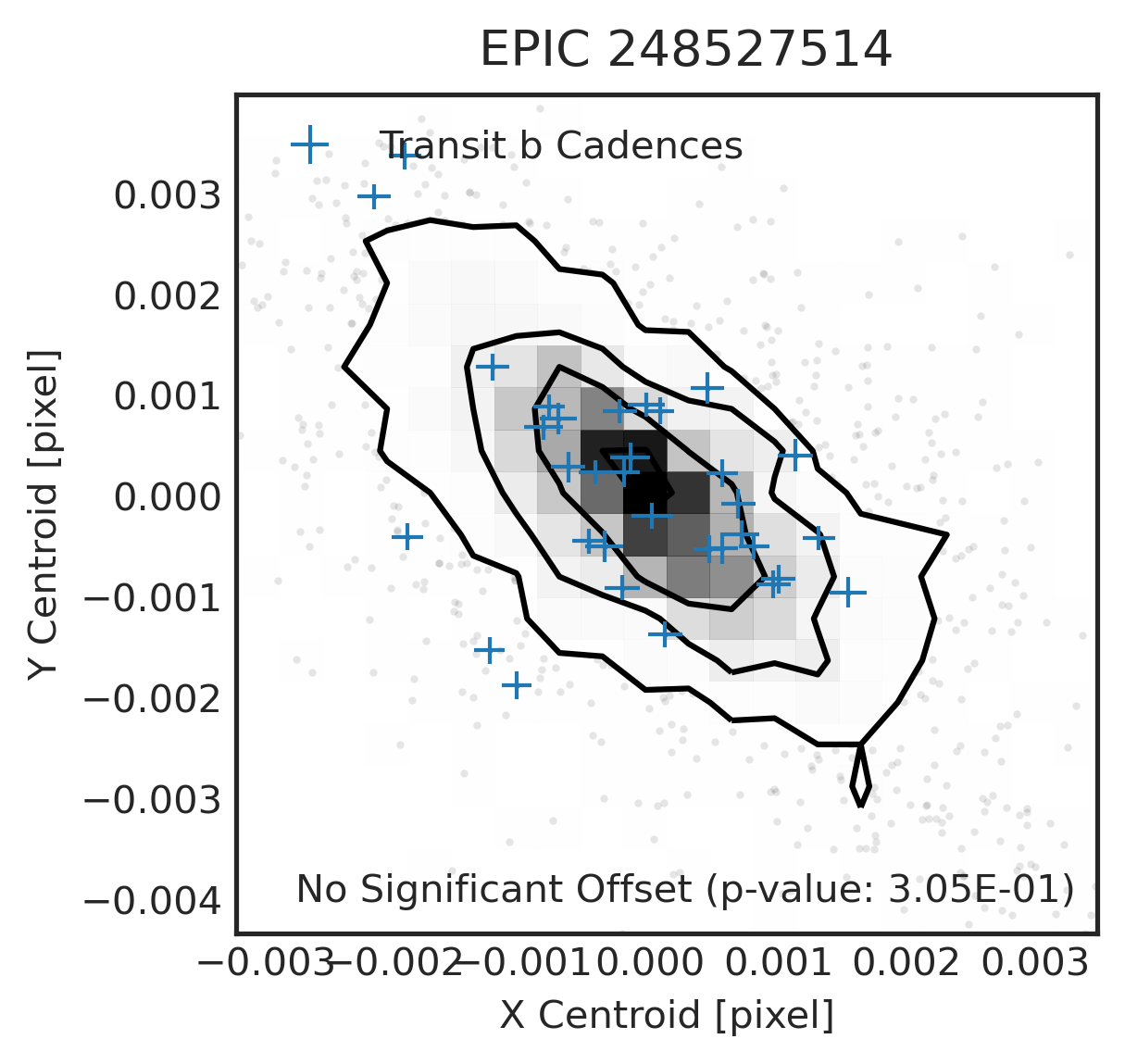}
\includegraphics[width=0.245\textwidth]{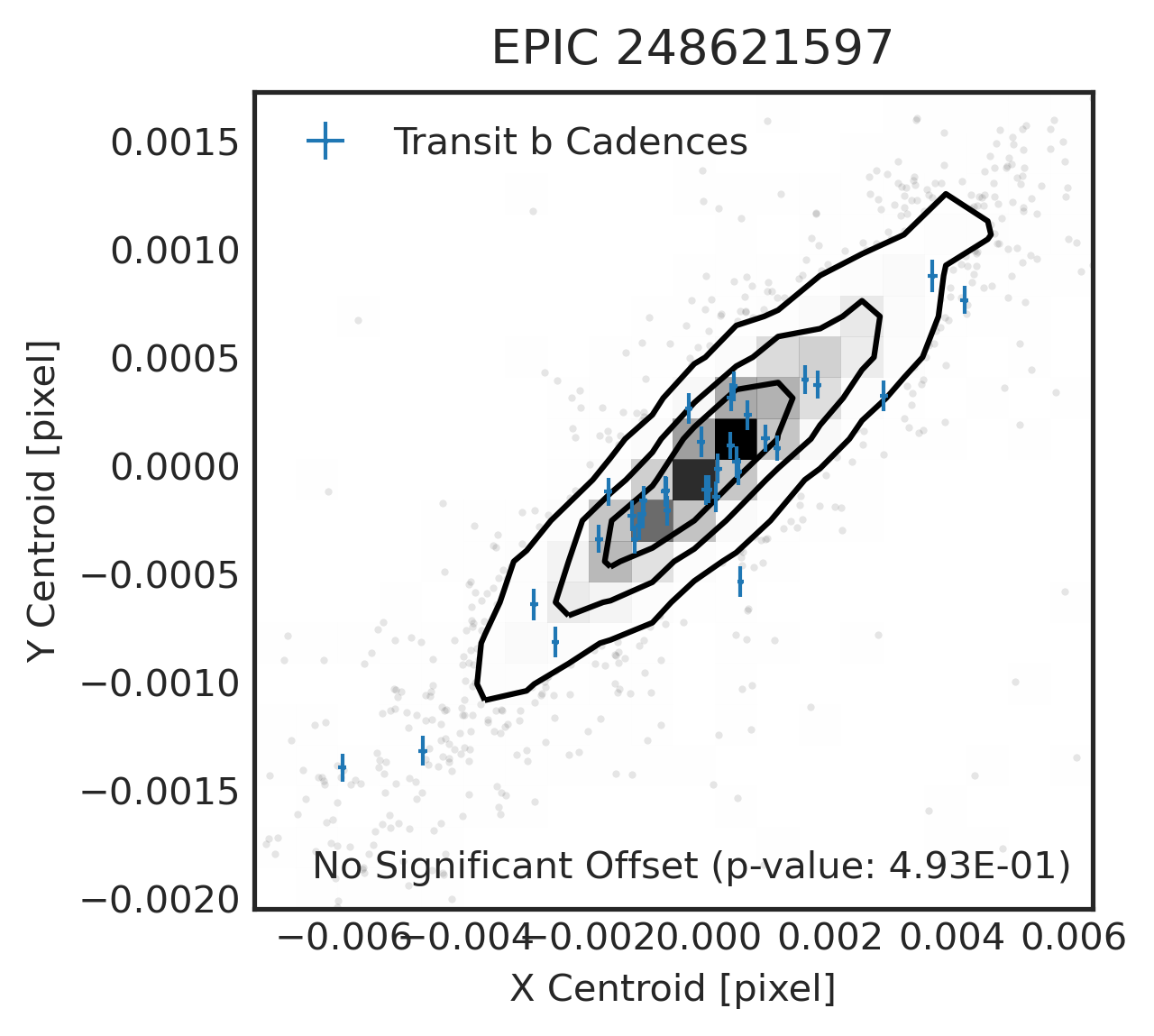}
\includegraphics[width=0.245\textwidth]{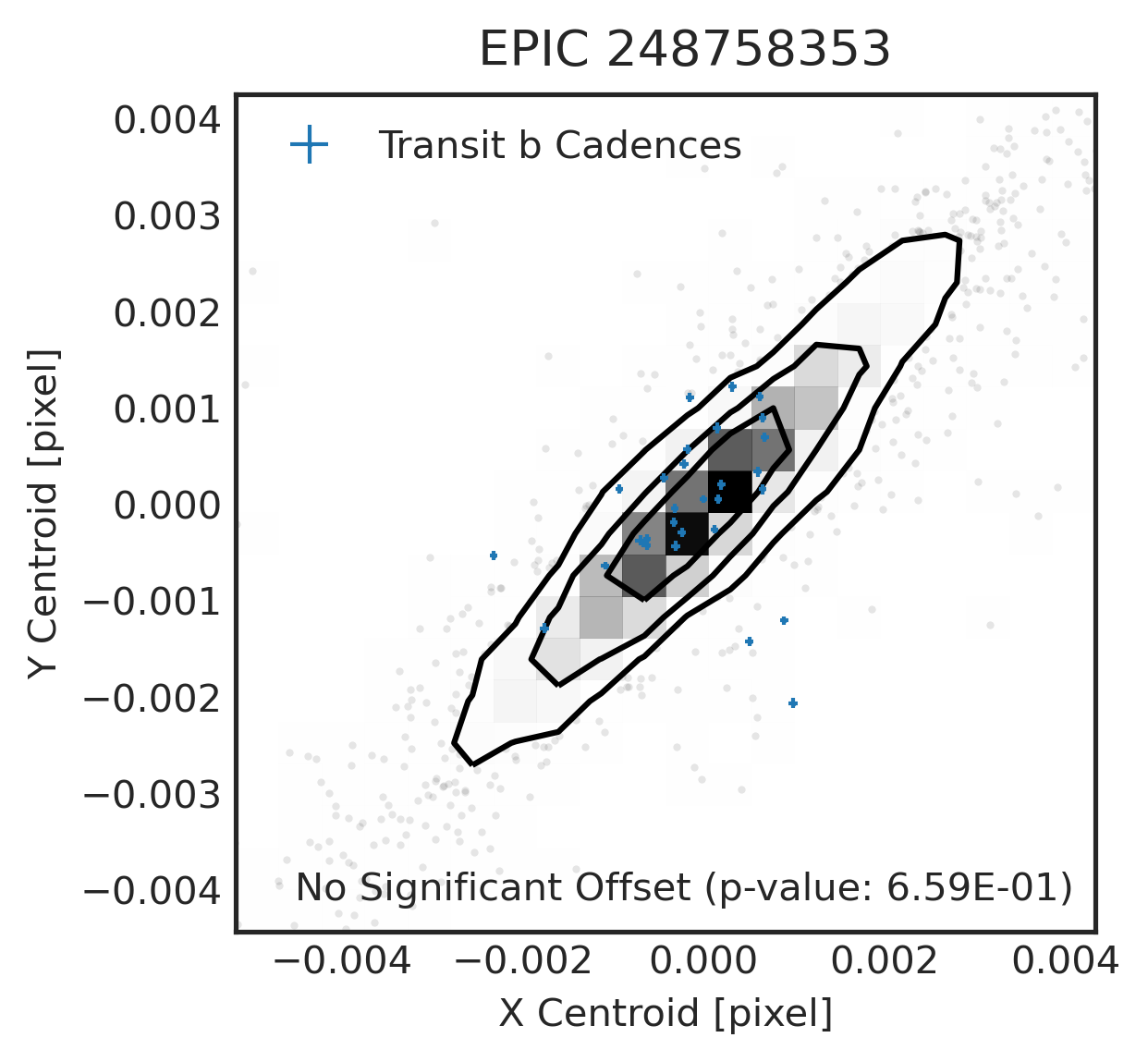}
\includegraphics[width=0.245\textwidth]{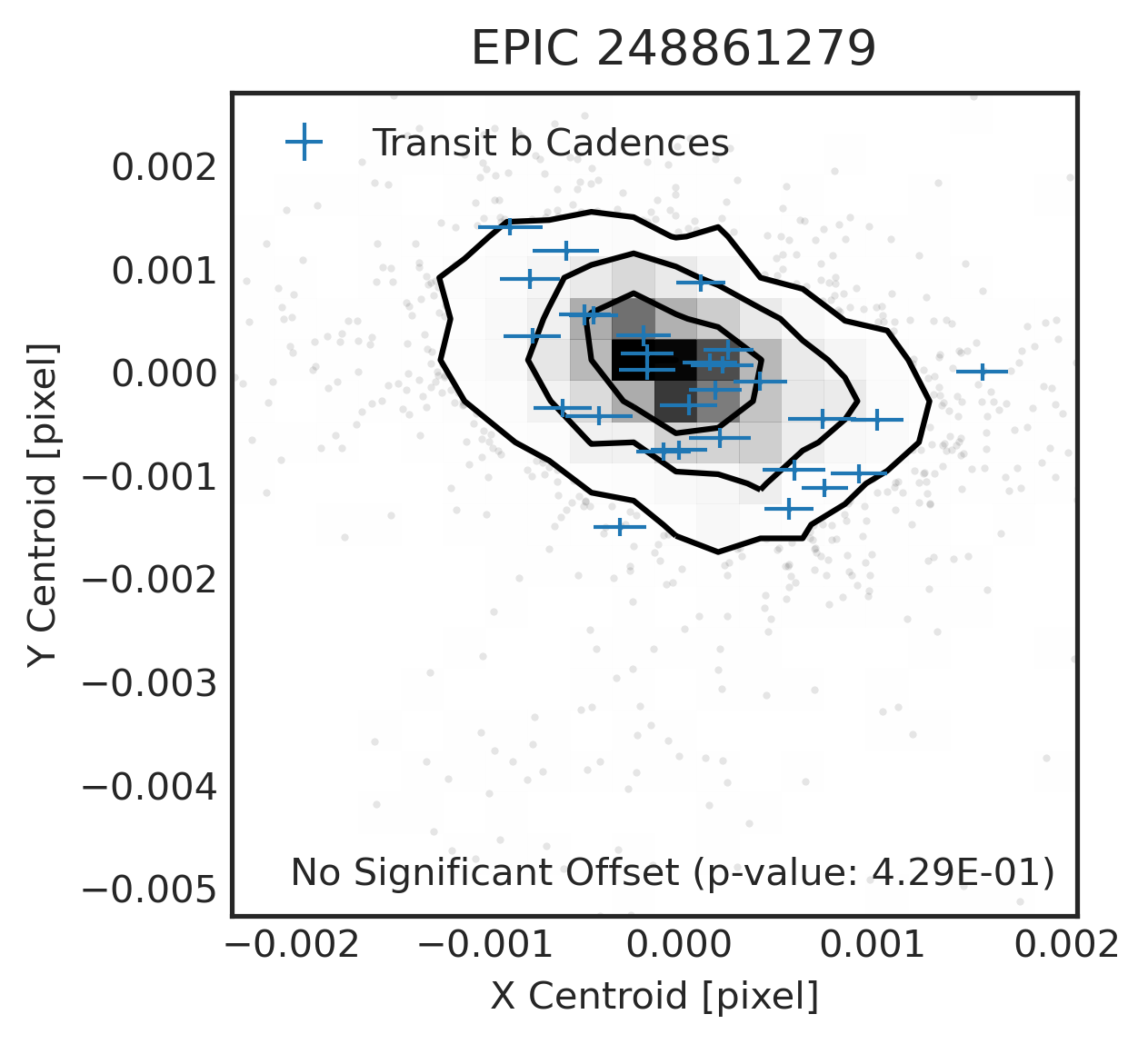}
\includegraphics[width=0.245\textwidth]{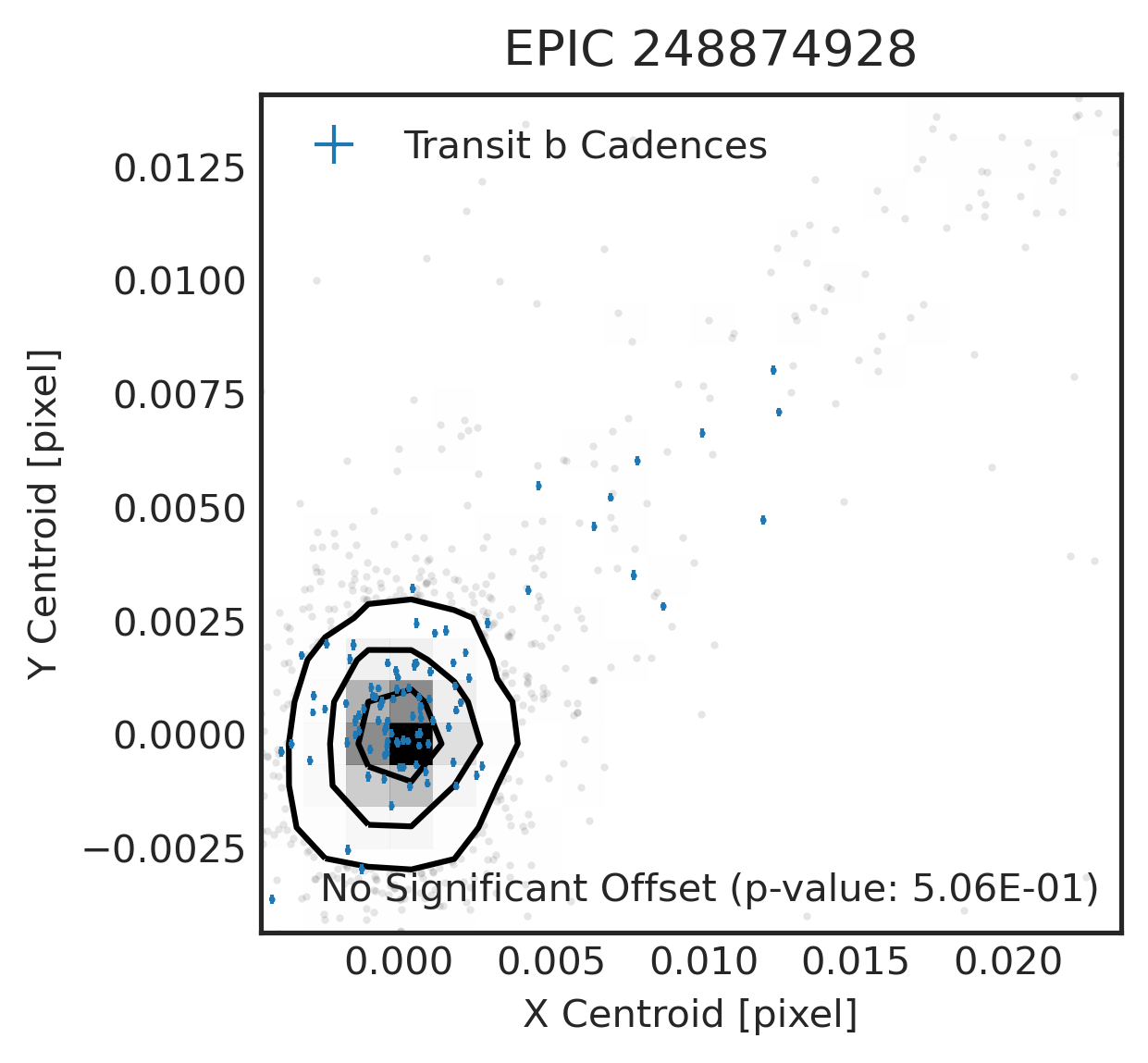}
\includegraphics[width=0.245\textwidth]{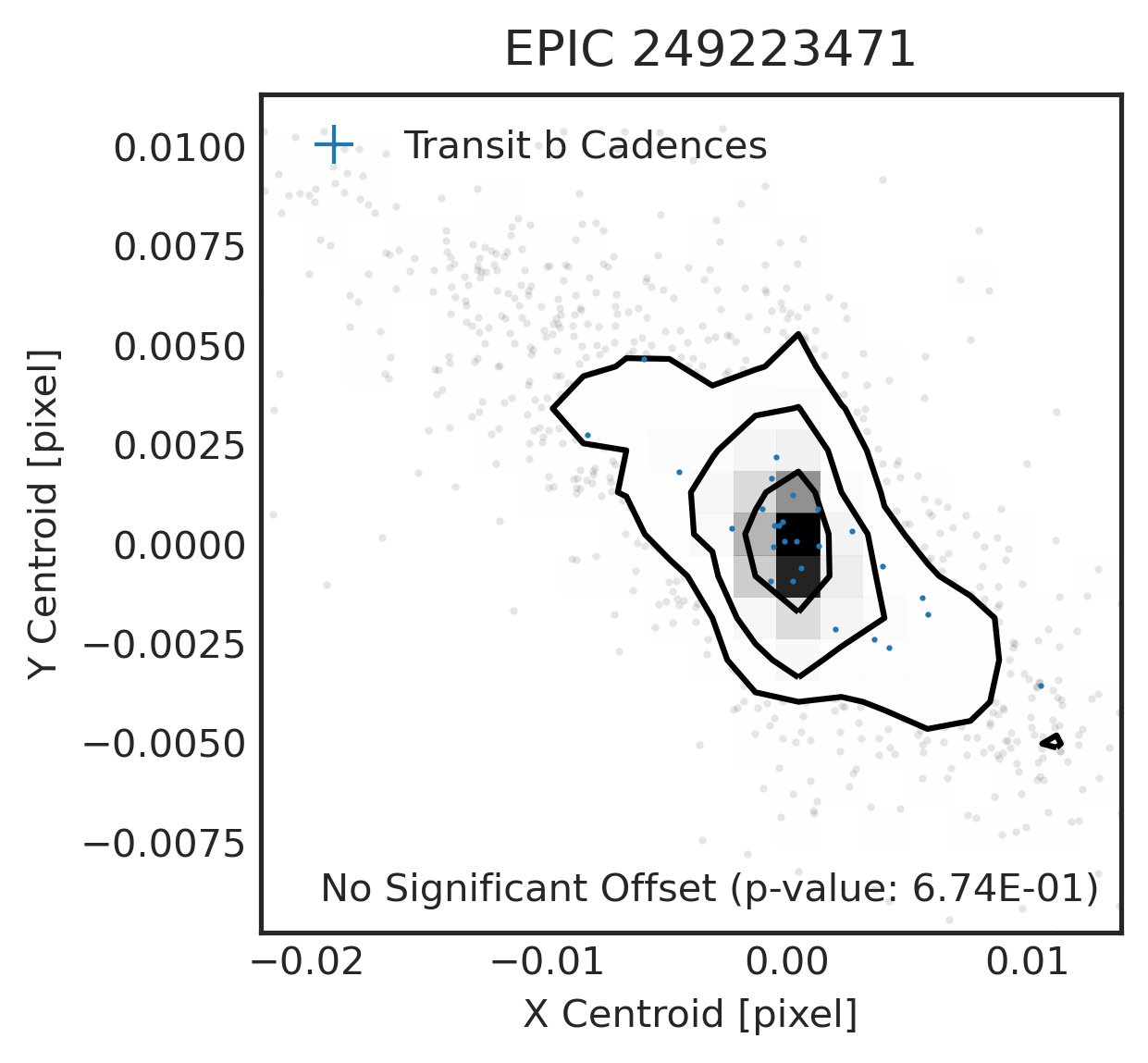}
\includegraphics[width=0.245\textwidth]{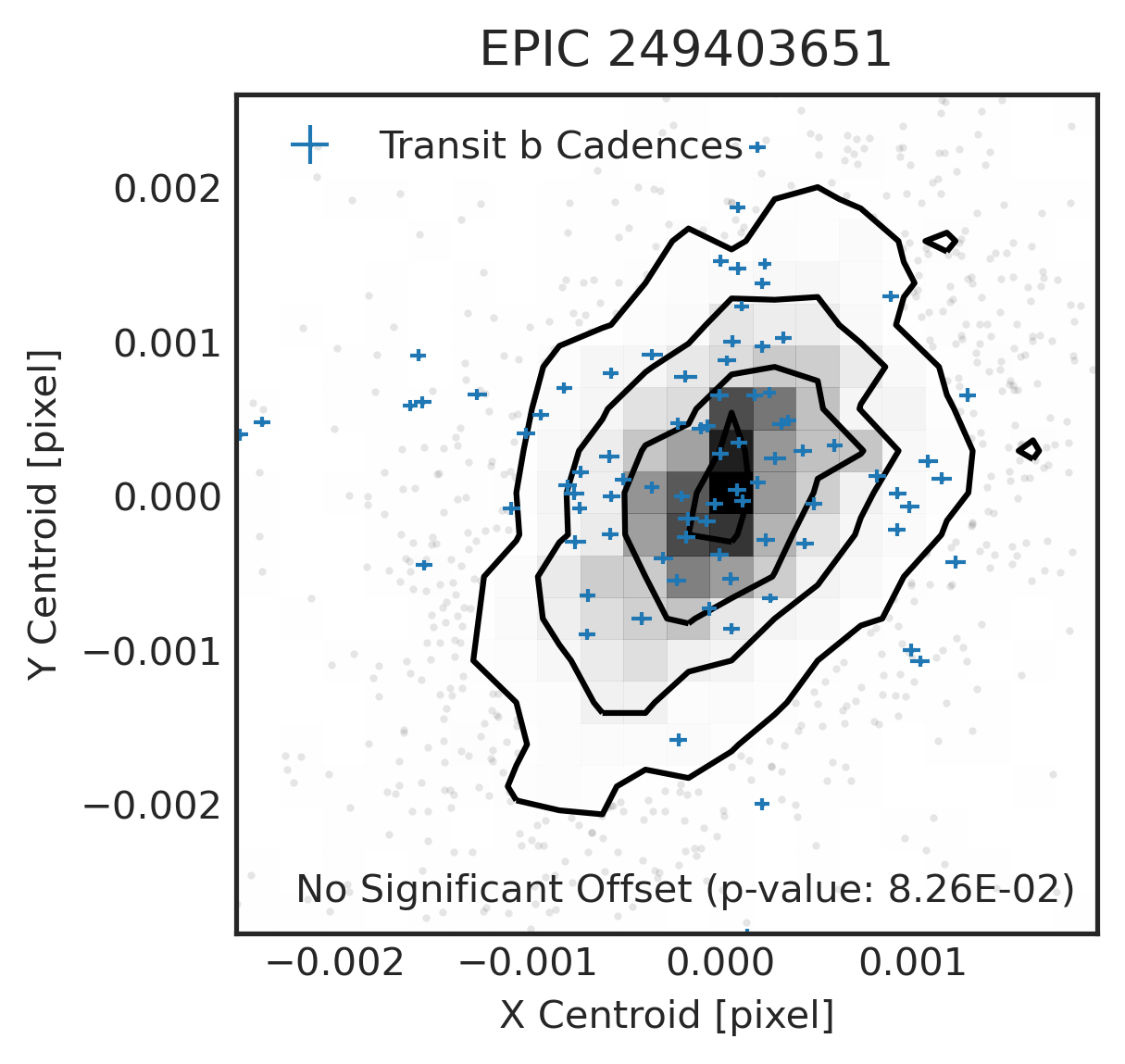}
\includegraphics[width=0.245\textwidth]{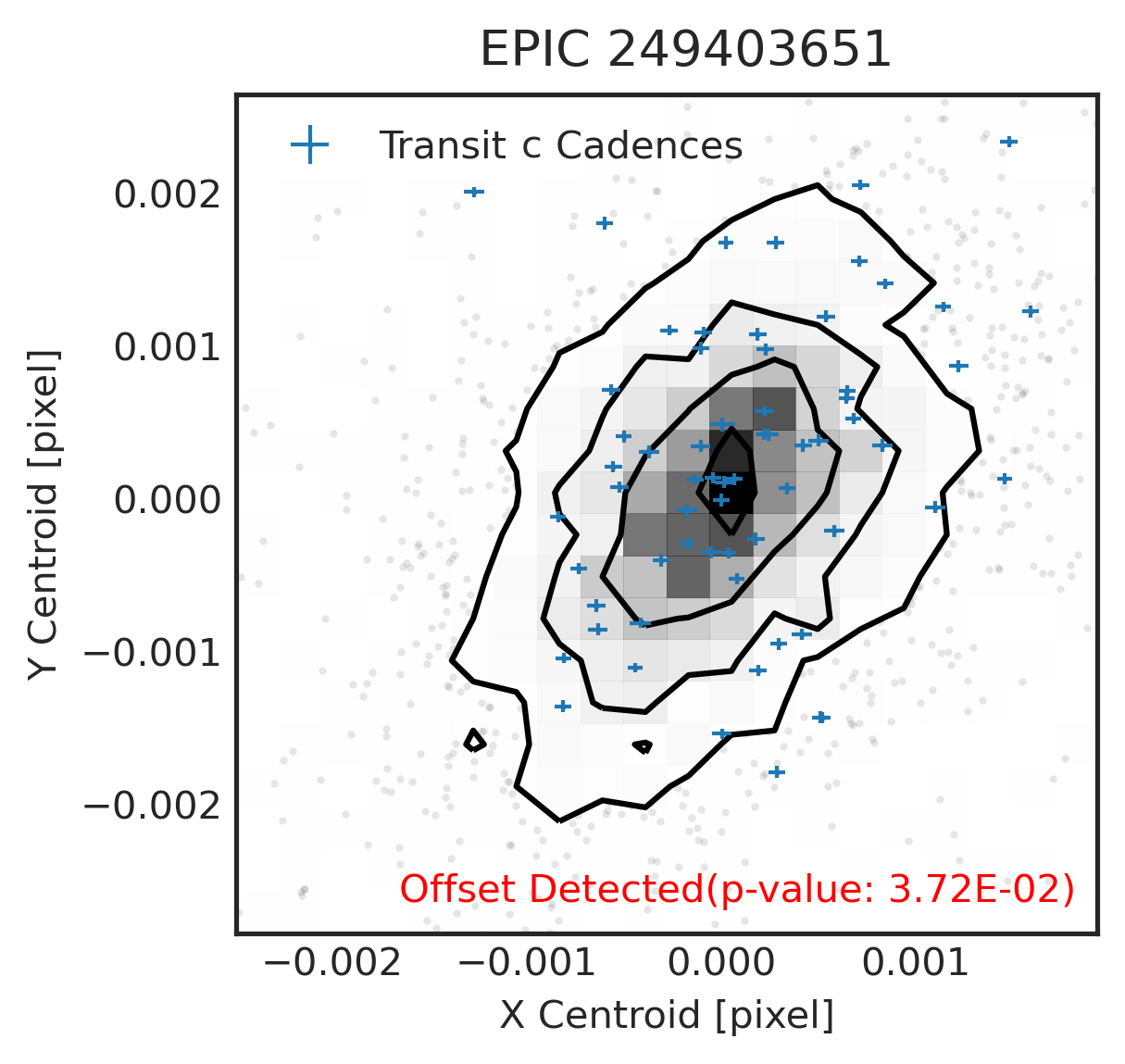}
\includegraphics[width=0.245\textwidth]{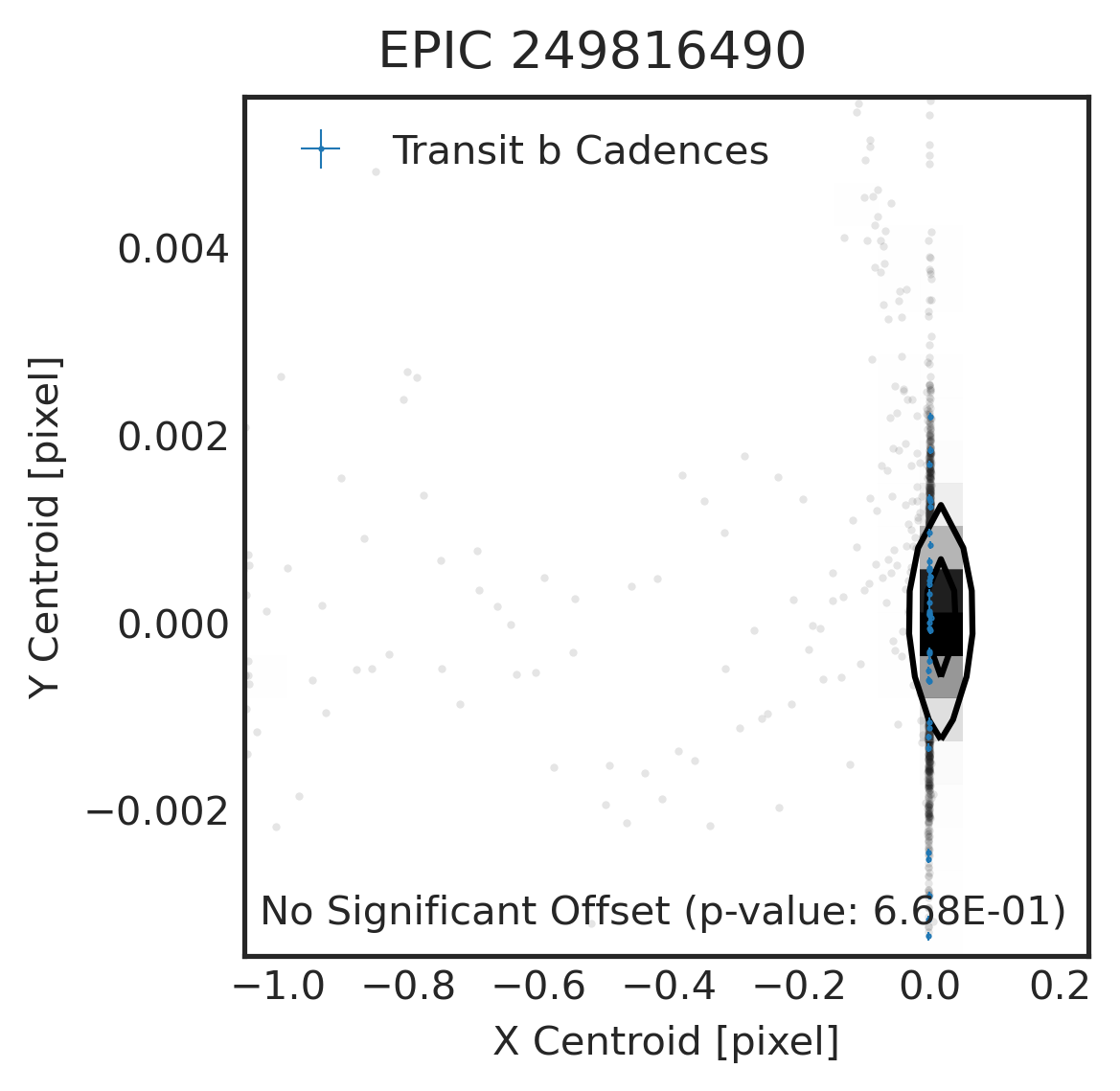}
\includegraphics[width=0.245\textwidth]{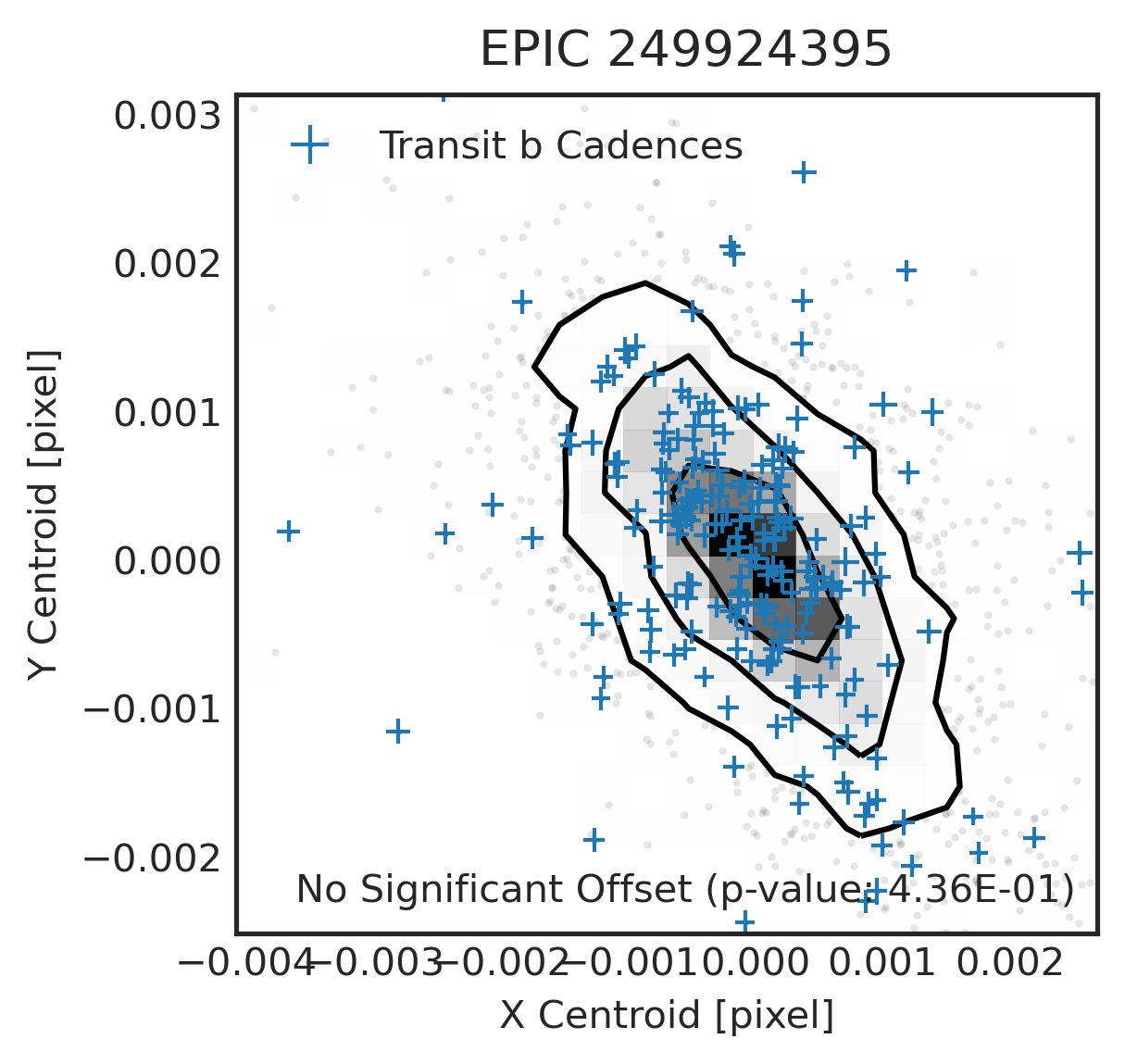}
\caption{Validated planet centroid plots. Description as for Figure \ref{fig:centroidplots1}.}
\label{fig:centroidplots3}
\end{figure}

We consider planet candidates with reconnaissance spectra and high-resolution imaging that show no evidence of stellar companions, a \texttt{vespa} FPP of $<1$\%, and a \texttt{centroid} value of $>0.05$ to be statistically validated. In total, examining \totalchecked\ candidates in \totalcheckedsystems\ systems, we validate \totalval\ planets in \totalsys\ systems. The properties of the new host stars are shown in the top panel of Figure \ref{fig:hrdiagram}, and the properties of the new planets are shown in the bottom panel. It total we validate \totalval\ new planets in \totalsys\ systems, ranging in period from the ultra-short period \epicthreeninetwobalias\ (0.673~d) and \epiconefourzerobalias\ (0.760~d) to  \epicthreefivethreebalias\ (33.590~d), and in size from the Earth-sized \epiceightthreeninecalias\ (1.075~\rearth) and \epictwoseventwobalias\ (1.191~\rearth) to the hot Jupiter \epicthreefiveninebalias\ (13.024~\rearth). We validate 36 new single-planet systems, six new two-planet systems, two new three-planet systems, one new five-planet system, and one new planet in a system with a previously validated planet. The phase-folded light curves with the best-fitting transit models are shown in Figures \ref{fig:phasedlcs1} to \ref{fig:phasedlcs4}, and the centroid plots are in Figures \ref{fig:centroidplots1} to \ref{fig:centroidplots3}.

\begin{figure}
\centering
\includegraphics[width=0.75\textwidth]{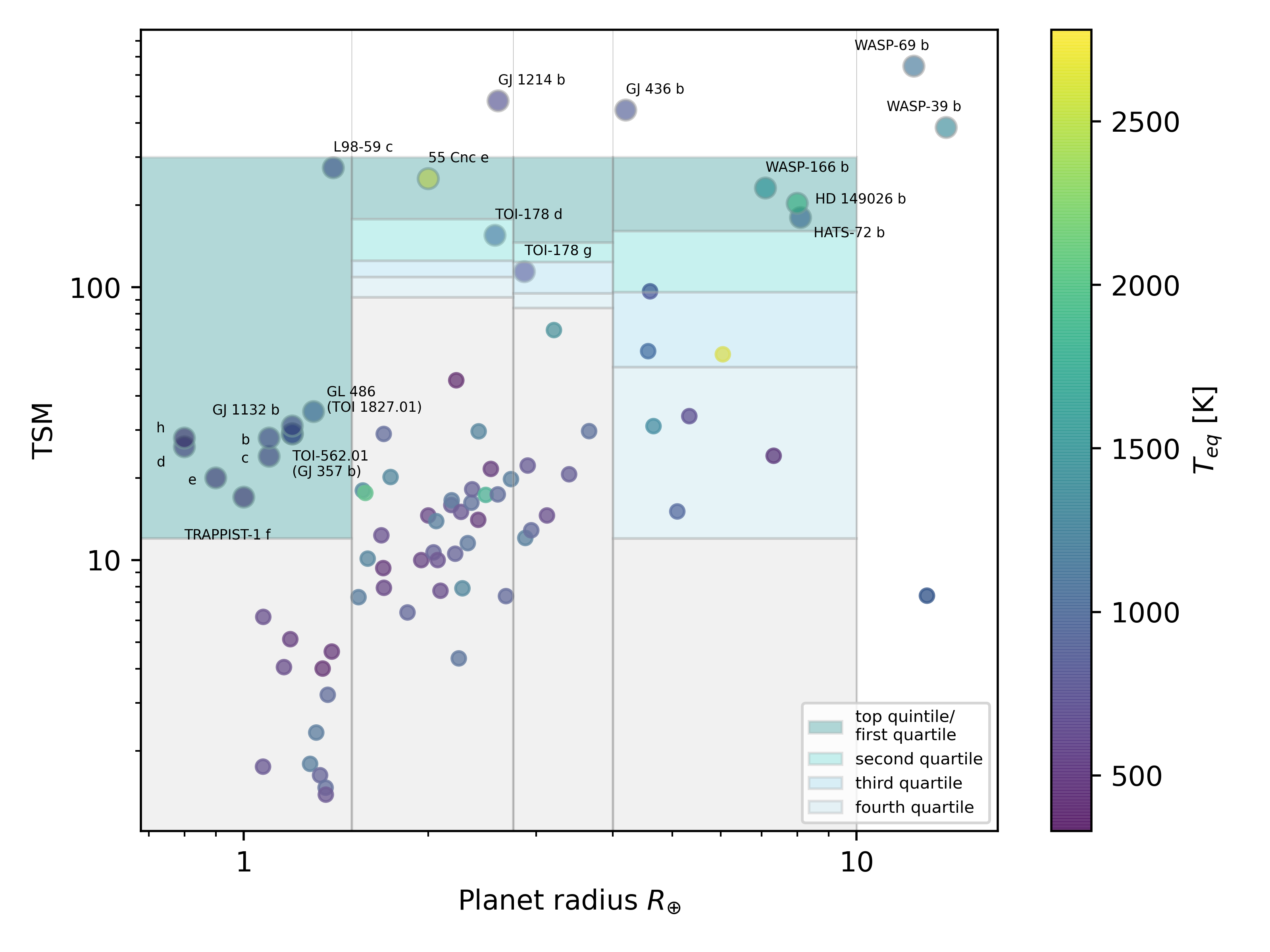}
\includegraphics[width=0.75\textwidth]{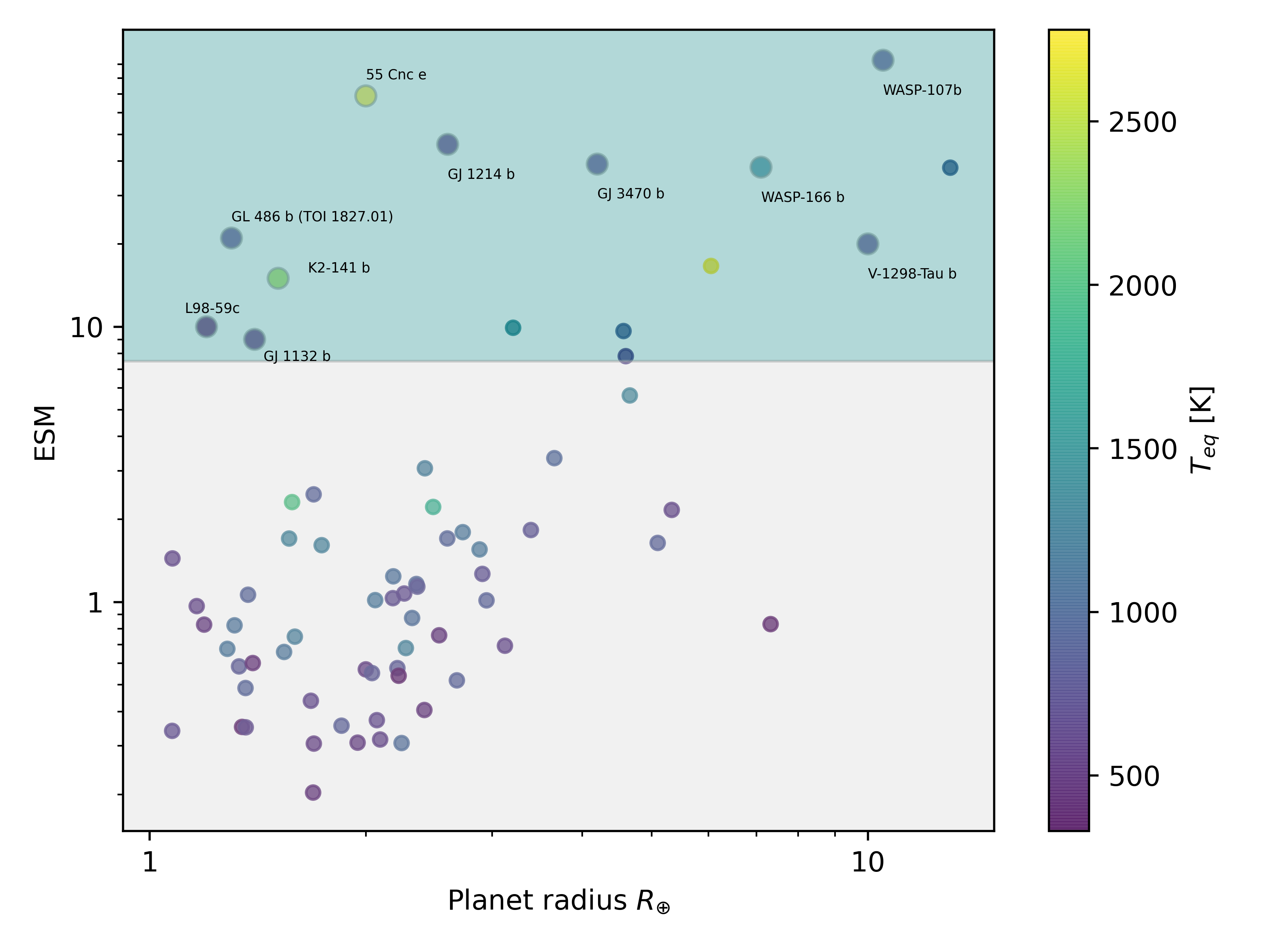}
\caption{Transmission spectroscopy values (top) and emission spectroscopy values (bottom) for the newly validated planets (small points) and planets slated for observation in JWST Cycle 1 (large points), color-coded by their equilibrium temperature. In each case the shaded regions indicate areas of interest as identified by \citet{Kempton2018}.}
\label{fig:tsm_esm}
\end{figure}

In this section we discuss some of the most interesting new systems; Appendix A describes the remaining validated planets in detail. For each planet, we estimate the planet mass using the \citet{Chen2017} mass-radius relationship. Although this is an empirical relationship which is uncorrected for observational biases, it has the advantage of covering the entire radius range of the planets validated here; our mass estimates and metrics subsequently derived from those mass estimates are provided as guideposts towards interesting systems, and should not be considered robust characterization of the planets' properties. From the mass estimate, we calculate the maximum semi-amplitude of the radial velocity variations that the planet would induce on the host star. In addition, we estimate the transmission and emission spectroscopy metrics \citep[TSM and ESM;][]{Kempton2018}. These metrics were designed to rank planets by their putative NASA James Webb Space Telescope (JWST) transmission and emission spectroscopy S/N respectively, taking into account the brightness of the target star and either the cloud-free scale height of the atmosphere (TSM) or the brightness temperature of the planet (ESM). This allows us to comment on the respective favorability of our newly validated planets as JWST targets compared to the thresholds set out in \citet{Kempton2018}. Figure \ref{fig:tsm_esm} shows the distribution of the TSM and ESM values of our newly validated planets compared with the JWST Cycle 1 exoplanet target list; there are a number of new planets, particularly larger than 3~\rearth, that may be interesting targets for future atmospheric follow-up. Where there are multiple planets in a system, we check for low-order mean motion resonances and, where found, estimate transit timing variations. Table \ref{tab:tess} shows which of the newly validated planets were also observed by the NASA Transiting Exoplanet Survey Satellite (\emph{TESS}) mission, and in which sector, determined using the Web TESS Viewing Tool\footnote{\url{https://heasarc.gsfc.nasa.gov/cgi-bin/tess/webtess/wtv.py}}. Although \emph{TESS} has smaller aperture telescopes (10-cm) than \emph{K2} (1-m), and each \emph{TESS} sector covers a shorter duration (27 days) than a \emph{K2} campaign (50--90 days), the additional data may be useful to, for instance, increase the observing baseline for potential transit timing variations.

\begin{longtable}[c]{lllllll}
 \caption{Estimated values and metrics for the newly validated planets.\label{tab:estimated}}\\

 \hline
 Planet & Candidate ID & $T_{\rm{eq}}$ [K] & $M_p$ [\mearth] & $K$ [m/s] & TSM & ESM\\ 
 \hline
 \endfirsthead

 \hline
 \multicolumn{6}{c}{Table \ref{tab:estimated} \emph{(continued)}}\\
 \hline
 Planet & Candidate ID & $T_{\rm{eq}}$ [K] & $M_p$ [\mearth] & $K$ [m/s] & TSM & ESM\\ 
 \hline
 \endhead

 \hline
 \endfoot

 \hline\hline
 \endlastfoot
\epiczerooneonesixbalias  &  204750116.01 &  $\sim$710   &   $8.65\pm5.47$   &  $2.20\pm1.50$    & $22.2$   &   $1.26$\\
\epicsixsixfourbalias     &  205111664.01 & $\sim$770   &   $6.34\pm3.66$   &$1.83\pm1.15$&   $18.2$   &   $1.14$\\
\epicnineeightonebalias   &  206055981.01 & $\sim$450   &   $6.47\pm4.02$   &$1.82\pm1.22$&   $14.6$   &   $0.41$\\
\epicsixeighttwobalias    &  206135682.01 &  $\sim$800   &   $2.34\pm1.34$   &$1.06\pm0.66$&   $1.7$   &   $0.58$\\
\epicsixeighttwocalias    &  206135682.02 &  $\sim$650   &   $2.49\pm1.31$   &$0.91\pm0.53$&   $1.4$   &   $0.35$\\
\epicsixeighttwodalias   &  206135682.03 &  $\sim$500   &   $4.64\pm2.73$   &$1.32\pm0.85$&   $10.9$   &   $0.31$\\
\epicninefivesevenbalias & 206146957.01 &  $\sim$1080   &   $2.2\pm1.3$   & $0.96\pm0.62$    &        $2.3$   &   $0.82$\\
\epictwoeightsixcalias &  206317286.01  &   $\sim$570   &   $5.37\pm3.13$   &$1.49\pm0.95$&   $7.6$   &   $0.32$\\
\epicfiveeightzerobalias  &  210797580.01 &  $\sim$1420   &   $10.42\pm6.6$        &    $5.22\pm3.55$ &   $71.1$   &   $9.92$\\
\epicthreefiveninebalias   &  211399359.01 & $\sim$1000   &   $\sim$1500 &    $\sim$700  &   $8.0$   &   $37.84$\\
\epicninenineninebalias   &  211490999.01 &  $\sim$870   &   $9.2\pm5.37$   & $3.07\pm1.80$    &        $12.6$   &   $1.01$\\
\epiczerofivefourbalias & 211539054.01 &  $\sim$1130   &   $5.15\pm2.85$        &    $1.34\pm0.81$ &   $13.9$   &   $1.02$\\
\epictwooneonesixbalias   &  211732116.01 &  $\sim$1210   &   $3.62\pm1.87$        &    $1.72\pm0.90$ &   $9.7$   &   $0.75$\\
\epictwooneonesixcalias   &  211732116.02 &  $\sim$790   &   $5.71\pm3.48$   &$1.77\pm1.08$&   $10.8$   &   $0.58$\\
\epicthreeoneeightbalias  & 212006318.01 &  $\sim$1000   &   $5.83\pm3.37$        &    $1.42\pm0.89$ &   $4.3$   &   $0.31$\\
\epicthreeeightthreebalias  &  212222383.01 &  $\sim$1240   &   $4.1\pm2.36$   &$1.56\pm0.98$    &        $20.6$   &   $1.61$\\
\epiconeoneeightbalias   &  212530118.01 &  $\sim$530   &   $3.93\pm2.34$   &$1.22\pm0.79$&   $7.7$   &   $0.31$\\
\epiceighttwoeightbalias  &  212575828.01 & $\sim$1130   &   $9.02\pm5.52$        &    $4.78\pm3.18$ &   $11.6$   &   $1.55$\\
\epiczerosixninebalias  &  214173069.01 & $\sim$680   &   $5.59\pm3.33$   &$1.92\pm1.24$&   $16.6$   &   $1.03$\\
\epicfivefourfivebalias  &  214419545.01 & $\sim$1110   &   $7.96\pm4.6$   &$2.42\pm1.52$&   $19.4$   &   $1.79$\\
\epiceightthreeninebalias  &  217192839.02 &  $\sim$670   &   $1.28\pm0.83$   &$0.39\pm0.27$    &        $1.7$   &   $0.34$\\
\epiceightthreeninecalias   &  217192839.01 &  $\sim$530   &   $5.04\pm3.01$   &$1.95\pm1.27$    &        $14.9$   &   $0.57$\\
\epiceightthreeninedalias   & 217192839.03 &  $\sim$440   &   $3.85\pm2.25$   &$0.99\pm0.63$&   $9.7$   &   $0.20$\\
\epiceightninefivebalias   &  217977895.01 &  $\sim$610   &   $5.15\pm3.23$   &$1.25\pm0.84$&   $10.2$   &   $0.37$\\
\epicsixzerotwobalias   &  218668602.01 & $\sim$1270   &   $3.26\pm1.66$        &    $1.63\pm0.92$&   $18.4$   &   $1.70$\\
\epictwoseventwobalias   &  220221272.05 & $\sim$600   &   $1.28\pm0.83$   &$0.72\pm0.47$&   $6.3$   &   $1.44$\\
\epictwoseventwocalias   &  220221272.03 & $\sim$480   &   $1.72\pm0.86$   &$1.22\pm0.61$    &        $5.1$   &   $0.83$\\
\epictwoseventwodalias   &  220221272.02 & $\sim$410   &   $2.71\pm1.4$   & $2.25\pm1.16$   &        $4.4$   &   $0.60$\\
\epictwoseventwoealias   &  220221272.04 & $\sim$370   &   $2.39\pm1.36$   &$1.50\pm0.85$&   $4.1$   &   $0.35$\\
\epictwoseventwofalias   &  220221272.01 & $\sim$330   &   $5.71\pm3.33$   &$5.85\pm3.41$&   $45.6$   &   $0.54$\\
\epicfoursevensevenbalias   &  220459477.01 & $\sim$1100   &   $3.2\pm1.79$   &$1.61\pm0.99$    &        $7.5$   &   $0.66$\\
\epiceightsevenfourbalias   &  220510874.01 &  $\sim$1010   &   $6.21\pm3.66$        &    $2.16\pm1.28$ &   $11.5$   &   $0.87$\\
\epictwothreethreebalias   &  220696233.01 & $\sim$340   &   $45.46\pm31.68$        &   $13.80\pm9.65$&   $23.1$   &   $0.83$\\
\epicfourfivefivebalias  &  245943455.01 &  $\sim$970   &   $13.1\pm8.36$   &$5.08\pm3.48$&   $30.3$   &   $3.33$\\
\epiczerofoureightbalias&  245991048.01 &  $\sim$1050   &   $5.65\pm3.3$   &$2.04\pm1.30$&   $16.4$   &   $1.24$\\
\epiczerofoureightcalias   &  245991048.02 &  $\sim$780   &   $5.26\pm2.73$   &$1.42\pm0.82$&   $10.7$   &   $0.55$\\
\epicninesevensevenbalias  &  245995977.01 & $\sim$1310   &   $20.24\pm12.4$        &   $10.17\pm6.73$  &   $30.3$   &   $5.63$\\
\epicthreeonefourbalias   &  246074314.01 & $\sim$900   &   $2.49\pm1.71$   &$1.15\pm0.84$&   $3.3$   &   $1.06$\\
\epicthreenineeightbalias  &  246084398.01 & $\sim$840   &   $4.45\pm2.52$   &$1.28\pm0.79$&   $6.4$   &   $0.36$\\
\epiczerofourninebalias  &  246429049.01 & $\sim$950   &   $6.34\pm4.06$   &$2.02\pm1.39$&   $15.6$   &   $1.16$\\
\epiczerofourzerobalias  &  246876040.01 & $\sim$890   &   $3.93\pm2.03$   &$2.39\pm1.37$&   $29.6$   &   $2.46$\\
\epiceightoneninebalias  &  246891819.01 &  $\sim$910   &   $7.64\pm4.5$   & $2.99\pm1.91$    &        $16.4$   &   $1.70$\\
\epiceightoneninecalias   &  246891819.02 & $\sim$750   &   $11.81\pm7.02$        &    $3.82\pm2.47$ &   $19.8$   &   $1.83$\\
\epicthreeninetwobalias   &  246953392.01 & $\sim$1990   &   $3.4\pm1.94$   &$2.56\pm1.60$&   $18.4$   &   $2.31$\\
\epicthreeninetwocalias   &  246953392.02 & $\sim$590   &   $10.0\pm6.01$   &$2.23\pm1.46$&   $14.6$   &   $0.69$\\
\epiczerozerothreebalias   &  247383003.01 &  $\sim$1210   &   $6.74\pm4.06$        &    $2.94\pm1.92$ &   $28.4$   &   $3.06$\\
\epicthreefivezerobalias   &  248463350.01 &  $\sim$1240   &   $6.08\pm3.59$        &    $2.17\pm1.29$ &   $7.7$   &   $0.68$\\
\epicthreefivezerobalias   &  248463350.02 &  $\sim$870   &   $23.41\pm13.1$        &    $5.85\pm3.29$ &   $14.5$   &   $1.64$\\
\epiconefourzerobalias   &  248472140.01 & $\sim$2780   &   $32.28\pm22.16$        &  $26.62\pm18.49$  &   $51.7$   &   $16.63$\\
\epicthreezerosevenbalias  &  248518307.01 & $\sim$570   &   $1.65\pm0.8$   & $1.29\pm0.63$    &        $4.0$   &   $0.97$\\
\epicfiveonefourbalias   &  248527514.01 & $\sim$660   &   $5.83\pm3.73$   &$2.22\pm1.53$&   $15.3$   &   $1.07$\\
\epicfiveninesevenbalias   &  248621597.01 & $\sim$870   &   $7.8\pm4.74$   & $2.00\pm1.32$   &        $7.4$   &   $0.52$\\
\epicthreefivethreebalias   &  248758353.01 &  $\sim$580   &   $24.91\pm16.38$        &   $5.81\pm4.08$ &   $33.0$   &   $2.16$\\
\epictwosevenninebalias   &  248861279.01 & $\sim$450   &   $7.33\pm4.06$   &$2.95\pm1.64$&   $19.9$   &   $0.76$\\
\epicninetwoeightbalias  &  248874928.01 &  $\sim$980   &   $18.63\pm10.87$        &   $7.72\pm4.90$  &   $59.5$   &   $9.65$\\
\epicfoursevenonebalias  &  249223471.01 &  $\sim$720   &   $20.04\pm12.4$        &    $5.27\pm3.51$ &   $89.8$   &   $7.82$\\

\epicsixfiveonebalias   &  249403651.01 &  $\sim$1110   &   $2.11\pm1.14$        &    $0.86\pm0.51$ &   $1.8$   &   $0.68$\\
\epicsixfiveonecalias   &  249403651.02 & $\sim$900   &   $2.49\pm1.4$   & $0.83\pm0.51$    &        $1.5$   &   $0.49$\\
\epicfourninezerobalias   &  249816490.01 & $\sim$610   &   $3.89\pm1.99$   &$1.04\pm0.59$&   $12.4$   &   $0.44$\\
\epicthreeninefivebalias  &  249924395.01 &  $\sim$1800   &   $6.74\pm4.06$        &    $3.77\pm2.45$ &   $17.7$   &   $2.21$\\

\end{longtable}

\begin{longtable}[c]{ll|ll}
 \caption{\emph{TESS} sectors during which the host stars of the validated planets are observed.\label{tab:tess}}\\

 \hline
 Planet & \emph{TESS} Sector/s & Planet & \emph{TESS} Sector/s\\ 
 \hline
 \endfirsthead

 \hline
 \multicolumn{4}{c}{Table \ref{tab:tess} \emph{(continued)}}\\
 \hline
 Planet & \emph{TESS} Sector/s & Planet & \emph{TESS} Sector/s\\ 
 \hline
 \endhead

 \hline
 \endfoot

 \hline\hline
 \endlastfoot
K2-367/EPIC 206055981 &S42&                K2-390/EPIC 245995977& S2, S29, S42\\
K2-368/EPIC 206135682 &S42&                K2-391/EPIC 246074314& S42\\
K2-369/EPIC 206146957 &S42&                K2-392/EPIC 246084398& S42\\
EPIC 206317286 &S42&                       K2-393/EPIC 246429049& S42, S43, S44\\
K2-370/EPIC 210797580& S43, S44&           K2-395/EPIC 246891819& S43, S44\\
K2-371/EPIC 211399359& S7, S34, S44--S46&  K2-396/EPIC 246953392& S43, S44\\
K2-372/EPIC 211490999& S44--S46&           K2-397/EPIC 247383003& S43, S44, S45\\
K2-373/EPIC 211539054& S7, S34, S44--S46&  K2-398/EPIC 248463350& S45, S46\\
K2-374/EPIC 211732116& S44--S46&           K2-399/EPIC 248472140& S35, S45, S46\\
K2-375/EPIC 212006318& S44--S46&           K2-400/EPIC 248518307& S8, S35, S45, S46\\
K2-376/EPIC 212222383& S21, S44--S46&      K2-401/EPIC 248527514& S8, S35, S45, S46\\
K2-384/EPIC 220221272& S3, S30, S42, S43&  K2-402/EPIC 248621597& S45, S46\\
K2-385/EPIC 220459477& S42&                K2-403/EPIC 248758353& S45, S46\\
K2-386/EPIC 220510874& S42, S43&           K2-404/EPIC 248861279& S45, S46\\
K2-387/EPIC 220696233& S42, S45&           K2-405/EPIC 248874928& S45, S46\\
K2-388/EPIC 245943455& S2, S29, S42&       K2-406/EPIC 249223471& S11, S38\\
K2-389/EPIC 245991048& S2, S29, S42&       
\end{longtable}

\subsection{EPIC 211399359}
\label{sec:threefivenine}




\epicthreefiveninebalias\ is a gas giant (13.02$^{+0.921}_{-0.868}$ \rearth) planet orbiting a moderately faint ($V$=14.65 mag, $K_s$=12.39 mag) K0V star (0.74\rsun, 0.96\msun). This target was observed in Campaigns 5 and 18, and a short-period candidate, EPIC 211399359.01, was identified by \citet{pop16} and \citet{bar16} that we validate here. Planet b orbits at a distance of 0.041~au, with a period of 3.114905~days and an equilibrium temperature of $\sim$1000~K. The host star has a clean, single-lined Keck/HIRES spectrum, and Keck/NIRC2 and Gemini/NIRI AO imaging which show no contaminating stellar companions. The \texttt{vespa} FPP value is $1.89 \times 10^{-13}$, without requiring use of the available contrast curves. The \texttt{centroid} p-value is 0.1241, which is consistent with the source of the transiting signal being on the target star. This planet falls into the range of \citet{Chen2017} with a degenerate mass-radius relation; taking a representative mass of $\sim$1500 \mearth, we find a TSM of 8.0. This lies below the fourth quintile threshold of 96 for giant planets suggested by \citet{Kempton2018}, however, the ESM value is $\sim$38, well above the 7.5 threshold suggested in that paper (the predicted value of a single secondary eclipse of GJ 1132 b in the JWST MIRI LRS bandpass), indicating that it is potentially a good target for emission spectroscopy measurements.

\subsection{EPIC 206135682}

\epicsixeighttwoalias\ is a moderately faint ($V$=13.54 mag, $K_s$=11.04 mag) K5V star (0.66~\rsun, 0.75~\msun) orbited by a compact system of three super-Earth planets that was observed in Campaign 3. \epicsixeighttwobalias\ is a 1.332~\rearth\ planet, orbiting at a distance of 0.052~au, with a period of 5.025598~days and an equilibrium temperature of $\sim$800~K; it was first noted as EPIC 206135682.01 by \citet{van16} and \citet{kru19}. It has a \texttt{vespa} FPP value of $1.90 \times 10^{-05}$ and a \texttt{centroid} p-value of 0.2070. Using the \citet{Chen2017} mass-radius relation predicts a mass of $\sim$2.3~\mearth. \epicsixeighttwocalias\ is a 1.361~\rearth\ planet, orbiting at a distance of 0.081~au, with a period of 9.660186~days and an equilibrium temperature of $\sim$650~K; it was first noted as EPIC 206135682.03 by \citet{kru19}. It has a \texttt{vespa} FPP value of $2.46 \times 10^{05}$ and a \texttt{centroid} p-value of 0.1385. Using the \citet{Chen2017} mass-radius relation predicts a mass of $\sim$2.5~\mearth. \epicsixeighttwodalias\ is a 1.948~\rearth\ planet, orbiting at a distance of 0.132~au, with a period of 20.201018~days and an equilibrium temperature of $\sim$500~K; it was first noted as EPIC 206135682.02 by \citet{kru19}. It has a \texttt{vespa} FPP value of $3.44 \times 10^{05}$ and a \texttt{centroid} p-value of 0.3671. Using the \citet{Chen2017} mass-radius relation predicts a mass of $\sim$4.6~\mearth. These mass estimates for the three planets result in challenging estimated RV semi-amplitudes ($K$ = 1.1 m/s, 0.91 m/s and 1.3 m/s for planets b, c, and d respectively), and TSM ($\sim$1.7, $\sim$1.4, $\sim$11) and ESM ($\sim$0.6, $\sim$0.3, $\sim$0.3) values indicating they would be challenging atmospheric targets. We note that \citet{kru19} also detected a longer period candidate at 37.8 days that was not recovered by our pipeline.

Planets b and d lie close to the 4:1 mean motion resonance, having a period ratio of 4.0196. Planet c, in between, lies interior to the 2:1 resonance with b, and exterior to the 2:1 resonance with d. The \texttt{TTVFaster} code of \citet{Agol2016} predicts transit timing variation (TTV) amplitudes for planets b, c, and d of 31 seconds, 89 seconds, and 18 seconds respectively, assuming circular orbits. This precision is well below the timing precision of individual K2 transits, given the 30-minute observing cadence, however may be accessible with higher cadence follow-up observations. If the orbits are non-circular, the amplitudes of these TTVs will increase; the amplitudes listed above should be considered lower limits.

\subsection{EPIC 210797580}



\epicfiveeightzerobalias\ is a sub-Neptune (3.207~\rearth) planet orbiting a bright ($V$=11.11 mag, $K_s$=9.24 mag) G8V star (0.97~\rsun, 0.98~\msun), observed in Campaign 13. It was first noted in Paper IV as EPIC 210797580.01; it orbits the star at a distance of 0.032~au, with a period of 2.140840~days and an equilibrium temperature of $\sim$1420~K. This makes it the second hottest, second brightest known planet in the 3--4~\rearth\ size range after K2-100~b. The host star has a clean, single-lined FLWO/TRES spectrum, and Keck/NIRC2 and WIYN/NESSI imaging which show no contaminating stellar companions. The \texttt{vespa} FPP value is $1.77 \times 10^{-03}$, using the three available contrast curves. The \texttt{centroid} p-value is 0.4204, which is consistent with the source of the transiting signal being on the target star. Using the \citet{Chen2017} mass-radius relation predicts a mass of $\sim$10.4~\mearth. This would result in a measurable RV semi-amplitude (\emph{K}$\sim$5.2 m/s), but predicts a TSM ($\sim$71) that is below the recommended level of interest for atmospheric follow-up recommended by \citet{Kempton2018}. However, due to the planet's high temperature, the ESM is predicted to be $\sim$10, above 7.5 (the predicted value of a single secondary eclipse of GJ 1132 b in the JWST MIRI LRS bandpass), indicating that it is potentially a good target for emission spectroscopy measurements.

\subsection{EPIC 217192839}

\epiceightthreeninealias\ is a multi-planet system consisting of an Earth-sized planet (b; 1.075~\rearth), a sub-Neptune (c; 2.000~\rearth), a super-Earth (d; 1.689~\rearth), and a sub-Earth-sized candidate (.04; 0.758~\rearth). The host star is a moderately faint ($V$=13.01 mag, $K_s$=10.30 mag) K5V star (0.68~\rsun, 0.75~\msun), which was observed in Campaign 7. Planet b, noted as candidate EPIC 217192839.02 in \citet{Mayo2018} and \citet{kru19}, orbits the star at a distance of 0.071~au, with a period of 7.938933355~days and an equilibrium temperature of $\sim$670~K; at 1.075~\rearth, it is the smallest planet validated in this paper. Planet c, noted as candidate EPIC 217192839.01 in \citet{Mayo2018}, \citet{Petigura18a}, \citet{kru19}, and \citet{zin19c}, orbits the star at a distance of 0.113~au, with a period of 16.034656~days and an equilibrium temperature of $\sim$530~K. Planet d, noted as candidate EPIC 217192839.03 in \citet{Mayo2018} and \citet{kru19}, orbits the star at a distance of 0.160~au, with a period of 26.803023~days and an equilibrium temperature of $\sim$440~K. In addition, our pipeline identified a tentative fourth candidate, EPIC 217192839.04, with a period of 40.934141~days and a radius of 0.758~\rearth. Using the \citet{Chen2017} mass-radius relation results in masses of $\sim$1.3 \mearth, $\sim$5.0 \mearth, and $\sim$3.9 \mearth\ for planets b, c, and d respectively, with challenging radial velocity semi-amplitudes and low prospects for atmospheric follow-up.

\epiceightthreeninealias\ has a clean, single-lined FLWO/TRES spectrum, and both Gemini/NIRI AO and Gemini/DSSI speckle imaging which show no contaminating stellar companions. The \texttt{vespa} FPP values for planets b, c, and d are $8.22 \times 10^{-03}$, $1.46 \times 10^{-05}$, and $1.16 \times 10^{-03}$ respectively, using the available contrast curves. The \texttt{centroid} p-values are 0.2215, 0.3334, and 0.2891, which are consistent with the source of the transiting signal being on the target star. The tentative fourth candidate at $\sim$41 days does not meet our validation criteria (having an FPP of 2.56\%), likely due to its low SNR.

\begin{figure}
\centering
\includegraphics[width=0.325\textwidth]{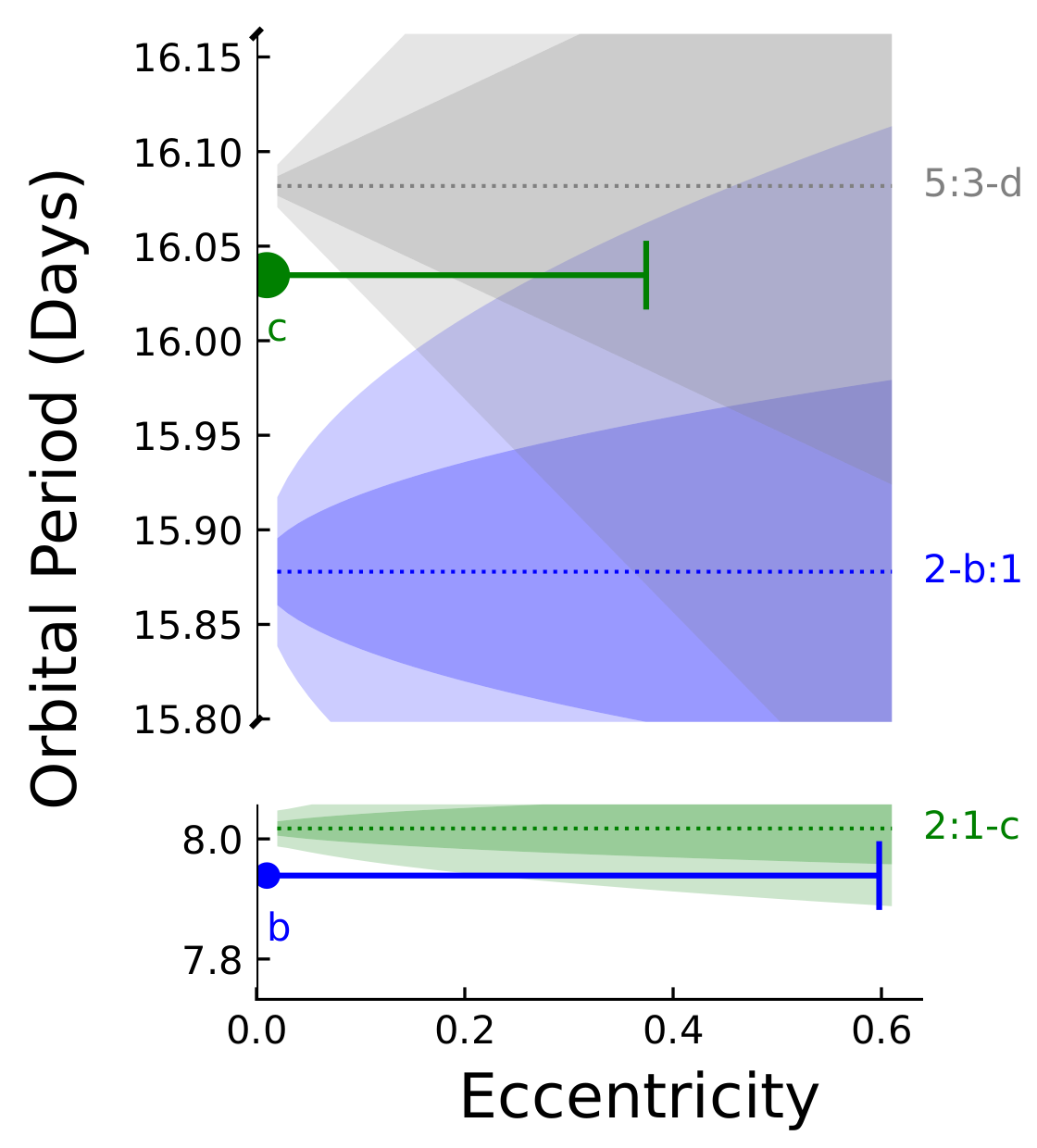}
\caption{Resonance locations in the \epiceightthreeninealias\ system. The orbital period is shown on the y-axis, and eccentricity is on the x-axis; note that the y-axes of the two subplots are discontinuous and not to scale. The location of planets b and c are shown with green and blue circles respectively, and the solid horizontal lines extend to the eccentricity at which each planet would cross the next planet's orbit. The shaded regions show, as a function of eccentricity, the location and analytically estimated widths of the 2:1 mean motion resonance for planets b and c, and the 5:3 mean-motion resonance for planets c and d; the label ``5:3-d'' indicates, for instance, that a test particle at that location would complete 3 orbits in the same amount of time that planet c takes to complete 5 orbits. The shaded regions surrounding each resonance line are the resonance widths corresponding to the lower (dark shading) and upper (light shading) planet mass limits, as propagated from the uncertainty on the planets' radii.}
\label{fig:eightthreenine_resonance}
\end{figure}

Planets c and d lie within 0.3\% of the 5:3 mean motion resonance. Figure \ref{fig:eightthreenine_resonance} shows the location and analytically estimated widths of the 5:3 and 7:4 mean-motion resonances for \epiceightthreeninealias\ c and d, using the program from \citet{Volk2020}\footnote{https://github.com/katvolk/analytical-resonance-widths}, as in \citet{HardegreeUllman2021}. For any modestly non-zero eccentricity, the planets are in resonance; having no constraints on the eccentricity from the light curve itself, we content ourselves with claiming they are near-resonant. Assuming circular orbits, the \texttt{TTVFaster} package \citep{Agol2016} predicts transit timing variations with amplitudes of 1.4 minutes for planet c and 1.8 minutes for planet d; increasing the eccentricity to 0.2 increases these to 24 and 54 minutes. Planets c and d are also both in the radius valley \citep{ful17}, making an attempt to constrain their bulk density with masses derived by transit timing variations a potentially valuable prospect.

\subsection{EPIC 220221272}

\epictwoseventwoalias\ is a moderately faint ($Kepmag$=14.26 mag, $K_s$=11.29 mag) M4V star (0.348~\rsun, 0.330~\msun) that was observed in Campaign 8. It hosts a compact system of five planets, ranging from Earth- to sub-Neptune-sized, including multiple pairs close to or in mean motion resonances. Planet b has a period of 2.231527~days and a radius of 1.076~\rearth; planet c has a period of 4.194766~days and a radius of 1.191~\rearth; planet d has a period of 6.679581~days and a radius of 1.392~\rearth; planet e has a period of 9.715043~days and a radius of 1.345~\rearth; and planet f has a period of 13.627490~days and a radius of 2.222~\rearth. The planets range in equilibrium temperature from 330--600~K, and, using the \citet{Chen2017} mass-radius relation, in mass from $\sim$1.3--5.7 \mearth. The host star has a clean, single-lined IRTF/SpeX spectrum, and WIYN/NESSI speckle imaging at both 562nm and 832nm that show no contaminating stellar companions. The \texttt{vespa} FPP values are $7.20 \times 10^{-3}$, $6.57 \times 10^{-4}$, $3.14 \times 10^{-4}$, $3.85 \times 10^{-3}$, and $1.12 \times 10^{-4}$ for planets b--f respectively using the available contrast curves and the Campaign 8 multiplicity boost. The \texttt{centroid} p-values are 0.6315, 0.1135, 0.5203, 0.4207, and 0.5460 respectively.

\begin{figure}
\centering
\includegraphics[width=0.4\textwidth]{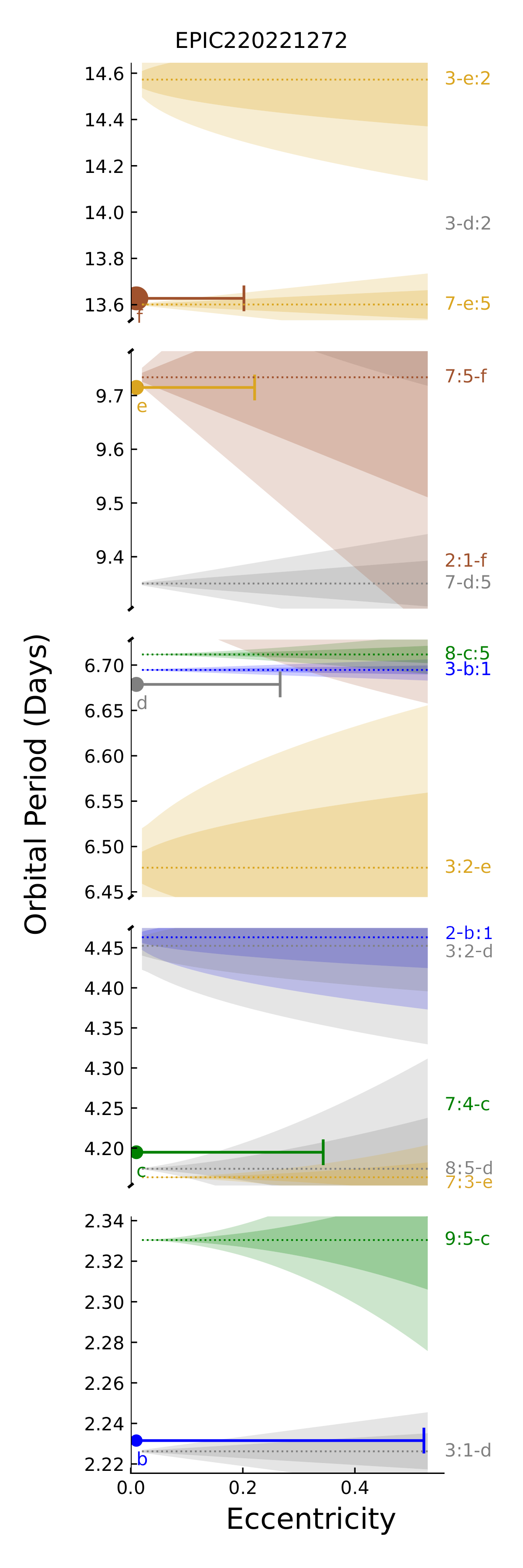}
\caption{The location and analytically estimated widths of the low-order mean-motion resonances in the \epictwoseventwoalias\ system. See Figure \ref{fig:eightthreenine_resonance} for a description.}
\label{fig:twoseventwo_resonances}
\end{figure}

Figure \ref{fig:twoseventwo_resonances} shows the proximity of the observed period ratios to the location and analytically estimated widths of the resonances in the system; planets b and c are close to the 3:1 resonance, planets c and d are close to the 8:5 resonance, and planets e and f are in the 7:5 resonance. This is reminiscent of other compact systems with chains of resonances and near-resonances, such as TRAPPIST-1 \citep{Gillon2017} and K2-138 \citep{Christiansen2018}. Assuming circular orbits, the \texttt{TTVFaster} package \citep{Agol2016} predicts transit timing variations with amplitudes of 0.28, 0.83, 3.33, 12.26, and 3.97 minutes for planets b--f respectively. Planet e, being somewhat smaller than planet f in the 7:5 resonance, is predicted to exhibit the largest transit timing variations, which would be detectable for a number of facilities \citep[e.g. Palomar/WIRC][]{Vis2020}, but difficult to discern in the 30-minute cadence \emph{K2} data. Given the magnitude of the target and the small nature of the planets, these would be challenging radial velocity measurements, with predicted amplitudes ranging from 0.7--5.9~m/s, but potentially a worthwhile investment. Although this is a slightly fainter target than TRAPPIST-1 ($K_s$=11.29 mag compared to $K_s$=10.30), and a larger host star, the transmission spectroscopy metric for planet f ($\sim$46) is higher than the values for the most promising TRAPPIST-1 planets (20--26 for planets c, d, and e). Indeed, the \epictwoseventwoalias\ system is one of the closest analogs to the TRAPPIST-1 system that has been found, as the only other M dwarfs hosting five or more transiting planets---Kepler-186 \citep{Quintana2014} and Kepler-296 \citep{Barclay2015}---are early M dwarfs (0.472~\rsun\ and 0.480~\rsun\ respectively), where \epictwoseventwoalias\ is a mid-M dwarf (0.348~\rsun) and TRAPPIST-1 is a late-M dwarf (0.119~\rsun).


\subsection{EPIC 220696233}



\epictwothreethreebalias\ is a sub-Saturn (7.322~\rearth) planet orbiting a faint ($V$=16.22 mag, $K_s$=12.29 mag) M1V star that was observed in Campaign 8. It orbits the star at a distance of 0.152~au, with a period of 28.735327~days. With an equilibrium temperature of $\sim$340~K, it is the coolest planet validated in this work. It was previously noted as EPIC 220696233.01 by \citet{Petigura18a} and \citet{kru19}. The host star has a clean, single-lined IRTF/SpeX spectrum, and both Palomar AO imaging and WIYN/NESSI speckle imaging that show no contaminating stellar companions. The \texttt{vespa} FPP value is $3.33 \times 10^{-10}$ using the available contrast curves. The \texttt{centroid} p-value is 0.3918, which is consistent with the source of the transiting signal being on the target star. Using the \citet{Chen2017} mass-radius relation predicts a mass of $\sim$45~\mearth. This would result in a measurable RV semi-amplitude (\emph{K}$\sim$14 m/s), although the star is very faint. The estimated planet mass produces a predicted TSM value ($\sim$23) in the fourth quartile of interest for atmospheric follow-up as defined by \citet{Kempton2018}, and an ESM value ($\sim$0.8) that is below the recommended threshold. However, we note that \epictwothreethreebalias\ occupies a relatively unpopulated region of period-radius parameter space (see Figure \ref{fig:hrdiagram}). There are few cool ($<500$~K) gas giant planets known, and even fewer orbiting cool host stars. The TSM value is one of the most favourable for known planets in this size and temperature regime.

\subsection{EPIC 245995977}



\epicninesevensevenalias\ is a moderately faint ($V$=13.57 mag, $K_s$=11.72 mag) G4V star (0.97~\rsun, 0.79~\msun) that was observed in Campaign 12. It is orbited by a Neptune-size planet on the edge of the hot Neptune desert that was first noted in Paper IV. \epicninesevensevenbalias\ is a 4.662~\rearth\ planet, orbiting at a distance of 0.040~au, with a period of 3.312792~days and an equilibrium temperature of $\sim$1310~K. The host star has a clean, single-lined Keck/HIRES spectrum, and Gemini/NIRI and Palomar/PHARO AO imaging that show no contaminating stellar companions. The \texttt{vespa} FPP value is $8.68 \times 10^{-8}$, using the available contrast curves. The \texttt{centroid} p-value is 0.6208, which is consistent with the source of the transiting signal being on the target star. Using the \citet{Chen2017} mass-radius relation predicts a mass of $\sim$20\mearth. This would result in a readily measurable RV semi-amplitude (\emph{K}$\sim$10.2m/s), and predicts a TSM value ($\sim$30) that puts \epicninesevensevenbalias\ in the fourth quartile for atmospheric follow-up priority according to the thresholds in \citet{Kempton2018}. The ESM value ($\sim$5.7) is comparable to but below the 7.5 threshold for atmospheric follow-up recommended by \citet{Kempton2018}. 

\subsection{EPIC 246876040}



\epiczerofourzeroalias\ is a moderately bright ($V$=12.72 mag, $K_s$=9.57 mag) G8V star (0.63~\rsun, 0.48~\msun) that was observed in Campaign 13. It is orbited by a super-Earth-size planet, \epiczerofourzerobalias\, which was previously noted as EPIC 246876040.01 by \citet{zin19c}. It is a 1.692~\rearth\ planet, orbiting at a distance of 0.045~au, with a period of 5.095717~days and an equilibrium temperature of $\sim$890~K. The host star has a clean, single-lined FLWO/TRES spectrum, and WIYN/NESSI speckle imaging at 562nm and 832nm that show no contaminating stellar companions. The \texttt{vespa} FPP value is $4.38 \times 10^{-4}$, using the available contrast curves. The \texttt{centroid} p-value is 0.2143, which is consistent with the source of the transiting signal being on the target star. Using the \citet{Chen2017} mass-radius relation predicts a mass of $\sim$3.9\mearth. This would result in a measurable RV semi-amplitude (\emph{K}$\sim$2.4m/s), but predicts TSM ($\sim$30) and ESM ($\sim$2.4) values that indicate atmospheric follow-up would be challenging, compared to the thresholds recommended by \citet{Kempton2018}. Given its relatively small size and relatively small host star, the TSM value does make it somewhat comparable to other close-in super-Earth planets, such as LTT 1445 A b \cite[TSM$\sim$40;][]{Winters2019} and L 98-59 c \cite[TSM$\sim$34;][]{Demangeon2021}.

\subsection{EPIC 248463350}




\epicthreefivezeroalias\ is a moderately faint ($V$=13.15 mag, $K_s$=11.72 mag) G2V star (1.24~\rsun, 0.95~\msun) that was observed in Campaign 14. It is orbited by two rather disparately sized planets that were first noted in Paper IV. \epicthreefivezerobalias\ is a 2.274~\rearth\ sub-Neptune-sized planet, orbiting at a distance of 0.066~au, with a period of 6.393025~days and an equilibrium temperature of $\sim$1240~K. \epicthreefivezerocalias\ is a 5.098~\rearth\ Neptune-sized planet, orbiting at a distance of 0.136~au, with a period of 18.787839~days and an equilibrium temperature of $\sim$870~K. The host star has clean, single-lined Keck/HIRES and FLWO/TRES spectra, and Keck/NIRC2 AO imaging and WIYN/NESSI speckle imaging at 562nm and 832nm that show no contaminating stellar companions. Planets b and c have \texttt{vespa} FPP values of $4.80 \times 10^{-4}$ and $2.43 \times 10^{-4}$ respectively, using the available contrast curves and the C14 multiplicity boost. The \texttt{centroid} p-values are 0.3000 and 0.3974 respectively, which is consistent with the source of the transiting signals being on the target star. For \epicthreefivezerobalias\, the \citet{Chen2017} mass-radius relation predicts a mass of $\sim$6.1\mearth. This would result in a measurable RV semi-amplitude (\emph{K}$\sim$2.2m/s), but predicts TSM ($\sim$7.7) and ESM ($\sim$0.7) values that indicate atmospheric follow-up would be challenging, compared to the thresholds recommended by \citet{Kempton2018}. For \epicthreefivezerocalias\, the \citet{Chen2017} mass-radius relation predicts a mass of $\sim$23\mearth. This would result in a measurable RV semi-amplitude (\emph{K}$\sim$5.9m/s), and predicts a TSM value ($\sim$15) that puts \epicthreefivezerocalias\ in the fourth quartile of atmospheric follow-up priority according to the thresholds in \citet{Kempton2018}. The ESM value ($\sim$1.6) indicates that atmospheric follow-up would be challenging. Given the proximity of the planets to the 3:1 mean motion resonance we investigated further with the analytic code of \citet{Volk2020}, but do not find the planets to plausibly be in resonance. 

\subsection{EPIC 248472140}



\epiconefourzerobalias\ is a ultra-short period sub-Saturn (6.049~\rearth) planet orbiting a moderately bright ($V$=12.48 mag, $K_s$=10.93 mag) F9V star (1.54~\rsun, 0.78~\msun), observed in Campaign 14. It was first noted in Paper IV; it orbits the star at a very close distance of 0.015~au, less than twice the stellar radius above the stellar surface, with a period of 0.759989~days. With a correspondingly very high equilibrium temperature of $\sim$2780~K, it is the hottest validated planet in this analysis, and the hottest planet between the size of Neptune and Jupiter discovered to date. It's uniqueness is illustrated in the right panel of Figure \ref{fig:hrdiagram}; the single nearby planet in the figure is Kepler 1520~b, which is 500~K cooler and 3.5 magnitudes fainter. This means that \epiconefourzerobalias\ may be an excellent probe of the ways in which atmospheres of planets in this size range respond to incredibly high insolation fluxes. \citet{Dai2021} note that ultra-short-period (USP) Neptunes are more similar to USP Jupiters, in having `lonely' system configurations than smaller USP planets; there is no evidence for additional transiting planets in the light curve of \epiconefourzeroalias. \citet{Dai2021} also note that USP Neptunes are more likely to orbit metal-rich stars than their smaller USP counterparts; \epiconefourzeroalias\ has a significantly super-solar metallicity of [Fe/H]$=0.252\pm0.036$ dex.

\epiconefourzeroalias\ has a clean, single-lined FLWO/TRES spectrum, and Keck/NIRC2 AO imaging that shows no contaminating stellar companions. The \emph{Gaia} RUWE value is 5.89, which is higher than expected for a single star, but that is not corroborated by the follow-up observations. The \texttt{vespa} FPP value is $7.80 \times 10^{-4}$ using the available contrast curve. The \texttt{centroid} p-value is 0.1608, which is consistent with the source of the transiting signal being on the target star. Using the \citet{Chen2017} mass-radius relation predicts a mass of $\sim$32~\mearth, resulting in a readily measurable RV semi-amplitude (\emph{K}$\sim$27 m/s). Although the predicted TSM ($\sim$52) falls below the recommended threshold in \citet{Kempton2018}, the predicted ESM ($\sim$16.6) is well above the threshold of 7 due to the high equilibrium temperature of the planet. We encourage additional characterization of this unique USP sub-Saturn.

\subsection{EPIC 248758353}



\epicthreefivethreealias\ is a moderately bright ($V$=12.36 mag, $K_s$=10.58 mag) G4V star (0.89~\rsun, 0.78~\msun) that was observed in Campaign 14. It is orbited by a warm sub-Saturn-size planet that was first noted in Paper IV. \epicthreefivethreebalias\ is a 5.333~\rearth\ planet, orbiting at a distance of 0.188~au, with a period of 33.589979~days and an equilibrium temperature of $\sim$575~K. The host star has a clean, single-lined FLWO/TRES spectrum, and WIYN/NESSI speckle imaging at 562nm and 832nm that show no contaminating stellar companions. The \texttt{vespa} FPP value is $1.05 \times 10^{-5}$, using the available contrast curves. The \texttt{centroid} p-value is 0.6589, which is consistent with the source of the transiting signal being on the target star. Using the \citet{Chen2017} mass-radius relation predicts a mass of $\sim$25\mearth. This would result in a measurable RV semi-amplitude (\emph{K}$\sim$5.8m/s). A predicted TSM of $\sim$33 places \epicthreefivethreebalias in the fourth quartile of large ($>$4 \rearth) planets as defined by \citet{Kempton2018}. The ESM ($\sim$2.1) value indicates emission spectroscopy would be challenging, compared to the threshold of 7.5 recommended by \citet{Kempton2018}.

\subsection{EPIC 248874928}






\epicninetwoeightbalias\ is a Neptune-sized (4.569~\rearth) planet orbiting a moderately bright ($V$=12.86 mag, $K_s$=10.55 mag) K0V star (0.77~\rsun, 1.03~\msun) observed in Campaign 14. It was first noted in Paper IV; it orbits the star at a distance of 0.045~au, with a period of 3.435471~days and an equilibrium temperature of $\sim$980~K. The host star has a clean, single-lined FLWO/TRES spectrum, and both Keck/NIRC2 AO imaging and WIYN speckle imaging that show no contaminating stellar companions. The \texttt{vespa} FPP value is $7.80 \times 10^{-4}$ using the available contrast curves. The \texttt{centroid} p-value is 0.5057, which is consistent with the source of the transiting signal being on the target star. Using the \citet{Chen2017} mass-radius relation predicts a mass of $\sim$19~\mearth. This would result in a measurable RV semi-amplitude (\emph{K}$\sim$7.7 m/s), and predicts a TSM ($\sim$60) value that puts \epicninetwoeightbalias\ in the third quartile ($>51$) for atmospheric follow-up recommended by \citet{Kempton2018}. The ESM value ($\sim$9.6) is comparable to the GJ~1132~b value of 7.5, indicating that \epicninetwoeightbalias\ may also be a favorable planet for further emission spectroscopy characterization.

\subsection{EPIC 249223471}


\epicfoursevenonealias\ is a bright ($V$=9.47 mag, $K_s$=8.03 mag) G4V star (0.96~\rsun, 0.80~\msun) that was observed in Campaign 15. It is orbited by a Neptune-size planet that was first noted in Paper IV. \epicfoursevenonebalias\ is a 4.602~\rearth\ planet, orbiting at a distance of 0.145~au, with a period of 22.549406~days and an equilibrium temperature of $\sim$720~K. The host star has clean, single-lined Keck/HIRES and FLWO/TRES spectra, and Keck/NIRC2 AO imaging and Gemini/DSSI speckle imaging at 692nm and 880nm that shows no contaminating stellar companions. The \texttt{vespa} FPP value is $1.30 \times 10^{-3}$, using the available contrast curves. The \texttt{centroid} p-value is 0.6741, which is consistent with the source of the transiting signal being on the target star. Using the \citet{Chen2017} mass-radius relation predicts a mass of $\sim$20\mearth. This would result in a measurable RV semi-amplitude (\emph{K}$\sim$5.3m/s). A predicted TSM of $\sim$90 places \epicthreefivethreebalias\ near the top of the third quartile of large ($>$4 \rearth) planets as defined by \citet{Kempton2018}. The predicted ESM ($\sim$7.7) value is comparable to the benchmark GJ~1132~b value of 7.5; \epicthreefivethreebalias\ may therefore be a high quality target for future atmospheric observations.

\subsection{EPIC 249816490}



\begin{figure}
\centering
\includegraphics[width=0.75\textwidth]{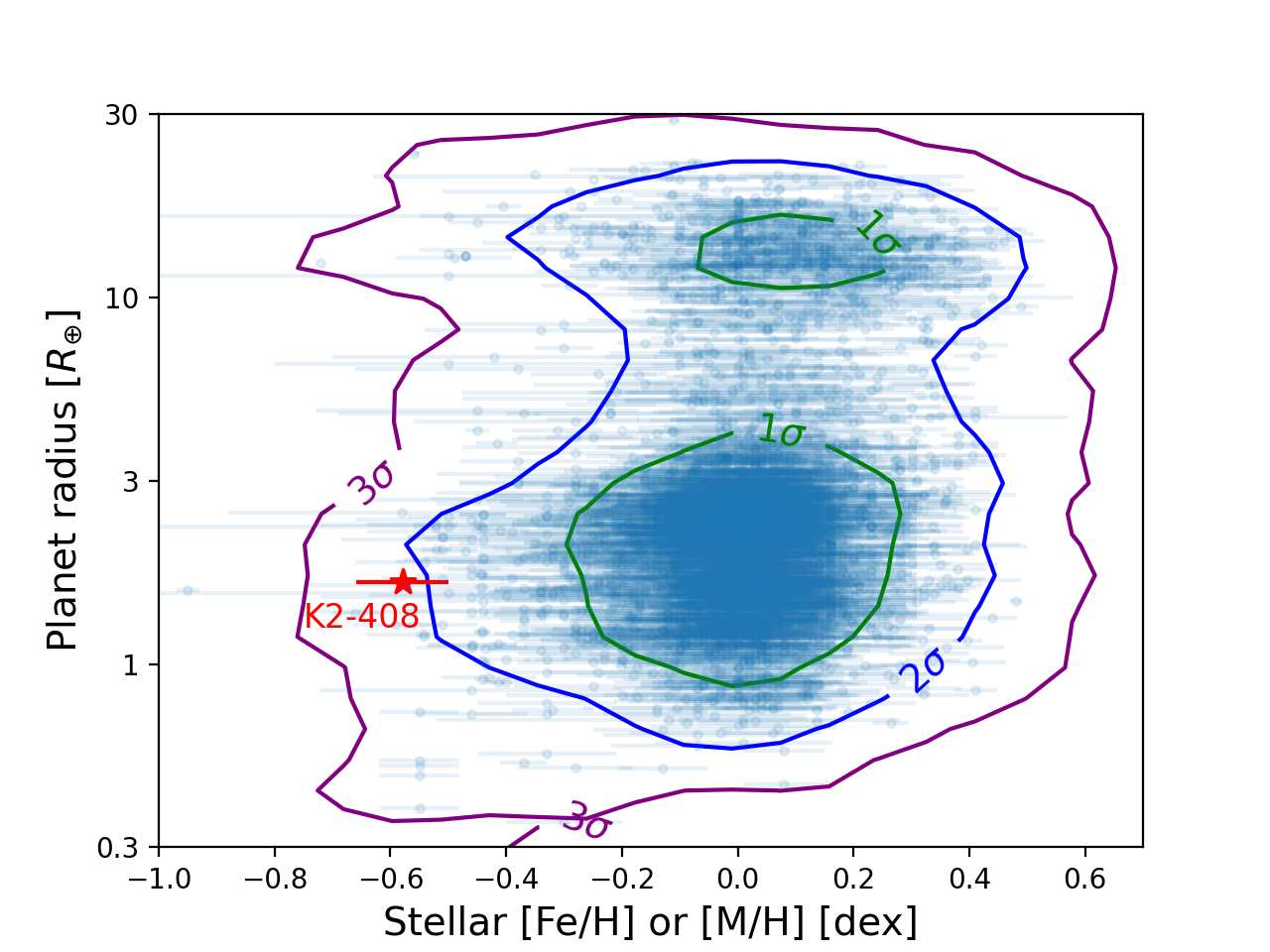}
\caption{The metallicity distribution of the host stars of planets with measured radii (3411 planets at the NASA Exoplanet Archive). The green, blue, and purple contours enclose 68\%, 95\%, and 99.7\% of the data respectively. \epicfourninezerobalias\ has one of the lowest well-constrained metallicities measured to date for an exoplanet host star.}
\label{fig:metallicitydistribution}
\end{figure}

\epicfourninezerobalias\ is a super-Earth (1.677~\rearth) planet orbiting a bright ($V$=11.93 mag, $K_s$=9.98 mag) G7V star (0.78~\rsun, 0.81\msun), observed in Campaign 15. It was first noted in Paper IV; it orbits the star at a distance of 0.14~au, with a period of 20.978959~days and an equilibrium temperature of $\sim$610~K. The host star has a clean, single-lined FLWO/TRES spectrum, and Gemini/DSSI imaging which shows no contaminating stellar companions. The \texttt{vespa} FPP value is $5.19 \times 10^{-4}$ using the available contrast curve. The \texttt{centroid} p-value is 0.6675, which is consistent with the source of the transiting signal being on the target star. Using the \citet{Chen2017} mass-radius relation predicts a mass of $\sim$3.9~\mearth. This would result in a challenging RV semi-amplitude (\emph{K}$\sim$1.0 m/s), and predicts TSM ($\sim$12) and ESM ($\sim$0.4) values that indicate atmospheric follow-up would be challenging, compared to the thresholds recommended by \citet{Kempton2018}. However, the FLWO/TRES spectrum points to the stellar host having a significantly sub-stellar metallicity ([Fe/H] = $-0.579\pm0.080$ dex), which makes this one of the most metal-poor stars known to host a super-Earth (see Figure \ref{fig:metallicitydistribution}). Understanding the lowest metallicity protoplanetary disks that can give rise to planet formation is critical input into planet formation theories \cite[e.g.][]{Dawson2015,Lee2015}. \epicfourninezerobalias\ represents a unique opportunity for the next generation of extreme precision radial velocity instruments to measure the density of a super-Earth orbiting a metal-poor star. The only planets smaller than 2~\rearth\ orbiting stars at such low spectroscopically-confirmed metallicity with measured radii and masses are the three planets in the comparably bright L 98-59 system ([Fe/H]$=-0.46\pm0.26$ dex, $V$=11.69 mag, $K_s$=7.10), and two K2 planets around fainter targets (K2-344 b: [Fe/H]$=-0.95\pm0.02$ dex, $V$=13.44 mag, $K_s$=9.66 mag, and K2-349 b: [Fe/H]$=-0.66\pm0.04$ dex, $V$=14.40 mag, $K_s$=12.20 mag \citep{deLeon2021}). We encourage further follow-up of this system.

\section{Planet candidates that did not pass our validation tests}
\label{sec:candidates}

Table \ref{tab:validation} also lists the planet candidates that did not pass our set of validation tests. These include: 

\begin{enumerate}
    \item Stars that were revealed with high resolution imaging to have nearby companions:
        \begin{itemize}
            \item EPIC 205029914
            \item EPIC 206192335
            \item EPIC 246947582
            \item EPIC 220294712
            \item EPIC 245955351
            \item EPIC 211428897
        \end{itemize}
    \item Stars that were revealed with reconnaissance spectra to be double-lined spectroscopic binaries or to have velocities inconsistent with planetary companions:
        \begin{itemize}
            \item EPIC 211830293
            \item EPIC 212705192
            \item EPIC 219388192 \citep{Nowak2017}
            \item EPIC 211830293
        \end{itemize}
    \item Candidate signals that failed our \texttt{vespa} FPP threshold, likely due to being v-shaped:
        \begin{itemize}
            \item EPIC 205944181.01
            \item EPIC 211784767.01
            \item EPIC 212585579.01
            \item EPIC 216892056.01
            \item EPIC 220571481.01
            \item EPIC 247724061.01
            \item EPIC 248222323.01
            \item EPIC 248827616.01
            \item EPIC 211800191.01
            \item EPIC 212797028.01
            \item EPIC 216334329.01
        \end{itemize}
    \item Candidate signals that failed our \texttt{vespa} FPP threshold, likely due to distortion by incompletely removed stellar rotation and pulsation signals:
        \begin{itemize}
            \item EPIC 211965883.01
            \item EPIC 249827330.01
            \item EPIC 248639411.01
            \item EPIC 248740016.01
        \end{itemize}
    \item Candidate signals that failed our \texttt{vespa} FPP threshold, likely due to being too low SNR: 
        \begin{itemize}
            \item EPIC 212730483.01
            \item EPIC 226042826.01
        \end{itemize}
    \item Candidate signals that failed our \texttt{vespa} FPP threshold, likely due to a potential secondary eclipse: 
        \begin{itemize}
            \item EPIC 247698108.01
            \item EPIC 212351868.01
        \end{itemize}
    \item Candidate signals that marginally failed our FPP threshold (having FPP values 0.01--0.10) but otherwise looked promising: 
        \begin{itemize}
            \item EPIC 205999468.01
            \item EPIC 213817056.01
            \item EPIC 220400100.01
            \item EPIC 205957328.01
        \end{itemize}
    \item Candidate signals that passed our other tests but failed the \texttt{centroid} test ($p<0.05$) as not conclusively being on the target star:
        \begin{itemize}
            \item EPIC 249865296.01
            \item EPIC 247164043.01
            \item EPIC 211711685.01
        \end{itemize}
\end{enumerate}

We discuss several of these in more detail below.

\begin{figure}
\centering
\includegraphics[width=0.75\textwidth]{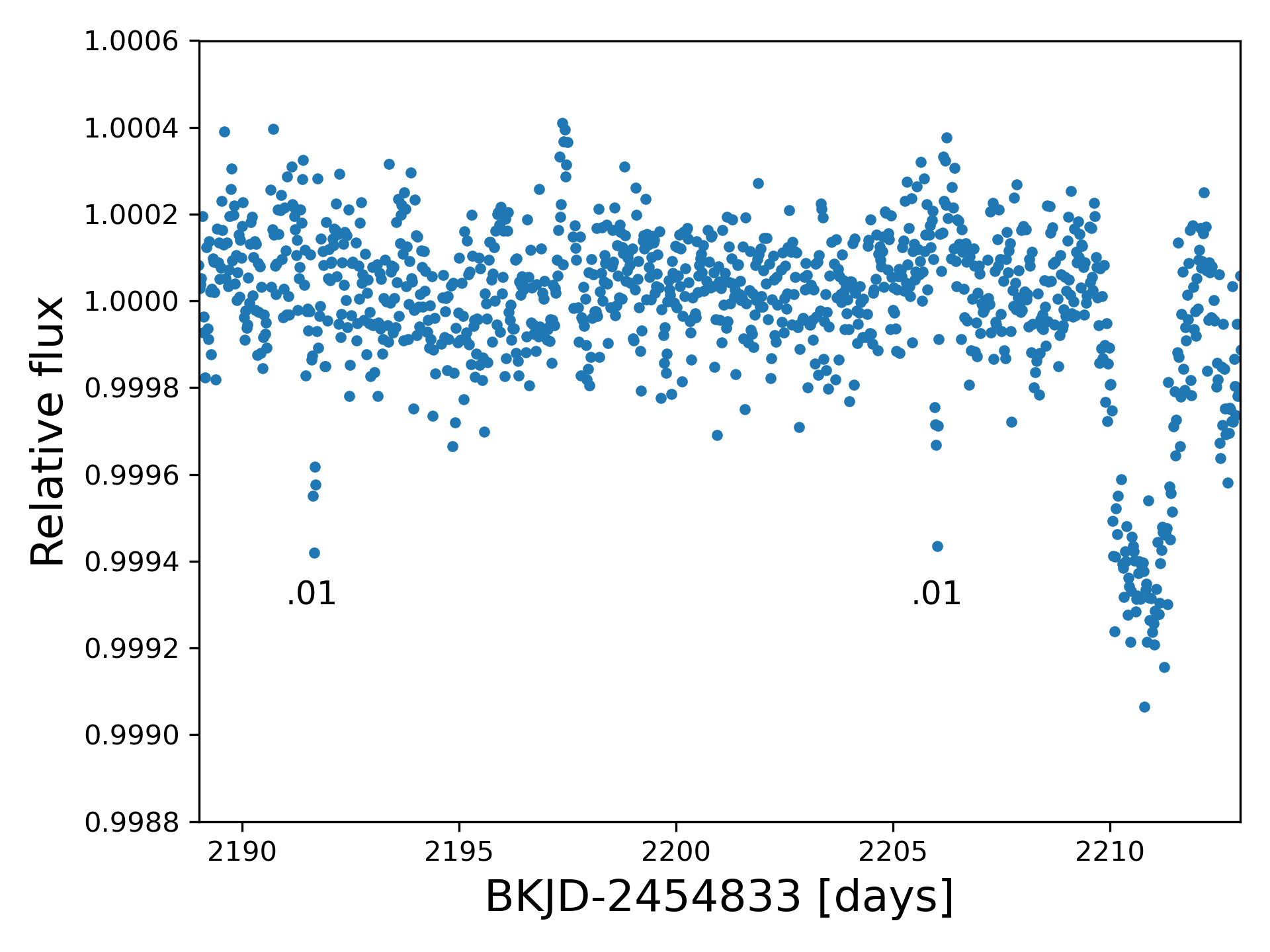}
\caption{A portion of the light curve of EPIC 205957328, showing two transits of the short period candidate EPIC 205957328.01, and a potential single transit of a longer period candidate centered at $\sim$2211 days.}
\label{fig:epic205957328_longtransit}
\end{figure}

EPIC 205957328.01 has a 14.35-day period sub-Neptune candidate, noted by \citet{van16}, \citet{Mayo2018}, and \citet{kru19}, with a radius of $\sim$1.9~\rearth. It does not pass our validation threshold, with an FPP of 0.022 (possibly due to one transit which has an inconsistent depth with the remaining transits), but we note that the light curve also shows a potential single transit, at BJD$=$2457043.5, of a much longer period candidate, EPIC 205957328.02, shown in Figure \ref{fig:epic205957328_longtransit}. The duration of the putative event is $\sim$46.3 hours, implying an orbital period of $>$80 years. The depth of the event is consistent with a sub-Neptune planet with a radius of $\sim$2.3~\rearth. This putative candidate would be challenging to confirm, given the timescale to observe a second transit or obtain sufficient orbital coverage with radial velocity observations.

The light curve of EPIC 211428897 shows five candidate signals, with periods ranging from 1.61--6.27~days, reported by \citet{pop16} (EPIC 211428897.01), \citet{pet18} (EPIC 211428897.01 and .02), \citet{Dressing2017} (EPIC 211428897.01, .02, and .03), and \citet{kru19} (EPIC 211428897.01, .02, .03, .04, and .05). High-resolution imaging of EPIC 211428897, which is relatively red ($V$=14.1, $K_s$=9.6) and was originally classified as an M2 dwarf \citep{Rod2019}, shows two nearly equal brightness stars separated by $\sim1^{\prime\prime}$. Given that the transits are relatively shallow, ranging in depth from 500--700~ppm, they are likely still planetary in nature after accounting for the dilution regardless of which star they orbit. We note that the candidates may not all be orbiting one of the stars; the work of disentangling which signals are associated with which host star is left for a future analysis.

Similarly, EPIC 245955351 shows two candidate signals, first reported in Paper IV, with periods of 7.98 and 18.43 days. The target, originally classified as a G6V star \citep{HardegreeUllman2021}, has a much fainter companion $\sim3^{\prime\prime}$ away that is revealed by high-resolution AO imaging. If the signals are associated with the bright star, they likely remain planetary in nature, and indeed have low FPP values ($4.72 \times 10^{-3}$ and $8.34 \times 10^{-5}$ respectively). However, if either or both of them are orbiting the fainter companion, the dilution is such that the radius of the transiting object would become too large to be planetary. Given the relatively deep signals ($\sim$1000~ppm and $\sim$3000~ppm), ground-based, seeing-limited photometry could be used to determine whether the signals were on the bright star. 

EPIC 249559552 shows two candidate signals, first reported in Paper IV, with periods of 7.82 and 19.52 days, very close to the 5:2 resonance. The target, originally classified as a K3V star \citep{HardegreeUllman2021}, has a much fainter companion $\sim0.4^{\prime\prime}$ away that is revealed by speckle imaging. As above, if the signals are associated with the bright star, they likely remain planetary in nature, however either or both may be on the nearby faint companion, which would preclude their planetary nature. In this case, the two stars are close enough on the sky that ground-based photometry would likely be insufficient to disentangle them, and something like a high precision spectrograph behind an AO system would be necessary (e.g., PARVI on the Palomar Hale Telescope, iLocator on the Large Binocular Telescope, or HISPEC on Keck)

\begin{figure}
\centering
\includegraphics[width=0.33\textwidth]{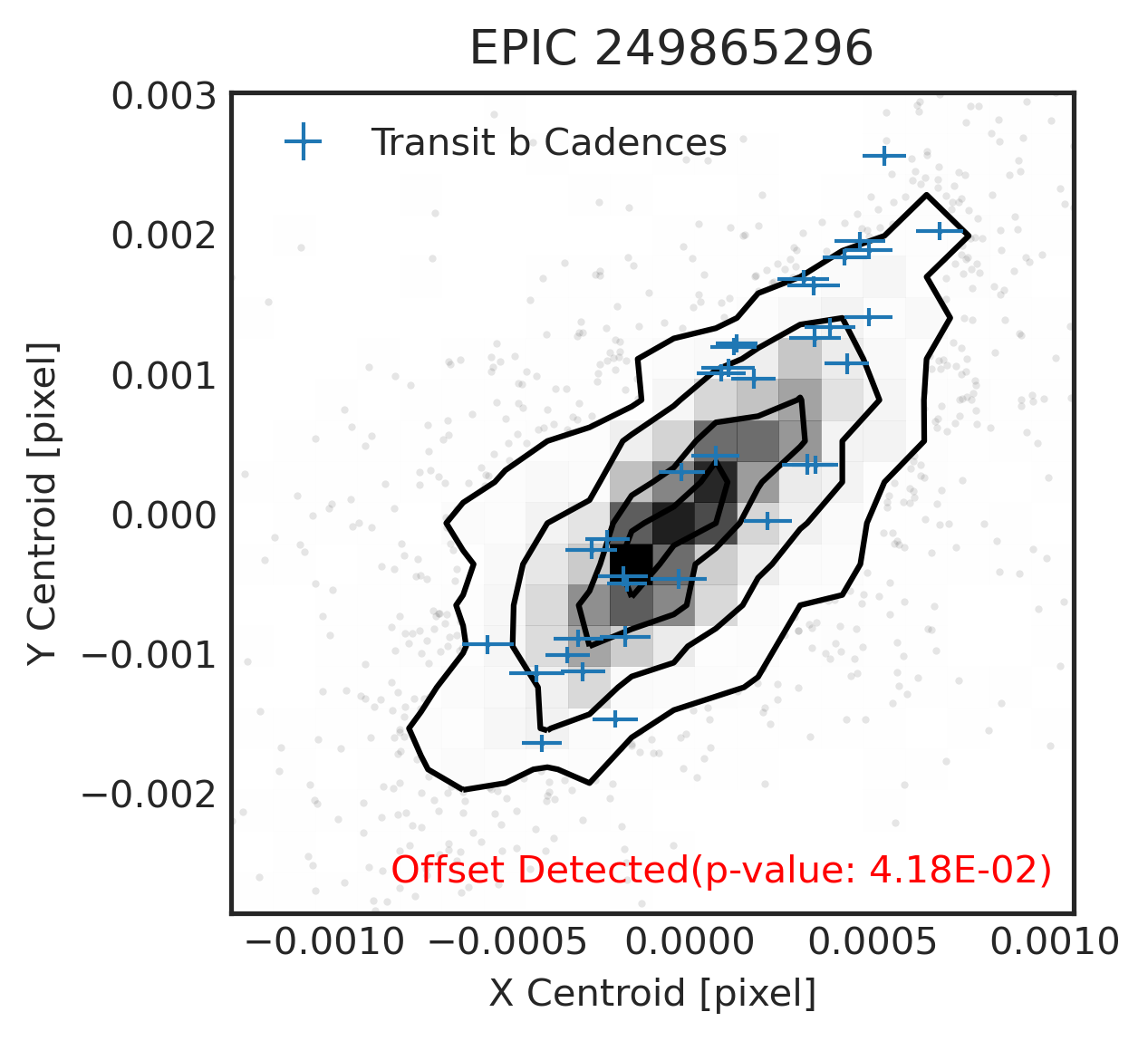}
\includegraphics[width=0.32\textwidth]{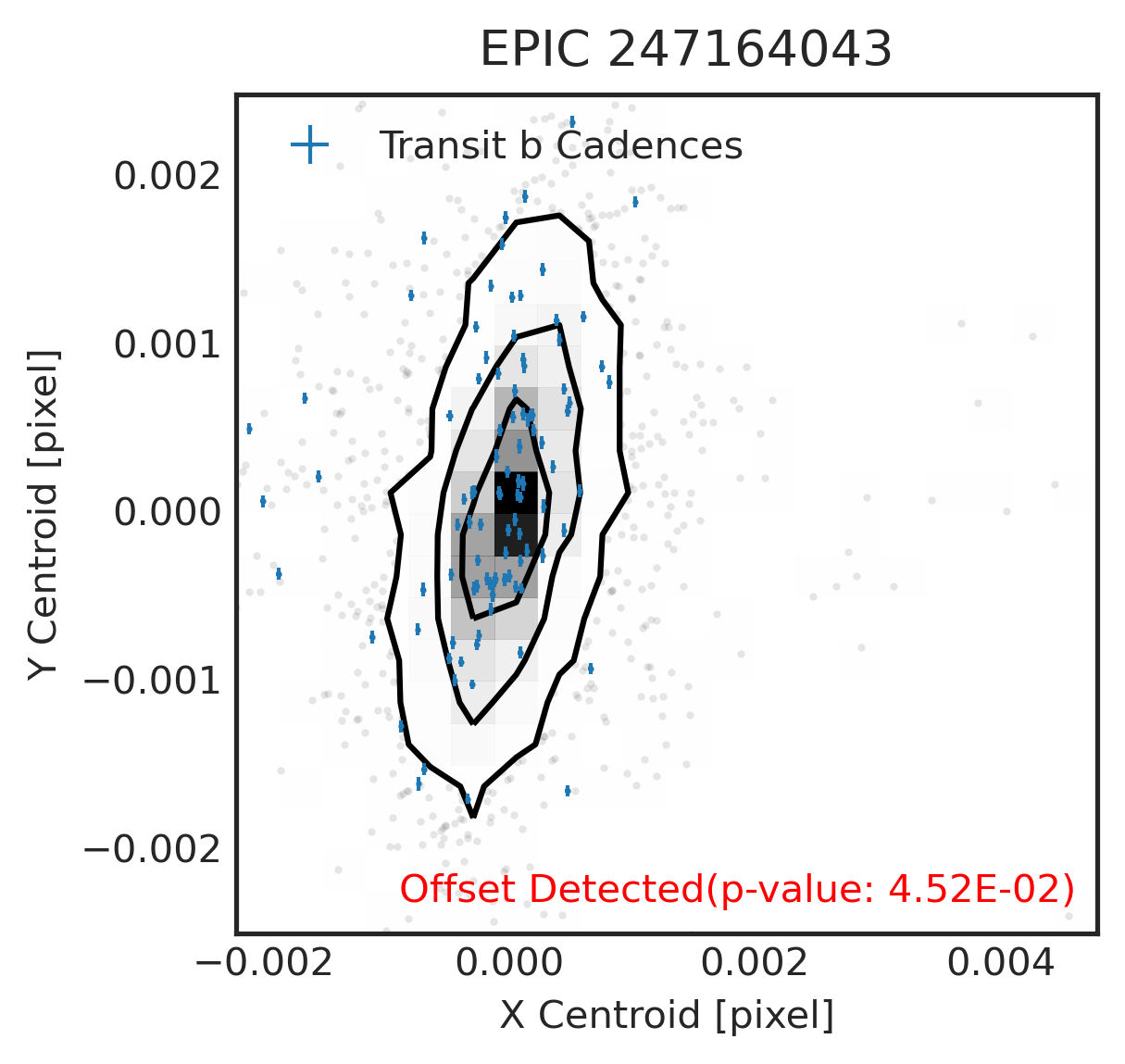}
\includegraphics[width=0.32\textwidth]{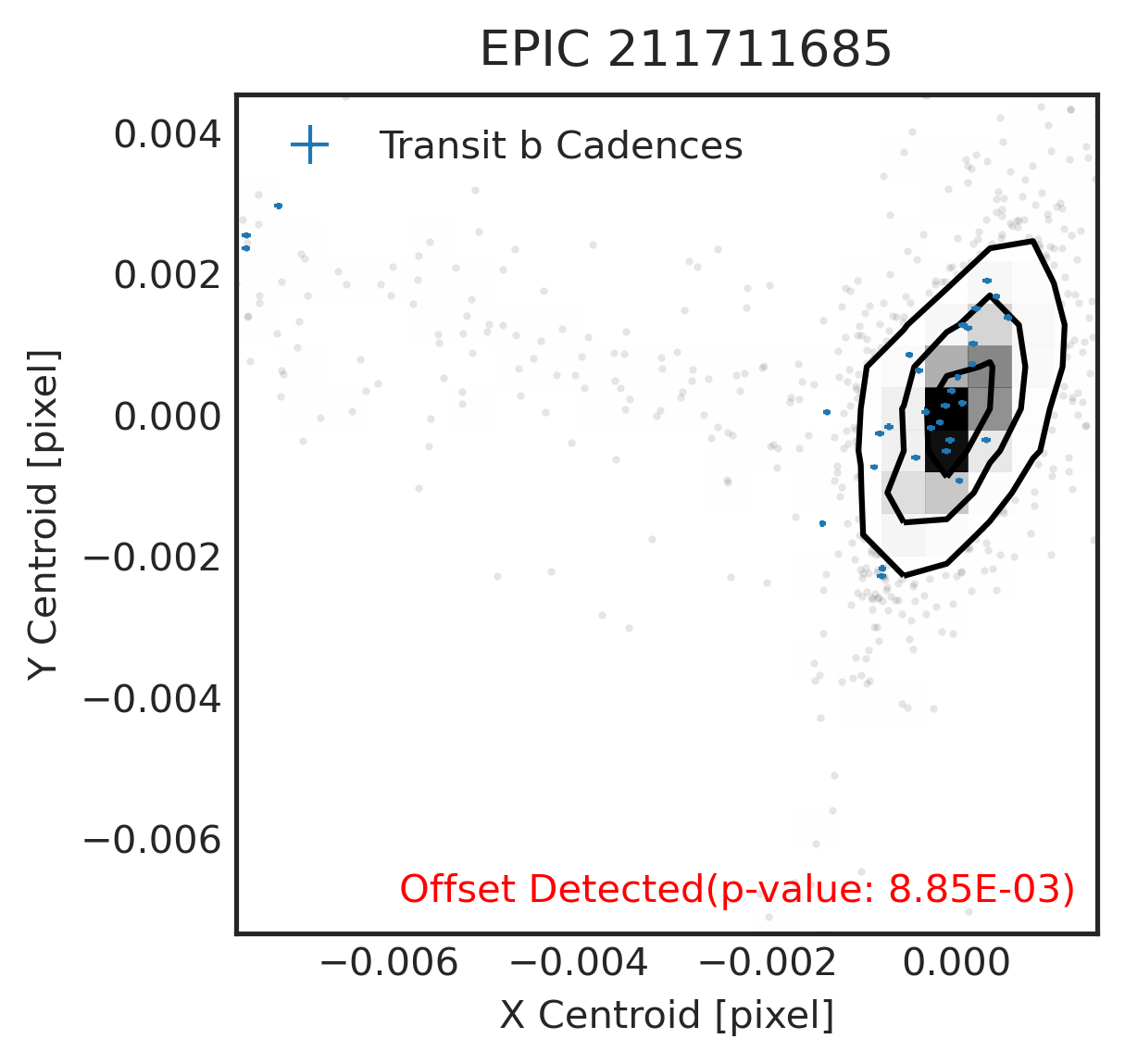}
\caption{Centroid plots for candidates with significant (p-value$<$0.05) centroid offsets during transit. The contours show the $1\sigma$, $2\sigma$, and $3\sigma$ contours for the centroid locations for out-of-transit cadences; the grey points show the remaining out-of-transit centroid locations that fall outside those contours. The blue points are the centroid locations of the in-transit cadences.}
\label{fig:centroid_fails}
\end{figure}

Finally, three candidates passed all our validation tests except the \texttt{centroid} test. Their centroid plots are shown in Figure \ref{fig:centroid_fails}; faint grey points are the centroid positions for out-of-transit observations in the light curve, blue points with error bars are the centroid positions for the in-transit observations. EPIC 249865296.01 and EPIC 247164043.01 have marginally significant offsets (p$=$0.0418 and 0.0452 respectively), and the distribution of their in-transit points do not clearly point in an obvious direction for a potential contaminating source, however they fail to pass our threshold and so we do not validate them. EPIC 211711685.01 has a very significant centroid offset (p=0.0089). Interestingly, most of the in-transit centroid positions for this candidate are in-family with the out-of-transit centroid positions, but a small number of in-transit points lie significantly offset from the location of the target star. As a result, we do not validate that the transiting signal is a planet on EPIC 21171168, however identifying the true source of the transiting signal is beyond the scope of this paper.

In addition to the above candidates, which had sufficient follow-up observations for us to attempt validation, we note a number of relatively bright ($Kepmag<13$) candidates in Table \ref{tab:needfollowup} that remain in our program. They pass visual inspection as clean planet candidates but still require either reconnaissance spectra or high-resolution imaging to proceed. We did not attempt to validate them, but highlight them for observers hoping to prioritize candidates for further follow-up. 


\begin{longrotatetable}
\begin{deluxetable*}{cccccccccc}
\tablewidth{700pt}
\tablecaption{Bright \textit{K2} planet candidates from Paper IV that require additional follow-up observations for validation. `Y' indicates that reconnaissance spectra or high-resolution imaging have already been obtained (many are available on ExoFOP); `-' indicates that they are still needed.\label{tab:needfollowup}}
\tabletypesize{\scriptsize}
\tablehead{
\colhead{Candidate ID} & 
\colhead{Camp} &
\colhead{Period} &
\colhead{$R_p$} &
\colhead{$T_{\rm mid}$} &
\colhead{$R_p/R_s$} &
\colhead{$Kepmag$} & 
\colhead{RUWE} & 
\colhead{Spectrum} & 
\colhead{Imaging} \\ 
\colhead{} & 
\colhead{} & 
\colhead{[d]} & 
\colhead{[$R_{\oplus}$]} & 
\colhead{[2454833-JD]} & 
\colhead{} & 
\colhead{[mag]} & 
\colhead{} &
\colhead{} &
\colhead{}
}
\startdata
201445732.01&   1 &     5.60203111 &    1.916  &        1978.88960 &    0.0149 &        11.95 & 0.88 & - & Y \\
204308061.01&   2 &     22.6435563 &    1.297  &        2067.43865 &    0.0143 &        10.17 & 0.92 & - & - \\
204888276.01&   2 &     16.5604729 &    1.217  &        2067.53685 &    0.0299 &        12.54 & -    & Y & - \\
205483055.01&   2 &     22.970936  &    2.394  &        2063.81798 &    0.0219 &        12.93 & 0.90 & - & - \\
205489894.01&   2 &     14.6573703 &    1.693  &        2074.07547 &    0.0313 &        12.34 & 1.14 & Y & - \\
205943325.01&   3 &     27.2701915 &    2.500  &        2151.27679 &    0.0142 &        11.28 & 0.99 & Y & - \\
206133795.01&   3 &     13.8506388 &    2.882  &        2147.80162 &    0.0152 &        11.63 & 1.04 & - & Y \\
206329937.01&   3 &     10.4990647 &    20.92  &        2151.04875 &    0.0338 &        11.79 & 1.23 & - & - \\
210550063.01&   4 &     2.16604744 &    2.611  &        2229.8188  &    0.0208 &        11.15 & 1.31 & - & Y \\
210678858.02&   4 &     14.8489834 &    2.466  &        2237.22147 &    0.0287 &        12.74 & 1.08 & - & Y \\
210945680.01&   4 &     20.6006563 &    2.166  &        2242.51612 &    0.0187 &        11.32 & 0.99 & - & Y \\
211505089.01&   5 &     24.443649  &    1.737  &        2313.9657  &    0.0222 &        12.94 & 1.48 & Y & - \\
211897272.01&   5 &     11.2356761 &    2.706  &        2310.56874 &    0.0266 &        12.57 & 1.16 & - & - \\
212346723.01&   6 &     15.9483772 &    2.345  &        2397.08637 &    0.0171 &        12.99 & 0.96 & - & - \\
212652418.01&   6 &     19.1374173 &    2.802  &        2401.96834 &    0.0191 &        12.27 & 1.11 & - & Y \\
212691727.01&   6 &     6.43038282 &    4.309  &        2386.5697  &    0.0374 &        12.66 & 0.92 & - & Y \\
212833814.01&   6 &     27.8083498 &    2.144  &        2403.11012 &    0.0130 &        10.72 & 0.97 & - & Y \\
216363472.01&   7 &     8.69447821 &    1.401  &        2468.88249 &    0.0129 &        11.87 & 1.01 & - & Y \\
218332621.01&   7 &     9.81158865 &    4.245  &        2469.06837 &    0.0207 &        12.75 & 0.87 & - & Y \\
218701083.01&   7 &     5.04906822 &    2.594  &        2471.91113 &    0.0161 &        12.49 & 0.93 & - & Y \\
246067130.01&   12&     6.78653927 &    1.200  &        2906.07556 &    0.0113 &        11.74 & 0.87 & - & Y \\
246777408.01&   13&     2.46563986 &    30.19  &        2988.02272 &    0.1107 &        11.21 & 1.21 & - & - \\
248865761.01&   14&     13.0710405 &    2.285  &        3085.12075 &    0.0168 &        12.94 & 1.06 & Y & - \\
248869142.01&   14&     28.8483394 &    3.657  &        3086.94431 &    0.0191 &        10.39 & 0.98 & - & - \\
249373875.01&   15&     7.52210942 &    2.249  &        3159.14609 &    0.0149 &        12.34 & 1.23 & - & - \\
249418609.01&   15&     15.3206064 &    1.770  &        3170.06807 &    0.0145 &        12.35 & 1.13 & - & - \\
249476992.01&   15&     7.9436039  &    1.992  &        3161.75698 &    0.0210 &        12.79 & 1.21 & - & Y \\
249703125.01&   15&     10.3073874 &    2.602  &        3157.55787 &    0.0272 &        12.88 & 1.26 & - & Y \\
249731291.01&   15&     10.2053773 &    6.310  &        3160.91941 &    0.0245 &        12.92 & 0.97 & - & Y \\
249731291.02&   15&     23.3367386 &    5.824  &        3179.76842 &    0.0226 &        12.92 & 0.97 & - & Y \\
249822095.01&   15&     1.69671578 &    5.738  &        3157.01732 &    0.0264 &        12.80 & 1.13 & - & Y \\
250106132.01&   15&     22.1223653 &    3.715  &        3157.32569 &    0.0351 &        11.19 & 0.90 & Y & - \\
211396431.01&   16&     8.51418384 &    38.30  &        3266.73236 &    0.0418 &        11.77 & 1.18 & - & Y \\
251304634.01&   16&     4.39250294 &    1.110  &        3264.36688 &    0.0152 &        12.25 & 1.02 & Y & - \\
251399501.01&   16&     11.6349883 &    2.605  &        3272.15823 &    0.0218 &        12.10 & 1.25 & Y & - \\
212479997.01&   17&     8.9317272  &    9.390  &        3348.03092 &    0.0478 &        10.80 & 1.07 & - & - \\
211408558.01&   18&     9.96768882 &    2.005  &        3427.65262 &    0.0178 &        11.51 & 1.13 & - & Y \\
\enddata
\end{deluxetable*}
\end{longrotatetable}

\section{Conclusions}
\label{sec:conclusions}

In total, we validate \totalval\ new \emph{K2} planets in \totalsys\ systems, out of 91 total candidates. This is a validation rate of 66\%, having already down-selected the full catalog in Paper IV to the most likely planetary signals by visual inspection. The most promising new planets for transmission spectroscopy are \epictwoseventwofalias, \epictwothreethreebalias, \epicninesevensevenbalias, \epicthreefivethreebalias, and \epicthreefivezerocalias, while the most promising new planets for emission spectroscopy are \epicthreefiveninebalias, \epicfiveeightzerobalias, and \epiconefourzerobalias. \epicfoursevenonebalias and \epicninetwoeightbalias show interesting TSM and ESM values, making them potentially valuable atmospheric characterization targets. In addition to these new planets, in Table \ref{tab:needfollowup} we highlight 37 planet candidates orbiting bright host stars that pass visual inspection and warrant further follow-up. 

We note that many additional planets remain in the \emph{K2} data. Besides the planet candidates listed above, Paper IV also includes 109 planet candidates that pass visual inspection orbiting stars $13<Kp<15$, and 90 planet candidates that pass visual inspection orbiting stars $Kp>15$, that require additional follow-up observations. There are also hundreds of additional \emph{K2} planet candidates at the NASA Exoplanet Archive; although Paper IV presented a large, uniform catalog of planet candidates, comparison with larger catalogs such as \citet{kru19} show that our catalog was incomplete. Given the plethora of archival data from multiple missions now available, there remain strong reasons to pursue \emph{K2} planet confirmation. There are significant ongoing and upcoming opportunities to exploit and expand upon the \emph{K2} data, including: (i) a re-observation of the ecliptic plane by the TESS mission \citep{Ricker2015} during its first extended mission; (ii) opportunities for measuring transit timing variations or searching for additional transits with CHEOPS \citep{Broeg2013}; (iii) a new generation of extreme precision radial velocity instruments to map out the composition of different classes and sub-classes of planets; and (iv) opportunities for characterizing exoplanet atmospheres to an unprecedented level of detail with JWST and high-resolution ground-based spectroscopy for statistical samples of planets.

\section{Acknowledgements}

The analysis described here were performed on the UCLA Hoffman2 shared computing cluster and using the resources provided by the Bhaumik Institute. This research has made use of the NASA Exoplanet Archive and the Exoplanet Follow-up Observation Program website, which are operated by the California Institute of Technology, under contract with the National Aeronautics and Space Administration under the Exoplanet Exploration Program. This publication makes use of data products collected by the \emph{Kepler} mission and obtained from the MAST data archive at the Space Telescope Science Institute (STScI); the K2SFF apertures can be accessed via \dataset[https://doi.org/10.17909/T9BC75]{https://doi.org/10.17909/T9BC75}, and the K2 target pixel files via \dataset[https://doi.org/10.17909/T9K30X]{https://doi.org/10.17909/T9K30X}. STScI is operated by the Association of Universities for Research in Astronomy, Inc., under NASA contract NAS5–26555. Support to MAST for these data is provided by the NASA Office of Space Science via grant NAG5–7584 and by other grants and contracts. Funding for the \emph{Kepler} mission was provided by the NASA Science Mission Directorate. This publication makes use of data products from the Two Micron All Sky Survey, which is a joint project of the University of Massachusetts and the Infrared Processing and Analysis Center/California Institute of Technology, funded by the National Aeronautics and Space Administration and the National Science Foundation. This dataset is made available by the Infrared Science Archive (IRSA) at IPAC, which is operated by the California Institute of Technology under contract with the National Aeronautics and Space Administration. The specific data products can be accessed via \dataset[https://doi.org/10.26131/IRSA2]{https://doi.org/10.26131/IRSA2}.

Funding for this work for CH is provided by grant number 80NSSC20K0874, through NASA ROSES. M.T. is supported by JSPS KAKENHI grant Nos.18H05442,15H02063, and 22000005. Support for this work was provided by NASA through the NASA Hubble Fellowship grant \#51497.001 awarded by the Space Telescope Science Institute, which is operated by the Association of Universities for Research in Astronomy, Inc., for NASA, under contract NAS5-26555. The material is based upon work supported by NASA under award number 80GSFC21M0002 and by the GSFC Sellers Exoplanet Environments Collaboration (SEEC), which is funded by the NASA Planetary Science Division’s Internal Scientist Funding Mode.

The observations in the paper made use of the NN-EXPLORE Exoplanet and Stellar Speckle Imager (NESSI). NESSI was funded by the NASA Exoplanet Exploration Program and the NASA Ames Research Center. NESSI was built at the Ames Research Center by Steve B. Howell, Nic Scott, Elliott P. Horch, and Emmett Quigley. The authors are honored to be permitted to conduct observations on Iolkam Du'ag (Kitt Peak), a mountain within the Tohono O'odham Nation with particular significance to the Tohono O'odham people. 

This work has made use of data from the European Space Agency (ESA) mission {\it Gaia} (\url{https://www.cosmos.esa.int/gaia}), processed by the {\it Gaia} Data Processing and Analysis Consortium (DPAC, \url{https://www.cosmos.esa.int/web/gaia/dpac/consortium}). Funding for the DPAC has been provided by national institutions, in particular the institutions participating in the {\it Gaia} Multilateral Agreement.

This work made use of the gaia-kepler.fun crossmatch database created by Megan Bedell.

\software{{\tt EVEREST} \citep{lug16,lug18}, {\tt TERRA} \citep{pet13b}, {\tt EDI-Vetter} \citep{zin20a}, {\tt \emph{K2}SFF} \citep{van14}, {\tt PyMC3} \citep{sal15}, {\tt Exoplanet} \citep{for19}, {\tt RoboVetter} \citep{tho18}, {\tt batman} \cite{kre15}, {\tt emcee} \citep{for13}}


\bibliography{paper}{}
\bibliographystyle{aasjournal}



\begin{appendices}
\label{app:validatedplanets} 

\section{Validated planets}

\subsection{EPIC 204750116}






\epiczerooneonesixbalias\ is a sub-Neptune (2.906~\rearth) planet orbiting a bright ($V$=11.37 mag, $K_s$=9.62 mag) G4V star (1.05~\rsun, 0.82~\msun) observed in Campaign 2. It was noted as EPIC 204750116.01 by \citet{van16}, \citet{riz18}, \citet{Mayo2018}, and \citet{kru19}, and it orbits the star at a distance of 0.150~au, with a period of 23.448172~days and an equilibrium temperature of 710~K. The host star has a clean, single-lined FLWO/TRES spectrum, and Keck/NIRC2 imaging which shows no contaminating stellar companions. The \texttt{vespa} FPP value is 0.0016, without requiring use of the available contrast curves. The \texttt{centroid} p-value is 0.5219, which is consistent with the source of the transiting signal being on the target star. Using the \citet{Chen2017} mass-radius relation predicts a mass of $\sim$8.7~\mearth. This would result in a measurable RV semi-amplitude (\emph{K}$\sim$2.2 m/s), but predicts TSM ($\sim$22) and ESM ($\sim$1.3) values that indicate atmospheric follow-up would be challenging, compared to the thresholds recommended by \citet{Kempton2018}. We note that our pipeline was able to marginally detect the shorter period candidate, EPIC 204750116.02, reported by \cite{kru19}, but with insufficient signal-to-noise to attempt validation, and it remains a candidate.

\subsection{EPIC 205111664}



\epicsixsixfouralias\ is a moderately bright ($V$=12.49 mag, $K_s$=9.88 mag) G4V star (0.94~\rsun, 0.83~\msun) that was observed in Campaign 2. It is orbited by a sub-Neptune-sized planet that was previously noted as EPIC 205111664.01 by \citet{van16} and \citet{kru19}. \epicsixsixfourbalias\ is a 2.359~\rearth\ planet, orbiting at a distance of 0.116~au, with a period of 15.937039~days and an equilibrium temperature of $\sim$770~K. The host star has a clean, single-lined FLWO/TRES spectrum, and Keck/NIRC2 AO imaging that shows no contaminating stellar companions. The \texttt{vespa} FPP value is $1.40 \times 10^{-4}$, without requiring use of the contrast curve. The \texttt{centroid} p-value is 0.4962, which is consistent with the source of the transiting signal being on the target star. Using the \citet{Chen2017} mass-radius relation predicts a mass of $\sim$6.3~\mearth. This would result in a challenging RV semi-amplitude (\emph{K}$\sim$1.8 m/s), and predicts TSM ($\sim$18) and ESM ($\sim$1.1) values that indicate atmospheric follow-up would be challenging, compared to the thresholds recommended by \citet{Kempton2018}.

\subsection{EPIC 206055981}



\epicnineeightonealias\ is a moderately faint ($V$=13.80 mag, $K_s$=10.96 mag) K5V star (0.63~\rsun, 0.76~\msun) that was observed in Campaign 3. It is orbited by a sub-Neptune-size planet, \epicnineeightonebalias\, which was previously identified as EPIC 206055981.01 by \citet{van16} and \citet{kru19}. It is a 2.413~\rearth\ planet, orbiting at a distance of 0.134~au, with a period of 20.645099~days and a relatively cool (for our sample) equilibrium temperature of $\sim$450~K. The host star has clean, single-lined Keck/HIRES and IRTF/SpeX spectra, and WIYN/NESSI speckle imaging at 562nm and 832nm that show no contaminating stellar companions. The \texttt{vespa} FPP value is 0.000721, using the available contrast curves. The \texttt{centroid} p-value is 0.5842, which is consistent with the source of the transiting signal being on the target star. Using the \citet{Chen2017} mass-radius relation predicts a mass of $\sim$6.5\mearth. This would result in a challenging RV semi-amplitude (\emph{K}$\sim$1.8m/s), especially given the faint host star, and predicts TSM ($\sim$15) and ESM ($\sim$0.4) values that indicate atmospheric follow-up would be challenging, compared to the thresholds recommended by \citet{Kempton2018}.

\subsection{EPIC 206146957}



\epicninefivesevenalias\ is a bright ($V$=11.79 mag, $K_s$=9.93 mag) G3V star (0.88~\rsun, 0.74~\msun) that was observed in Campaign 3. It is orbited by a super-Earth-size planet, \epicninefivesevenbalias\, which was previously identified as EPIC 206146957.01 by \citet{van16}, \citet{Mayo2018}, and \citet{kru19}. It is a 1.313~\rearth\ planet, orbiting at a distance of 0.057~au, with a period of 5.761612~days and an equilibrium temperature of $\sim$1080~K. The host star has a clean, single-lined FLWO/TRES spectra, and Palomar/PHARO AO imaging that shows no contaminating stellar companions. The \texttt{vespa} FPP value is $3.02 \times 10^{-5}$, using the available contrast curve. The \texttt{centroid} p-value is 0.4584, which is consistent with the source of the transiting signal being on the target star. Using the \citet{Chen2017} mass-radius relation predicts a mass of $\sim$2.2\mearth. This would result in a challenging RV semi-amplitude (\emph{K}$\sim$1.0m/s), and predicts TSM ($\sim$2.3) and ESM ($\sim$0.8) values that indicate atmospheric follow-up would be challenging, compared to the thresholds recommended by \citet{Kempton2018}.

\subsection{EPIC 206317286}





\epictwoeightsixalias\ (K2-303) is a moderately faint ($V$=14.05 mag, $K_s$=11.64 mag) K3V star (0.76~\rsun, 0.83~\msun) that was observed in Campaign 3. It was previously found to host a short period (1.58~d) Earth-sized (0.96~\rearth) planet, \epictwoeightsixbalias\ (EPIC 206317286.02) \citep{Heller2019}. An additional longer period candidate, EPIC 206317286.01, was noted by \citet{van16} and \citet{kru19}, and was recovered by our pipeline. \epictwoeightsixcalias\ is a 2.464~\rearth\ sub-Neptune-sized planet, orbiting at a distance of 0.124~au, with a period of 17.515472~days and an equilibrium temperature of $\sim$570~K. The host star has a clean, single-lined IRTF/SpeX spectrum. \citet{Heller2019} validated \epictwoeightsixbalias\ in part by using Pan-STARRS imaging (with a resolution down to 200 mas/pixel) to eliminate background flux contaminants; we therefore did not obtain additional high-resolution imaging. The \texttt{vespa} FPP value of \epictwoeightsixcalias\ is $8.16 \times 10^{-5}$ using the Campaign 3 multiplicity boost. The \texttt{centroid} p-value is 0.1811, which is consistent with the source of the transiting signal being on the target star. Using the \citet{Chen2017} mass-radius relation predicts a mass of planet c of $\sim$5.4~\mearth. This would result in a challenging RV semi-amplitude (\emph{K}$\sim$1.5 m/s), and predicts TSM ($\sim$8.0) and ESM ($\sim$0.3) values that indicate atmospheric follow-up would be challenging, compared to the thresholds recommended by \citet{Kempton2018}.

\subsection{EPIC 211490999}



\epicninenineninealias\ is a moderately faint ($V$=13.60 mag, $K_s$=11.87 mag) G5V star (0.92~\rsun, 0.85~\msun) that was observed in Campaigns 5, 16, and 18. It is orbited by a sub-Neptune-sized planet, \epicninenineninebalias\, previously noted as EPIC 211490999.01 by \citet{pop16}, \citet{Petigura18a}, \citet{kru19}, and \citet{deLeon2021}. It is a 2.945~\rearth\ planet, orbiting at a distance of 0.085~au, with a period of 9.844077~days and an equilibrium temperature of $\sim$870~K. The host star has clean, single-lined Keck/HIRES and McDonald/TS23 spectra, and Gemini/NIRI AO and WIYN/NESSI speckle imaging at 562nm and 832nm that show no contaminating stellar companions. The \texttt{vespa} FPP value is $3.39 \times 10^{-3}$, using the available contrast curves. The \texttt{centroid} p-value is 0.4752, which is consistent with the source of the transiting signal being on the target star. Using the \citet{Chen2017} mass-radius relation predicts a mass of $\sim$9.2~\mearth. This would result in a measurable RV semi-amplitude (\emph{K}$\sim$3.1 m/s), but predicts TSM ($\sim$13) and ESM ($\sim$1.0) values that indicate atmospheric follow-up would be challenging, compared to the thresholds recommended by \citet{Kempton2018}.

\subsection{EPIC 211539054}



\epiczerofivefouralias\ is a bright ($V$=10.47 mag, $K_s$=9.16 mag) G1V star (1.49~\rsun, 1.16~\msun) that was observed in Campaign 5. It is orbited by a sub-Neptune-size planet, \epiczerofivefourbalias\, which was first noted by \citet{kru19} as EPIC 211539054.01. It is a 2.061~\rearth\ planet, orbiting at a distance of 0.102~au, with a period of 11.020660~days and an equilibrium temperature of $\sim$1130~K. The host star has a clean, single-lined FLWO/TRES spectra, and Palomar/PHARO AO imaging that shows no contaminating stellar companions. The \texttt{vespa} FPP value is 0.00213, using the available contrast curve. The \texttt{centroid} p-value is 0.3779, which is consistent with the source of the transiting signal being on the target star. Using the \citet{Chen2017} mass-radius relation predicts a mass of $\sim$5.2\mearth. This would result in a challenging RV semi-amplitude (\emph{K}$\sim$1.3m/s), and predicts TSM ($\sim$14) and ESM ($\sim$1.0) values that indicate atmospheric follow-up would be challenging, compared to the thresholds recommended by \citet{Kempton2018}.

\subsection{EPIC 211732116}




\epictwooneonesixalias\ is a moderately bright ($V$=12.80 mag, $K_s$=11.29 mag) G3V star (0.91~\rsun, 0.73~\msun) that was observed in Campaign 16. It is orbited by a super-Earth and a sub-Neptune, which were first noted in Paper IV. \epictwooneonesixbalias\ (EPIC 211732116.01) is a 2.597~\rearth\ planet, orbiting at a distance of 0.055~au, with a period of 4.521953~days and a high equilibrium temperature of $\sim$1210~K. \epictwooneonesixcalias\ (EPIC 211732116.02) is a 2.213~\rearth\ planet, orbiting at a distance of 0.114~au, with a period of 16.434450~days and an equilibrium temperature of $\sim$790~K. The host star has a clean, single-lined FLWO/TRES spectrum, and Keck/NIRC2 AO imaging that shows no contaminating stellar companions. The \texttt{vespa} FPP values of planets b and c are $6.80 \times 10^{-5}$ and $2.42 \times 10^{-3}$ respectively, using the available contrast curve and the multiplicity boost. The \texttt{centroid} p-values are 0.5403 and 0.5749 respectively, consistent with the source of the transiting signals being on the target star. Using the \citet{Chen2017} mass-radius relation predicts masses of $\sim$3.6~\mearth\ and $\sim$5.7~\mearth\ for planets b and c respectively. These mass estimates for the two planets result in challenging RV semi-amplitudes ($K$ = 1.7 m/s, and 1.8 m/s), and TSM ($\sim$10, $\sim$11) and ESM ($\sim$0.7, $\sim$0.6) values below the thresholds of interest recommended by \citet{Kempton2018}.

\subsection{EPIC 212006318}



\epicthreeoneeightalias\ is a moderately faint ($V$=13.04 mag, $K_s$=11.56 mag) G2V star (1.54~\rsun, 1.12~\msun) that was observed in Campaign 3. It is orbited by a sub-Neptune-sized planet, \epicthreeoneeightbalias, which was previously noted as EPIC 212006318.01 by \citet{pop16}, \citet{Mayo2018}, \citet{kru19}, and \citet{deLeon2021}. \epicthreeoneeightbalias\ is a 2.243~\rearth\ planet, orbiting at a distance of 0.121~au, with a period of 14.453895~days and an equilibrium temperature of $\sim$1000~K. The host star has clean, single-lined Keck/HIRES and FLWO/TRES spectra, and Palomar/PHARO AO and WIYN/NESSI speckle imaging at 562nm and 832nm that show no contaminating stellar companions. The \texttt{vespa} FPP value is $4.11 \times 10^{-3}$, without using the available contrast curves. The \texttt{centroid} p-value is 0.1458, which is consistent with the source of the transiting signal being on the target star. Using the \citet{Chen2017} mass-radius relation predicts a mass of $\sim$5.8\mearth. This would result in a challenging RV semi-amplitude (\emph{K}$\sim$1.4 m/s), and predicts TSM ($\sim$4.3) and ESM ($\sim$0.31) values that indicate atmospheric follow-up would be challenging, compared to the thresholds recommended by \citet{Kempton2018}.

\subsection{EPIC 212222383}



\epicthreeeightthreealias\ is a bright ($V$=10.42 mag, $K_s$=9.01 mag) G1V star (1.15~\rsun, 0.91~\msun) that was observed in Campaign 16. It is orbited by a super-Earth-size planet that was first noted in Paper IV as EPIC 212222383.01. \epicthreeeightthreebalias\ is a 1.736~\rearth\ planet, orbiting at a distance of 0.061~au, with a period of 5.776475~days and an equilibrium temperature of $\sim$1240~K. The host star has a clean, single-lined FLWO/TRES spectrum, and Palomar/PHARO AO imaging that shows no contaminating stellar companions. The \texttt{vespa} FPP value is 0.00170, using the available contrast curve. The \texttt{centroid} p-value is 0.6183, which is consistent with the source of the transiting signal being on the target star. Using the \citet{Chen2017} mass-radius relation predicts a mass of $\sim$4.1\mearth. This would result in a challenging RV semi-amplitude (\emph{K}$\sim$1.6m/s), and predicts TSM ($\sim$21) and ESM ($\sim$1.6) values that indicate atmospheric follow-up would be challenging, compared to the thresholds recommended by \citet{Kempton2018}.

\subsection{EPIC 212530118}



\epiconeoneeightalias\ is a moderately faint ($V$=14.05 mag, $K_s$=10.99 mag) K7V star (0.69~\rsun, 0.82~\msun) that was observed in Campaign 6. It is orbited by a super-Earth-sized planet that was previously noted as EPIC 212530118.01 by \citet{bar16} and \citet{kru19}. \epiconeoneeightbalias\ is a 1.693~\rearth\ planet, orbiting at a distance of 0.100~au, with a period of 12.932235~days and an equilibrium temperature of $\sim$530~K. The host star has a clean, single-lined IRTF/SpeX spectrum, and WIYN/NESSI speckle imaging at 692nm and 880nm that show no contaminating stellar companions. The \texttt{vespa} FPP value is $1.14 \times 10^{-3}$, using the available contrast curves. The \texttt{centroid} p-value is 0.5774, which is consistent with the source of the transiting signal being on the target star. Using the \citet{Chen2017} mass-radius relation predicts a mass of $\sim$3.9\mearth. This would result in a challenging RV semi-amplitude (\emph{K}$\sim$1.2 m/s), and predicts TSM ($\sim$7.7) and ESM ($\sim$0.3) values that indicate atmospheric follow-up would be challenging, compared to the thresholds recommended by \citet{Kempton2018}.

\subsection{EPIC 212575828}



\epiceighttwoeightalias\ is a faint ($V$=15.79 mag, $K_s$=13.39 mag) K3V star (0.73~\rsun, 0.92~\msun), observed in Campaign 6, that is orbited by a sub-Neptune-sized planet. \epiceighttwoeightbalias, previously noted by \citet{pop16}, \citet{Crossfield2016}, and \citet{kru19} as EPIC 212575828.01, is a 2.880~\rearth\ planet, orbiting at a distance of 0.031~au, with a period of 2.060438~days and an equilibrium temperature of $\sim$1135~K. The host star has a clean, single-lined Hale/TripleSpec spectrum, and Gemini/DSSI speckle imaging at both 692nm and 880nm that show no contaminating stellar companions. The \texttt{vespa} FPP value is $3.78 \times 10^{-3}$, using the available contrast curves. The \texttt{centroid} p-value is 0.4968, which is consistent with the source of the transiting signal being on the target star. Using the \citet{Chen2017} mass-radius relation predicts a mass of $\sim$9.0~\mearth. This would result in a measurable RV semi-amplitude (\emph{K}$\sim$4.5 m/s), albeit around a faint target, but predicts TSM ($\sim$12) and ESM ($\sim$1.6) values that indicate atmospheric follow-up would be challenging, compared to the thresholds recommended by \citet{Kempton2018}.

\subsection{EPIC 214173069}





\epiczerosixninealias\ is a moderately faint ($V$=13.21 mag, $K_s$=10.51 mag) K3V star (0.75~\rsun, 0.86~\msun) that was observed in Campaign 7. It is orbited by a sub-Neptune-sized planet, \epiczerosixninebalias\, which was first noted as EPIC 214173069.01 by \citet{kru19}. It is a 2.181~\rearth\ planet, orbiting at a distance of 0.079~au, with a period of 8.777247~days and an equilibrium temperature of $\sim$680~K. The host star has clean, single-lined IRTF/SpeX and FLWO/TRES spectra, and Gemini/DSSI speckle imaging at 692nm and 880nm that show no contaminating stellar companions. The \texttt{vespa} FPP value is $5.38 \times 10^{-3}$, using the available contrast curves. The \texttt{centroid} p-value is 0.8487, which is consistent with the source of the transiting signal being on the target star. Using the \citet{Chen2017} mass-radius relation predicts a mass of $\sim$5.6\mearth. This would result in a challenging RV semi-amplitude (\emph{K}$\sim$1.9 m/s), and predicts TSM ($\sim$17) and ESM ($\sim$0.8) values that indicate atmospheric follow-up would be challenging, compared to the thresholds recommended by \citet{Kempton2018}. We note that our pipeline did not recover the smaller (1.4\rearth), longer period (14.1297-day) candidate, EPIC 214173069.02, reported by \citet{kru19}.

\subsection{EPIC 214419545}



\epicfivefourfivealias\ is a bright ($V$=11.65 mag, $K_s$=9.68 mag) F8V star (1.32~\rsun, 0.99~\msun) that was observed in Campaign 7. It is orbited by a sub-Neptune-sized planet, \epicfivefourfivebalias, which was first noted as EPIC 214419545.01 by \citet{kru19}. It is a 2.729~\rearth\ planet, orbiting at a distance of 0.087~au, with a period of 9.401312~days and an equilibrium temperature of $\sim$1110~K. The host star has a clean, single-lined FLWO/TRES spectrum, and Keck/NIRC2 AO imaging that show no contaminating stellar companions. The \texttt{vespa} FPP value is $8.82 \times 10^{-3}$, without using the available contrast curves. The \texttt{centroid} p-value is 0.6540, which is consistent with the source of the transiting signal being on the target star. Using the \citet{Chen2017} mass-radius relation predicts a mass of $\sim$8.0\mearth. This would result in a measurable RV semi-amplitude (\emph{K}$\sim$2.4 m/s), but predicts TSM ($\sim$19) and ESM ($\sim$1.8) values that indicate atmospheric follow-up would be challenging, compared to the thresholds recommended by \citet{Kempton2018}.

\subsection{EPIC 217977895}




\epiceightninefivebalias\ is a sub-Neptune (2.072~\rearth) planet in the small planet radius valley, orbiting a moderately faint ($V$=12.98 mag, $K_s$=11.05 mag) G8V star (0.81~\rsun, 0.92~\msun). It was observed in Campaign 7, and was first noted in \citet{Mayo2018} and \citet{kru19} as EPIC 217977895.01. It orbits the star at a distance of 0.148~au, with a period of 21.700156~days and an equilibrium temperature of 615~K. The host star has a clean, single-lined FLWO/TRES spectrum, and Gemini/DSSI speckle imaging at 692~nm and 880~nm which show no contaminating stellar companions. The \texttt{vespa} FPP value is $2.62 \times 10^{-3}$ using the available contrast curves. The \texttt{centroid} p-value is 0.3721, which is consistent with the source of the transiting signal being on the target star. Using the \citet{Chen2017} mass-radius relation predicts a mass of $\sim$5.2~\mearth. This would result in a challenging RV semi-amplitude (\emph{K}$\sim$1.3 m/s), and predicts TSM ($\sim$10) and ESM ($\sim$0.4) values that indicate atmospheric follow-up would be challenging, compared to the thresholds recommended by \citet{Kempton2018}.

\subsection{EPIC 218668602}



\epicsixzerotwoalias\ is a moderately bright ($V$=12.68 mag, $K_s$=10.46 mag) K0V star (0.80~\rsun, 1.06~\msun) that was observed in Campaign 7. It is orbited by a super-Earth-sized planet, \epicsixzerotwobalias\, which was previously noted by \citet{Mayo2018} and \citet{kru19} at EPIC 218668602.01. It is a 1.564~\rearth\ planet, orbiting at a distance of 0.030~au, with a period of 1.865962~days and an equilibrium temperature of $\sim$1270~K. The host star has a clean, single-lined FLWO/TRES spectrum, and Keck/NIRC2 AO imaging that show no contaminating stellar companions. The \texttt{vespa} FPP value is $2.12 \times 10^{-3}$, without using the available contrast curves. The \texttt{centroid} p-value is 0.1391, which is consistent with the source of the transiting signal being on the target star. Using the \citet{Chen2017} mass-radius relation predicts a mass of $\sim$3.3\mearth. This would result in a challenging RV semi-amplitude (\emph{K}$\sim$1.6 m/s), but predicts TSM ($\sim$18) and ESM ($\sim$1.7) values that indicate atmospheric follow-up would be challenging, compared to the thresholds recommended by \citet{Kempton2018}.

\subsection{EPIC 220459477}



\epicfoursevensevenbalias\ is a super-Earth (1.539~\rearth) planet orbiting a moderately faint ($V$=14.65 mag, $K_s$=12.24 mag) K3V star that was observed in Campaign 8. It was first noted as EPIC 220459477.01 by \citet{kru19}. It orbits the star at a distance of 0.034~au, with a period of 2.380867~days and an equilibrium temperature of $\sim$1100~K. The host star has a clean, single-lined TripleSpec spectrum, and WIYN/NESSI speckle imaging at both 562nm and 832nm that show no contaminating stellar companions. The \texttt{vespa} FPP value is $1.09 \times 10^{-5}$ using the available contrast curves. The \texttt{centroid} p-value is 0.1107, which is consistent with the source of the transiting signal being on the target star. Using the \citet{Chen2017} mass-radius relation predicts a mass of $\sim$3.2~\mearth. This would result in a challenging RV semi-amplitude (\emph{K}$\sim$1.6 m/s), especially given the faintness of the target, and predicts TSM ($\sim$8) and ESM ($\sim$0.7) values that indicate atmospheric follow-up would be challenging, compared to the thresholds recommended by \citet{Kempton2018}.

\subsection{EPIC 220510874}



\epiceightsevenfouralias\ is a moderately faint ($V$=13.17 mag, $K_s$=11.52 mag) F9V star (0.97~\rsun, 0.91~\msun) that was observed in Campaign 8. It is orbited by a sub-Neptune-sized planet, \epiceightsevenfourbalias\, which was previously noted as EPIC 220510874.01 by \citet{kru19} and \citet{zin19c}. \epiceightsevenfourbalias\ is a 2.320~\rearth\ planet, orbiting at a distance of 0.073~au, with a period of 7.473223~days and an equilibrium temperature of $\sim$1010~K. The host star has a clean, single-lined McDonald TS23 spectrum, and WIYN/NESSI speckle imaging at 562nm and 832nm that show no contaminating stellar companions. The \texttt{vespa} FPP value is $3.34 \times 10^{-5}$, using the available contrast curves. The \texttt{centroid} p-value is 0.6971, which is consistent with the source of the transiting signal being on the target star. Using the \citet{Chen2017} mass-radius relation predicts a mass of $\sim$6.2\mearth. This would result in a measurable RV semi-amplitude (\emph{K}$\sim$2.2 m/s), but predicts TSM ($\sim$12) and ESM ($\sim$0.9) values that indicate atmospheric follow-up would be challenging, compared to the thresholds recommended by \citet{Kempton2018}.

\subsection{EPIC 245943455}



\epicfourfivefivebalias\ is a sub-Neptune (3.660~\rearth) planet orbiting a moderately bright ($V$=12.82 mag, $K_s$=10.93 mag) G8V star (0.90~\rsun, 0.84~\msun). The target was observed in Campaign 12, and the transiting signal was previously noted in \citet{dat19}. The planet orbits the star at a distance of 0.063~au, with a period of 6.339069~days and an equilibrium temperature of $\sim$970~K. The host star has a clean, single-lined Keck/HIRES spectrum, and Palomar/PHARO, Gemini/NIRI, and WIYN/NESSI imaging which all show no contaminating stellar companions. The \texttt{vespa} FPP value is $8.02 \times 10^{-8}$, using the available contrast curves. The \texttt{centroid} p-value is 0.0938, which is moderately inconsistent with the source of the transiting signal being on the target star at the $\sim$2-$\sigma$ level, but does not fall below our $p=0.05$ threshold for revoking a planet's validation status. In \citet{dat19}, they note that transit depth changes as a function of aperture size for this signal, which may indicate a contaminating background star. Although this planet passes our statistical thresholds for validation, it may warrant additional ground-based time series photometry to confirm the transit is indeed on the target star. Using the \citet{Chen2017} mass-radius relation predicts a mass of $\sim$13.1~\mearth. This would result in a measurable RV semi-amplitude (\emph{K}$\sim$5.1 m/s), but predicts TSM ($\sim$30a) and ESM ($\sim$3.4) values that indicate atmospheric follow-up would be challenging, compared to the thresholds recommended by \citet{Kempton2018}.

\subsection{EPIC 245991048}

\epiczerofoureightalias\ is a moderately bright ($V$=12.30 mag, $K_s$=10.18 mag) G3V star (1.08~\rsun, 0.80~\msun). It was observed in Campaign 12 and is orbited by two sub-Neptunes, which were first noted Paper IV. \epiczerofoureightbalias\ is a 2.118~\rearth\ planet, orbiting at a distance of 0.076~au, with a period of 8.583566~days and an equilibrium temperature of $\sim$1050~K. Using the \citet{Chen2017} mass-radius relation predicts a mass of $\sim$5.7~\mearth. This would result in a measurable RV semi-amplitude (\emph{K}$\sim$2.0 m/s), but predicts TSM ($\sim$16) and ESM ($\sim$1.3) values that indicate atmospheric follow-up would be challenging, compared to the thresholds recommended by \citet{Kempton2018}. \epiczerofoureightcalias\ is a 2.040~\rearth\ planet, orbiting at a distance of 0.138~au, with a period of 20.851337~days and an equilibrium temperature of $\sim$780~K. Using the \citet{Chen2017} mass-radius relation predicts a mass of $\sim$5.3~\mearth. This would result in a challenging RV semi-amplitude (\emph{K}$\sim$1.4 m/s), and predicts TSM ($\sim$11) and ESM ($\sim$0.6) values that indicate atmospheric follow-up would be challenging, compared to the thresholds recommended by \citet{Kempton2018}. The host star has a clean, single-lined FLWO/TRES spectrum, as well as Gemini/NIRI and Palomar/PHARO AO imaging and WIYN/NESSI speckle imaging that show no contaminating stellar companions. The \texttt{vespa} FPP value for planet b is $6.41 \times 10^{-5}$ using the available contrast curves and the application of the Campaign 12 multiplicity boost. The \texttt{centroid} p-value for planet b of is 0.2457, which is consistent with the source of the transiting signal being on the target star. The \texttt{vespa} FPP value for planet c is $5.75 \times 10^{-5}$ using the available contrast curves and the application of the Campaign 12 multiplicity boost. The \texttt{centroid} p-value for planet c of is 0.2053, which is consistent with the source of the transiting signal being on the target star.

\subsection{EPIC 246074314}




\epicthreeonefourbalias\ is a near-Earth-sized (1.371~\rearth) planet orbiting a moderately faint ($Kep$=12.36 mag, $K_s$=10.71 mag) G7V star (0.57~\rsun, 0.76~\msun), observed in Campaign 12. It was first noted in Paper IV; it orbits the star at a distance of 0.049~au, with a period of 4.622654~days and an equilibrium temperature of $\sim$900~K. The host star has a clean, single-lined FLWO/TRES spectrum, and WIYN/NESSI speckle imaging at 562~nm and 832~nm which show no contaminating stellar companions. The \texttt{vespa} FPP value is $9.87 \times 10^{-5}$ using the available contrast curves. The \texttt{centroid} p-value is 0.3581, which is consistent with the source of the transiting signal being on the target star. Using the \citet{Chen2017} mass-radius relation predicts a mass of $\sim$2.5~\mearth. This would result in a challenging RV semi-amplitude (\emph{K}$\sim$1.2 m/s), and predicts TSM ($\sim$3.3) and ESM ($\sim$1.1) values that indicate atmospheric follow-up would be challenging, compared to the thresholds recommended by \citet{Kempton2018}.

\subsection{EPIC 246084398}



\epicthreenineeightbalias\ is a super-Earth (1.850~\rearth) planet orbiting a moderately bright ($V$=12.82 mag, $K_s$=11.24 mag) G3V star (1.06~\rsun, 0.84~\msun), observed in Campaign 12. It was first noted in Paper IV; it orbits the star at a distance of 0.063~au, with a period of 15.399~days and an equilibrium temperature of $\sim$840~K. The host star has a clean, single-lined FLWO/TRES spectrum, and WIYN/NESSI speckle imaging at 562~nm and 832~nm which show no contaminating stellar companions. The \texttt{vespa} FPP value is $2.12 \times 10^{-3}$, using the available contrast curves. The \texttt{centroid} p-value is 0.5614, which is consistent with the source of the transiting signal being on the target star. Using the \citet{Chen2017} mass-radius relation predicts a mass of $\sim$4.5~\mearth. This would result in a challenging RV semi-amplitude (\emph{K}$\sim$1.3 m/s), and predicts TSM ($\sim$6.4) and ESM ($\sim$0.4) values that indicate atmospheric follow-up would be challenging, compared to the thresholds recommended by \citet{Kempton2018}.

\subsection{EPIC 246429049}



\epiczerofourninealias\ is a bright ($V$=11.84 mag, $K_s$=10.28 mag) F9V star (1.07~\rsun, 0.88~\msun) that was observed in Campaign 12. It is orbited by a sub-Neptune-sized planet, \epiczerofourninebalias\, which was previously noted by \citet{zin19c} as EPIC 246429049.01. It is a 2.352~\rearth\ planet, orbiting at a distance of 0.089~au, with a period of 10.413181~days and an equilibrium temperature of $\sim$950~K. The host star has a clean, single-lined FLWO/TRES spectrum, and WIYN/NESSI speckle imaging at 562nm and 832nm that show no contaminating stellar companions. The \texttt{vespa} FPP value is $9.00 \times 10^{-3}$, using the available contrast curves. The \texttt{centroid} p-value is 0.3512, which is consistent with the source of the transiting signal being on the target star. Using the \citet{Chen2017} mass-radius relation predicts a mass of $\sim$6.3\mearth. This would result in a measurable RV semi-amplitude (\emph{K}$\sim$2.0m/s), but predicts TSM ($\sim$16) and ESM ($\sim$1.1) values that indicate atmospheric follow-up would be challenging, compared to the thresholds recommended by \citet{Kempton2018}.

\subsection{EPIC 246891819}




\epiceightoneninealias\ is a moderately faint ($V$=14.67 mag, $K_s$=11.37 mag) K0V star (0.78~\rsun, 0.95~\msun) observed in Campaign 13 that is orbited by two close-in sub-Neptune planets, which were first noted in Paper IV. \epiceightoneninebalias\ is a 2.597~\rearth\ planet, orbiting at a distance of 0.055~au, with a period of 4.803352~days and an equilibrium temperature of $\sim$910~K. \epiceightoneninecalias\ is a 3.395~\rearth\ planet, orbiting at a distance of 0.080~au, with a period of 8.491282~days and an equilibrium temperature of $\sim$750~K. The host star has a clean, single-lined TripleSpec spectrum, and Keck/NIRC2 AO imaging that shows no contaminating stellar companions. The \texttt{vespa} FPP values of planets b and c are $6.35 \times 10^{-10}$ and $1.74 \times 10^{-5}$ respectively, using the available contrast curve and the Campaign 13 multiplicity boost. The \texttt{centroid} p-values are 0.4413 and 0.4640 respectively, consistent with the source of the transiting signals being on the target star. Using the \citet{Chen2017} mass-radius relation predicts masses of $\sim$7.6~\mearth\ and $\sim$11.8~\mearth\ for planets b and c respectively. These mass estimates for the two planets result in potentially measurable RV semi-amplitudes ($K$ = 3.0 m/s, and 3.8 m/s), but TSM ($\sim$16, $\sim$20) and ESM ($\sim$1.7, $\sim$1.9) values below the thresholds of interest recommended by \citet{Kempton2018}. 

\subsection{EPIC 246953392}

\epicthreeninetwoalias\ is a moderately faint ($V$=13.24 mag, $K_s$=10.81 mag) G9V star (0.92~\rsun, 0.95~\msun) observed in Campaign 13. It is orbited by an ultra-short period super-Earth, \epicthreeninetwobalias, first reported in Paper IV as EPIC 246953392.01, and a longer-period sub-Neptune, \epicthreeninetwocalias (EPIC 246953392.02), reported here for the first time. \epicthreeninetwobalias\ orbits the star at a distance of 0.015~au, with an ultra-short period of 0.673862~days. With a high equilibrium temperature of $\sim$1990~K, it is the second hottest of the validated planets in this analysis. \epicthreeninetwocalias\ orbits the star at a distance of 0.168~au, with a period of 25.760509~days, and an equilibrium temperature of $\sim$590~K. The host star has a clean, single-lined FLWO/TRES spectrum, and WIYN/NESSI speckle imaging at both 562nm and 832nm that show no contaminating stellar companions. The \texttt{vespa} FPP values of planets b and c are $4.16 \times 10^{-4}$ and $1.07 \times 10^{-5}$ using the available contrast curves and Campaign 13 multiplicity boost. The \texttt{centroid} p-values are 0.4124 and 0.3421, which is consistent with the sources of the transiting signals being on the target star. Using the \citet{Chen2017} mass-radius relation predicts masses of $\sim$3.4~\mearth\ and $\sim$10.0~\mearth\ for planets b and c respectively. This would result in measurable RV semi-amplitudes (\emph{K}$\sim$2.6 m/s and $\sim$2.2 m/s), but predicts TSM ($\sim$18, $\sim$15) and ESM ($\sim$2.3, $\sim$0.7) values that indicate atmospheric follow-up would be challenging, compared to the thresholds recommended by \citet{Kempton2018}.

\subsection{EPIC 247383003}



\epiczerozerothreealias\ is a bright ($V$=11.83 mag, $K_s$=9.95 mag) G7V star (0.97~\rsun, 0.94~\msun) that was observed in Campaign 13. It is orbited by a sub-Neptune-size planet that was first noted in Paper IV. \epiczerozerothreebalias\ is a 2.417~\rearth\ planet, orbiting at a distance of 0.045~au, with a period of 3.572326~days and an equilibrium temperature of $\sim$1210~K. The host star has a clean, single-lined FLWO/TRES spectrum, and Keck/NIRC2 AO imaging and WIYN/NESSI speckle imaging at 562nm and 832nm that show no contaminating stellar companions. The \texttt{vespa} FPP value is 0.00799, using the available contrast curves. The \texttt{centroid} p-value is 0.5226, which is consistent with the source of the transiting signal being on the target star. Using the \citet{Chen2017} mass-radius relation predicts a mass of $\sim$6.7\mearth. This would result in a measurable RV semi-amplitude (\emph{K}$\sim$2.9m/s), but predicts TSM ($\sim$28) and ESM ($\sim$3.0) values that indicate atmospheric follow-up would be challenging, compared to the thresholds recommended by \citet{Kempton2018}.

\subsection{EPIC 248518307}



\epicthreezerosevenalias\ is a faint ($V$=15.13 mag, $K_s$=10.41 mag) M3V star (0.38~\rsun, 0.37~\msun) that was observed in Campaign 14. It is orbited by an Earth-size planet that was first noted in Paper IV as EPIC 248518307.01. \epicthreezerosevenbalias\ is a 1.163~\rearth\ planet, orbiting at a distance of 0.035~au, with a period of 3.865053~days and an equilibrium temperature of $\sim$570~K. The host star has a clean, single-lined Hale/TripleSpec spectrum, and WIYN/NESSI speckle imaging at 562nm and 832nm that shows no contaminating stellar companions. The \texttt{vespa} FPP value is $1.35 \times 10^{-3}$, using the available contrast curve. The \texttt{centroid} p-value is 0.5332, which is consistent with the source of the transiting signal being on the target star. Using the \citet{Chen2017} mass-radius relation predicts a mass of $\sim$1.7\mearth. This would result in a challenging RV semi-amplitude (\emph{K}$\sim$1.3m/s), and predicts TSM ($\sim$4.0) and ESM ($\sim$1.0) values that indicate atmospheric follow-up would be challenging, compared to the thresholds recommended by \citet{Kempton2018}.

\subsection{EPIC 248527514}


\epicfiveonefouralias\ is a moderately faint ($V$=13.72 mag, $K_s$=11.11 mag) K5V star (0.68~\rsun, 0.87~\msun) that was observed in Campaign 14. It is orbited by a sub-Neptune that was first noted in Paper IV. \epicfiveonefourbalias\ is a 2.262~\rearth\ planet, orbiting at a distance of 0.064~au, with a period of 6.293099~days and an equilibrium temperature of $\sim$660~K. The host star has a clean, single-lined TripleSpec spectrum, and WIYN/NESSI speckle imaging at both 562nm and 832nm that show no contaminating stellar companions. We note that \citet{Dressing2019} find that \epicfiveonefouralias\ lies above the main sequence and may therefore be an unresolved eclipsing binary; however the clean spectrum and high-resolution imaging, as well as the low \emph{Gaia} RUWE value of 1.09, all point to a single host star. The \texttt{vespa} FPP value is $3.78 \times 10^{-3}$, using the available contrast curves. The \texttt{centroid} p-value is 0.3051, which is consistent with the source of the transiting signal being on the target star. Using the \citet{Chen2017} mass-radius relation predicts a mass of $\sim$5.8~\mearth. This would result in a measurable RV semi-amplitude (\emph{K}$\sim$2.2 m/s), but predicts TSM ($\sim$15) and ESM ($\sim$1.1) values that indicate atmospheric follow-up would be challenging, compared to the thresholds recommended by \citet{Kempton2018}.

\subsection{EPIC 248621597}




\epicfiveninesevenbalias\ is a sub-Neptune (2.678~\rearth) planet orbiting a moderately faint ($V$=13.03 mag, $K_s$=11.38 mag) G2V star (1.28~\rsun, 0.94~\msun), observed in Campaign 14. It was first noted in Paper IV; it orbits the star at a distance of 0.128~au, with a period of 17.274740~days and an equilibrium temperature of $\sim$870~K. The host star has a clean, single-lined FLWO/TRES spectrum, and WIYN/NESSI speckle imaging at 562~nm and 832~nm which show no contaminating stellar companions. The \texttt{vespa} FPP value is $2.07 \times 10^{-4}$ using the available contrast curves. The \texttt{centroid} p-value is 0.4934, which is consistent with the source of the transiting signal being on the target star. Using the \citet{Chen2017} mass-radius relation predicts a mass of $\sim$7.8~\mearth. This would result in a measurable RV semi-amplitude (\emph{K}$\sim$2.0 m/s), and predicts TSM ($\sim$7.4) and ESM ($\sim$0.5) values that indicate atmospheric follow-up would be challenging, compared to the thresholds recommended by \citet{Kempton2018}.

\subsection{EPIC 248861279}




\epictwosevenninebalias\ is a sub-Neptune (2.530~\rearth) planet orbiting a moderately faint ($V$=14.56 mag, $K_s$=10.75 mag) early M dwarf (M1V, 0.55~\rsun, 0.55~\msun), observed in Campaign 14. It was first noted in Paper IV; it orbits the star at a distance of 0.089~au, with a period of 13.115365~days. With a relatively low equilibrium temperature of $\sim$450~K, it is one of the coolest validated planets in this analysis. The host star has a clean, single-lined TripleSpec spectrum, and both Keck/NIRC2 AO imaging and WIYN/NESSI speckle imaging which show no contaminating stellar companions. The \texttt{vespa} FPP value is $8.62 \times 10^{-4}$ using the available contrast curves. The \texttt{centroid} p-value is 0.4292, which is consistent with the source of the transiting signal being on the target star. Using the \citet{Chen2017} mass-radius relation predicts a mass of $\sim$7.3~\mearth. This would result in a measurable RV semi-amplitude (\emph{K}$\sim$3.0 m/s), and predicts TSM ($\sim$20) and ESM ($\sim$0.8) values that indicate atmospheric follow-up would be challenging, compared to the thresholds recommended by \citet{Kempton2018}.

\subsection{EPIC 249403651}




\epicsixfiveonealias\ is a bright ($V$=11.97 mag, $K_s$=10.21 mag) early G4V star (0.93~\rsun, 0.88~\msun) that was observed in Campaign 15 and is orbited by two small, close-in planets. The two candidates were first noted in Paper IV; \epicsixfiveonebalias\ is a super-Earth (1.283~\rearth) planet, orbiting at a distance of 0.054~au, with a period of 4.941907~days and an equilibrium temperature of $\sim$1105~K, and \epicsixfiveonecalias\ is a super-Earth (1.360~\rearth) planet, orbiting at a distance of 0.083~au, with a period of 9.224530~days and an equilibrium temperature of $\sim$900~K. The host star has a clean, single-lined FLWO/TRES spectrum, and Gemini/DSSI speckle imaging at 692nm and 880nm which show no contaminating stellar companions. The \texttt{vespa} FPP values for planets b and c are $3.23 \times 10^{-5}$ and $1.03 \times 10^{-4}$ respectively, incorporating the available contrast curves and Campaign 15 multiplicity boost. The \texttt{centroid} p-value for planet b is 0.0826, which is marginally consistent with the source of the transiting signal being on the target star. The \texttt{centroid} p-value for planet c is 0.0372, which is marginally inconsistent with the source of the transiting signal being on the target star given our threshold of 0.05. However, visual inspection of the centroid plots shows no systematic offset, with the transit cadences that fall outside the significance contours spread around a large fraction of the exterior of the contours, and the presence of multiple signals in the same light curve gives credence to their planetary nature. We consider planet c to be validated, but note that additional ground-based time series photometry, while challenging given the shallow depths of these transits, would provide additional assurance. From the \citet{Chen2017} mass-radius relation, we estimate masses of $\sim$2.1 \mearth\ and $\sim$2.5 \mearth\ for planets b and c respectively, leading to very low radial velocity semi-amplitude estimates and atmospheric spectroscopy metrics.

\subsection{EPIC 249924395}




\epicthreeninefivebalias\ is a sub-Neptune (2.482~\rearth) planet orbiting a moderately faint ($V$=12.76 mag, $K_s$=11.04 mag) G4V star (1.23~\rsun, 0.89~\msun), observed in Campaign 15. It was first noted in Paper IV; it orbits the star at a distance of 0.029~au, with a period of 1.908084~days and a very high equilibrium temperature of $\sim$1800~K. The host star has a clean, single-lined FLWO/TRES spectrum, and Gemini/DSSI speckle imaging which shows no contaminating stellar companions. Ground-based time-series photometry of \epicthreeninefivealias\ during a predicted transit with the 0.6-m ULMT found no detectable event on the target star (consistent with the shallow K2 detection) and no nearby binaries eclipsing at the predicted transit time within $\sim$120'' down to $\Delta$mag = 6.5, eliminating many blended eclipsing binary scenarios. The \texttt{vespa} FPP value is $3.92 \times 10^{-5}$ using the available contrast curves. The \texttt{centroid} p-value is 0.4360, which is consistent with the source of the transiting signal being on the target star. Using the \citet{Chen2017} mass-radius relation predicts a mass of $\sim$6.7~\mearth. This would result in a measurable RV semi-amplitude (\emph{K}$\sim$3.8 m/s), but given the relative faintness of the target, predicts TSM ($\sim$18) and ESM ($\sim$2.2) values that indicate atmospheric follow-up would be challenging, compared to the thresholds recommended by \citet{Kempton2018}.

\end{appendices}

\end{document}